\documentclass[reprint,showpacs,superscriptaddress,english,twocolumn,notitlepage,longbibliography,nofootinbib]{revtex4-1}
\usepackage[utf8]{inputenc}
\usepackage{color}
\usepackage{array}
\usepackage{verbatim}
\usepackage{multirow}
\usepackage{amsmath}
\usepackage{amssymb}
\usepackage{dsfont}
\usepackage{graphicx}
\usepackage{esint}
\usepackage[unicode=true,
 bookmarks=true,bookmarksnumbered=true,bookmarksopen=false,
 breaklinks=true,pdfborder={0 0 0},backref=false,colorlinks=true]
 {hyperref}
\usepackage{enumitem}
\usepackage{titlesec}
\usepackage{cleveref}
\usepackage{longtable}
\usepackage{bbm}
\usepackage[normalem]{ulem}
\usepackage{listings}
\usepackage{xcolor,colortbl}

\usepackage{listings}
\definecolor{mygreen}{RGB}{28,172,0} 
\definecolor{mylilas}{RGB}{170,55,241}
\hypersetup{
citecolor={blue},
urlcolor={blue}
}

\usepackage{cleveref}
\crefname{appendix}{Appendix}{Appendices}
\crefname{equation}{Eq.}{Eqs.}
\crefname{figure}{Fig.}{Figs.}
\crefname{table}{Table}{Tables}
\crefname{section}{Section}{Sections}
\crefname{mythe}{Theorem}{Theorems}
\crefname{mydef}{Definition}{Definitions}
\newcommand{\dummylabel}[2]{\def\@currentlabel{#2}\label{#1}}

\renewcommand{\paragraph}[1]{\vspace{0.2cm}{\bf \textit{#1}}}
\def\ie{{\it i.e.},\ }
\def\eg{{\it e.g.},\ }

\definecolor{Gray}{gray}{0.85}
\newcolumntype{a}{>{\columncolor{Gray}}c}

\usepackage{simpler-wick}
\allowdisplaybreaks[1] 

\newcommand{\mcl}{\mathcal}

\newcommand{\mrm}{\mathrm}

\newcommand{\td}{\widetilde}
\newcommand{\ovl}{\overline}

\def\pare#1{\left( #1 \right)}

\def\bra#1{\langle #1 |}
\def\ket#1{| #1 \rangle}

\def\inn#1{\langle #1 \rangle}

\def\nono{\nonumber}

\def\up{\uparrow}
\def\down{\downarrow}

\def\Tr{\mrm{Tr}}

\def\ee{\epsilon}

\def\PH{{\mathcal{P}}}

\def\qq{\mathbf{q}}
\def\kk{\mathbf{k}}
\def\KK{\mathbf{K}}
\def\DKK{\Delta\mathbf{K}}

\def\RR{\mathbf{R}}
\def\tt{\mathbf{t}}
\def\rr{\mathbf{r}}
\def\GG{\mathbf{G}}
\def\QQ{\mathbf{Q}}
\def\aa{\mathbf{a}}
\def\bb{\mathbf{b}}

\def\spin{{\varsigma}}

\def\hH{{ \hat{H} }}
\def\hrho{ \hat{\rho} }
\def\hg{\hat{g}}

\def\UC{{\hat{\Theta}}}
\def\UF{{\hat{\Sigma}}}

\def\mK{{\mathcal{K}}}

\def\mJ{{\mathcal{J}}}
\def\mK{{\mathcal{K}}}

\newcounter{SMmark}
\makeatletter

\@addtoreset{equation}{SMmark}
\@addtoreset{figure}{SMmark}
\@addtoreset{table}{SMmark}
\@addtoreset{page}{SMmark}
\@addtoreset{section}{SMmark}

\makeatother

\begin{document}

\title{MATBG as Topological Heavy Fermion: I. Exact Mapping and Correlated Insulators}
\author{Zhi-Da Song}
\email{songzd@pku.edu.cn}
\affiliation{International Center for Quantum Materials, School of Physics, Peking University, Beijing 100871, China}
\affiliation{Department of Physics, Princeton University, Princeton, New Jersey 08544, USA}
\author{B. Andrei Bernevig}
\affiliation{Department of Physics, Princeton University, Princeton, New Jersey 08544, USA}
\affiliation{Donostia International Physics Center, P. Manuel de Lardizabal 4, 20018 Donostia-San Sebastian, Spain}
\affiliation{IKERBASQUE, Basque Foundation for Science, Bilbao, Spain}

\begin{abstract}
Magic-angle ($\theta=1.05^\circ$) twisted bilayer graphene (MATBG) has shown two seemingly contradictory characters: the localization and quantum-dot-like behavior in STM experiments, and delocalization in transport experiments.
We construct a model, which naturally captures the two aspects, from the Bistritzer-MacDonald (BM) model in a {\it first principle spirit}. 
A set of local flat-band orbitals ($f$) centered at the AA-stacking regions are responsible to the localization.
A set of extended {\it topological} semi-metallic conduction bands ($c$), which are at small energetic separation from the local orbitals, are responsible to the delocalization and transport. 
The topological flat bands of the BM model appear as a result of the hybridization of $f$- and $c$-electrons.
This model then provides a new perspective for the strong correlation physics, which is now described as strongly correlated $f$-electrons coupled to nearly free $c$-electrons - we hence name our model as the {\it topological heavy fermion model}. 
Using this model, we obtain the U(4) and U(4)$\times$U(4) symmetries of Refs. \cite{kang_strong_2019,bultinck_ground_2020,vafek2020hidden,Biao-TBG3,Biao-TBG4} as well as the correlated insulator phases and their energies. 
Simple rules for the ground states and their Chern numbers are derived. 
Moreover, features such as the large dispersion of the charge $\pm1$ excitations \cite{TBG5,vafek2021lattice,bultinck_ground_2020}, and the minima of the charge gap at the $\Gamma_M$ point can now, for the first time, be understood both qualitatively and quantitatively in a simple physical picture. 
Our mapping opens the prospect of using heavy-fermion physics machinery to the superconducting physics of MATBG.
\end{abstract}

\maketitle

{\it Introduction ---}
Since the initial experimental discovery of the correlated insulator phases \cite{cao_correlated_2018} and superconductivity \cite{cao_unconventional_2018} in MATBG \cite{bistritzer_moire_2011}, extensive experimental
\cite{lu2019superconductors, yankowitz2019tuning, sharpe_emergent_2019, saito_independent_2020, stepanov_interplay_2020, liu2021tuning, arora_2020, serlin_QAH_2019, cao_strange_2020, polshyn_linear_2019,  xie2019spectroscopic, choi_imaging_2019, kerelsky_2019_stm, jiang_charge_2019,  wong_cascade_2020, zondiner_cascade_2020,  nuckolls_chern_2020, choi2020tracing, saito2020,das2020symmetry, wu_chern_2020,park2020flavour, saito2020isospin,rozen2020entropic, lu2020fingerprints}
and theoretical 
\cite{tarnopolsky_origin_2019, zou2018, liu2019pseudo, Efimkin2018TBG, kang_symmetry_2018, song_all_2019,po_faithful_2019, hejazi_multiple_2019,padhi2018doped,lian2020, hejazi_landau_2019, padhi2020transport, xu2018topological,  koshino_maximally_2018, ochi_possible_2018, xux2018, guinea2018, venderbos2018, you2019,  wu_collective_2020, Lian2019TBG,Wu2018TBG-BCS, isobe2018unconventional,liu2018chiral, bultinck2020, zhang2019nearly, liu2019quantum,  wux2018b, thomson2018triangular,  dodaro2018phases, gonzalez2019kohn, yuan2018model,kang_strong_2019,bultinck_ground_2020,seo_ferro_2019, hejazi2020hybrid, khalaf_charged_2020,po_origin_2018,xie_superfluid_2020,julku_superfluid_2020, hu2019_superfluid, kang_nonabelian_2020, soejima2020efficient, pixley2019, knig2020spin, christos2020superconductivity,lewandowski2020pairing,Kwan2020Twisted, Parameswaran2021Exciton, xie_HF_2020,liu2020theories, cea_band_2020,zhang_HF_2020,liu2020nematic, daliao_VBO_2019,daliao2020correlation, classen2019competing, kennes2018strong, eugenio2020dmrg, huang2020deconstructing, huang2019antiferromagnetically,guo2018pairing, ledwith2020, repellin_EDDMRG_2020,abouelkomsan2020,repellin_FCI_2020, vafek2020hidden, fernandes_nematic_2020, Wilson2020TBG, wang2020chiral, TBG1,Song-TBG2,Biao-TBG3,Biao-TBG4,TBG5,TBG6, zhang2021momentum,vafek2021lattice,cha2021strange,Chichinadze2020Nematic,sheffer2021chiral,kang2021cascades,hofmann2021fermionic,Calder2020Interactions,Thomson2021Recovery}
efforts have been made to understand the nature of these exotic phases. 
Theoretical challenges for understanding the correlation physics come from both the strong interaction compared to relatively small band width as well as from the topology \cite{liu2019pseudo,song_all_2019,po_faithful_2019,Song-TBG2,tarnopolsky_origin_2019,hejazi_multiple_2019}, which forbids a symmetric lattice description of the problem.  
The two flat bands of MATBG posses strong topology in the presence of  $C_{2z}T$ (time-reversal followed by $C_{2z}$ rotation) and particle-hole ($P$) symmetries \cite{Song-TBG2}, which supersedes the earlier, $C_{2z}T$ symmetry-protected fragile topology \cite{song_all_2019,po_faithful_2019}. This strong topology extends to the entire continuum BM model, and implies  the absence of a lattice model for any number of bands.
The topology is also responsible to exotic phases such as  quantum anomalous Hall states \cite{bultinck2020,liu2020theories,Biao-TBG4,wu_collective_2020,bultinck_ground_2020,Parameswaran2021Exciton} and fractional Chern states \cite{sheffer2021chiral,repellin_FCI_2020,abouelkomsan2020,ledwith2020}.

Two types of complementary strategies have been proposed to resolve the problem of the lattice description. One is to construct extended Hubbard models \cite{po_origin_2018,kang_symmetry_2018,kang_strong_2019,koshino_maximally_2018,po_faithful_2019,vafek2021lattice,zou2018,xux2018,yuan2018model}, where either $C_{2z}T$ \cite{kang_symmetry_2018,kang_strong_2019,koshino_maximally_2018,vafek2021lattice,yuan2018model} or $P$ \cite{po_faithful_2019} becomes non-local in real space.
The other is to adopt a full momentum-space formalism \cite{Biao-TBG4,TBG5,TBG6,zhang2021momentum,bultinck_ground_2020,cea_band_2020,zhang_HF_2020,hofmann2021fermionic}, where locality becomes hidden. 
(Besides the two strategies, some phenomenological models are also proposed \cite{Efimkin2018TBG,wux2018b,xu2018topological,thomson2018triangular,classen2019competing,eugenio2020dmrg,repellin_EDDMRG_2020,fernandes_nematic_2020}.)
The real and momentum space strong coupling models elucidated the nature of the correlated insulator states: they are ferromagnets - sometimes carrying Chern numbers -  in a large U(4) or U(4)$\times$U(4) symmetry space that contains spin, valley and band quantum number \cite{kang_strong_2019, bultinck_ground_2020,Biao-TBG3}. 
The dispersion of the excitations above the correlated insulators  \cite{TBG5,vafek2021lattice,bultinck_ground_2020} - where superconductivity appears upon doping - is, despite being exact - not physically understood.

In the current manuscript, nevertheless, we find it possible to write down a fully symmetric model that has a simple real space picture, which, remarkably and elegantly, solves the aforementioned puzzles. We reformulate and map the interacting MATBG as an effective topological heavy fermion system, which consists of local orbitals ($f$) centered at the AA-stacking regions and delocalized topological conduction bands ($c$).
The $f$-electrons are so localized that they have an almost zero kinetic energy ($\sim 0.1$meV) but a strong on-site Coulomb repulsion that we compute to be $\sim60$meV.  
The $c$-electrons carry the symmetry anomaly and have unbounded kinetic energies. 
The actual flat bands of the BM model are from a hybridization ($\sim$20meV) between the $f$- and $c$-bands.
The interacting Hamiltonian also couples the $f$ and $c$ electrons through the presence of several types of interactions.  
Using this model, the ground states \cite{po_origin_2018,padhi2018doped,bultinck_ground_2020,kang_strong_2019,liu2020theories,ochi_possible_2018,venderbos2018,dodaro2018phases,seo_ferro_2019,pixley2019,Xie2020TBG,cea_band_2020,zhang_HF_2020} and their topologies can be understood in a simple, physical picture.
The quasi-particle excitation bandwidth can even be analytically determined. 



\begin{figure}[t]
\centering
\includegraphics[width=1.0\linewidth]{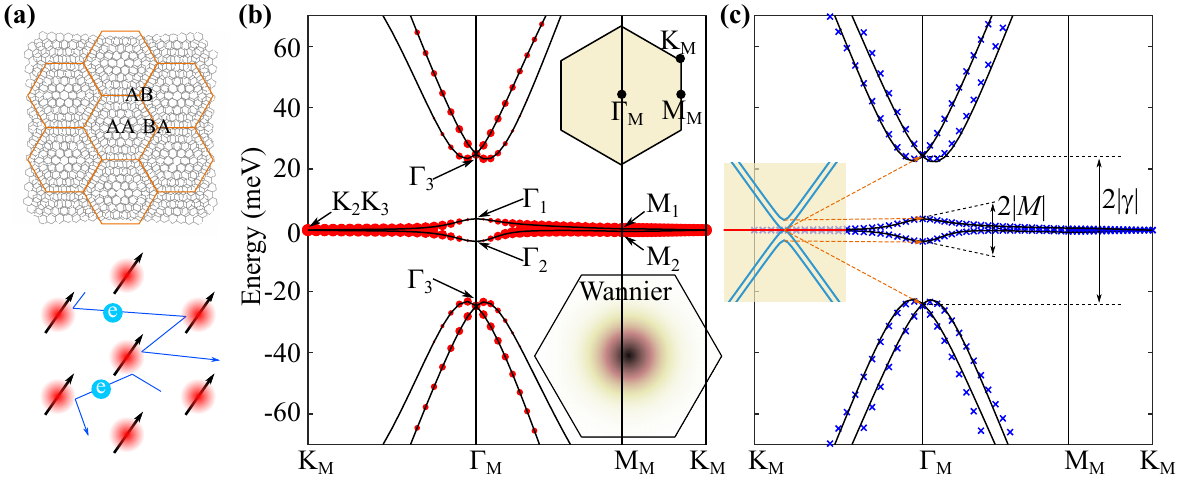}
\caption{Topological heavy fermion model. 
(a) A sketch of the moir\'e unit cell of MATBG and its heavy fermion analog, where the local moments and itinerant electrons are formed by the effective $f$-orbitals at the AA-stacking regions and topological conduction bands ($c$), respectively.   
(b) The band structure of the BM model at the magic angle $\theta=1.05^\circ$, where the moir\'e BZ and high symmetry momenta are illustrated in the upper inset panel. The overlaps between the Bloch states and the trial WFs are represented by the red circles. The density profile of the constructed maximally localized WFs ($f$-orbitals) is shown in the lower inset panel. 
(c) Bands given by the topological heavy fermion model (black lines) compared to the BM bands (blue crosses). 
The $c$- (blue) and $f$-bands (red) in the decoupled limit, where $\gamma=v_\star'=0$, are shown in the inset. Orange dashed lines indicate evolution of energy levels as $f$-$c$ coupling is turned on. 
\label{fig:Model} 
}
\end{figure}

{\it Topological heavy fermion model ---}
The single-valley BM model has the symmetry of the magnetic space group $P6'2'2$, generated by $C_{3z}$, $C_{2x}$, $C_{2z}T$, and moir\'e translations. 
(See Refs.~\cite{song_all_2019,SM} for this group and its irreducible representations - irreps.)
The energy bands in the valley $\eta=+$ of the BM model are shown in \cref{fig:Model}(b), where the bands are labeled by their irreps.   
Refs.~\cite{song_all_2019,po_faithful_2019} showed that the irreps formed by the two flat bands, \ie $\Gamma_1\oplus\Gamma_2$; $M_1\oplus M_2$; $K_2K_3$, are not consistent with any local orbitals (band representations \cite{bradlyn_topological_2017}) and indicate a fragile \cite{po_fragile_2018,ahn_failure_2019,Slager2019WL,cano_fragile_2018} topological obstruction to a two-band lattice model. 
Here we resolve the fragile topology by involving higher energy bands.
Suppose we can ``borrow'' a $\Gamma_3$ irrep from higher ($\sim$20meV) energy bands and use it to replace the  $\Gamma_1\oplus\Gamma_2$ states; then the replaced irreps - $\Gamma_3$, $M_1\oplus M_2$, $K_2K_3$ - are consistent with $p_x\pm i p_y$ orbitals located at the triangular lattice. 
We hence introduce two trial Gaussian-type Wannier functions (WFs) that transform as $p_x\pm i p_y$ orbitals under the crystalline symmetries. 
As indicated by the overlaps between the trial WFs and the Bloch bands (\cref{fig:Model}(a)), the trial WFs are supported by the flat band states at $\kk$ away from $\Gamma_M$ and by the lowest higher energy band states around $\Gamma_M$.
Feeding the overlaps into the program Wannier90 \cite{Wannier90-1,Wannier90-2,Wannier90-3}, we obtain the corresponding maximally localized WFs, density profile of which is shown in \cref{fig:Model}(b) \cite{SM}. 
(Similar local states are also discussed using different methods in Refs.~\cite{liu2019pseudo,Calder2020Interactions}.)
These WFs are extremely localized - their nearest neighbor hoppings are about 0.1meV - and span 96\% percent of the flat bands.

To recover the irreps and topology of the middle two bands, we have to take into account the remaining 4\% states, without which the localized electrons could not form a superconductor.
To do this, we define the projector into the WFs as $\mathbb{P}$, the projector into the lowest six bands (per spin valley) as $\mathbb{I}$, and divide the low energy BM Hamiltonian $H_{BM}$ into four parts: $H^{(f)}=\mathbb{P} H_{BM} \mathbb{P}$, $H^{(c)} = \mathbb{Q} H_{BM} \mathbb{Q}$, $H^{(fc)}=\mathbb{P} H_{BM} \mathbb{Q}$, $H^{(cf)}=H^{(fc)\dagger}$, where $ \mathbb{Q} = \mathbb{I} - \mathbb{P}$, $H^{(c)}$ is the remaining Hamiltonian, and $H^{(fc)}+h.c.$ is the coupling between WFs and the remaining states.
As the couplings between WFs are extremely weak ($\sim$0.1meV) we find $H^{(f)}\approx 0$.  
Since the two states in $\mathbb{P}$ form $\Gamma_3$ at $\Gamma_M$, the four states in $\mathbb{Q}$ must form $\Gamma_3\oplus \Gamma_1\oplus \Gamma_2$ at $\Gamma_M$ due to the irrep counting. 
Due to the crystalline and $P$ symmetries, $H^{(c)}$ in the valley $\eta$ takes the form \cite{SM} 
{\small
\begin{equation} \label{eq:Hc-maintext}
    H^{(c,\eta)}(\kk) = \begin{pmatrix}
    0_{2\times 2} & v_\star(\eta k_x\sigma_0 + i k_y\sigma_z) \\
    v_\star(\eta k_x\sigma_0 - i k_y\sigma_z) & M \sigma_x
    \end{pmatrix}
\end{equation}}
to linear order of $\kk$, where the first two-by-two block is spanned by the $\Gamma_3$ states and the second two-by-two block is spanned by the $\Gamma_1\oplus \Gamma_2$ states. 
The $\Gamma_1$ and $\Gamma_2$ states are split by the $M$ term (blue bands in \cref{fig:Model}(c)), while the $\Gamma_3$ states form a quadratic touching at $\kk=0$, which is shown in Ref.~\cite{SM} responsible to the symmetry anomaly \cite{Song-TBG2} jointly protected by $C_{2z}T$ and $P$. 
The coupling $H^{(fc)}$ in the valley $\eta$ has the form
{\small
\begin{equation} \label{eq:Hfc-maintext}
    H^{(fc,\eta)}(\kk) = \begin{pmatrix}
     \gamma \sigma_0 + v_\star'(\eta k_x\sigma_x + k_y\sigma_y), & 0_{2\times 2} 
    \end{pmatrix}\ ,
\end{equation}}
where the second block is computed to be extremely small and hence is omitted and written as $0_{2\times2}$.  
$H^{(fc,\eta)}$ will gap $H^{(c,\eta)}$, and hence provides for both the single particle gap and for the flat band topology of the BM model. 
Using a set of usually adopt parameters for MATBG, we find $v_\star = -4.303 \mrm{eV}\cdot\mrm{\mathring{A}}$, $M=3.697$meV, $\gamma=-24.75$meV, $v_\star'=1.622 \mrm{eV}\cdot\mrm{\mathring{A}}$. 

Since the WFs and the remaining ``$c$'' degrees of freedom have localized and plane-wave-like wave functions, respectively, we make the analogy with local orbitals and conduction bands in heavy fermion systems. 
We refer to them as local $f$-orbitals and (topological) conduction $c$-bands, respectively. 
We use $f_{\RR \alpha\eta s}$ ($\alpha=1,2$, $\eta=\pm$, $s=\up,\down$) to represent the annihilation operator of the $\alpha$-th WF of the valley $\eta$ and spin $s$ at the moir\'e unit cell $\RR$.  We use $c_{\kk a\eta s}$ ($a=1,2,3,4$) to represent the annihilation operator of the $a$-th conduction band basis of the valley $\eta$ and spin $s$ at the moir\'e momentum $\kk$.
The single-particle Hamiltonian can be written as
{\small
\begin{align} \label{eq:H0-maintext}
\hH_0 =& \sum_{|\kk|<\Lambda_c} \sum_{a a'\eta s} H^{(c,\eta)}_{aa'}(\kk) c_{\kk a\eta s}^\dagger c_{\kk a'\eta s} + \frac1{\sqrt{N}} \sum_{\substack{|\kk|<\Lambda_c\\ \RR}} \sum_{\alpha a \eta s} \Big(\nono\\
 &  e^{i\kk\cdot\RR -\frac{|\kk|^2\lambda^2}2} H^{(fc,\eta)}_{\alpha a} (\kk)  f_{\RR \alpha\eta s}^\dagger c_{\kk a\eta s} + h.c. \Big) \ ,
\end{align}}
where $\Lambda_c$  is the momentum cutoff for the $c$-electrons, $a_M$ is the moir\'e lattice constant, $N$ is the number of moir\'e unit cell in the system, and $\lambda$, which is found to be $0.3375a_M$, is a damping factor proportional to the size of WFs.
We plot the band structure of $\hH_0$ in \cref{fig:Model}(c), where the splitting of the two $\Gamma_3$ states is given by $2|\gamma|$ and the bandwidth of the two flat bands is given by $2M \approx 7.4 \text{meV}$. 
The spectrum of $\hH_0$ matches very well with the BM model (\cref{fig:Model}(a)) in the energy range [-70meV, 70meV].

{\it The U(4) symmetry ---}
The projected model of MATBG \cite{kang_strong_2019, bultinck_ground_2020, Biao-TBG3} is found to possess a U(4) symmetry if the kinetic energy of the flat bands is omitted.
In the heavy fermion basis, this U(4) symmetry can be realized by imposing the flat band condition, \ie $M=0$. 
(Note that $2|M|$ is the bandwidth of the flat bands.)
The U(4) moments of the $f$-electrons, $\Gamma_3$ $c$-electrons, and $\Gamma_1\oplus\Gamma_2$ $c$-electrons are given by \cite{SM}
{\small
\begin{equation}  \label{eq:U4op-maintext}
\begin{aligned}
\UF_{\mu\nu}^{(f,\xi)}(\RR) =& \frac{\delta_{\xi, (-1)^{\alpha-1}\eta}}2 A^{\mu\nu}_{\alpha\eta s, \alpha'\eta's'} f_{\RR \alpha \eta s}^\dagger f_{\RR \alpha' \eta' s'} \\ 
\UF_{\mu\nu}^{(c\prime,\xi)}(\qq) =& \frac{\delta_{\xi, (-1)^{a-1}\eta}}{2N} A^{\mu\nu}_{a\eta s, a'\eta's'} c_{\kk+\qq a \eta s}^\dagger c_{\kk a' \eta' s'},\; (a=1,2) \\
\UF_{\mu\nu}^{(c\prime\prime,\xi)}(\qq) =& \frac{\delta_{\xi, (-1)^{a-1}\eta}}{2N} B^{\mu\nu}_{a\eta s, a'\eta's'} c_{\kk+\qq a \eta s}^\dagger c_{\kk a' \eta' s'},\; (a=3,4)
\end{aligned}
\end{equation}}
respectively, where repeated indices should be summed over and $A^{\mu\nu},B^{\mu\nu}$ ($\mu,\nu=0,x,y,z$) are eight-by-eight matrices 
\begin{equation} \label{eq:flat-U4-maintext}
\begin{aligned}
A^{\mu\nu} =& \{ \sigma_0 \tau_0 \spin_\nu, \sigma_y \tau_x \spin_\nu, \sigma_y \tau_y \spin_\nu, \sigma_0 \tau_z \spin_\nu \} \\
B^{\mu\nu} =& \{ \sigma_0 \tau_0 \spin_\nu, - \sigma_y \tau_x \spin_\nu, - \sigma_y \tau_y \spin_\nu, \sigma_0 \tau_z \spin_\nu \}\ ,
\end{aligned}
\end{equation}
with $\sigma_{0,x,y,z}$, $\tau_{0,x,y,z}$, $\spin_{0,x,y,z}$ being the Pauli or identity matrices for the orbital, valley, and spin degrees of freedom, respectively.  
The $\pm1$ valued index $\xi$, equal to $(-1)^{\alpha-1}\eta$ or $(-1)^{a-1}\eta$ in the moments, labels different fundamental representations of the U(4) group.
The global U(4) rotations are generated by $\UF_{\mu\nu} = \sum_{\xi=\pm1} \UF_{\mu\nu}^{(f,\xi)} + \UF_{\mu\nu}^{(c\prime,\xi)} + \UF_{\mu\nu}^{(c\prime\prime,\xi)}$. 
Unlike the U(4) rotations found in  Refs.~\cite{kang_strong_2019, bultinck_ground_2020, Biao-TBG3}, which only commute the projected Hamiltonian into the flat bands, the U(4) rotations here commute with the full Hamiltonian. 
(Generators of the U(4) or U(4)$\times$U(4) symmetry in the first chiral limit \cite{tarnopolsky_origin_2019,wang2020chiral} is also given in Ref.~\cite{SM}.)



{\it Interaction Hamiltonian ---} To obtain the interaction Hamiltonian in the heavy fermion basis, we can first express the density operator $\rho(\rr)$ of the BM model in terms of $f_{\RR\alpha\eta s}$ and $c_{\kk a\eta s}$, and then substitute it into the Coulomb interaction, $\rho(\rr) V(\rr-\rr') \rho(\rr')$.
By evaluating the Coulomb integrals, we obtain the interaction Hamiltonian resembling a periodic Anderson model with extra $f$-$c$ exchange interactions \cite{SM}, 
\begin{equation}
   \hH_I= \hH_{U_1} + \hH_J +  \hH_{U_2} + \hH_V + \hH_W\ .
\end{equation}
$\hH_{U_1} =  \frac{U_1}2 \sum_{\RR} :\rho^f_{\RR}: :\rho^f_{\RR}:$ are the on-site interactions of $f$-electrons, where $\rho^f_{\RR} = \sum_{\alpha\eta s} f_{\RR \alpha\eta s}^\dagger f_{\RR \alpha\eta s}$ is the $f$-electrons density and the colon symbols represent the normal ordered operator with respect to the normal state. 
{\small
\begin{equation} \label{eq:HJ-maintext}
\hH_J = - J\sum_{\RR\qq}\sum_{\mu\nu} \sum_{\xi=\pm} e^{-i\qq\cdot\RR}:\UF_{\mu\nu}^{(f,\xi)}(\RR): :\UF_{\mu\nu}^{(c\prime\prime,\xi)}(\qq):
\end{equation}}
is a ferromagnetic exchange coupling between U(4) moments of $f$-electrons and $\Gamma_1\oplus\Gamma_2$ $c$-electrons. 
Using practical parameters for MATBG, we obtain $U_1=57.95$meV and $J=16.38$meV.
The other three terms in $\hH_I$ are: 
$H_{U_2}$ - repulsion ($\sim2.3$meV) between nearest neighbor $f$-electrons,
$H_V$ - repulsion ($\sim48$meV) between $c$-electrons, 
$H_W$ - repulsion ($\sim47$meV) between $c$- and $f$-electrons. 

As a widely adopted approximation in heavy fermion materials,  $\hH_{U_2} + \hH_V + \hH_W$ can be decoupled in the Hartree channel due to the delocalized and localized natures of $c$- and $f$-electrons. 
Hence these terms only effectively shift the band energies of $f$- and $c$-bands.
Then, $U_1$ - the on-site repulsion of the $f$-electrons - is by far the largest energy scale of the problem - more than twice the hybridization ($\gamma$) and three times the exchange ($J$). 
In Hartree-Fock (HF) calculations $U_1$ is found to be the source of spontaneous symmetry-breakings. 

\begin{figure}[t]
\centering
\includegraphics[width=\linewidth]{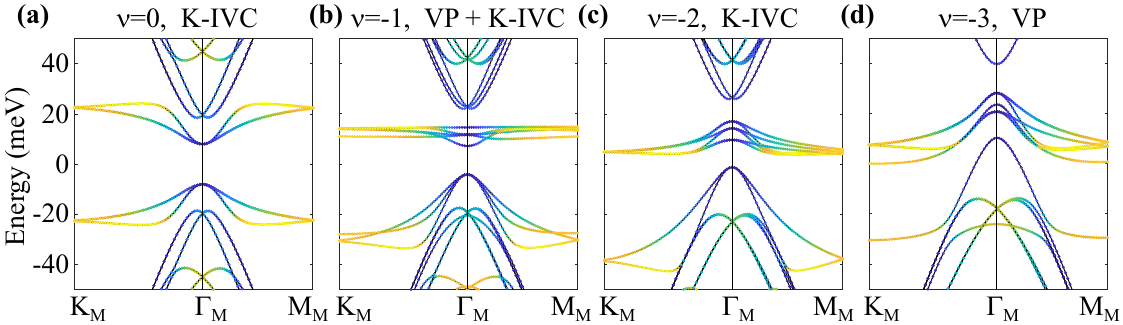}
\caption{The self-consistent HF bands upon the ground states at the fillings $\nu=0,-1,-2,-3$. The color of the bands represent the contributing components, wherein yellow represents the $f$-electron states and blue represents the $c$-electron states. 
}
\label{fig:HFbands-maintext}
\end{figure}

{\it Ground states ---}
Since $U_1$ is much larger than the couplings ($\gamma$, $J$, $v_\star' (\eta k_x \sigma_x + k_y\sigma_y)$) between $f$- and $c$-electrons, a reasonable guess of the ground states would be product states of $f$-multiplets and the (gapless point) Fermi liquid state ($\ket{\text{FS}}$) of the half-filled $c$-electrons. 
We call such product states ``the parent states''. 
{\it E.g.}, the parent valley-polarized (VP) state at the charge neutrality ($\nu=0$) is 
\begin{equation}\label{eq:VP-maintext}
\ket{\text{VP}^{\nu=0}_0} = \prod_{\RR} \prod_{\alpha=1,2} \prod_{s=\up\down} f_{\RR, \alpha, +, s}^\dagger \ket{\text{FS}}\ . 
\end{equation}
The parent Kramers inter-valley-coherent (K-IVC) state is a U(4)-rotation of $\ket{\text{VP}^{\nu=0}_0}$ along the $\tau_x$-direction
{\small
\begin{align}\label{eq:K-IVC-maintext}
& \ket{\text{K-IVC}^{\nu=0}_0} =   e^{-i\frac{\pi}2 \UF_{x0}} \ket{\text{VP}^{\nu=0}_0} \nono\\
=& \prod_{\RR} \prod_{s=\up\down} \frac12
    (f_{\RR, 1, +, s}^\dagger + f_{\RR, 2, -, s}^\dagger ) 
    (-f_{\RR, 1, -, s}^\dagger + f_{\RR, 2, +, s}^\dagger )
    \ket{\text{FS}}\ . 
\end{align}}
Parent states at other integer fillings ($\nu=0,\pm1,\pm2,\pm3$) can be similarly constructed \cite{SM}.
They would be ground states of the Hamiltonian if $\gamma$, $J$, $v_\star'$ terms vanished;
hybridization of $f$- and $c$-electrons will develop, \ie $\inn{f^\dagger c}\neq 0$, otherwise. 
The determination of ground states by self-consistent HF calculation with initial states given by the parent states is given in Ref.~\cite{SM}.
The numerically found HF ground states at the integer fillings (\cref{fig:HFbands-maintext}) are fully consistent with those in Ref.~\cite{Biao-TBG4}. 

\begin{figure}[t]
\centering
\includegraphics[width=\linewidth]{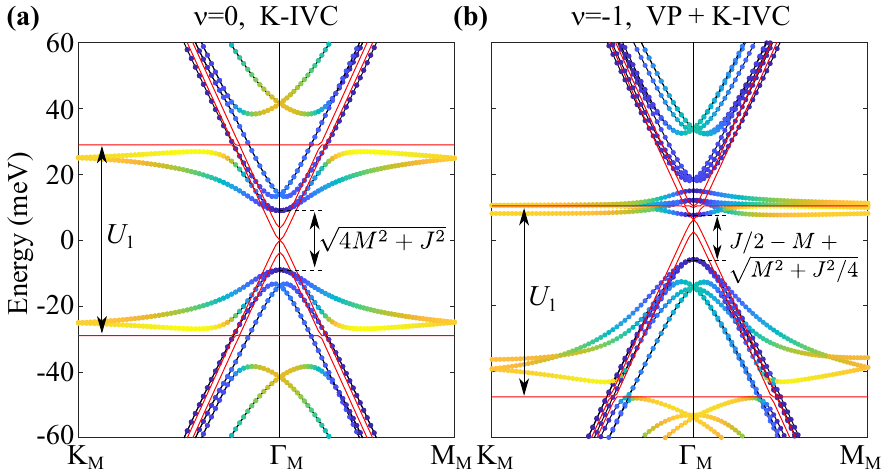}
\caption{The one-shot HF bands of the ground states at the fillings $\nu=0,-1$.
The red solid bands are the quasi-particle bands of the decoupled Hamiltonian, where $\gamma=v_\star'=J=0$.
The horizontal and dispersive red bands are of the $f$- and $c$-electrons, respectively.
The touching point of the dispersive red bands at $\Gamma_M$ is quadratic, while since $M$ is small, it may look like linear. 
The one-shot bands can be understood as a result of hybridization between $f$- and $c$-electrons.
}
\label{fig:excitation-maintext}
\end{figure}

The parent states are so good initial states for the HF calculations  that the one-shot HF is already qualitatively same as the self-consistent HF (see \cref{fig:excitation-maintext}). 
Thanks to the simplicity of the heavy fermion model, the one-shot energies can be analytically estimated and we are able to derive two rules for the ground states \cite{SM}.  
\emph{First}, in the parent state, $f$-electrons at each site tend to be symmetric under permutation of U(4) indices to save the Coulomb energy (Hunds' rule). 
Both \cref{eq:VP-maintext,eq:K-IVC-maintext} satisfy the first rule.
{\it Second}, for U(4)-related states at a given integer filling $\nu$, the state that minimizes $\hH_M + \hH_J$ is the ground state, where $\hH_M$ is the U(4)-breaking $M$ term in $\hH_0$ (\cref{eq:Hc-maintext}). 
This energy can be estimated by the lowest $\nu+4$ levels of the mean field Hamiltonian $H^{(\Gamma_1\oplus\Gamma_2)}$ spanned by the $\Gamma_1\oplus\Gamma_2$ basis of the $c$-bands at $\kk=0$, which reads (up to constants)
\begin{equation} \label{eq:H-mean-field-maintext}
H^{(\Gamma_1\oplus\Gamma_2)} = M\sigma_x \tau_0 \spin_0 - \frac{J}2  (\tau_z \ovl{O}^{fT} \tau_z + \sigma_z \ovl{O}^{fT} \sigma_z )\ .
\end{equation}
Here $\ovl{O}^f_{\alpha\eta s, \alpha'\eta's'} = \inn{f_{\RR \alpha\eta s}^\dagger f_{\RR\alpha'\eta's'}} - \frac12 \delta_{\alpha\alpha'} \delta_{\eta\eta'} \delta_{ss'}$ is the density matrix of the local $f$-orbitals with homogeneity assumed.
We have assumed that, for all the integer fillings, the eight lowest energy levels (closest to the gap) are always contributed by the $\Gamma_1\oplus\Gamma_2$ $c$-band basis - which are part of the flat bands - hence we only need to look at the $\Gamma_1\oplus\Gamma_2$ subspace. 
This assumption is fully justified by previous studies based on two-band projected Hamiltonian, where only $\Gamma_1\oplus\Gamma_2$ basis exists, and will become clearer after we discuss the charge $\pm1$ excitations. 



We now apply the second rule to \cref{eq:VP-maintext} and (\ref{eq:K-IVC-maintext}) to determine the one-shot state of lowest energy. 
$\ovl{O}^f$ matrices given by \cref{eq:VP-maintext,eq:K-IVC-maintext} are $\frac12\sigma_0\tau_z\spin_0$ and $-\frac12\sigma_y\tau_y\spin_0$, respectively; the resulted lowest four levels of $H^{(\Gamma_1\oplus\Gamma_2)}$ are 
$\pm M -J/2$ (each 2-fold) and $-\sqrt{M^2+J^2/4}$ (4-fold), respectively. 
It is direct to verify that the latter (K-IVC) has a lower energy.
Applying the two rules to parent states at other fillings, we obtain consistent results with the full numerical calculations in Refs.~\cite{Biao-TBG4,TBG6}. 
We also obtain an analytical expression for the Chern numbers of ground states \cite{SM}.


{\it Charge $\pm 1$ excitations ---}
As shown in \cref{fig:HFbands-maintext,fig:excitation-maintext} and in Refs. \cite{TBG5,bultinck_ground_2020,vafek2021lattice,cea_band_2020,zhang_HF_2020,kang2021cascades}, at $\kk$ away from $\Gamma_M$, the quasi-particle bands have a large gap ($\sim U_1$) and are relatively flat; at $\kk$ around $\Gamma_M$, the bands have significant dip. Such features are found related to the topology of the two flat-bands \cite{TBG5, vafek2021lattice} but have not yet been quantitatively understood.
The heavy fermion model provides a natural explanation to these features. 
We first consider the decoupled limit ($\gamma=v_\star'=J=0$) at $\nu=0$, where the $f$-electron bands are flat and have a (charge) gap $U_1$, and the $c$-electron bands are given by $H^{(c,\eta)}$  (\cref{fig:excitation-maintext}(a)). 
Tuning on $\gamma,v_\star',J$  then yields the one-shot quasi-particle bands.
At $\kk=0$, $\gamma$ gaps out the $\Gamma_3$ $c$-bands, and $J$ further gaps out the $\Gamma_1\oplus\Gamma_2$ $c$-bands.  
As the splitting of $\Gamma_1 - \Gamma_2$ is smaller than that of the $\Gamma_3$, the lowest excitations will carry $\Gamma_1, \Gamma_2$ representations, matching Refs.~\cite{TBG5, vafek2021lattice, bultinck_ground_2020}  and, according to the discussion after \cref{eq:H-mean-field-maintext}, equals to $2\sqrt{M^2+J^2/4}$ and $|J-2M|$ for K-IVC and VP states, respectively. 
At $\kk\neq0$, the $v_\star'$ term hybridizes the flat $f$-bands and dispersive $c$-bands. 
For large $\kk$, where the $c$-bands have very high kinetic energies, the hybridization is relatively weak and the gap is still approximately $U_1$. 
Thus the shape of the quasi-particle bands is explained, and its bandwidth is approximately given by $(U_1- J)/2$ when $M$ is small. 
As discussed in Ref.~\cite{SM}, the feature that the larger ($\sim U_1$) and smaller ($\sim J$) gaps are contributed by $f$- and $c$-electrons, respectively, is reflected in the STM spectra and Landau levels at different regions (AA or AB sites) of MATBG.

At nonzero fillings, the quasi-particle bands can also be understood as hybridized flat $f$-bands and dispersive $c$-bands, except that the $f$- and $c$-bands may feel different effective chemical potentials due to the density-density interactions between them. 
For example, at $\nu=-1$, the upper branch of the $f$-bands is shifted to an energy close to the quadratic touching of the $c$-bands (\cref{fig:excitation-maintext}(b)) \cite{SM}. 
Thus one of the hybridized bands is extremely flat. 


{\it Discussion ---}
The coexistence of quantum-dot-like behavior \cite{xie2019spectroscopic,wong_cascade_2020} and superconductivity \cite{cao_unconventional_2018,lu2019superconductors,yankowitz2019tuning,saito_independent_2020,stepanov_interplay_2020} may now be understood - they come from two different types ($f$ and $c$) of carriers. 
In fact, inspired by the pomeranchuk effect experiments \cite{saito2020isospin,rozen2020entropic} and strange metal behavior \cite{cao_strange_2020,polshyn_linear_2019}, authors of Refs.~\cite{saito2020isospin,rozen2020entropic,cha2021strange} also conjecture the possibility of coexistence of local momenta and itinerant electrons. 
(The heavy fermion theory analog may also exist in other twisted materials \cite{Ramires2021Emulating}.)
Our paper \emph{derives} and shows the exact mapping of MATBG to such a heavy-fermion type model.  
As such, the machinery of heavy fermions \cite{si2010heavy-review,gegenwart2008quantum,Coleman1984,Coleman2010Topological,tsvelick1983exact,Millis2006,Dai2013Correlated,Dai2014Topological,Kotliar2006Electronic,Kivelson1992Mapping,Kim2018Two-Stage,Nagaosa1994Kondo,Fradkin1996Kondo,Philip1997Kondo,Sudip1995Kondo} can now be applied, for the first time, to MATBG. 
We speculate that it will lead to pairing \cite{guinea2018,you2019,isobe2018unconventional,liu2018chiral,gonzalez2019kohn,khalaf_charged_2020,christos2020superconductivity,knig2020spin,kennes2018strong,huang2019antiferromagnetically,guo2018pairing,Chichinadze2020Nematic,Lian2019TBG,Wu2018TBG-BCS,lewandowski2020pairing} in nontrivial gap channels.

\begin{acknowledgements}
We thank X. Dai, O Vafek, A. Yazdani, P. Coleman, EA. Kim, Q. Si, R Fernandez, P. Jarillo-Herrero and D. Efetov for discussions.
B. A. B.'s work was primarily supported by the DOE Grant No. DE-SC0016239, the
Schmidt Fund for Innovative Research, Simons Investigator Grant No. 404513, and the Packard Foundation. 
Z.-D. S. was supported by ONR No. N00014-20-1-2303, the National Key Research and Development Program of China (No. 2021YFA1401900), NSF-MRSEC No. DMR-1420541, and Gordon and Betty Moore Foundation through Grant GBMF8685 towards the Princeton theory program. 
B.A.B. also acknowledges support from the European Research Council (ERC) under the European Union's Horizon 2020 research and innovation program (Grant Agreement No. 101020833).
\end{acknowledgements}



\bibliography{ref}

\begin{thebibliography}{146}%
\makeatletter
\providecommand \@ifxundefined [1]{%
 \@ifx{#1\undefined}
}%
\providecommand \@ifnum [1]{%
 \ifnum #1\expandafter \@firstoftwo
 \else \expandafter \@secondoftwo
 \fi
}%
\providecommand \@ifx [1]{%
 \ifx #1\expandafter \@firstoftwo
 \else \expandafter \@secondoftwo
 \fi
}%
\providecommand \natexlab [1]{#1}%
\providecommand \enquote  [1]{``#1''}%
\providecommand \bibnamefont  [1]{#1}%
\providecommand \bibfnamefont [1]{#1}%
\providecommand \citenamefont [1]{#1}%
\providecommand \href@noop [0]{\@secondoftwo}%
\providecommand \href [0]{\begingroup \@sanitize@url \@href}%
\providecommand \@href[1]{\@@startlink{#1}\@@href}%
\providecommand \@@href[1]{\endgroup#1\@@endlink}%
\providecommand \@sanitize@url [0]{\catcode `\\12\catcode `\$12\catcode
  `\&12\catcode `\#12\catcode `\^12\catcode `\_12\catcode `\%12\relax}%
\providecommand \@@startlink[1]{}%
\providecommand \@@endlink[0]{}%
\providecommand \url  [0]{\begingroup\@sanitize@url \@url }%
\providecommand \@url [1]{\endgroup\@href {#1}{\urlprefix }}%
\providecommand \urlprefix  [0]{URL }%
\providecommand \Eprint [0]{\href }%
\providecommand \doibase [0]{http://dx.doi.org/}%
\providecommand \selectlanguage [0]{\@gobble}%
\providecommand \bibinfo  [0]{\@secondoftwo}%
\providecommand \bibfield  [0]{\@secondoftwo}%
\providecommand \translation [1]{[#1]}%
\providecommand \BibitemOpen [0]{}%
\providecommand \bibitemStop [0]{}%
\providecommand \bibitemNoStop [0]{.\EOS\space}%
\providecommand \EOS [0]{\spacefactor3000\relax}%
\providecommand \BibitemShut  [1]{\csname bibitem#1\endcsname}%
\let\auto@bib@innerbib\@empty
\bibitem [{\citenamefont {Kang}\ and\ \citenamefont
  {Vafek}(2019)}]{kang_strong_2019}%
  \BibitemOpen
  \bibfield  {author} {\bibinfo {author} {\bibfnamefont {Jian}\ \bibnamefont
  {Kang}}\ and\ \bibinfo {author} {\bibfnamefont {Oskar}\ \bibnamefont
  {Vafek}},\ }\bibfield  {title} {\enquote {\bibinfo {title} {Strong {Coupling}
  {Phases} of {Partially} {Filled} {Twisted} {Bilayer} {Graphene} {Narrow}
  {Bands}},}\ }\href {\doibase 10.1103/PhysRevLett.122.246401} {\bibfield
  {journal} {\bibinfo  {journal} {Physical Review Letters}\ }\textbf {\bibinfo
  {volume} {122}},\ \bibinfo {pages} {246401} (\bibinfo {year}
  {2019})}\BibitemShut {NoStop}%
\bibitem [{\citenamefont {Bultinck}\ \emph
  {et~al.}(2020{\natexlab{a}})\citenamefont {Bultinck}, \citenamefont {Khalaf},
  \citenamefont {Liu}, \citenamefont {Chatterjee}, \citenamefont {Vishwanath},\
  and\ \citenamefont {Zaletel}}]{bultinck_ground_2020}%
  \BibitemOpen
  \bibfield  {author} {\bibinfo {author} {\bibfnamefont {Nick}\ \bibnamefont
  {Bultinck}}, \bibinfo {author} {\bibfnamefont {Eslam}\ \bibnamefont
  {Khalaf}}, \bibinfo {author} {\bibfnamefont {Shang}\ \bibnamefont {Liu}},
  \bibinfo {author} {\bibfnamefont {Shubhayu}\ \bibnamefont {Chatterjee}},
  \bibinfo {author} {\bibfnamefont {Ashvin}\ \bibnamefont {Vishwanath}}, \ and\
  \bibinfo {author} {\bibfnamefont {Michael~P.}\ \bibnamefont {Zaletel}},\
  }\bibfield  {title} {\enquote {\bibinfo {title} {Ground state and hidden
  symmetry of magic-angle graphene at even integer filling},}\ }\href {\doibase
  10.1103/PhysRevX.10.031034} {\bibfield  {journal} {\bibinfo  {journal} {Phys.
  Rev. X}\ }\textbf {\bibinfo {volume} {10}},\ \bibinfo {pages} {031034}
  (\bibinfo {year} {2020}{\natexlab{a}})}\BibitemShut {NoStop}%
\bibitem [{\citenamefont {Vafek}\ and\ \citenamefont
  {Kang}(2020)}]{vafek2020hidden}%
  \BibitemOpen
  \bibfield  {author} {\bibinfo {author} {\bibfnamefont {Oskar}\ \bibnamefont
  {Vafek}}\ and\ \bibinfo {author} {\bibfnamefont {Jian}\ \bibnamefont
  {Kang}},\ }\bibfield  {title} {\enquote {\bibinfo {title} {Renormalization
  group study of hidden symmetry in twisted bilayer graphene with coulomb
  interactions},}\ }\href {\doibase 10.1103/PhysRevLett.125.257602} {\bibfield
  {journal} {\bibinfo  {journal} {Phys. Rev. Lett.}\ }\textbf {\bibinfo
  {volume} {125}},\ \bibinfo {pages} {257602} (\bibinfo {year}
  {2020})}\BibitemShut {NoStop}%
\bibitem [{\citenamefont {Bernevig}\ \emph
  {et~al.}(2021{\natexlab{a}})\citenamefont {Bernevig}, \citenamefont {Song},
  \citenamefont {Regnault},\ and\ \citenamefont {Lian}}]{Biao-TBG3}%
  \BibitemOpen
  \bibfield  {author} {\bibinfo {author} {\bibfnamefont {B.~Andrei}\
  \bibnamefont {Bernevig}}, \bibinfo {author} {\bibfnamefont {Zhi-Da}\
  \bibnamefont {Song}}, \bibinfo {author} {\bibfnamefont {Nicolas}\
  \bibnamefont {Regnault}}, \ and\ \bibinfo {author} {\bibfnamefont {Biao}\
  \bibnamefont {Lian}},\ }\bibfield  {title} {\enquote {\bibinfo {title}
  {Twisted bilayer graphene. {III}. {Interacting} {Hamiltonian} and exact
  symmetries},}\ }\href {\doibase 10.1103/PhysRevB.103.205413} {\bibfield
  {journal} {\bibinfo  {journal} {Physical Review B}\ }\textbf {\bibinfo
  {volume} {103}},\ \bibinfo {pages} {205413} (\bibinfo {year}
  {2021}{\natexlab{a}})},\ \bibinfo {note} {publisher: American Physical
  Society}\BibitemShut {NoStop}%
\bibitem [{\citenamefont {Lian}\ \emph {et~al.}(2021)\citenamefont {Lian},
  \citenamefont {Song}, \citenamefont {Regnault}, \citenamefont {Efetov},
  \citenamefont {Yazdani},\ and\ \citenamefont {Bernevig}}]{Biao-TBG4}%
  \BibitemOpen
  \bibfield  {author} {\bibinfo {author} {\bibfnamefont {Biao}\ \bibnamefont
  {Lian}}, \bibinfo {author} {\bibfnamefont {Zhi-Da}\ \bibnamefont {Song}},
  \bibinfo {author} {\bibfnamefont {Nicolas}\ \bibnamefont {Regnault}},
  \bibinfo {author} {\bibfnamefont {Dmitri~K.}\ \bibnamefont {Efetov}},
  \bibinfo {author} {\bibfnamefont {Ali}\ \bibnamefont {Yazdani}}, \ and\
  \bibinfo {author} {\bibfnamefont {B.~Andrei}\ \bibnamefont {Bernevig}},\
  }\bibfield  {title} {\enquote {\bibinfo {title} {Twisted bilayer graphene.
  {IV}. {Exact} insulator ground states and phase diagram},}\ }\href {\doibase
  10.1103/PhysRevB.103.205414} {\bibfield  {journal} {\bibinfo  {journal}
  {Physical Review B}\ }\textbf {\bibinfo {volume} {103}},\ \bibinfo {pages}
  {205414} (\bibinfo {year} {2021})},\ \bibinfo {note} {publisher: American
  Physical Society}\BibitemShut {NoStop}%
\bibitem [{\citenamefont {Bernevig}\ \emph
  {et~al.}(2021{\natexlab{b}})\citenamefont {Bernevig}, \citenamefont {Lian},
  \citenamefont {Cowsik}, \citenamefont {Xie}, \citenamefont {Regnault},\ and\
  \citenamefont {Song}}]{TBG5}%
  \BibitemOpen
  \bibfield  {author} {\bibinfo {author} {\bibfnamefont {B.~Andrei}\
  \bibnamefont {Bernevig}}, \bibinfo {author} {\bibfnamefont {Biao}\
  \bibnamefont {Lian}}, \bibinfo {author} {\bibfnamefont {Aditya}\ \bibnamefont
  {Cowsik}}, \bibinfo {author} {\bibfnamefont {Fang}\ \bibnamefont {Xie}},
  \bibinfo {author} {\bibfnamefont {Nicolas}\ \bibnamefont {Regnault}}, \ and\
  \bibinfo {author} {\bibfnamefont {Zhi-Da}\ \bibnamefont {Song}},\ }\bibfield
  {title} {\enquote {\bibinfo {title} {Twisted bilayer graphene. {V}. {Exact}
  analytic many-body excitations in {Coulomb} {Hamiltonians}: {Charge} gap,
  {Goldstone} modes, and absence of {Cooper} pairing},}\ }\href {\doibase
  10.1103/PhysRevB.103.205415} {\bibfield  {journal} {\bibinfo  {journal}
  {Physical Review B}\ }\textbf {\bibinfo {volume} {103}},\ \bibinfo {pages}
  {205415} (\bibinfo {year} {2021}{\natexlab{b}})},\ \bibinfo {note}
  {publisher: American Physical Society}\BibitemShut {NoStop}%
\bibitem [{\citenamefont {Vafek}\ and\ \citenamefont
  {Kang}(2021)}]{vafek2021lattice}%
  \BibitemOpen
  \bibfield  {author} {\bibinfo {author} {\bibfnamefont {Oskar}\ \bibnamefont
  {Vafek}}\ and\ \bibinfo {author} {\bibfnamefont {Jian}\ \bibnamefont
  {Kang}},\ }\bibfield  {title} {\enquote {\bibinfo {title} {Lattice model for
  the coulomb interacting chiral limit of the magic angle twisted bilayer
  graphene: symmetries, obstructions and excitations},}\ }\href
  {https://arxiv.org/abs/2106.05670} {\bibfield  {journal} {\bibinfo  {journal}
  {arXiv preprint arXiv:2106.05670}\ } (\bibinfo {year} {2021})}\BibitemShut
  {NoStop}%
\bibitem [{\citenamefont {Cao}\ \emph {et~al.}(2018{\natexlab{a}})\citenamefont
  {Cao}, \citenamefont {Fatemi}, \citenamefont {Demir}, \citenamefont {Fang},
  \citenamefont {Tomarken}, \citenamefont {Luo}, \citenamefont
  {Sanchez-Yamagishi}, \citenamefont {Watanabe}, \citenamefont {Taniguchi},
  \citenamefont {Kaxiras}, \citenamefont {Ashoori},\ and\ \citenamefont
  {Jarillo-Herrero}}]{cao_correlated_2018}%
  \BibitemOpen
  \bibfield  {author} {\bibinfo {author} {\bibfnamefont {Yuan}\ \bibnamefont
  {Cao}}, \bibinfo {author} {\bibfnamefont {Valla}\ \bibnamefont {Fatemi}},
  \bibinfo {author} {\bibfnamefont {Ahmet}\ \bibnamefont {Demir}}, \bibinfo
  {author} {\bibfnamefont {Shiang}\ \bibnamefont {Fang}}, \bibinfo {author}
  {\bibfnamefont {Spencer~L.}\ \bibnamefont {Tomarken}}, \bibinfo {author}
  {\bibfnamefont {Jason~Y.}\ \bibnamefont {Luo}}, \bibinfo {author}
  {\bibfnamefont {Javier~D.}\ \bibnamefont {Sanchez-Yamagishi}}, \bibinfo
  {author} {\bibfnamefont {Kenji}\ \bibnamefont {Watanabe}}, \bibinfo {author}
  {\bibfnamefont {Takashi}\ \bibnamefont {Taniguchi}}, \bibinfo {author}
  {\bibfnamefont {Efthimios}\ \bibnamefont {Kaxiras}}, \bibinfo {author}
  {\bibfnamefont {Ray~C.}\ \bibnamefont {Ashoori}}, \ and\ \bibinfo {author}
  {\bibfnamefont {Pablo}\ \bibnamefont {Jarillo-Herrero}},\ }\bibfield  {title}
  {\enquote {\bibinfo {title} {Correlated insulator behaviour at half-filling
  in magic-angle graphene superlattices},}\ }\href {\doibase
  10.1038/nature26154} {\bibfield  {journal} {\bibinfo  {journal} {Nature}\
  }\textbf {\bibinfo {volume} {556}},\ \bibinfo {pages} {80--84} (\bibinfo
  {year} {2018}{\natexlab{a}})}\BibitemShut {NoStop}%
\bibitem [{\citenamefont {Cao}\ \emph {et~al.}(2018{\natexlab{b}})\citenamefont
  {Cao}, \citenamefont {Fatemi}, \citenamefont {Fang}, \citenamefont
  {Watanabe}, \citenamefont {Taniguchi}, \citenamefont {Kaxiras},\ and\
  \citenamefont {Jarillo-Herrero}}]{cao_unconventional_2018}%
  \BibitemOpen
  \bibfield  {author} {\bibinfo {author} {\bibfnamefont {Yuan}\ \bibnamefont
  {Cao}}, \bibinfo {author} {\bibfnamefont {Valla}\ \bibnamefont {Fatemi}},
  \bibinfo {author} {\bibfnamefont {Shiang}\ \bibnamefont {Fang}}, \bibinfo
  {author} {\bibfnamefont {Kenji}\ \bibnamefont {Watanabe}}, \bibinfo {author}
  {\bibfnamefont {Takashi}\ \bibnamefont {Taniguchi}}, \bibinfo {author}
  {\bibfnamefont {Efthimios}\ \bibnamefont {Kaxiras}}, \ and\ \bibinfo {author}
  {\bibfnamefont {Pablo}\ \bibnamefont {Jarillo-Herrero}},\ }\bibfield  {title}
  {\enquote {\bibinfo {title} {Unconventional superconductivity in magic-angle
  graphene superlattices},}\ }\href {\doibase 10.1038/nature26160} {\bibfield
  {journal} {\bibinfo  {journal} {Nature}\ }\textbf {\bibinfo {volume} {556}},\
  \bibinfo {pages} {43--50} (\bibinfo {year} {2018}{\natexlab{b}})}\BibitemShut
  {NoStop}%
\bibitem [{\citenamefont {Bistritzer}\ and\ \citenamefont
  {MacDonald}(2011)}]{bistritzer_moire_2011}%
  \BibitemOpen
  \bibfield  {author} {\bibinfo {author} {\bibfnamefont {Rafi}\ \bibnamefont
  {Bistritzer}}\ and\ \bibinfo {author} {\bibfnamefont {Allan~H.}\ \bibnamefont
  {MacDonald}},\ }\bibfield  {title} {\enquote {\bibinfo {title} {Moiré bands
  in twisted double-layer graphene},}\ }\href {\doibase
  10.1073/pnas.1108174108} {\bibfield  {journal} {\bibinfo  {journal}
  {Proceedings of the National Academy of Sciences}\ }\textbf {\bibinfo
  {volume} {108}},\ \bibinfo {pages} {12233--12237} (\bibinfo {year}
  {2011})}\BibitemShut {NoStop}%
\bibitem [{\citenamefont {Lu}\ \emph {et~al.}(2019)\citenamefont {Lu},
  \citenamefont {Stepanov}, \citenamefont {Yang}, \citenamefont {Xie},
  \citenamefont {Aamir}, \citenamefont {Das}, \citenamefont {Urgell},
  \citenamefont {Watanabe}, \citenamefont {Taniguchi}, \citenamefont {Zhang},
  \citenamefont {Bachtold}, \citenamefont {MacDonald},\ and\ \citenamefont
  {Efetov}}]{lu2019superconductors}%
  \BibitemOpen
  \bibfield  {author} {\bibinfo {author} {\bibfnamefont {Xiaobo}\ \bibnamefont
  {Lu}}, \bibinfo {author} {\bibfnamefont {Petr}\ \bibnamefont {Stepanov}},
  \bibinfo {author} {\bibfnamefont {Wei}\ \bibnamefont {Yang}}, \bibinfo
  {author} {\bibfnamefont {Ming}\ \bibnamefont {Xie}}, \bibinfo {author}
  {\bibfnamefont {Mohammed~Ali}\ \bibnamefont {Aamir}}, \bibinfo {author}
  {\bibfnamefont {Ipsita}\ \bibnamefont {Das}}, \bibinfo {author}
  {\bibfnamefont {Carles}\ \bibnamefont {Urgell}}, \bibinfo {author}
  {\bibfnamefont {Kenji}\ \bibnamefont {Watanabe}}, \bibinfo {author}
  {\bibfnamefont {Takashi}\ \bibnamefont {Taniguchi}}, \bibinfo {author}
  {\bibfnamefont {Guangyu}\ \bibnamefont {Zhang}}, \bibinfo {author}
  {\bibfnamefont {Adrian}\ \bibnamefont {Bachtold}}, \bibinfo {author}
  {\bibfnamefont {Allan~H.}\ \bibnamefont {MacDonald}}, \ and\ \bibinfo
  {author} {\bibfnamefont {Dmitri~K.}\ \bibnamefont {Efetov}},\ }\bibfield
  {title} {\enquote {\bibinfo {title} {Superconductors, orbital magnets and
  correlated states in magic-angle bilayer graphene},}\ }\href
  {https://www.nature.com/articles/s41586-019-1695-0} {\bibfield  {journal}
  {\bibinfo  {journal} {Nature}\ }\textbf {\bibinfo {volume} {574}},\ \bibinfo
  {pages} {653--657} (\bibinfo {year} {2019})}\BibitemShut {NoStop}%
\bibitem [{\citenamefont {Yankowitz}\ \emph {et~al.}(2019)\citenamefont
  {Yankowitz}, \citenamefont {Chen}, \citenamefont {Polshyn}, \citenamefont
  {Zhang}, \citenamefont {Watanabe}, \citenamefont {Taniguchi}, \citenamefont
  {Graf}, \citenamefont {Young},\ and\ \citenamefont
  {Dean}}]{yankowitz2019tuning}%
  \BibitemOpen
  \bibfield  {author} {\bibinfo {author} {\bibfnamefont {Matthew}\ \bibnamefont
  {Yankowitz}}, \bibinfo {author} {\bibfnamefont {Shaowen}\ \bibnamefont
  {Chen}}, \bibinfo {author} {\bibfnamefont {Hryhoriy}\ \bibnamefont
  {Polshyn}}, \bibinfo {author} {\bibfnamefont {Yuxuan}\ \bibnamefont {Zhang}},
  \bibinfo {author} {\bibfnamefont {K}~\bibnamefont {Watanabe}}, \bibinfo
  {author} {\bibfnamefont {T}~\bibnamefont {Taniguchi}}, \bibinfo {author}
  {\bibfnamefont {David}\ \bibnamefont {Graf}}, \bibinfo {author}
  {\bibfnamefont {Andrea~F}\ \bibnamefont {Young}}, \ and\ \bibinfo {author}
  {\bibfnamefont {Cory~R}\ \bibnamefont {Dean}},\ }\bibfield  {title} {\enquote
  {\bibinfo {title} {Tuning superconductivity in twisted bilayer graphene},}\
  }\href
  {https://science.sciencemag.org/content/363/6431/1059.abstract?casa_token=lunm_WUYs2YAAAAA:VYMQ9xKAP9yNieasreqWu0I0g8sN82wxfevMxLMfsegLO9RZtKOt45kmqcsGZAKERIiy2VDY21ejfWs}
  {\bibfield  {journal} {\bibinfo  {journal} {Science}\ }\textbf {\bibinfo
  {volume} {363}},\ \bibinfo {pages} {1059--1064} (\bibinfo {year}
  {2019})}\BibitemShut {NoStop}%
\bibitem [{\citenamefont {Sharpe}\ \emph {et~al.}(2019)\citenamefont {Sharpe},
  \citenamefont {Fox}, \citenamefont {Barnard}, \citenamefont {Finney},
  \citenamefont {Watanabe}, \citenamefont {Taniguchi}, \citenamefont
  {Kastner},\ and\ \citenamefont {Goldhaber-Gordon}}]{sharpe_emergent_2019}%
  \BibitemOpen
  \bibfield  {author} {\bibinfo {author} {\bibfnamefont {Aaron~L.}\
  \bibnamefont {Sharpe}}, \bibinfo {author} {\bibfnamefont {Eli~J.}\
  \bibnamefont {Fox}}, \bibinfo {author} {\bibfnamefont {Arthur~W.}\
  \bibnamefont {Barnard}}, \bibinfo {author} {\bibfnamefont {Joe}\ \bibnamefont
  {Finney}}, \bibinfo {author} {\bibfnamefont {Kenji}\ \bibnamefont
  {Watanabe}}, \bibinfo {author} {\bibfnamefont {Takashi}\ \bibnamefont
  {Taniguchi}}, \bibinfo {author} {\bibfnamefont {M.~A.}\ \bibnamefont
  {Kastner}}, \ and\ \bibinfo {author} {\bibfnamefont {David}\ \bibnamefont
  {Goldhaber-Gordon}},\ }\bibfield  {title} {\enquote {\bibinfo {title}
  {Emergent ferromagnetism near three-quarters filling in twisted bilayer
  graphene},}\ }\href {\doibase 10.1126/science.aaw3780} {\bibfield  {journal}
  {\bibinfo  {journal} {Science}\ }\textbf {\bibinfo {volume} {365}},\ \bibinfo
  {pages} {605–608} (\bibinfo {year} {2019})}\BibitemShut {NoStop}%
\bibitem [{\citenamefont {Saito}\ \emph {et~al.}(2020)\citenamefont {Saito},
  \citenamefont {Ge}, \citenamefont {Watanabe}, \citenamefont {Taniguchi},\
  and\ \citenamefont {Young}}]{saito_independent_2020}%
  \BibitemOpen
  \bibfield  {author} {\bibinfo {author} {\bibfnamefont {Yu}~\bibnamefont
  {Saito}}, \bibinfo {author} {\bibfnamefont {Jingyuan}\ \bibnamefont {Ge}},
  \bibinfo {author} {\bibfnamefont {Kenji}\ \bibnamefont {Watanabe}}, \bibinfo
  {author} {\bibfnamefont {Takashi}\ \bibnamefont {Taniguchi}}, \ and\ \bibinfo
  {author} {\bibfnamefont {Andrea~F.}\ \bibnamefont {Young}},\ }\bibfield
  {title} {\enquote {\bibinfo {title} {Independent superconductors and
  correlated insulators in twisted bilayer graphene},}\ }\href {\doibase
  10.1038/s41567-020-0928-3} {\bibfield  {journal} {\bibinfo  {journal} {Nature
  Physics}\ }\textbf {\bibinfo {volume} {16}},\ \bibinfo {pages} {926–930}
  (\bibinfo {year} {2020})}\BibitemShut {NoStop}%
\bibitem [{\citenamefont {Stepanov}\ \emph {et~al.}(2020)\citenamefont
  {Stepanov}, \citenamefont {Das}, \citenamefont {Lu}, \citenamefont
  {Fahimniya}, \citenamefont {Watanabe}, \citenamefont {Taniguchi},
  \citenamefont {Koppens}, \citenamefont {Lischner}, \citenamefont {Levitov},\
  and\ \citenamefont {Efetov}}]{stepanov_interplay_2020}%
  \BibitemOpen
  \bibfield  {author} {\bibinfo {author} {\bibfnamefont {Petr}\ \bibnamefont
  {Stepanov}}, \bibinfo {author} {\bibfnamefont {Ipsita}\ \bibnamefont {Das}},
  \bibinfo {author} {\bibfnamefont {Xiaobo}\ \bibnamefont {Lu}}, \bibinfo
  {author} {\bibfnamefont {Ali}\ \bibnamefont {Fahimniya}}, \bibinfo {author}
  {\bibfnamefont {Kenji}\ \bibnamefont {Watanabe}}, \bibinfo {author}
  {\bibfnamefont {Takashi}\ \bibnamefont {Taniguchi}}, \bibinfo {author}
  {\bibfnamefont {Frank H.~L.}\ \bibnamefont {Koppens}}, \bibinfo {author}
  {\bibfnamefont {Johannes}\ \bibnamefont {Lischner}}, \bibinfo {author}
  {\bibfnamefont {Leonid}\ \bibnamefont {Levitov}}, \ and\ \bibinfo {author}
  {\bibfnamefont {Dmitri~K.}\ \bibnamefont {Efetov}},\ }\bibfield  {title}
  {\enquote {\bibinfo {title} {Untying the insulating and superconducting
  orders in magic-angle graphene},}\ }\href {\doibase
  10.1038/s41586-020-2459-6} {\bibfield  {journal} {\bibinfo  {journal}
  {Nature}\ }\textbf {\bibinfo {volume} {583}},\ \bibinfo {pages} {375–378}
  (\bibinfo {year} {2020})}\BibitemShut {NoStop}%
\bibitem [{\citenamefont {Liu}\ \emph {et~al.}(2021{\natexlab{a}})\citenamefont
  {Liu}, \citenamefont {Wang}, \citenamefont {Watanabe}, \citenamefont
  {Taniguchi}, \citenamefont {Vafek},\ and\ \citenamefont
  {Li}}]{liu2021tuning}%
  \BibitemOpen
  \bibfield  {author} {\bibinfo {author} {\bibfnamefont {Xiaoxue}\ \bibnamefont
  {Liu}}, \bibinfo {author} {\bibfnamefont {Zhi}\ \bibnamefont {Wang}},
  \bibinfo {author} {\bibfnamefont {Kenji}\ \bibnamefont {Watanabe}}, \bibinfo
  {author} {\bibfnamefont {Takashi}\ \bibnamefont {Taniguchi}}, \bibinfo
  {author} {\bibfnamefont {Oskar}\ \bibnamefont {Vafek}}, \ and\ \bibinfo
  {author} {\bibfnamefont {JIA}\ \bibnamefont {Li}},\ }\bibfield  {title}
  {\enquote {\bibinfo {title} {Tuning electron correlation in magic-angle
  twisted bilayer graphene using coulomb screening},}\ }\href
  {https://www.science.org/doi/abs/10.1126/science.abb8754} {\bibfield
  {journal} {\bibinfo  {journal} {Science}\ }\textbf {\bibinfo {volume}
  {371}},\ \bibinfo {pages} {1261--1265} (\bibinfo {year}
  {2021}{\natexlab{a}})}\BibitemShut {NoStop}%
\bibitem [{\citenamefont {Arora}\ \emph {et~al.}(2020)\citenamefont {Arora},
  \citenamefont {Polski}, \citenamefont {Zhang}, \citenamefont {Thomson},
  \citenamefont {Choi}, \citenamefont {Kim}, \citenamefont {Lin}, \citenamefont
  {Wilson}, \citenamefont {Xu}, \citenamefont {Chu},\ and\ \citenamefont
  {et~al.}}]{arora_2020}%
  \BibitemOpen
  \bibfield  {author} {\bibinfo {author} {\bibfnamefont {Harpreet~Singh}\
  \bibnamefont {Arora}}, \bibinfo {author} {\bibfnamefont {Robert}\
  \bibnamefont {Polski}}, \bibinfo {author} {\bibfnamefont {Yiran}\
  \bibnamefont {Zhang}}, \bibinfo {author} {\bibfnamefont {Alex}\ \bibnamefont
  {Thomson}}, \bibinfo {author} {\bibfnamefont {Youngjoon}\ \bibnamefont
  {Choi}}, \bibinfo {author} {\bibfnamefont {Hyunjin}\ \bibnamefont {Kim}},
  \bibinfo {author} {\bibfnamefont {Zhong}\ \bibnamefont {Lin}}, \bibinfo
  {author} {\bibfnamefont {Ilham~Zaky}\ \bibnamefont {Wilson}}, \bibinfo
  {author} {\bibfnamefont {Xiaodong}\ \bibnamefont {Xu}}, \bibinfo {author}
  {\bibfnamefont {Jiun-Haw}\ \bibnamefont {Chu}}, \ and\ \bibinfo {author}
  {\bibnamefont {et~al.}},\ }\bibfield  {title} {\enquote {\bibinfo {title}
  {Superconductivity in metallic twisted bilayer graphene stabilized by
  wse2},}\ }\href {\doibase 10.1038/s41586-020-2473-8} {\bibfield  {journal}
  {\bibinfo  {journal} {Nature}\ }\textbf {\bibinfo {volume} {583}},\ \bibinfo
  {pages} {379–384} (\bibinfo {year} {2020})}\BibitemShut {NoStop}%
\bibitem [{\citenamefont {Serlin}\ \emph {et~al.}(2019)\citenamefont {Serlin},
  \citenamefont {Tschirhart}, \citenamefont {Polshyn}, \citenamefont {Zhang},
  \citenamefont {Zhu}, \citenamefont {Watanabe}, \citenamefont {Taniguchi},
  \citenamefont {Balents},\ and\ \citenamefont {Young}}]{serlin_QAH_2019}%
  \BibitemOpen
  \bibfield  {author} {\bibinfo {author} {\bibfnamefont {M.}~\bibnamefont
  {Serlin}}, \bibinfo {author} {\bibfnamefont {C.~L.}\ \bibnamefont
  {Tschirhart}}, \bibinfo {author} {\bibfnamefont {H.}~\bibnamefont {Polshyn}},
  \bibinfo {author} {\bibfnamefont {Y.}~\bibnamefont {Zhang}}, \bibinfo
  {author} {\bibfnamefont {J.}~\bibnamefont {Zhu}}, \bibinfo {author}
  {\bibfnamefont {K.}~\bibnamefont {Watanabe}}, \bibinfo {author}
  {\bibfnamefont {T.}~\bibnamefont {Taniguchi}}, \bibinfo {author}
  {\bibfnamefont {L.}~\bibnamefont {Balents}}, \ and\ \bibinfo {author}
  {\bibfnamefont {A.~F.}\ \bibnamefont {Young}},\ }\bibfield  {title} {\enquote
  {\bibinfo {title} {Intrinsic quantized anomalous hall effect in a moiré
  heterostructure},}\ }\href {\doibase 10.1126/science.aay5533} {\bibfield
  {journal} {\bibinfo  {journal} {Science}\ }\textbf {\bibinfo {volume}
  {367}},\ \bibinfo {pages} {900–903} (\bibinfo {year} {2019})}\BibitemShut
  {NoStop}%
\bibitem [{\citenamefont {Cao}\ \emph {et~al.}(2020)\citenamefont {Cao},
  \citenamefont {Chowdhury}, \citenamefont {Rodan-Legrain}, \citenamefont
  {Rubies-Bigorda}, \citenamefont {Watanabe}, \citenamefont {Taniguchi},
  \citenamefont {Senthil},\ and\ \citenamefont
  {Jarillo-Herrero}}]{cao_strange_2020}%
  \BibitemOpen
  \bibfield  {author} {\bibinfo {author} {\bibfnamefont {Yuan}\ \bibnamefont
  {Cao}}, \bibinfo {author} {\bibfnamefont {Debanjan}\ \bibnamefont
  {Chowdhury}}, \bibinfo {author} {\bibfnamefont {Daniel}\ \bibnamefont
  {Rodan-Legrain}}, \bibinfo {author} {\bibfnamefont {Oriol}\ \bibnamefont
  {Rubies-Bigorda}}, \bibinfo {author} {\bibfnamefont {Kenji}\ \bibnamefont
  {Watanabe}}, \bibinfo {author} {\bibfnamefont {Takashi}\ \bibnamefont
  {Taniguchi}}, \bibinfo {author} {\bibfnamefont {T.}~\bibnamefont {Senthil}},
  \ and\ \bibinfo {author} {\bibfnamefont {Pablo}\ \bibnamefont
  {Jarillo-Herrero}},\ }\bibfield  {title} {\enquote {\bibinfo {title} {Strange
  metal in magic-angle graphene with near planckian dissipation},}\ }\href
  {\doibase 10.1103/PhysRevLett.124.076801} {\bibfield  {journal} {\bibinfo
  {journal} {Phys. Rev. Lett.}\ }\textbf {\bibinfo {volume} {124}},\ \bibinfo
  {pages} {076801} (\bibinfo {year} {2020})}\BibitemShut {NoStop}%
\bibitem [{\citenamefont {Polshyn}\ \emph {et~al.}(2019)\citenamefont
  {Polshyn}, \citenamefont {Yankowitz}, \citenamefont {Chen}, \citenamefont
  {Zhang}, \citenamefont {Watanabe}, \citenamefont {Taniguchi}, \citenamefont
  {Dean},\ and\ \citenamefont {Young}}]{polshyn_linear_2019}%
  \BibitemOpen
  \bibfield  {author} {\bibinfo {author} {\bibfnamefont {Hryhoriy}\
  \bibnamefont {Polshyn}}, \bibinfo {author} {\bibfnamefont {Matthew}\
  \bibnamefont {Yankowitz}}, \bibinfo {author} {\bibfnamefont {Shaowen}\
  \bibnamefont {Chen}}, \bibinfo {author} {\bibfnamefont {Yuxuan}\ \bibnamefont
  {Zhang}}, \bibinfo {author} {\bibfnamefont {K.}~\bibnamefont {Watanabe}},
  \bibinfo {author} {\bibfnamefont {T.}~\bibnamefont {Taniguchi}}, \bibinfo
  {author} {\bibfnamefont {Cory~R.}\ \bibnamefont {Dean}}, \ and\ \bibinfo
  {author} {\bibfnamefont {Andrea~F.}\ \bibnamefont {Young}},\ }\bibfield
  {title} {\enquote {\bibinfo {title} {Large linear-in-temperature resistivity
  in twisted bilayer graphene},}\ }\href {\doibase 10.1038/s41567-019-0596-3}
  {\bibfield  {journal} {\bibinfo  {journal} {Nature Physics}\ }\textbf
  {\bibinfo {volume} {15}},\ \bibinfo {pages} {1011–1016} (\bibinfo {year}
  {2019})}\BibitemShut {NoStop}%
\bibitem [{\citenamefont {Xie}\ \emph {et~al.}(2019)\citenamefont {Xie},
  \citenamefont {Lian}, \citenamefont {J{\"a}ck}, \citenamefont {Liu},
  \citenamefont {Chiu}, \citenamefont {Watanabe}, \citenamefont {Taniguchi},
  \citenamefont {Bernevig},\ and\ \citenamefont
  {Yazdani}}]{xie2019spectroscopic}%
  \BibitemOpen
  \bibfield  {author} {\bibinfo {author} {\bibfnamefont {Yonglong}\
  \bibnamefont {Xie}}, \bibinfo {author} {\bibfnamefont {Biao}\ \bibnamefont
  {Lian}}, \bibinfo {author} {\bibfnamefont {Berthold}\ \bibnamefont
  {J{\"a}ck}}, \bibinfo {author} {\bibfnamefont {Xiaomeng}\ \bibnamefont
  {Liu}}, \bibinfo {author} {\bibfnamefont {Cheng-Li}\ \bibnamefont {Chiu}},
  \bibinfo {author} {\bibfnamefont {Kenji}\ \bibnamefont {Watanabe}}, \bibinfo
  {author} {\bibfnamefont {Takashi}\ \bibnamefont {Taniguchi}}, \bibinfo
  {author} {\bibfnamefont {B~Andrei}\ \bibnamefont {Bernevig}}, \ and\ \bibinfo
  {author} {\bibfnamefont {Ali}\ \bibnamefont {Yazdani}},\ }\bibfield  {title}
  {\enquote {\bibinfo {title} {Spectroscopic signatures of many-body
  correlations in magic-angle twisted bilayer graphene},}\ }\href
  {https://www.nature.com/articles/s41586-019-1422-x} {\bibfield  {journal}
  {\bibinfo  {journal} {Nature}\ }\textbf {\bibinfo {volume} {572}},\ \bibinfo
  {pages} {101--105} (\bibinfo {year} {2019})}\BibitemShut {NoStop}%
\bibitem [{\citenamefont {Choi}\ \emph {et~al.}(2019)\citenamefont {Choi},
  \citenamefont {Kemmer}, \citenamefont {Peng}, \citenamefont {Thomson},
  \citenamefont {Arora}, \citenamefont {Polski}, \citenamefont {Zhang},
  \citenamefont {Ren}, \citenamefont {Alicea}, \citenamefont {Refael},\ and\
  \citenamefont {et~al.}}]{choi_imaging_2019}%
  \BibitemOpen
  \bibfield  {author} {\bibinfo {author} {\bibfnamefont {Youngjoon}\
  \bibnamefont {Choi}}, \bibinfo {author} {\bibfnamefont {Jeannette}\
  \bibnamefont {Kemmer}}, \bibinfo {author} {\bibfnamefont {Yang}\ \bibnamefont
  {Peng}}, \bibinfo {author} {\bibfnamefont {Alex}\ \bibnamefont {Thomson}},
  \bibinfo {author} {\bibfnamefont {Harpreet}\ \bibnamefont {Arora}}, \bibinfo
  {author} {\bibfnamefont {Robert}\ \bibnamefont {Polski}}, \bibinfo {author}
  {\bibfnamefont {Yiran}\ \bibnamefont {Zhang}}, \bibinfo {author}
  {\bibfnamefont {Hechen}\ \bibnamefont {Ren}}, \bibinfo {author}
  {\bibfnamefont {Jason}\ \bibnamefont {Alicea}}, \bibinfo {author}
  {\bibfnamefont {Gil}\ \bibnamefont {Refael}}, \ and\ \bibinfo {author}
  {\bibnamefont {et~al.}},\ }\bibfield  {title} {\enquote {\bibinfo {title}
  {Electronic correlations in twisted bilayer graphene near the magic angle},}\
  }\href {\doibase 10.1038/s41567-019-0606-5} {\bibfield  {journal} {\bibinfo
  {journal} {Nature Physics}\ }\textbf {\bibinfo {volume} {15}},\ \bibinfo
  {pages} {1174–1180} (\bibinfo {year} {2019})}\BibitemShut {NoStop}%
\bibitem [{\citenamefont {Kerelsky}\ \emph {et~al.}(2019)\citenamefont
  {Kerelsky}, \citenamefont {McGilly}, \citenamefont {Kennes}, \citenamefont
  {Xian}, \citenamefont {Yankowitz}, \citenamefont {Chen}, \citenamefont
  {Watanabe}, \citenamefont {Taniguchi}, \citenamefont {Hone}, \citenamefont
  {Dean},\ and\ \citenamefont {et~al.}}]{kerelsky_2019_stm}%
  \BibitemOpen
  \bibfield  {author} {\bibinfo {author} {\bibfnamefont {Alexander}\
  \bibnamefont {Kerelsky}}, \bibinfo {author} {\bibfnamefont {Leo~J.}\
  \bibnamefont {McGilly}}, \bibinfo {author} {\bibfnamefont {Dante~M.}\
  \bibnamefont {Kennes}}, \bibinfo {author} {\bibfnamefont {Lede}\ \bibnamefont
  {Xian}}, \bibinfo {author} {\bibfnamefont {Matthew}\ \bibnamefont
  {Yankowitz}}, \bibinfo {author} {\bibfnamefont {Shaowen}\ \bibnamefont
  {Chen}}, \bibinfo {author} {\bibfnamefont {K.}~\bibnamefont {Watanabe}},
  \bibinfo {author} {\bibfnamefont {T.}~\bibnamefont {Taniguchi}}, \bibinfo
  {author} {\bibfnamefont {James}\ \bibnamefont {Hone}}, \bibinfo {author}
  {\bibfnamefont {Cory}\ \bibnamefont {Dean}}, \ and\ \bibinfo {author}
  {\bibnamefont {et~al.}},\ }\bibfield  {title} {\enquote {\bibinfo {title}
  {Maximized electron interactions at the magic angle in twisted bilayer
  graphene},}\ }\href {\doibase 10.1038/s41586-019-1431-9} {\bibfield
  {journal} {\bibinfo  {journal} {Nature}\ }\textbf {\bibinfo {volume} {572}},\
  \bibinfo {pages} {95–100} (\bibinfo {year} {2019})}\BibitemShut {NoStop}%
\bibitem [{\citenamefont {Jiang}\ \emph {et~al.}(2019)\citenamefont {Jiang},
  \citenamefont {Lai}, \citenamefont {Watanabe}, \citenamefont {Taniguchi},
  \citenamefont {Haule}, \citenamefont {Mao},\ and\ \citenamefont
  {Andrei}}]{jiang_charge_2019}%
  \BibitemOpen
  \bibfield  {author} {\bibinfo {author} {\bibfnamefont {Yuhang}\ \bibnamefont
  {Jiang}}, \bibinfo {author} {\bibfnamefont {Xinyuan}\ \bibnamefont {Lai}},
  \bibinfo {author} {\bibfnamefont {Kenji}\ \bibnamefont {Watanabe}}, \bibinfo
  {author} {\bibfnamefont {Takashi}\ \bibnamefont {Taniguchi}}, \bibinfo
  {author} {\bibfnamefont {Kristjan}\ \bibnamefont {Haule}}, \bibinfo {author}
  {\bibfnamefont {Jinhai}\ \bibnamefont {Mao}}, \ and\ \bibinfo {author}
  {\bibfnamefont {Eva~Y.}\ \bibnamefont {Andrei}},\ }\bibfield  {title}
  {\enquote {\bibinfo {title} {Charge order and broken rotational symmetry in
  magic-angle twisted bilayer graphene},}\ }\href {\doibase
  10.1038/s41586-019-1460-4} {\bibfield  {journal} {\bibinfo  {journal}
  {Nature}\ }\textbf {\bibinfo {volume} {573}},\ \bibinfo {pages} {91–95}
  (\bibinfo {year} {2019})}\BibitemShut {NoStop}%
\bibitem [{\citenamefont {Wong}\ \emph {et~al.}(2020)\citenamefont {Wong},
  \citenamefont {Nuckolls}, \citenamefont {Oh}, \citenamefont {Lian},
  \citenamefont {Xie}, \citenamefont {Jeon}, \citenamefont {Watanabe},
  \citenamefont {Taniguchi}, \citenamefont {Bernevig},\ and\ \citenamefont
  {Yazdani}}]{wong_cascade_2020}%
  \BibitemOpen
  \bibfield  {author} {\bibinfo {author} {\bibfnamefont {Dillon}\ \bibnamefont
  {Wong}}, \bibinfo {author} {\bibfnamefont {Kevin~P.}\ \bibnamefont
  {Nuckolls}}, \bibinfo {author} {\bibfnamefont {Myungchul}\ \bibnamefont
  {Oh}}, \bibinfo {author} {\bibfnamefont {Biao}\ \bibnamefont {Lian}},
  \bibinfo {author} {\bibfnamefont {Yonglong}\ \bibnamefont {Xie}}, \bibinfo
  {author} {\bibfnamefont {Sangjun}\ \bibnamefont {Jeon}}, \bibinfo {author}
  {\bibfnamefont {Kenji}\ \bibnamefont {Watanabe}}, \bibinfo {author}
  {\bibfnamefont {Takashi}\ \bibnamefont {Taniguchi}}, \bibinfo {author}
  {\bibfnamefont {B.~Andrei}\ \bibnamefont {Bernevig}}, \ and\ \bibinfo
  {author} {\bibfnamefont {Ali}\ \bibnamefont {Yazdani}},\ }\bibfield  {title}
  {\enquote {\bibinfo {title} {Cascade of electronic transitions in magic-angle
  twisted bilayer graphene},}\ }\href {\doibase 10.1038/s41586-020-2339-0}
  {\bibfield  {journal} {\bibinfo  {journal} {Nature}\ }\textbf {\bibinfo
  {volume} {582}},\ \bibinfo {pages} {198–202} (\bibinfo {year}
  {2020})}\BibitemShut {NoStop}%
\bibitem [{\citenamefont {Zondiner}\ \emph {et~al.}(2020)\citenamefont
  {Zondiner}, \citenamefont {Rozen}, \citenamefont {Rodan-Legrain},
  \citenamefont {Cao}, \citenamefont {Queiroz}, \citenamefont {Taniguchi},
  \citenamefont {Watanabe}, \citenamefont {Oreg}, \citenamefont {von Oppen},
  \citenamefont {Stern},\ and\ \citenamefont {et~al.}}]{zondiner_cascade_2020}%
  \BibitemOpen
  \bibfield  {author} {\bibinfo {author} {\bibfnamefont {U.}~\bibnamefont
  {Zondiner}}, \bibinfo {author} {\bibfnamefont {A.}~\bibnamefont {Rozen}},
  \bibinfo {author} {\bibfnamefont {D.}~\bibnamefont {Rodan-Legrain}}, \bibinfo
  {author} {\bibfnamefont {Y.}~\bibnamefont {Cao}}, \bibinfo {author}
  {\bibfnamefont {R.}~\bibnamefont {Queiroz}}, \bibinfo {author} {\bibfnamefont
  {T.}~\bibnamefont {Taniguchi}}, \bibinfo {author} {\bibfnamefont
  {K.}~\bibnamefont {Watanabe}}, \bibinfo {author} {\bibfnamefont
  {Y.}~\bibnamefont {Oreg}}, \bibinfo {author} {\bibfnamefont {F.}~\bibnamefont
  {von Oppen}}, \bibinfo {author} {\bibfnamefont {Ady}\ \bibnamefont {Stern}},
  \ and\ \bibinfo {author} {\bibnamefont {et~al.}},\ }\bibfield  {title}
  {\enquote {\bibinfo {title} {Cascade of phase transitions and dirac revivals
  in magic-angle graphene},}\ }\href {\doibase 10.1038/s41586-020-2373-y}
  {\bibfield  {journal} {\bibinfo  {journal} {Nature}\ }\textbf {\bibinfo
  {volume} {582}},\ \bibinfo {pages} {203–208} (\bibinfo {year}
  {2020})}\BibitemShut {NoStop}%
\bibitem [{\citenamefont {Nuckolls}\ \emph {et~al.}(2020)\citenamefont
  {Nuckolls}, \citenamefont {Oh}, \citenamefont {Wong}, \citenamefont {Lian},
  \citenamefont {Watanabe}, \citenamefont {Taniguchi}, \citenamefont
  {Bernevig},\ and\ \citenamefont {Yazdani}}]{nuckolls_chern_2020}%
  \BibitemOpen
  \bibfield  {author} {\bibinfo {author} {\bibfnamefont {Kevin~P}\ \bibnamefont
  {Nuckolls}}, \bibinfo {author} {\bibfnamefont {Myungchul}\ \bibnamefont
  {Oh}}, \bibinfo {author} {\bibfnamefont {Dillon}\ \bibnamefont {Wong}},
  \bibinfo {author} {\bibfnamefont {Biao}\ \bibnamefont {Lian}}, \bibinfo
  {author} {\bibfnamefont {Kenji}\ \bibnamefont {Watanabe}}, \bibinfo {author}
  {\bibfnamefont {Takashi}\ \bibnamefont {Taniguchi}}, \bibinfo {author}
  {\bibfnamefont {B~Andrei}\ \bibnamefont {Bernevig}}, \ and\ \bibinfo {author}
  {\bibfnamefont {Ali}\ \bibnamefont {Yazdani}},\ }\bibfield  {title} {\enquote
  {\bibinfo {title} {Strongly correlated chern insulators in magic-angle
  twisted bilayer graphene},}\ }\href
  {https://www.nature.com/articles/s41586-020-3028-8} {\bibfield  {journal}
  {\bibinfo  {journal} {Nature}\ }\textbf {\bibinfo {volume} {588}},\ \bibinfo
  {pages} {610--615} (\bibinfo {year} {2020})}\BibitemShut {NoStop}%
\bibitem [{\citenamefont {Choi}\ \emph {et~al.}(2021)\citenamefont {Choi},
  \citenamefont {Kim}, \citenamefont {Peng}, \citenamefont {Thomson},
  \citenamefont {Lewandowski}, \citenamefont {Polski}, \citenamefont {Zhang},
  \citenamefont {Arora}, \citenamefont {Watanabe}, \citenamefont {Taniguchi}
  \emph {et~al.}}]{choi2020tracing}%
  \BibitemOpen
  \bibfield  {author} {\bibinfo {author} {\bibfnamefont {Youngjoon}\
  \bibnamefont {Choi}}, \bibinfo {author} {\bibfnamefont {Hyunjin}\
  \bibnamefont {Kim}}, \bibinfo {author} {\bibfnamefont {Yang}\ \bibnamefont
  {Peng}}, \bibinfo {author} {\bibfnamefont {Alex}\ \bibnamefont {Thomson}},
  \bibinfo {author} {\bibfnamefont {Cyprian}\ \bibnamefont {Lewandowski}},
  \bibinfo {author} {\bibfnamefont {Robert}\ \bibnamefont {Polski}}, \bibinfo
  {author} {\bibfnamefont {Yiran}\ \bibnamefont {Zhang}}, \bibinfo {author}
  {\bibfnamefont {Harpreet~Singh}\ \bibnamefont {Arora}}, \bibinfo {author}
  {\bibfnamefont {Kenji}\ \bibnamefont {Watanabe}}, \bibinfo {author}
  {\bibfnamefont {Takashi}\ \bibnamefont {Taniguchi}},  \emph {et~al.},\
  }\bibfield  {title} {\enquote {\bibinfo {title} {Correlation-driven
  topological phases in magic-angle twisted bilayer graphene},}\ }\href
  {\doibase 10.1038/s41586-020-03159-7} {\bibfield  {journal} {\bibinfo
  {journal} {Nature}\ }\textbf {\bibinfo {volume} {589}},\ \bibinfo {pages}
  {536--541} (\bibinfo {year} {2021})}\BibitemShut {NoStop}%
\bibitem [{\citenamefont {Saito}\ \emph
  {et~al.}(2021{\natexlab{a}})\citenamefont {Saito}, \citenamefont {Ge},
  \citenamefont {Rademaker}, \citenamefont {Watanabe}, \citenamefont
  {Taniguchi}, \citenamefont {Abanin},\ and\ \citenamefont
  {Young}}]{saito2020}%
  \BibitemOpen
  \bibfield  {author} {\bibinfo {author} {\bibfnamefont {Yu}~\bibnamefont
  {Saito}}, \bibinfo {author} {\bibfnamefont {Jingyuan}\ \bibnamefont {Ge}},
  \bibinfo {author} {\bibfnamefont {Louk}\ \bibnamefont {Rademaker}}, \bibinfo
  {author} {\bibfnamefont {Kenji}\ \bibnamefont {Watanabe}}, \bibinfo {author}
  {\bibfnamefont {Takashi}\ \bibnamefont {Taniguchi}}, \bibinfo {author}
  {\bibfnamefont {Dmitry~A}\ \bibnamefont {Abanin}}, \ and\ \bibinfo {author}
  {\bibfnamefont {Andrea~F}\ \bibnamefont {Young}},\ }\bibfield  {title}
  {\enquote {\bibinfo {title} {Hofstadter subband ferromagnetism and
  symmetry-broken chern insulators in twisted bilayer graphene},}\ }\href
  {https://www.nature.com/articles/s41567-020-01129-4} {\bibfield  {journal}
  {\bibinfo  {journal} {Nature Physics}\ }\textbf {\bibinfo {volume} {17}},\
  \bibinfo {pages} {478--481} (\bibinfo {year}
  {2021}{\natexlab{a}})}\BibitemShut {NoStop}%
\bibitem [{\citenamefont {Das}\ \emph {et~al.}(2021)\citenamefont {Das},
  \citenamefont {Lu}, \citenamefont {Herzog-Arbeitman}, \citenamefont {Song},
  \citenamefont {Watanabe}, \citenamefont {Taniguchi}, \citenamefont
  {Bernevig},\ and\ \citenamefont {Efetov}}]{das2020symmetry}%
  \BibitemOpen
  \bibfield  {author} {\bibinfo {author} {\bibfnamefont {Ipsita}\ \bibnamefont
  {Das}}, \bibinfo {author} {\bibfnamefont {Xiaobo}\ \bibnamefont {Lu}},
  \bibinfo {author} {\bibfnamefont {Jonah}\ \bibnamefont {Herzog-Arbeitman}},
  \bibinfo {author} {\bibfnamefont {Zhi-Da}\ \bibnamefont {Song}}, \bibinfo
  {author} {\bibfnamefont {Kenji}\ \bibnamefont {Watanabe}}, \bibinfo {author}
  {\bibfnamefont {Takashi}\ \bibnamefont {Taniguchi}}, \bibinfo {author}
  {\bibfnamefont {B~Andrei}\ \bibnamefont {Bernevig}}, \ and\ \bibinfo {author}
  {\bibfnamefont {Dmitri~K}\ \bibnamefont {Efetov}},\ }\bibfield  {title}
  {\enquote {\bibinfo {title} {Symmetry-broken chern insulators and rashba-like
  landau-level crossings in magic-angle bilayer graphene},}\ }\href
  {https://www.nature.com/articles/s41567-021-01186-3} {\bibfield  {journal}
  {\bibinfo  {journal} {Nature Physics}\ }\textbf {\bibinfo {volume} {17}},\
  \bibinfo {pages} {710--714} (\bibinfo {year} {2021})}\BibitemShut {NoStop}%
\bibitem [{\citenamefont {Wu}\ \emph {et~al.}(2021)\citenamefont {Wu},
  \citenamefont {Zhang}, \citenamefont {Watanabe}, \citenamefont {Taniguchi},\
  and\ \citenamefont {Andrei}}]{wu_chern_2020}%
  \BibitemOpen
  \bibfield  {author} {\bibinfo {author} {\bibfnamefont {Shuang}\ \bibnamefont
  {Wu}}, \bibinfo {author} {\bibfnamefont {Zhenyuan}\ \bibnamefont {Zhang}},
  \bibinfo {author} {\bibfnamefont {K}~\bibnamefont {Watanabe}}, \bibinfo
  {author} {\bibfnamefont {T}~\bibnamefont {Taniguchi}}, \ and\ \bibinfo
  {author} {\bibfnamefont {Eva~Y}\ \bibnamefont {Andrei}},\ }\bibfield  {title}
  {\enquote {\bibinfo {title} {Chern insulators, van hove singularities and
  topological flat bands in magic-angle twisted bilayer graphene},}\ }\href
  {https://www.nature.com/articles/s41563-020-00911-2} {\bibfield  {journal}
  {\bibinfo  {journal} {Nature materials}\ }\textbf {\bibinfo {volume} {20}},\
  \bibinfo {pages} {488--494} (\bibinfo {year} {2021})}\BibitemShut {NoStop}%
\bibitem [{\citenamefont {Park}\ \emph {et~al.}(2021)\citenamefont {Park},
  \citenamefont {Cao}, \citenamefont {Watanabe}, \citenamefont {Taniguchi},\
  and\ \citenamefont {Jarillo-Herrero}}]{park2020flavour}%
  \BibitemOpen
  \bibfield  {author} {\bibinfo {author} {\bibfnamefont {Jeong~Min}\
  \bibnamefont {Park}}, \bibinfo {author} {\bibfnamefont {Yuan}\ \bibnamefont
  {Cao}}, \bibinfo {author} {\bibfnamefont {Kenji}\ \bibnamefont {Watanabe}},
  \bibinfo {author} {\bibfnamefont {Takashi}\ \bibnamefont {Taniguchi}}, \ and\
  \bibinfo {author} {\bibfnamefont {Pablo}\ \bibnamefont {Jarillo-Herrero}},\
  }\bibfield  {title} {\enquote {\bibinfo {title} {Flavour hund’s coupling,
  chern gaps and charge diffusivity in moir{\'e} graphene},}\ }\href
  {https://www.nature.com/articles/s41586-021-03366-w} {\bibfield  {journal}
  {\bibinfo  {journal} {Nature}\ }\textbf {\bibinfo {volume} {592}},\ \bibinfo
  {pages} {43--48} (\bibinfo {year} {2021})}\BibitemShut {NoStop}%
\bibitem [{\citenamefont {Saito}\ \emph
  {et~al.}(2021{\natexlab{b}})\citenamefont {Saito}, \citenamefont {Yang},
  \citenamefont {Ge}, \citenamefont {Liu}, \citenamefont {Taniguchi},
  \citenamefont {Watanabe}, \citenamefont {Li}, \citenamefont {Berg},\ and\
  \citenamefont {Young}}]{saito2020isospin}%
  \BibitemOpen
  \bibfield  {author} {\bibinfo {author} {\bibfnamefont {Yu}~\bibnamefont
  {Saito}}, \bibinfo {author} {\bibfnamefont {Fangyuan}\ \bibnamefont {Yang}},
  \bibinfo {author} {\bibfnamefont {Jingyuan}\ \bibnamefont {Ge}}, \bibinfo
  {author} {\bibfnamefont {Xiaoxue}\ \bibnamefont {Liu}}, \bibinfo {author}
  {\bibfnamefont {Takashi}\ \bibnamefont {Taniguchi}}, \bibinfo {author}
  {\bibfnamefont {Kenji}\ \bibnamefont {Watanabe}}, \bibinfo {author}
  {\bibfnamefont {JIA}\ \bibnamefont {Li}}, \bibinfo {author} {\bibfnamefont
  {Erez}\ \bibnamefont {Berg}}, \ and\ \bibinfo {author} {\bibfnamefont
  {Andrea~F}\ \bibnamefont {Young}},\ }\bibfield  {title} {\enquote {\bibinfo
  {title} {Isospin pomeranchuk effect in twisted bilayer graphene},}\ }\href
  {\doibase 10.1038/s41586-021-03409-2} {\bibfield  {journal} {\bibinfo
  {journal} {Nature}\ }\textbf {\bibinfo {volume} {592}},\ \bibinfo {pages}
  {220--224} (\bibinfo {year} {2021}{\natexlab{b}})}\BibitemShut {NoStop}%
\bibitem [{\citenamefont {Rozen}\ \emph {et~al.}(2021)\citenamefont {Rozen},
  \citenamefont {Park}, \citenamefont {Zondiner}, \citenamefont {Cao},
  \citenamefont {Rodan-Legrain}, \citenamefont {Taniguchi}, \citenamefont
  {Watanabe}, \citenamefont {Oreg}, \citenamefont {Stern}, \citenamefont {Berg}
  \emph {et~al.}}]{rozen2020entropic}%
  \BibitemOpen
  \bibfield  {author} {\bibinfo {author} {\bibfnamefont {Asaf}\ \bibnamefont
  {Rozen}}, \bibinfo {author} {\bibfnamefont {Jeong~Min}\ \bibnamefont {Park}},
  \bibinfo {author} {\bibfnamefont {Uri}\ \bibnamefont {Zondiner}}, \bibinfo
  {author} {\bibfnamefont {Yuan}\ \bibnamefont {Cao}}, \bibinfo {author}
  {\bibfnamefont {Daniel}\ \bibnamefont {Rodan-Legrain}}, \bibinfo {author}
  {\bibfnamefont {Takashi}\ \bibnamefont {Taniguchi}}, \bibinfo {author}
  {\bibfnamefont {Kenji}\ \bibnamefont {Watanabe}}, \bibinfo {author}
  {\bibfnamefont {Yuval}\ \bibnamefont {Oreg}}, \bibinfo {author}
  {\bibfnamefont {Ady}\ \bibnamefont {Stern}}, \bibinfo {author} {\bibfnamefont
  {Erez}\ \bibnamefont {Berg}},  \emph {et~al.},\ }\bibfield  {title} {\enquote
  {\bibinfo {title} {Entropic evidence for a pomeranchuk effect in magic-angle
  graphene},}\ }\href {https://www.nature.com/articles/s41586-021-03319-3}
  {\bibfield  {journal} {\bibinfo  {journal} {Nature}\ }\textbf {\bibinfo
  {volume} {592}},\ \bibinfo {pages} {214--219} (\bibinfo {year}
  {2021})}\BibitemShut {NoStop}%
\bibitem [{\citenamefont {Lu}\ \emph {et~al.}(2021)\citenamefont {Lu},
  \citenamefont {Lian}, \citenamefont {Chaudhary}, \citenamefont {Piot},
  \citenamefont {Romagnoli}, \citenamefont {Watanabe}, \citenamefont
  {Taniguchi}, \citenamefont {Poggio}, \citenamefont {MacDonald}, \citenamefont
  {Bernevig},\ and\ \citenamefont {Efetov}}]{lu2020fingerprints}%
  \BibitemOpen
  \bibfield  {author} {\bibinfo {author} {\bibfnamefont {Xiaobo}\ \bibnamefont
  {Lu}}, \bibinfo {author} {\bibfnamefont {Biao}\ \bibnamefont {Lian}},
  \bibinfo {author} {\bibfnamefont {Gaurav}\ \bibnamefont {Chaudhary}},
  \bibinfo {author} {\bibfnamefont {Benjamin~A.}\ \bibnamefont {Piot}},
  \bibinfo {author} {\bibfnamefont {Giulio}\ \bibnamefont {Romagnoli}},
  \bibinfo {author} {\bibfnamefont {Kenji}\ \bibnamefont {Watanabe}}, \bibinfo
  {author} {\bibfnamefont {Takashi}\ \bibnamefont {Taniguchi}}, \bibinfo
  {author} {\bibfnamefont {Martino}\ \bibnamefont {Poggio}}, \bibinfo {author}
  {\bibfnamefont {Allan~H.}\ \bibnamefont {MacDonald}}, \bibinfo {author}
  {\bibfnamefont {B.~Andrei}\ \bibnamefont {Bernevig}}, \ and\ \bibinfo
  {author} {\bibfnamefont {Dmitri~K.}\ \bibnamefont {Efetov}},\ }\bibfield
  {title} {\enquote {\bibinfo {title} {Multiple flat bands and topological
  hofstadter butterfly in twisted bilayer graphene close to the second magic
  angle},}\ }\href {\doibase 10.1073/pnas.2100006118} {\bibfield  {journal}
  {\bibinfo  {journal} {Proceedings of the National Academy of Sciences}\
  }\textbf {\bibinfo {volume} {118}} (\bibinfo {year} {2021}),\
  10.1073/pnas.2100006118}\BibitemShut {NoStop}%
\bibitem [{\citenamefont {Tarnopolsky}\ \emph {et~al.}(2019)\citenamefont
  {Tarnopolsky}, \citenamefont {Kruchkov},\ and\ \citenamefont
  {Vishwanath}}]{tarnopolsky_origin_2019}%
  \BibitemOpen
  \bibfield  {author} {\bibinfo {author} {\bibfnamefont {Grigory}\ \bibnamefont
  {Tarnopolsky}}, \bibinfo {author} {\bibfnamefont {Alex~Jura}\ \bibnamefont
  {Kruchkov}}, \ and\ \bibinfo {author} {\bibfnamefont {Ashvin}\ \bibnamefont
  {Vishwanath}},\ }\bibfield  {title} {\enquote {\bibinfo {title} {Origin of
  {Magic} {Angles} in {Twisted} {Bilayer} {Graphene}},}\ }\href {\doibase
  10.1103/PhysRevLett.122.106405} {\bibfield  {journal} {\bibinfo  {journal}
  {Physical Review Letters}\ }\textbf {\bibinfo {volume} {122}},\ \bibinfo
  {pages} {106405} (\bibinfo {year} {2019})}\BibitemShut {NoStop}%
\bibitem [{\citenamefont {Zou}\ \emph {et~al.}(2018)\citenamefont {Zou},
  \citenamefont {Po}, \citenamefont {Vishwanath},\ and\ \citenamefont
  {Senthil}}]{zou2018}%
  \BibitemOpen
  \bibfield  {author} {\bibinfo {author} {\bibfnamefont {Liujun}\ \bibnamefont
  {Zou}}, \bibinfo {author} {\bibfnamefont {Hoi~Chun}\ \bibnamefont {Po}},
  \bibinfo {author} {\bibfnamefont {Ashvin}\ \bibnamefont {Vishwanath}}, \ and\
  \bibinfo {author} {\bibfnamefont {T.}~\bibnamefont {Senthil}},\ }\bibfield
  {title} {\enquote {\bibinfo {title} {Band structure of twisted bilayer
  graphene: Emergent symmetries, commensurate approximants, and wannier
  obstructions},}\ }\href {\doibase 10.1103/PhysRevB.98.085435} {\bibfield
  {journal} {\bibinfo  {journal} {Phys. Rev. B}\ }\textbf {\bibinfo {volume}
  {98}},\ \bibinfo {pages} {085435} (\bibinfo {year} {2018})}\BibitemShut
  {NoStop}%
\bibitem [{\citenamefont {Liu}\ \emph {et~al.}(2019{\natexlab{a}})\citenamefont
  {Liu}, \citenamefont {Liu},\ and\ \citenamefont {Dai}}]{liu2019pseudo}%
  \BibitemOpen
  \bibfield  {author} {\bibinfo {author} {\bibfnamefont {Jianpeng}\
  \bibnamefont {Liu}}, \bibinfo {author} {\bibfnamefont {Junwei}\ \bibnamefont
  {Liu}}, \ and\ \bibinfo {author} {\bibfnamefont {Xi}~\bibnamefont {Dai}},\
  }\bibfield  {title} {\enquote {\bibinfo {title} {Pseudo landau level
  representation of twisted bilayer graphene: Band topology and implications on
  the correlated insulating phase},}\ }\href
  {https://journals.aps.org/prb/abstract/10.1103/PhysRevB.99.155415} {\bibfield
   {journal} {\bibinfo  {journal} {Physical Review B}\ }\textbf {\bibinfo
  {volume} {99}},\ \bibinfo {pages} {155415} (\bibinfo {year}
  {2019}{\natexlab{a}})}\BibitemShut {NoStop}%
\bibitem [{\citenamefont {Efimkin}\ and\ \citenamefont
  {MacDonald}(2018)}]{Efimkin2018TBG}%
  \BibitemOpen
  \bibfield  {author} {\bibinfo {author} {\bibfnamefont {Dmitry~K.}\
  \bibnamefont {Efimkin}}\ and\ \bibinfo {author} {\bibfnamefont {Allan~H.}\
  \bibnamefont {MacDonald}},\ }\bibfield  {title} {\enquote {\bibinfo {title}
  {Helical network model for twisted bilayer graphene},}\ }\href {\doibase
  10.1103/PhysRevB.98.035404} {\bibfield  {journal} {\bibinfo  {journal} {Phys.
  Rev. B}\ }\textbf {\bibinfo {volume} {98}},\ \bibinfo {pages} {035404}
  (\bibinfo {year} {2018})}\BibitemShut {NoStop}%
\bibitem [{\citenamefont {Kang}\ and\ \citenamefont
  {Vafek}(2018)}]{kang_symmetry_2018}%
  \BibitemOpen
  \bibfield  {author} {\bibinfo {author} {\bibfnamefont {Jian}\ \bibnamefont
  {Kang}}\ and\ \bibinfo {author} {\bibfnamefont {Oskar}\ \bibnamefont
  {Vafek}},\ }\bibfield  {title} {\enquote {\bibinfo {title} {Symmetry,
  {Maximally} {Localized} {Wannier} {States}, and a {Low}-{Energy} {Model} for
  {Twisted} {Bilayer} {Graphene} {Narrow} {Bands}},}\ }\href {\doibase
  10.1103/PhysRevX.8.031088} {\bibfield  {journal} {\bibinfo  {journal} {Phys.
  Rev. X}\ }\textbf {\bibinfo {volume} {8}},\ \bibinfo {pages} {031088}
  (\bibinfo {year} {2018})}\BibitemShut {NoStop}%
\bibitem [{\citenamefont {Song}\ \emph {et~al.}(2019)\citenamefont {Song},
  \citenamefont {Wang}, \citenamefont {Shi}, \citenamefont {Li}, \citenamefont
  {Fang},\ and\ \citenamefont {Bernevig}}]{song_all_2019}%
  \BibitemOpen
  \bibfield  {author} {\bibinfo {author} {\bibfnamefont {Zhida}\ \bibnamefont
  {Song}}, \bibinfo {author} {\bibfnamefont {Zhijun}\ \bibnamefont {Wang}},
  \bibinfo {author} {\bibfnamefont {Wujun}\ \bibnamefont {Shi}}, \bibinfo
  {author} {\bibfnamefont {Gang}\ \bibnamefont {Li}}, \bibinfo {author}
  {\bibfnamefont {Chen}\ \bibnamefont {Fang}}, \ and\ \bibinfo {author}
  {\bibfnamefont {B.~Andrei}\ \bibnamefont {Bernevig}},\ }\bibfield  {title}
  {\enquote {\bibinfo {title} {All {Magic} {Angles} in {Twisted} {Bilayer}
  {Graphene} are {Topological}},}\ }\href {\doibase
  10.1103/PhysRevLett.123.036401} {\bibfield  {journal} {\bibinfo  {journal}
  {Physical Review Letters}\ }\textbf {\bibinfo {volume} {123}},\ \bibinfo
  {pages} {036401} (\bibinfo {year} {2019})}\BibitemShut {NoStop}%
\bibitem [{\citenamefont {Po}\ \emph {et~al.}(2019)\citenamefont {Po},
  \citenamefont {Zou}, \citenamefont {Senthil},\ and\ \citenamefont
  {Vishwanath}}]{po_faithful_2019}%
  \BibitemOpen
  \bibfield  {author} {\bibinfo {author} {\bibfnamefont {Hoi~Chun}\
  \bibnamefont {Po}}, \bibinfo {author} {\bibfnamefont {Liujun}\ \bibnamefont
  {Zou}}, \bibinfo {author} {\bibfnamefont {T.}~\bibnamefont {Senthil}}, \ and\
  \bibinfo {author} {\bibfnamefont {Ashvin}\ \bibnamefont {Vishwanath}},\
  }\bibfield  {title} {\enquote {\bibinfo {title} {Faithful tight-binding
  models and fragile topology of magic-angle bilayer graphene},}\ }\href
  {\doibase 10.1103/PhysRevB.99.195455} {\bibfield  {journal} {\bibinfo
  {journal} {Physical Review B}\ }\textbf {\bibinfo {volume} {99}},\ \bibinfo
  {pages} {195455} (\bibinfo {year} {2019})}\BibitemShut {NoStop}%
\bibitem [{\citenamefont {Hejazi}\ \emph
  {et~al.}(2019{\natexlab{a}})\citenamefont {Hejazi}, \citenamefont {Liu},
  \citenamefont {Shapourian}, \citenamefont {Chen},\ and\ \citenamefont
  {Balents}}]{hejazi_multiple_2019}%
  \BibitemOpen
  \bibfield  {author} {\bibinfo {author} {\bibfnamefont {Kasra}\ \bibnamefont
  {Hejazi}}, \bibinfo {author} {\bibfnamefont {Chunxiao}\ \bibnamefont {Liu}},
  \bibinfo {author} {\bibfnamefont {Hassan}\ \bibnamefont {Shapourian}},
  \bibinfo {author} {\bibfnamefont {Xiao}\ \bibnamefont {Chen}}, \ and\
  \bibinfo {author} {\bibfnamefont {Leon}\ \bibnamefont {Balents}},\ }\bibfield
   {title} {\enquote {\bibinfo {title} {Multiple topological transitions in
  twisted bilayer graphene near the first magic angle},}\ }\href {\doibase
  10.1103/PhysRevB.99.035111} {\bibfield  {journal} {\bibinfo  {journal} {Phys.
  Rev. B}\ }\textbf {\bibinfo {volume} {99}},\ \bibinfo {pages} {035111}
  (\bibinfo {year} {2019}{\natexlab{a}})}\BibitemShut {NoStop}%
\bibitem [{\citenamefont {Padhi}\ \emph {et~al.}(2018)\citenamefont {Padhi},
  \citenamefont {Setty},\ and\ \citenamefont {Phillips}}]{padhi2018doped}%
  \BibitemOpen
  \bibfield  {author} {\bibinfo {author} {\bibfnamefont {Bikash}\ \bibnamefont
  {Padhi}}, \bibinfo {author} {\bibfnamefont {Chandan}\ \bibnamefont {Setty}},
  \ and\ \bibinfo {author} {\bibfnamefont {Philip~W}\ \bibnamefont
  {Phillips}},\ }\bibfield  {title} {\enquote {\bibinfo {title} {Doped twisted
  bilayer graphene near magic angles: proximity to wigner crystallization, not
  mott insulation},}\ }\href@noop {} {\bibfield  {journal} {\bibinfo  {journal}
  {Nano letters}\ }\textbf {\bibinfo {volume} {18}},\ \bibinfo {pages}
  {6175--6180} (\bibinfo {year} {2018})}\BibitemShut {NoStop}%
\bibitem [{\citenamefont {Lian}\ \emph {et~al.}(2020)\citenamefont {Lian},
  \citenamefont {Xie},\ and\ \citenamefont {Bernevig}}]{lian2020}%
  \BibitemOpen
  \bibfield  {author} {\bibinfo {author} {\bibfnamefont {Biao}\ \bibnamefont
  {Lian}}, \bibinfo {author} {\bibfnamefont {Fang}\ \bibnamefont {Xie}}, \ and\
  \bibinfo {author} {\bibfnamefont {B.~Andrei}\ \bibnamefont {Bernevig}},\
  }\bibfield  {title} {\enquote {\bibinfo {title} {Landau level of fragile
  topology},}\ }\href {\doibase 10.1103/PhysRevB.102.041402} {\bibfield
  {journal} {\bibinfo  {journal} {Phys. Rev. B}\ }\textbf {\bibinfo {volume}
  {102}},\ \bibinfo {pages} {041402} (\bibinfo {year} {2020})}\BibitemShut
  {NoStop}%
\bibitem [{\citenamefont {Hejazi}\ \emph
  {et~al.}(2019{\natexlab{b}})\citenamefont {Hejazi}, \citenamefont {Liu},\
  and\ \citenamefont {Balents}}]{hejazi_landau_2019}%
  \BibitemOpen
  \bibfield  {author} {\bibinfo {author} {\bibfnamefont {Kasra}\ \bibnamefont
  {Hejazi}}, \bibinfo {author} {\bibfnamefont {Chunxiao}\ \bibnamefont {Liu}},
  \ and\ \bibinfo {author} {\bibfnamefont {Leon}\ \bibnamefont {Balents}},\
  }\bibfield  {title} {\enquote {\bibinfo {title} {Landau levels in twisted
  bilayer graphene and semiclassical orbits},}\ }\href {\doibase
  10.1103/physrevb.100.035115} {\bibfield  {journal} {\bibinfo  {journal}
  {Physical Review B}\ }\textbf {\bibinfo {volume} {100}} (\bibinfo {year}
  {2019}{\natexlab{b}}),\ 10.1103/physrevb.100.035115}\BibitemShut {NoStop}%
\bibitem [{\citenamefont {Padhi}\ \emph {et~al.}(2020)\citenamefont {Padhi},
  \citenamefont {Tiwari}, \citenamefont {Neupert},\ and\ \citenamefont
  {Ryu}}]{padhi2020transport}%
  \BibitemOpen
  \bibfield  {author} {\bibinfo {author} {\bibfnamefont {Bikash}\ \bibnamefont
  {Padhi}}, \bibinfo {author} {\bibfnamefont {Apoorv}\ \bibnamefont {Tiwari}},
  \bibinfo {author} {\bibfnamefont {Titus}\ \bibnamefont {Neupert}}, \ and\
  \bibinfo {author} {\bibfnamefont {Shinsei}\ \bibnamefont {Ryu}},\ }\bibfield
  {title} {\enquote {\bibinfo {title} {Transport across twist angle domains in
  moiré graphene},}\ }\href@noop {} {\  (\bibinfo {year} {2020})},\ \Eprint
  {http://arxiv.org/abs/2005.02406} {arXiv:2005.02406 [cond-mat.mes-hall]}
  \BibitemShut {NoStop}%
\bibitem [{\citenamefont {Xu}\ and\ \citenamefont
  {Balents}(2018)}]{xu2018topological}%
  \BibitemOpen
  \bibfield  {author} {\bibinfo {author} {\bibfnamefont {Cenke}\ \bibnamefont
  {Xu}}\ and\ \bibinfo {author} {\bibfnamefont {Leon}\ \bibnamefont
  {Balents}},\ }\bibfield  {title} {\enquote {\bibinfo {title} {Topological
  superconductivity in twisted multilayer graphene},}\ }\href
  {https://journals.aps.org/prl/abstract/10.1103/PhysRevLett.121.087001}
  {\bibfield  {journal} {\bibinfo  {journal} {Physical review letters}\
  }\textbf {\bibinfo {volume} {121}},\ \bibinfo {pages} {087001} (\bibinfo
  {year} {2018})}\BibitemShut {NoStop}%
\bibitem [{\citenamefont {Koshino}\ \emph {et~al.}(2018)\citenamefont
  {Koshino}, \citenamefont {Yuan}, \citenamefont {Koretsune}, \citenamefont
  {Ochi}, \citenamefont {Kuroki},\ and\ \citenamefont
  {Fu}}]{koshino_maximally_2018}%
  \BibitemOpen
  \bibfield  {author} {\bibinfo {author} {\bibfnamefont {Mikito}\ \bibnamefont
  {Koshino}}, \bibinfo {author} {\bibfnamefont {Noah F.~Q.}\ \bibnamefont
  {Yuan}}, \bibinfo {author} {\bibfnamefont {Takashi}\ \bibnamefont
  {Koretsune}}, \bibinfo {author} {\bibfnamefont {Masayuki}\ \bibnamefont
  {Ochi}}, \bibinfo {author} {\bibfnamefont {Kazuhiko}\ \bibnamefont {Kuroki}},
  \ and\ \bibinfo {author} {\bibfnamefont {Liang}\ \bibnamefont {Fu}},\
  }\bibfield  {title} {\enquote {\bibinfo {title} {Maximally localized wannier
  orbitals and the extended hubbard model for twisted bilayer graphene},}\
  }\href {\doibase 10.1103/PhysRevX.8.031087} {\bibfield  {journal} {\bibinfo
  {journal} {Phys. Rev. X}\ }\textbf {\bibinfo {volume} {8}},\ \bibinfo {pages}
  {031087} (\bibinfo {year} {2018})}\BibitemShut {NoStop}%
\bibitem [{\citenamefont {Ochi}\ \emph {et~al.}(2018)\citenamefont {Ochi},
  \citenamefont {Koshino},\ and\ \citenamefont {Kuroki}}]{ochi_possible_2018}%
  \BibitemOpen
  \bibfield  {author} {\bibinfo {author} {\bibfnamefont {Masayuki}\
  \bibnamefont {Ochi}}, \bibinfo {author} {\bibfnamefont {Mikito}\ \bibnamefont
  {Koshino}}, \ and\ \bibinfo {author} {\bibfnamefont {Kazuhiko}\ \bibnamefont
  {Kuroki}},\ }\bibfield  {title} {\enquote {\bibinfo {title} {Possible
  correlated insulating states in magic-angle twisted bilayer graphene under
  strongly competing interactions},}\ }\href {\doibase
  10.1103/PhysRevB.98.081102} {\bibfield  {journal} {\bibinfo  {journal} {Phys.
  Rev. B}\ }\textbf {\bibinfo {volume} {98}},\ \bibinfo {pages} {081102}
  (\bibinfo {year} {2018})}\BibitemShut {NoStop}%
\bibitem [{\citenamefont {Xu}\ \emph {et~al.}(2018)\citenamefont {Xu},
  \citenamefont {Law},\ and\ \citenamefont {Lee}}]{xux2018}%
  \BibitemOpen
  \bibfield  {author} {\bibinfo {author} {\bibfnamefont {Xiao~Yan}\
  \bibnamefont {Xu}}, \bibinfo {author} {\bibfnamefont {K.~T.}\ \bibnamefont
  {Law}}, \ and\ \bibinfo {author} {\bibfnamefont {Patrick~A.}\ \bibnamefont
  {Lee}},\ }\bibfield  {title} {\enquote {\bibinfo {title} {Kekul\'e valence
  bond order in an extended hubbard model on the honeycomb lattice with
  possible applications to twisted bilayer graphene},}\ }\href {\doibase
  10.1103/PhysRevB.98.121406} {\bibfield  {journal} {\bibinfo  {journal} {Phys.
  Rev. B}\ }\textbf {\bibinfo {volume} {98}},\ \bibinfo {pages} {121406}
  (\bibinfo {year} {2018})}\BibitemShut {NoStop}%
\bibitem [{\citenamefont {Guinea}\ and\ \citenamefont
  {Walet}(2018)}]{guinea2018}%
  \BibitemOpen
  \bibfield  {author} {\bibinfo {author} {\bibfnamefont {Francisco}\
  \bibnamefont {Guinea}}\ and\ \bibinfo {author} {\bibfnamefont {Niels~R.}\
  \bibnamefont {Walet}},\ }\bibfield  {title} {\enquote {\bibinfo {title}
  {Electrostatic effects, band distortions, and superconductivity in twisted
  graphene bilayers},}\ }\href {\doibase 10.1073/pnas.1810947115} {\bibfield
  {journal} {\bibinfo  {journal} {Proceedings of the National Academy of
  Sciences}\ }\textbf {\bibinfo {volume} {115}},\ \bibinfo {pages}
  {13174--13179} (\bibinfo {year} {2018})}\BibitemShut {NoStop}%
\bibitem [{\citenamefont {Venderbos}\ and\ \citenamefont
  {Fernandes}(2018)}]{venderbos2018}%
  \BibitemOpen
  \bibfield  {author} {\bibinfo {author} {\bibfnamefont {J\"orn W.~F.}\
  \bibnamefont {Venderbos}}\ and\ \bibinfo {author} {\bibfnamefont {Rafael~M.}\
  \bibnamefont {Fernandes}},\ }\bibfield  {title} {\enquote {\bibinfo {title}
  {Correlations and electronic order in a two-orbital honeycomb lattice model
  for twisted bilayer graphene},}\ }\href {\doibase 10.1103/PhysRevB.98.245103}
  {\bibfield  {journal} {\bibinfo  {journal} {Phys. Rev. B}\ }\textbf {\bibinfo
  {volume} {98}},\ \bibinfo {pages} {245103} (\bibinfo {year}
  {2018})}\BibitemShut {NoStop}%
\bibitem [{\citenamefont {{You}}\ and\ \citenamefont
  {{Vishwanath}}(2019)}]{you2019}%
  \BibitemOpen
  \bibfield  {author} {\bibinfo {author} {\bibfnamefont {Y.-Z.}\ \bibnamefont
  {{You}}}\ and\ \bibinfo {author} {\bibfnamefont {A.}~\bibnamefont
  {{Vishwanath}}},\ }\bibfield  {title} {\enquote {\bibinfo {title}
  {{Superconductivity from Valley Fluctuations and Approximate SO(4) Symmetry
  in a Weak Coupling Theory of Twisted Bilayer Graphene}},}\ }\href {\doibase
  10.1038/s41535-019-0153-4} {\bibfield  {journal} {\bibinfo  {journal} {npj
  Quantum Materials}\ }\textbf {\bibinfo {volume} {4}},\ \bibinfo {pages} {16}
  (\bibinfo {year} {2019})}\BibitemShut {NoStop}%
\bibitem [{\citenamefont {Wu}\ and\ \citenamefont
  {Das~Sarma}(2020)}]{wu_collective_2020}%
  \BibitemOpen
  \bibfield  {author} {\bibinfo {author} {\bibfnamefont {Fengcheng}\
  \bibnamefont {Wu}}\ and\ \bibinfo {author} {\bibfnamefont {Sankar}\
  \bibnamefont {Das~Sarma}},\ }\bibfield  {title} {\enquote {\bibinfo {title}
  {Collective excitations of quantum anomalous hall ferromagnets in twisted
  bilayer graphene},}\ }\href {\doibase 10.1103/physrevlett.124.046403}
  {\bibfield  {journal} {\bibinfo  {journal} {Physical Review Letters}\
  }\textbf {\bibinfo {volume} {124}} (\bibinfo {year} {2020}),\
  10.1103/physrevlett.124.046403}\BibitemShut {NoStop}%
\bibitem [{\citenamefont {Lian}\ \emph {et~al.}(2019)\citenamefont {Lian},
  \citenamefont {Wang},\ and\ \citenamefont {Bernevig}}]{Lian2019TBG}%
  \BibitemOpen
  \bibfield  {author} {\bibinfo {author} {\bibfnamefont {Biao}\ \bibnamefont
  {Lian}}, \bibinfo {author} {\bibfnamefont {Zhijun}\ \bibnamefont {Wang}}, \
  and\ \bibinfo {author} {\bibfnamefont {B.~Andrei}\ \bibnamefont {Bernevig}},\
  }\bibfield  {title} {\enquote {\bibinfo {title} {Twisted bilayer graphene: A
  phonon-driven superconductor},}\ }\href {\doibase
  10.1103/PhysRevLett.122.257002} {\bibfield  {journal} {\bibinfo  {journal}
  {Phys. Rev. Lett.}\ }\textbf {\bibinfo {volume} {122}},\ \bibinfo {pages}
  {257002} (\bibinfo {year} {2019})}\BibitemShut {NoStop}%
\bibitem [{\citenamefont {Wu}\ \emph {et~al.}(2018)\citenamefont {Wu},
  \citenamefont {MacDonald},\ and\ \citenamefont {Martin}}]{Wu2018TBG-BCS}%
  \BibitemOpen
  \bibfield  {author} {\bibinfo {author} {\bibfnamefont {Fengcheng}\
  \bibnamefont {Wu}}, \bibinfo {author} {\bibfnamefont {A.~H.}\ \bibnamefont
  {MacDonald}}, \ and\ \bibinfo {author} {\bibfnamefont {Ivar}\ \bibnamefont
  {Martin}},\ }\bibfield  {title} {\enquote {\bibinfo {title} {Theory of
  phonon-mediated superconductivity in twisted bilayer graphene},}\ }\href
  {\doibase 10.1103/PhysRevLett.121.257001} {\bibfield  {journal} {\bibinfo
  {journal} {Phys. Rev. Lett.}\ }\textbf {\bibinfo {volume} {121}},\ \bibinfo
  {pages} {257001} (\bibinfo {year} {2018})}\BibitemShut {NoStop}%
\bibitem [{\citenamefont {Isobe}\ \emph {et~al.}(2018)\citenamefont {Isobe},
  \citenamefont {Yuan},\ and\ \citenamefont {Fu}}]{isobe2018unconventional}%
  \BibitemOpen
  \bibfield  {author} {\bibinfo {author} {\bibfnamefont {Hiroki}\ \bibnamefont
  {Isobe}}, \bibinfo {author} {\bibfnamefont {Noah F.~Q.}\ \bibnamefont
  {Yuan}}, \ and\ \bibinfo {author} {\bibfnamefont {Liang}\ \bibnamefont
  {Fu}},\ }\bibfield  {title} {\enquote {\bibinfo {title} {Unconventional
  superconductivity and density waves in twisted bilayer graphene},}\ }\href
  {\doibase 10.1103/PhysRevX.8.041041} {\bibfield  {journal} {\bibinfo
  {journal} {Phys. Rev. X}\ }\textbf {\bibinfo {volume} {8}},\ \bibinfo {pages}
  {041041} (\bibinfo {year} {2018})}\BibitemShut {NoStop}%
\bibitem [{\citenamefont {Liu}\ \emph {et~al.}(2018)\citenamefont {Liu},
  \citenamefont {Zhang}, \citenamefont {Chen},\ and\ \citenamefont
  {Yang}}]{liu2018chiral}%
  \BibitemOpen
  \bibfield  {author} {\bibinfo {author} {\bibfnamefont {Cheng-Cheng}\
  \bibnamefont {Liu}}, \bibinfo {author} {\bibfnamefont {Li-Da}\ \bibnamefont
  {Zhang}}, \bibinfo {author} {\bibfnamefont {Wei-Qiang}\ \bibnamefont {Chen}},
  \ and\ \bibinfo {author} {\bibfnamefont {Fan}\ \bibnamefont {Yang}},\
  }\bibfield  {title} {\enquote {\bibinfo {title} {Chiral spin density wave and
  d+ i d superconductivity in the magic-angle-twisted bilayer graphene},}\
  }\href {https://journals.aps.org/prl/abstract/10.1103/PhysRevLett.121.217001}
  {\bibfield  {journal} {\bibinfo  {journal} {Physical review letters}\
  }\textbf {\bibinfo {volume} {121}},\ \bibinfo {pages} {217001} (\bibinfo
  {year} {2018})}\BibitemShut {NoStop}%
\bibitem [{\citenamefont {Bultinck}\ \emph
  {et~al.}(2020{\natexlab{b}})\citenamefont {Bultinck}, \citenamefont
  {Chatterjee},\ and\ \citenamefont {Zaletel}}]{bultinck2020}%
  \BibitemOpen
  \bibfield  {author} {\bibinfo {author} {\bibfnamefont {Nick}\ \bibnamefont
  {Bultinck}}, \bibinfo {author} {\bibfnamefont {Shubhayu}\ \bibnamefont
  {Chatterjee}}, \ and\ \bibinfo {author} {\bibfnamefont {Michael~P.}\
  \bibnamefont {Zaletel}},\ }\bibfield  {title} {\enquote {\bibinfo {title}
  {Mechanism for anomalous hall ferromagnetism in twisted bilayer graphene},}\
  }\href {\doibase 10.1103/PhysRevLett.124.166601} {\bibfield  {journal}
  {\bibinfo  {journal} {Phys. Rev. Lett.}\ }\textbf {\bibinfo {volume} {124}},\
  \bibinfo {pages} {166601} (\bibinfo {year} {2020}{\natexlab{b}})}\BibitemShut
  {NoStop}%
\bibitem [{\citenamefont {Zhang}\ \emph {et~al.}(2019)\citenamefont {Zhang},
  \citenamefont {Mao}, \citenamefont {Cao}, \citenamefont {Jarillo-Herrero},\
  and\ \citenamefont {Senthil}}]{zhang2019nearly}%
  \BibitemOpen
  \bibfield  {author} {\bibinfo {author} {\bibfnamefont {Ya-Hui}\ \bibnamefont
  {Zhang}}, \bibinfo {author} {\bibfnamefont {Dan}\ \bibnamefont {Mao}},
  \bibinfo {author} {\bibfnamefont {Yuan}\ \bibnamefont {Cao}}, \bibinfo
  {author} {\bibfnamefont {Pablo}\ \bibnamefont {Jarillo-Herrero}}, \ and\
  \bibinfo {author} {\bibfnamefont {T}~\bibnamefont {Senthil}},\ }\bibfield
  {title} {\enquote {\bibinfo {title} {Nearly flat chern bands in moir{\'e}
  superlattices},}\ }\href
  {https://journals.aps.org/prb/abstract/10.1103/PhysRevB.99.075127} {\bibfield
   {journal} {\bibinfo  {journal} {Physical Review B}\ }\textbf {\bibinfo
  {volume} {99}},\ \bibinfo {pages} {075127} (\bibinfo {year}
  {2019})}\BibitemShut {NoStop}%
\bibitem [{\citenamefont {Liu}\ \emph {et~al.}(2019{\natexlab{b}})\citenamefont
  {Liu}, \citenamefont {Ma}, \citenamefont {Gao},\ and\ \citenamefont
  {Dai}}]{liu2019quantum}%
  \BibitemOpen
  \bibfield  {author} {\bibinfo {author} {\bibfnamefont {Jianpeng}\
  \bibnamefont {Liu}}, \bibinfo {author} {\bibfnamefont {Zhen}\ \bibnamefont
  {Ma}}, \bibinfo {author} {\bibfnamefont {Jinhua}\ \bibnamefont {Gao}}, \ and\
  \bibinfo {author} {\bibfnamefont {Xi}~\bibnamefont {Dai}},\ }\bibfield
  {title} {\enquote {\bibinfo {title} {Quantum valley hall effect, orbital
  magnetism, and anomalous hall effect in twisted multilayer graphene
  systems},}\ }\href
  {https://journals.aps.org/prx/abstract/10.1103/PhysRevX.9.031021} {\bibfield
  {journal} {\bibinfo  {journal} {Physical Review X}\ }\textbf {\bibinfo
  {volume} {9}},\ \bibinfo {pages} {031021} (\bibinfo {year}
  {2019}{\natexlab{b}})}\BibitemShut {NoStop}%
\bibitem [{\citenamefont {Wu}\ \emph {et~al.}(2019)\citenamefont {Wu},
  \citenamefont {Jian},\ and\ \citenamefont {Xu}}]{wux2018b}%
  \BibitemOpen
  \bibfield  {author} {\bibinfo {author} {\bibfnamefont {Xiao-Chuan}\
  \bibnamefont {Wu}}, \bibinfo {author} {\bibfnamefont {Chao-Ming}\
  \bibnamefont {Jian}}, \ and\ \bibinfo {author} {\bibfnamefont {Cenke}\
  \bibnamefont {Xu}},\ }\bibfield  {title} {\enquote {\bibinfo {title}
  {Coupled-wire description of the correlated physics in twisted bilayer
  graphene},}\ }\href {\doibase 10.1103/physrevb.99.161405} {\bibfield
  {journal} {\bibinfo  {journal} {Physical Review B}\ }\textbf {\bibinfo
  {volume} {99}} (\bibinfo {year} {2019}),\
  10.1103/physrevb.99.161405}\BibitemShut {NoStop}%
\bibitem [{\citenamefont {Thomson}\ \emph {et~al.}(2018)\citenamefont
  {Thomson}, \citenamefont {Chatterjee}, \citenamefont {Sachdev},\ and\
  \citenamefont {Scheurer}}]{thomson2018triangular}%
  \BibitemOpen
  \bibfield  {author} {\bibinfo {author} {\bibfnamefont {Alex}\ \bibnamefont
  {Thomson}}, \bibinfo {author} {\bibfnamefont {Shubhayu}\ \bibnamefont
  {Chatterjee}}, \bibinfo {author} {\bibfnamefont {Subir}\ \bibnamefont
  {Sachdev}}, \ and\ \bibinfo {author} {\bibfnamefont {Mathias~S.}\
  \bibnamefont {Scheurer}},\ }\bibfield  {title} {\enquote {\bibinfo {title}
  {Triangular antiferromagnetism on the honeycomb lattice of twisted bilayer
  graphene},}\ }\href {\doibase 10.1103/physrevb.98.075109} {\bibfield
  {journal} {\bibinfo  {journal} {Physical Review B}\ }\textbf {\bibinfo
  {volume} {98}} (\bibinfo {year} {2018}),\
  10.1103/physrevb.98.075109}\BibitemShut {NoStop}%
\bibitem [{\citenamefont {Dodaro}\ \emph {et~al.}(2018)\citenamefont {Dodaro},
  \citenamefont {Kivelson}, \citenamefont {Schattner}, \citenamefont {Sun},\
  and\ \citenamefont {Wang}}]{dodaro2018phases}%
  \BibitemOpen
  \bibfield  {author} {\bibinfo {author} {\bibfnamefont {John~F}\ \bibnamefont
  {Dodaro}}, \bibinfo {author} {\bibfnamefont {Steven~A}\ \bibnamefont
  {Kivelson}}, \bibinfo {author} {\bibfnamefont {Yoni}\ \bibnamefont
  {Schattner}}, \bibinfo {author} {\bibfnamefont {Xiao-Qi}\ \bibnamefont
  {Sun}}, \ and\ \bibinfo {author} {\bibfnamefont {Chao}\ \bibnamefont
  {Wang}},\ }\bibfield  {title} {\enquote {\bibinfo {title} {Phases of a
  phenomenological model of twisted bilayer graphene},}\ }\href
  {https://journals.aps.org/prb/abstract/10.1103/PhysRevB.98.075154} {\bibfield
   {journal} {\bibinfo  {journal} {Physical Review B}\ }\textbf {\bibinfo
  {volume} {98}},\ \bibinfo {pages} {075154} (\bibinfo {year}
  {2018})}\BibitemShut {NoStop}%
\bibitem [{\citenamefont {Gonzalez}\ and\ \citenamefont
  {Stauber}(2019)}]{gonzalez2019kohn}%
  \BibitemOpen
  \bibfield  {author} {\bibinfo {author} {\bibfnamefont {Jose}\ \bibnamefont
  {Gonzalez}}\ and\ \bibinfo {author} {\bibfnamefont {Tobias}\ \bibnamefont
  {Stauber}},\ }\bibfield  {title} {\enquote {\bibinfo {title} {Kohn-luttinger
  superconductivity in twisted bilayer graphene},}\ }\href
  {https://journals.aps.org/prl/abstract/10.1103/PhysRevLett.122.026801}
  {\bibfield  {journal} {\bibinfo  {journal} {Physical review letters}\
  }\textbf {\bibinfo {volume} {122}},\ \bibinfo {pages} {026801} (\bibinfo
  {year} {2019})}\BibitemShut {NoStop}%
\bibitem [{\citenamefont {Yuan}\ and\ \citenamefont
  {Fu}(2018)}]{yuan2018model}%
  \BibitemOpen
  \bibfield  {author} {\bibinfo {author} {\bibfnamefont {Noah~FQ}\ \bibnamefont
  {Yuan}}\ and\ \bibinfo {author} {\bibfnamefont {Liang}\ \bibnamefont {Fu}},\
  }\bibfield  {title} {\enquote {\bibinfo {title} {Model for the
  metal-insulator transition in graphene superlattices and beyond},}\ }\href
  {https://journals.aps.org/prb/abstract/10.1103/PhysRevB.98.045103} {\bibfield
   {journal} {\bibinfo  {journal} {Physical Review B}\ }\textbf {\bibinfo
  {volume} {98}},\ \bibinfo {pages} {045103} (\bibinfo {year}
  {2018})}\BibitemShut {NoStop}%
\bibitem [{\citenamefont {Seo}\ \emph {et~al.}(2019)\citenamefont {Seo},
  \citenamefont {Kotov},\ and\ \citenamefont {Uchoa}}]{seo_ferro_2019}%
  \BibitemOpen
  \bibfield  {author} {\bibinfo {author} {\bibfnamefont {Kangjun}\ \bibnamefont
  {Seo}}, \bibinfo {author} {\bibfnamefont {Valeri~N.}\ \bibnamefont {Kotov}},
  \ and\ \bibinfo {author} {\bibfnamefont {Bruno}\ \bibnamefont {Uchoa}},\
  }\bibfield  {title} {\enquote {\bibinfo {title} {Ferromagnetic mott state in
  twisted graphene bilayers at the magic angle},}\ }\href {\doibase
  10.1103/PhysRevLett.122.246402} {\bibfield  {journal} {\bibinfo  {journal}
  {Phys. Rev. Lett.}\ }\textbf {\bibinfo {volume} {122}},\ \bibinfo {pages}
  {246402} (\bibinfo {year} {2019})}\BibitemShut {NoStop}%
\bibitem [{\citenamefont {Hejazi}\ \emph {et~al.}(2021)\citenamefont {Hejazi},
  \citenamefont {Chen},\ and\ \citenamefont {Balents}}]{hejazi2020hybrid}%
  \BibitemOpen
  \bibfield  {author} {\bibinfo {author} {\bibfnamefont {Kasra}\ \bibnamefont
  {Hejazi}}, \bibinfo {author} {\bibfnamefont {Xiao}\ \bibnamefont {Chen}}, \
  and\ \bibinfo {author} {\bibfnamefont {Leon}\ \bibnamefont {Balents}},\
  }\bibfield  {title} {\enquote {\bibinfo {title} {Hybrid wannier chern bands
  in magic angle twisted bilayer graphene and the quantized anomalous hall
  effect},}\ }\href {\doibase 10.1103/PhysRevResearch.3.013242} {\bibfield
  {journal} {\bibinfo  {journal} {Phys. Rev. Research}\ }\textbf {\bibinfo
  {volume} {3}},\ \bibinfo {pages} {013242} (\bibinfo {year}
  {2021})}\BibitemShut {NoStop}%
\bibitem [{\citenamefont {Khalaf}\ \emph {et~al.}(2021)\citenamefont {Khalaf},
  \citenamefont {Chatterjee}, \citenamefont {Bultinck}, \citenamefont
  {Zaletel},\ and\ \citenamefont {Vishwanath}}]{khalaf_charged_2020}%
  \BibitemOpen
  \bibfield  {author} {\bibinfo {author} {\bibfnamefont {Eslam}\ \bibnamefont
  {Khalaf}}, \bibinfo {author} {\bibfnamefont {Shubhayu}\ \bibnamefont
  {Chatterjee}}, \bibinfo {author} {\bibfnamefont {Nick}\ \bibnamefont
  {Bultinck}}, \bibinfo {author} {\bibfnamefont {Michael~P}\ \bibnamefont
  {Zaletel}}, \ and\ \bibinfo {author} {\bibfnamefont {Ashvin}\ \bibnamefont
  {Vishwanath}},\ }\bibfield  {title} {\enquote {\bibinfo {title} {Charged
  skyrmions and topological origin of superconductivity in magic-angle
  graphene},}\ }\href {https://www.science.org/doi/10.1126/sciadv.abf5299}
  {\bibfield  {journal} {\bibinfo  {journal} {Science advances}\ }\textbf
  {\bibinfo {volume} {7}},\ \bibinfo {pages} {eabf5299} (\bibinfo {year}
  {2021})}\BibitemShut {NoStop}%
\bibitem [{\citenamefont {Po}\ \emph {et~al.}(2018{\natexlab{a}})\citenamefont
  {Po}, \citenamefont {Zou}, \citenamefont {Vishwanath},\ and\ \citenamefont
  {Senthil}}]{po_origin_2018}%
  \BibitemOpen
  \bibfield  {author} {\bibinfo {author} {\bibfnamefont {Hoi~Chun}\
  \bibnamefont {Po}}, \bibinfo {author} {\bibfnamefont {Liujun}\ \bibnamefont
  {Zou}}, \bibinfo {author} {\bibfnamefont {Ashvin}\ \bibnamefont
  {Vishwanath}}, \ and\ \bibinfo {author} {\bibfnamefont {T.}~\bibnamefont
  {Senthil}},\ }\bibfield  {title} {\enquote {\bibinfo {title} {Origin of
  {Mott} {Insulating} {Behavior} and {Superconductivity} in {Twisted} {Bilayer}
  {Graphene}},}\ }\href {\doibase 10.1103/PhysRevX.8.031089} {\bibfield
  {journal} {\bibinfo  {journal} {Physical Review X}\ }\textbf {\bibinfo
  {volume} {8}},\ \bibinfo {pages} {031089} (\bibinfo {year}
  {2018}{\natexlab{a}})}\BibitemShut {NoStop}%
\bibitem [{\citenamefont {Xie}\ \emph {et~al.}(2020)\citenamefont {Xie},
  \citenamefont {Song}, \citenamefont {Lian},\ and\ \citenamefont
  {Bernevig}}]{xie_superfluid_2020}%
  \BibitemOpen
  \bibfield  {author} {\bibinfo {author} {\bibfnamefont {Fang}\ \bibnamefont
  {Xie}}, \bibinfo {author} {\bibfnamefont {Zhida}\ \bibnamefont {Song}},
  \bibinfo {author} {\bibfnamefont {Biao}\ \bibnamefont {Lian}}, \ and\
  \bibinfo {author} {\bibfnamefont {B.~Andrei}\ \bibnamefont {Bernevig}},\
  }\bibfield  {title} {\enquote {\bibinfo {title} {Topology-bounded superfluid
  weight in twisted bilayer graphene},}\ }\href {\doibase
  10.1103/PhysRevLett.124.167002} {\bibfield  {journal} {\bibinfo  {journal}
  {Phys. Rev. Lett.}\ }\textbf {\bibinfo {volume} {124}},\ \bibinfo {pages}
  {167002} (\bibinfo {year} {2020})}\BibitemShut {NoStop}%
\bibitem [{\citenamefont {Julku}\ \emph {et~al.}(2020)\citenamefont {Julku},
  \citenamefont {Peltonen}, \citenamefont {Liang}, \citenamefont {Heikkilä},\
  and\ \citenamefont {Törmä}}]{julku_superfluid_2020}%
  \BibitemOpen
  \bibfield  {author} {\bibinfo {author} {\bibfnamefont {A.}~\bibnamefont
  {Julku}}, \bibinfo {author} {\bibfnamefont {T.~J.}\ \bibnamefont {Peltonen}},
  \bibinfo {author} {\bibfnamefont {L.}~\bibnamefont {Liang}}, \bibinfo
  {author} {\bibfnamefont {T.~T.}\ \bibnamefont {Heikkilä}}, \ and\ \bibinfo
  {author} {\bibfnamefont {P.}~\bibnamefont {Törmä}},\ }\bibfield  {title}
  {\enquote {\bibinfo {title} {Superfluid weight and
  berezinskii-kosterlitz-thouless transition temperature of twisted bilayer
  graphene},}\ }\href {\doibase 10.1103/physrevb.101.060505} {\bibfield
  {journal} {\bibinfo  {journal} {Physical Review B}\ }\textbf {\bibinfo
  {volume} {101}} (\bibinfo {year} {2020}),\
  10.1103/physrevb.101.060505}\BibitemShut {NoStop}%
\bibitem [{\citenamefont {Hu}\ \emph {et~al.}(2019)\citenamefont {Hu},
  \citenamefont {Hyart}, \citenamefont {Pikulin},\ and\ \citenamefont
  {Rossi}}]{hu2019_superfluid}%
  \BibitemOpen
  \bibfield  {author} {\bibinfo {author} {\bibfnamefont {Xiang}\ \bibnamefont
  {Hu}}, \bibinfo {author} {\bibfnamefont {Timo}\ \bibnamefont {Hyart}},
  \bibinfo {author} {\bibfnamefont {Dmitry~I.}\ \bibnamefont {Pikulin}}, \ and\
  \bibinfo {author} {\bibfnamefont {Enrico}\ \bibnamefont {Rossi}},\ }\bibfield
   {title} {\enquote {\bibinfo {title} {Geometric and conventional contribution
  to the superfluid weight in twisted bilayer graphene},}\ }\href {\doibase
  10.1103/PhysRevLett.123.237002} {\bibfield  {journal} {\bibinfo  {journal}
  {Phys. Rev. Lett.}\ }\textbf {\bibinfo {volume} {123}},\ \bibinfo {pages}
  {237002} (\bibinfo {year} {2019})}\BibitemShut {NoStop}%
\bibitem [{\citenamefont {Kang}\ and\ \citenamefont
  {Vafek}(2020)}]{kang_nonabelian_2020}%
  \BibitemOpen
  \bibfield  {author} {\bibinfo {author} {\bibfnamefont {Jian}\ \bibnamefont
  {Kang}}\ and\ \bibinfo {author} {\bibfnamefont {Oskar}\ \bibnamefont
  {Vafek}},\ }\bibfield  {title} {\enquote {\bibinfo {title} {Non-abelian dirac
  node braiding and near-degeneracy of correlated phases at odd integer filling
  in magic-angle twisted bilayer graphene},}\ }\href {\doibase
  10.1103/PhysRevB.102.035161} {\bibfield  {journal} {\bibinfo  {journal}
  {Phys. Rev. B}\ }\textbf {\bibinfo {volume} {102}},\ \bibinfo {pages}
  {035161} (\bibinfo {year} {2020})}\BibitemShut {NoStop}%
\bibitem [{\citenamefont {Soejima}\ \emph {et~al.}(2020)\citenamefont
  {Soejima}, \citenamefont {Parker}, \citenamefont {Bultinck}, \citenamefont
  {Hauschild},\ and\ \citenamefont {Zaletel}}]{soejima2020efficient}%
  \BibitemOpen
  \bibfield  {author} {\bibinfo {author} {\bibfnamefont {Tomohiro}\
  \bibnamefont {Soejima}}, \bibinfo {author} {\bibfnamefont {Daniel~E.}\
  \bibnamefont {Parker}}, \bibinfo {author} {\bibfnamefont {Nick}\ \bibnamefont
  {Bultinck}}, \bibinfo {author} {\bibfnamefont {Johannes}\ \bibnamefont
  {Hauschild}}, \ and\ \bibinfo {author} {\bibfnamefont {Michael~P.}\
  \bibnamefont {Zaletel}},\ }\bibfield  {title} {\enquote {\bibinfo {title}
  {Efficient simulation of moir\'e materials using the density matrix
  renormalization group},}\ }\href {\doibase 10.1103/PhysRevB.102.205111}
  {\bibfield  {journal} {\bibinfo  {journal} {Phys. Rev. B}\ }\textbf {\bibinfo
  {volume} {102}},\ \bibinfo {pages} {205111} (\bibinfo {year}
  {2020})}\BibitemShut {NoStop}%
\bibitem [{\citenamefont {Pixley}\ and\ \citenamefont
  {Andrei}(2019)}]{pixley2019}%
  \BibitemOpen
  \bibfield  {author} {\bibinfo {author} {\bibfnamefont {Jed~H.}\ \bibnamefont
  {Pixley}}\ and\ \bibinfo {author} {\bibfnamefont {Eva~Y.}\ \bibnamefont
  {Andrei}},\ }\bibfield  {title} {\enquote {\bibinfo {title} {Ferromagnetism
  in magic-angle graphene},}\ }\href {\doibase 10.1126/science.aay3409}
  {\bibfield  {journal} {\bibinfo  {journal} {Science}\ }\textbf {\bibinfo
  {volume} {365}},\ \bibinfo {pages} {543--543} (\bibinfo {year}
  {2019})}\BibitemShut {NoStop}%
\bibitem [{\citenamefont {K\"onig}\ \emph {et~al.}(2020)\citenamefont
  {K\"onig}, \citenamefont {Coleman},\ and\ \citenamefont
  {Tsvelik}}]{knig2020spin}%
  \BibitemOpen
  \bibfield  {author} {\bibinfo {author} {\bibfnamefont {E.~J.}\ \bibnamefont
  {K\"onig}}, \bibinfo {author} {\bibfnamefont {Piers}\ \bibnamefont
  {Coleman}}, \ and\ \bibinfo {author} {\bibfnamefont {A.~M.}\ \bibnamefont
  {Tsvelik}},\ }\bibfield  {title} {\enquote {\bibinfo {title} {Spin
  magnetometry as a probe of stripe superconductivity in twisted bilayer
  graphene},}\ }\href {\doibase 10.1103/PhysRevB.102.104514} {\bibfield
  {journal} {\bibinfo  {journal} {Phys. Rev. B}\ }\textbf {\bibinfo {volume}
  {102}},\ \bibinfo {pages} {104514} (\bibinfo {year} {2020})}\BibitemShut
  {NoStop}%
\bibitem [{\citenamefont {Christos}\ \emph {et~al.}(2020)\citenamefont
  {Christos}, \citenamefont {Sachdev},\ and\ \citenamefont
  {Scheurer}}]{christos2020superconductivity}%
  \BibitemOpen
  \bibfield  {author} {\bibinfo {author} {\bibfnamefont {Maine}\ \bibnamefont
  {Christos}}, \bibinfo {author} {\bibfnamefont {Subir}\ \bibnamefont
  {Sachdev}}, \ and\ \bibinfo {author} {\bibfnamefont {Mathias~S.}\
  \bibnamefont {Scheurer}},\ }\bibfield  {title} {\enquote {\bibinfo {title}
  {Superconductivity, correlated insulators, and
  wess{\textendash}zumino{\textendash}witten terms in twisted bilayer
  graphene},}\ }\href {\doibase 10.1073/pnas.2014691117} {\bibfield  {journal}
  {\bibinfo  {journal} {Proceedings of the National Academy of Sciences}\
  }\textbf {\bibinfo {volume} {117}},\ \bibinfo {pages} {29543--29554}
  (\bibinfo {year} {2020})}\BibitemShut {NoStop}%
\bibitem [{\citenamefont {Lewandowski}\ \emph {et~al.}(2021)\citenamefont
  {Lewandowski}, \citenamefont {Chowdhury},\ and\ \citenamefont
  {Ruhman}}]{lewandowski2020pairing}%
  \BibitemOpen
  \bibfield  {author} {\bibinfo {author} {\bibfnamefont {Cyprian}\ \bibnamefont
  {Lewandowski}}, \bibinfo {author} {\bibfnamefont {Debanjan}\ \bibnamefont
  {Chowdhury}}, \ and\ \bibinfo {author} {\bibfnamefont {Jonathan}\
  \bibnamefont {Ruhman}},\ }\bibfield  {title} {\enquote {\bibinfo {title}
  {Pairing in magic-angle twisted bilayer graphene: Role of phonon and plasmon
  umklapp},}\ }\href {\doibase 10.1103/PhysRevB.103.235401} {\bibfield
  {journal} {\bibinfo  {journal} {Phys. Rev. B}\ }\textbf {\bibinfo {volume}
  {103}},\ \bibinfo {pages} {235401} (\bibinfo {year} {2021})}\BibitemShut
  {NoStop}%
\bibitem [{\citenamefont {Kwan}\ \emph {et~al.}(2020)\citenamefont {Kwan},
  \citenamefont {Parameswaran},\ and\ \citenamefont
  {Sondhi}}]{Kwan2020Twisted}%
  \BibitemOpen
  \bibfield  {author} {\bibinfo {author} {\bibfnamefont {Yves~H.}\ \bibnamefont
  {Kwan}}, \bibinfo {author} {\bibfnamefont {S.~A.}\ \bibnamefont
  {Parameswaran}}, \ and\ \bibinfo {author} {\bibfnamefont {S.~L.}\
  \bibnamefont {Sondhi}},\ }\bibfield  {title} {\enquote {\bibinfo {title}
  {Twisted bilayer graphene in a parallel magnetic field},}\ }\href {\doibase
  10.1103/PhysRevB.101.205116} {\bibfield  {journal} {\bibinfo  {journal}
  {Phys. Rev. B}\ }\textbf {\bibinfo {volume} {101}},\ \bibinfo {pages}
  {205116} (\bibinfo {year} {2020})}\BibitemShut {NoStop}%
\bibitem [{\citenamefont {Kwan}\ \emph {et~al.}(2021)\citenamefont {Kwan},
  \citenamefont {Hu}, \citenamefont {Simon},\ and\ \citenamefont
  {Parameswaran}}]{Parameswaran2021Exciton}%
  \BibitemOpen
  \bibfield  {author} {\bibinfo {author} {\bibfnamefont {Yves~H.}\ \bibnamefont
  {Kwan}}, \bibinfo {author} {\bibfnamefont {Yichen}\ \bibnamefont {Hu}},
  \bibinfo {author} {\bibfnamefont {Steven~H.}\ \bibnamefont {Simon}}, \ and\
  \bibinfo {author} {\bibfnamefont {S.~A.}\ \bibnamefont {Parameswaran}},\
  }\bibfield  {title} {\enquote {\bibinfo {title} {Exciton band topology in
  spontaneous quantum anomalous hall insulators: Applications to twisted
  bilayer graphene},}\ }\href {\doibase 10.1103/PhysRevLett.126.137601}
  {\bibfield  {journal} {\bibinfo  {journal} {Phys. Rev. Lett.}\ }\textbf
  {\bibinfo {volume} {126}},\ \bibinfo {pages} {137601} (\bibinfo {year}
  {2021})}\BibitemShut {NoStop}%
\bibitem [{\citenamefont {Xie}\ and\ \citenamefont
  {MacDonald}(2020{\natexlab{a}})}]{xie_HF_2020}%
  \BibitemOpen
  \bibfield  {author} {\bibinfo {author} {\bibfnamefont {Ming}\ \bibnamefont
  {Xie}}\ and\ \bibinfo {author} {\bibfnamefont {A.~H.}\ \bibnamefont
  {MacDonald}},\ }\bibfield  {title} {\enquote {\bibinfo {title} {Nature of the
  correlated insulator states in twisted bilayer graphene},}\ }\href {\doibase
  10.1103/PhysRevLett.124.097601} {\bibfield  {journal} {\bibinfo  {journal}
  {Phys. Rev. Lett.}\ }\textbf {\bibinfo {volume} {124}},\ \bibinfo {pages}
  {097601} (\bibinfo {year} {2020}{\natexlab{a}})}\BibitemShut {NoStop}%
\bibitem [{\citenamefont {Liu}\ and\ \citenamefont
  {Dai}(2021)}]{liu2020theories}%
  \BibitemOpen
  \bibfield  {author} {\bibinfo {author} {\bibfnamefont {Jianpeng}\
  \bibnamefont {Liu}}\ and\ \bibinfo {author} {\bibfnamefont {Xi}~\bibnamefont
  {Dai}},\ }\bibfield  {title} {\enquote {\bibinfo {title} {Theories for the
  correlated insulating states and quantum anomalous hall effect phenomena in
  twisted bilayer graphene},}\ }\href {\doibase 10.1103/PhysRevB.103.035427}
  {\bibfield  {journal} {\bibinfo  {journal} {Phys. Rev. B}\ }\textbf {\bibinfo
  {volume} {103}},\ \bibinfo {pages} {035427} (\bibinfo {year}
  {2021})}\BibitemShut {NoStop}%
\bibitem [{\citenamefont {Cea}\ and\ \citenamefont
  {Guinea}(2020)}]{cea_band_2020}%
  \BibitemOpen
  \bibfield  {author} {\bibinfo {author} {\bibfnamefont {Tommaso}\ \bibnamefont
  {Cea}}\ and\ \bibinfo {author} {\bibfnamefont {Francisco}\ \bibnamefont
  {Guinea}},\ }\bibfield  {title} {\enquote {\bibinfo {title} {Band structure
  and insulating states driven by coulomb interaction in twisted bilayer
  graphene},}\ }\href {\doibase 10.1103/PhysRevB.102.045107} {\bibfield
  {journal} {\bibinfo  {journal} {Phys. Rev. B}\ }\textbf {\bibinfo {volume}
  {102}},\ \bibinfo {pages} {045107} (\bibinfo {year} {2020})}\BibitemShut
  {NoStop}%
\bibitem [{\citenamefont {Zhang}\ \emph {et~al.}(2020)\citenamefont {Zhang},
  \citenamefont {Jiang}, \citenamefont {Wang},\ and\ \citenamefont
  {Zhang}}]{zhang_HF_2020}%
  \BibitemOpen
  \bibfield  {author} {\bibinfo {author} {\bibfnamefont {Yi}~\bibnamefont
  {Zhang}}, \bibinfo {author} {\bibfnamefont {Kun}\ \bibnamefont {Jiang}},
  \bibinfo {author} {\bibfnamefont {Ziqiang}\ \bibnamefont {Wang}}, \ and\
  \bibinfo {author} {\bibfnamefont {Fuchun}\ \bibnamefont {Zhang}},\ }\bibfield
   {title} {\enquote {\bibinfo {title} {Correlated insulating phases of twisted
  bilayer graphene at commensurate filling fractions: A hartree-fock study},}\
  }\href {\doibase 10.1103/PhysRevB.102.035136} {\bibfield  {journal} {\bibinfo
   {journal} {Phys. Rev. B}\ }\textbf {\bibinfo {volume} {102}},\ \bibinfo
  {pages} {035136} (\bibinfo {year} {2020})}\BibitemShut {NoStop}%
\bibitem [{\citenamefont {Liu}\ \emph {et~al.}(2021{\natexlab{b}})\citenamefont
  {Liu}, \citenamefont {Khalaf}, \citenamefont {Lee},\ and\ \citenamefont
  {Vishwanath}}]{liu2020nematic}%
  \BibitemOpen
  \bibfield  {author} {\bibinfo {author} {\bibfnamefont {Shang}\ \bibnamefont
  {Liu}}, \bibinfo {author} {\bibfnamefont {Eslam}\ \bibnamefont {Khalaf}},
  \bibinfo {author} {\bibfnamefont {Jong~Yeon}\ \bibnamefont {Lee}}, \ and\
  \bibinfo {author} {\bibfnamefont {Ashvin}\ \bibnamefont {Vishwanath}},\
  }\bibfield  {title} {\enquote {\bibinfo {title} {Nematic topological
  semimetal and insulator in magic-angle bilayer graphene at charge
  neutrality},}\ }\href {\doibase 10.1103/PhysRevResearch.3.013033} {\bibfield
  {journal} {\bibinfo  {journal} {Phys. Rev. Research}\ }\textbf {\bibinfo
  {volume} {3}},\ \bibinfo {pages} {013033} (\bibinfo {year}
  {2021}{\natexlab{b}})}\BibitemShut {NoStop}%
\bibitem [{\citenamefont {Da~Liao}\ \emph {et~al.}(2019)\citenamefont
  {Da~Liao}, \citenamefont {Meng},\ and\ \citenamefont {Xu}}]{daliao_VBO_2019}%
  \BibitemOpen
  \bibfield  {author} {\bibinfo {author} {\bibfnamefont {Yuan}\ \bibnamefont
  {Da~Liao}}, \bibinfo {author} {\bibfnamefont {Zi~Yang}\ \bibnamefont {Meng}},
  \ and\ \bibinfo {author} {\bibfnamefont {Xiao~Yan}\ \bibnamefont {Xu}},\
  }\bibfield  {title} {\enquote {\bibinfo {title} {Valence bond orders at
  charge neutrality in a possible two-orbital extended hubbard model for
  twisted bilayer graphene},}\ }\href {\doibase 10.1103/PhysRevLett.123.157601}
  {\bibfield  {journal} {\bibinfo  {journal} {Phys. Rev. Lett.}\ }\textbf
  {\bibinfo {volume} {123}},\ \bibinfo {pages} {157601} (\bibinfo {year}
  {2019})}\BibitemShut {NoStop}%
\bibitem [{\citenamefont {Da~Liao}\ \emph {et~al.}(2021)\citenamefont
  {Da~Liao}, \citenamefont {Kang}, \citenamefont {Brei\o{}}, \citenamefont
  {Xu}, \citenamefont {Wu}, \citenamefont {Andersen}, \citenamefont
  {Fernandes},\ and\ \citenamefont {Meng}}]{daliao2020correlation}%
  \BibitemOpen
  \bibfield  {author} {\bibinfo {author} {\bibfnamefont {Yuan}\ \bibnamefont
  {Da~Liao}}, \bibinfo {author} {\bibfnamefont {Jian}\ \bibnamefont {Kang}},
  \bibinfo {author} {\bibfnamefont {Clara~N.}\ \bibnamefont {Brei\o{}}},
  \bibinfo {author} {\bibfnamefont {Xiao~Yan}\ \bibnamefont {Xu}}, \bibinfo
  {author} {\bibfnamefont {Han-Qing}\ \bibnamefont {Wu}}, \bibinfo {author}
  {\bibfnamefont {Brian~M.}\ \bibnamefont {Andersen}}, \bibinfo {author}
  {\bibfnamefont {Rafael~M.}\ \bibnamefont {Fernandes}}, \ and\ \bibinfo
  {author} {\bibfnamefont {Zi~Yang}\ \bibnamefont {Meng}},\ }\bibfield  {title}
  {\enquote {\bibinfo {title} {Correlation-induced insulating topological
  phases at charge neutrality in twisted bilayer graphene},}\ }\href {\doibase
  10.1103/PhysRevX.11.011014} {\bibfield  {journal} {\bibinfo  {journal} {Phys.
  Rev. X}\ }\textbf {\bibinfo {volume} {11}},\ \bibinfo {pages} {011014}
  (\bibinfo {year} {2021})}\BibitemShut {NoStop}%
\bibitem [{\citenamefont {Classen}\ \emph {et~al.}(2019)\citenamefont
  {Classen}, \citenamefont {Honerkamp},\ and\ \citenamefont
  {Scherer}}]{classen2019competing}%
  \BibitemOpen
  \bibfield  {author} {\bibinfo {author} {\bibfnamefont {Laura}\ \bibnamefont
  {Classen}}, \bibinfo {author} {\bibfnamefont {Carsten}\ \bibnamefont
  {Honerkamp}}, \ and\ \bibinfo {author} {\bibfnamefont {Michael~M.}\
  \bibnamefont {Scherer}},\ }\bibfield  {title} {\enquote {\bibinfo {title}
  {Competing phases of interacting electrons on triangular lattices in moir\'e
  heterostructures},}\ }\href {\doibase 10.1103/PhysRevB.99.195120} {\bibfield
  {journal} {\bibinfo  {journal} {Phys. Rev. B}\ }\textbf {\bibinfo {volume}
  {99}},\ \bibinfo {pages} {195120} (\bibinfo {year} {2019})}\BibitemShut
  {NoStop}%
\bibitem [{\citenamefont {Kennes}\ \emph {et~al.}(2018)\citenamefont {Kennes},
  \citenamefont {Lischner},\ and\ \citenamefont {Karrasch}}]{kennes2018strong}%
  \BibitemOpen
  \bibfield  {author} {\bibinfo {author} {\bibfnamefont {Dante~M.}\
  \bibnamefont {Kennes}}, \bibinfo {author} {\bibfnamefont {Johannes}\
  \bibnamefont {Lischner}}, \ and\ \bibinfo {author} {\bibfnamefont
  {Christoph}\ \bibnamefont {Karrasch}},\ }\bibfield  {title} {\enquote
  {\bibinfo {title} {Strong correlations and $d+\mathit{id}$ superconductivity
  in twisted bilayer graphene},}\ }\href {\doibase 10.1103/PhysRevB.98.241407}
  {\bibfield  {journal} {\bibinfo  {journal} {Phys. Rev. B}\ }\textbf {\bibinfo
  {volume} {98}},\ \bibinfo {pages} {241407} (\bibinfo {year}
  {2018})}\BibitemShut {NoStop}%
\bibitem [{\citenamefont {Eugenio}\ and\ \citenamefont
  {Dağ}(2020)}]{eugenio2020dmrg}%
  \BibitemOpen
  \bibfield  {author} {\bibinfo {author} {\bibfnamefont {P.~Myles}\
  \bibnamefont {Eugenio}}\ and\ \bibinfo {author} {\bibfnamefont {Ceren~B.}\
  \bibnamefont {Dağ}},\ }\bibfield  {title} {\enquote {\bibinfo {title} {{DMRG
  study of strongly interacting $\mathbb{Z}_2$ flatbands: a toy model inspired
  by twisted bilayer graphene}},}\ }\href {\doibase
  10.21468/SciPostPhysCore.3.2.015} {\bibfield  {journal} {\bibinfo  {journal}
  {SciPost Phys. Core}\ }\textbf {\bibinfo {volume} {3}},\ \bibinfo {pages}
  {15} (\bibinfo {year} {2020})}\BibitemShut {NoStop}%
\bibitem [{\citenamefont {Huang}\ \emph {et~al.}(2020)\citenamefont {Huang},
  \citenamefont {Hosur},\ and\ \citenamefont {Pal}}]{huang2020deconstructing}%
  \BibitemOpen
  \bibfield  {author} {\bibinfo {author} {\bibfnamefont {Yixuan}\ \bibnamefont
  {Huang}}, \bibinfo {author} {\bibfnamefont {Pavan}\ \bibnamefont {Hosur}}, \
  and\ \bibinfo {author} {\bibfnamefont {Hridis~K.}\ \bibnamefont {Pal}},\
  }\bibfield  {title} {\enquote {\bibinfo {title} {Quasi-flat-band physics in a
  two-leg ladder model and its relation to magic-angle twisted bilayer
  graphene},}\ }\href {\doibase 10.1103/PhysRevB.102.155429} {\bibfield
  {journal} {\bibinfo  {journal} {Phys. Rev. B}\ }\textbf {\bibinfo {volume}
  {102}},\ \bibinfo {pages} {155429} (\bibinfo {year} {2020})}\BibitemShut
  {NoStop}%
\bibitem [{\citenamefont {Huang}\ \emph {et~al.}(2019)\citenamefont {Huang},
  \citenamefont {Zhang},\ and\ \citenamefont
  {Ma}}]{huang2019antiferromagnetically}%
  \BibitemOpen
  \bibfield  {author} {\bibinfo {author} {\bibfnamefont {Tongyun}\ \bibnamefont
  {Huang}}, \bibinfo {author} {\bibfnamefont {Lufeng}\ \bibnamefont {Zhang}}, \
  and\ \bibinfo {author} {\bibfnamefont {Tianxing}\ \bibnamefont {Ma}},\
  }\bibfield  {title} {\enquote {\bibinfo {title} {Antiferromagnetically
  ordered mott insulator and d+id superconductivity in twisted bilayer
  graphene: a quantum monte carlo study},}\ }\href {\doibase
  https://doi.org/10.1016/j.scib.2019.01.026} {\bibfield  {journal} {\bibinfo
  {journal} {Science Bulletin}\ }\textbf {\bibinfo {volume} {64}},\ \bibinfo
  {pages} {310--314} (\bibinfo {year} {2019})}\BibitemShut {NoStop}%
\bibitem [{\citenamefont {Guo}\ \emph {et~al.}(2018)\citenamefont {Guo},
  \citenamefont {Zhu}, \citenamefont {Feng},\ and\ \citenamefont
  {Scalettar}}]{guo2018pairing}%
  \BibitemOpen
  \bibfield  {author} {\bibinfo {author} {\bibfnamefont {Huaiming}\
  \bibnamefont {Guo}}, \bibinfo {author} {\bibfnamefont {Xingchuan}\
  \bibnamefont {Zhu}}, \bibinfo {author} {\bibfnamefont {Shiping}\ \bibnamefont
  {Feng}}, \ and\ \bibinfo {author} {\bibfnamefont {Richard~T.}\ \bibnamefont
  {Scalettar}},\ }\bibfield  {title} {\enquote {\bibinfo {title} {Pairing
  symmetry of interacting fermions on a twisted bilayer graphene
  superlattice},}\ }\href {\doibase 10.1103/PhysRevB.97.235453} {\bibfield
  {journal} {\bibinfo  {journal} {Phys. Rev. B}\ }\textbf {\bibinfo {volume}
  {97}},\ \bibinfo {pages} {235453} (\bibinfo {year} {2018})}\BibitemShut
  {NoStop}%
\bibitem [{\citenamefont {Ledwith}\ \emph {et~al.}(2020)\citenamefont
  {Ledwith}, \citenamefont {Tarnopolsky}, \citenamefont {Khalaf},\ and\
  \citenamefont {Vishwanath}}]{ledwith2020}%
  \BibitemOpen
  \bibfield  {author} {\bibinfo {author} {\bibfnamefont {Patrick~J.}\
  \bibnamefont {Ledwith}}, \bibinfo {author} {\bibfnamefont {Grigory}\
  \bibnamefont {Tarnopolsky}}, \bibinfo {author} {\bibfnamefont {Eslam}\
  \bibnamefont {Khalaf}}, \ and\ \bibinfo {author} {\bibfnamefont {Ashvin}\
  \bibnamefont {Vishwanath}},\ }\bibfield  {title} {\enquote {\bibinfo {title}
  {Fractional chern insulator states in twisted bilayer graphene: An analytical
  approach},}\ }\href {\doibase 10.1103/PhysRevResearch.2.023237} {\bibfield
  {journal} {\bibinfo  {journal} {Phys. Rev. Research}\ }\textbf {\bibinfo
  {volume} {2}},\ \bibinfo {pages} {023237} (\bibinfo {year}
  {2020})}\BibitemShut {NoStop}%
\bibitem [{\citenamefont {Repellin}\ \emph {et~al.}(2020)\citenamefont
  {Repellin}, \citenamefont {Dong}, \citenamefont {Zhang},\ and\ \citenamefont
  {Senthil}}]{repellin_EDDMRG_2020}%
  \BibitemOpen
  \bibfield  {author} {\bibinfo {author} {\bibfnamefont {C\'ecile}\
  \bibnamefont {Repellin}}, \bibinfo {author} {\bibfnamefont {Zhihuan}\
  \bibnamefont {Dong}}, \bibinfo {author} {\bibfnamefont {Ya-Hui}\ \bibnamefont
  {Zhang}}, \ and\ \bibinfo {author} {\bibfnamefont {T.}~\bibnamefont
  {Senthil}},\ }\bibfield  {title} {\enquote {\bibinfo {title} {Ferromagnetism
  in narrow bands of moir\'e superlattices},}\ }\href {\doibase
  10.1103/PhysRevLett.124.187601} {\bibfield  {journal} {\bibinfo  {journal}
  {Phys. Rev. Lett.}\ }\textbf {\bibinfo {volume} {124}},\ \bibinfo {pages}
  {187601} (\bibinfo {year} {2020})}\BibitemShut {NoStop}%
\bibitem [{\citenamefont {Abouelkomsan}\ \emph {et~al.}(2020)\citenamefont
  {Abouelkomsan}, \citenamefont {Liu},\ and\ \citenamefont
  {Bergholtz}}]{abouelkomsan2020}%
  \BibitemOpen
  \bibfield  {author} {\bibinfo {author} {\bibfnamefont {Ahmed}\ \bibnamefont
  {Abouelkomsan}}, \bibinfo {author} {\bibfnamefont {Zhao}\ \bibnamefont
  {Liu}}, \ and\ \bibinfo {author} {\bibfnamefont {Emil~J.}\ \bibnamefont
  {Bergholtz}},\ }\bibfield  {title} {\enquote {\bibinfo {title} {Particle-hole
  duality, emergent fermi liquids, and fractional chern insulators in moir\'e
  flatbands},}\ }\href {\doibase 10.1103/PhysRevLett.124.106803} {\bibfield
  {journal} {\bibinfo  {journal} {Phys. Rev. Lett.}\ }\textbf {\bibinfo
  {volume} {124}},\ \bibinfo {pages} {106803} (\bibinfo {year}
  {2020})}\BibitemShut {NoStop}%
\bibitem [{\citenamefont {Repellin}\ and\ \citenamefont
  {Senthil}(2020)}]{repellin_FCI_2020}%
  \BibitemOpen
  \bibfield  {author} {\bibinfo {author} {\bibfnamefont {C\'ecile}\
  \bibnamefont {Repellin}}\ and\ \bibinfo {author} {\bibfnamefont
  {T.}~\bibnamefont {Senthil}},\ }\bibfield  {title} {\enquote {\bibinfo
  {title} {Chern bands of twisted bilayer graphene: Fractional chern insulators
  and spin phase transition},}\ }\href {\doibase
  10.1103/PhysRevResearch.2.023238} {\bibfield  {journal} {\bibinfo  {journal}
  {Phys. Rev. Research}\ }\textbf {\bibinfo {volume} {2}},\ \bibinfo {pages}
  {023238} (\bibinfo {year} {2020})}\BibitemShut {NoStop}%
\bibitem [{\citenamefont {Fernandes}\ and\ \citenamefont
  {Venderbos}(2020)}]{fernandes_nematic_2020}%
  \BibitemOpen
  \bibfield  {author} {\bibinfo {author} {\bibfnamefont {Rafael~M.}\
  \bibnamefont {Fernandes}}\ and\ \bibinfo {author} {\bibfnamefont {J{\"o}rn
  W.~F.}\ \bibnamefont {Venderbos}},\ }\bibfield  {title} {\enquote {\bibinfo
  {title} {Nematicity with a twist: Rotational symmetry breaking in a moir{\'e}
  superlattice},}\ }\href {\doibase 10.1126/sciadv.aba8834} {\bibfield
  {journal} {\bibinfo  {journal} {Science Advances}\ }\textbf {\bibinfo
  {volume} {6}} (\bibinfo {year} {2020}),\ 10.1126/sciadv.aba8834}\BibitemShut
  {NoStop}%
\bibitem [{\citenamefont {Wilson}\ \emph {et~al.}(2020)\citenamefont {Wilson},
  \citenamefont {Fu}, \citenamefont {Das~Sarma},\ and\ \citenamefont
  {Pixley}}]{Wilson2020TBG}%
  \BibitemOpen
  \bibfield  {author} {\bibinfo {author} {\bibfnamefont {Justin~H.}\
  \bibnamefont {Wilson}}, \bibinfo {author} {\bibfnamefont {Yixing}\
  \bibnamefont {Fu}}, \bibinfo {author} {\bibfnamefont {S.}~\bibnamefont
  {Das~Sarma}}, \ and\ \bibinfo {author} {\bibfnamefont {J.~H.}\ \bibnamefont
  {Pixley}},\ }\bibfield  {title} {\enquote {\bibinfo {title} {Disorder in
  twisted bilayer graphene},}\ }\href {\doibase
  10.1103/PhysRevResearch.2.023325} {\bibfield  {journal} {\bibinfo  {journal}
  {Phys. Rev. Research}\ }\textbf {\bibinfo {volume} {2}},\ \bibinfo {pages}
  {023325} (\bibinfo {year} {2020})}\BibitemShut {NoStop}%
\bibitem [{\citenamefont {Wang}\ \emph {et~al.}(2021)\citenamefont {Wang},
  \citenamefont {Zheng}, \citenamefont {Millis},\ and\ \citenamefont
  {Cano}}]{wang2020chiral}%
  \BibitemOpen
  \bibfield  {author} {\bibinfo {author} {\bibfnamefont {Jie}\ \bibnamefont
  {Wang}}, \bibinfo {author} {\bibfnamefont {Yunqin}\ \bibnamefont {Zheng}},
  \bibinfo {author} {\bibfnamefont {Andrew~J.}\ \bibnamefont {Millis}}, \ and\
  \bibinfo {author} {\bibfnamefont {Jennifer}\ \bibnamefont {Cano}},\
  }\bibfield  {title} {\enquote {\bibinfo {title} {Chiral approximation to
  twisted bilayer graphene: Exact intravalley inversion symmetry, nodal
  structure, and implications for higher magic angles},}\ }\href {\doibase
  10.1103/PhysRevResearch.3.023155} {\bibfield  {journal} {\bibinfo  {journal}
  {Phys. Rev. Research}\ }\textbf {\bibinfo {volume} {3}},\ \bibinfo {pages}
  {023155} (\bibinfo {year} {2021})}\BibitemShut {NoStop}%
\bibitem [{\citenamefont {Bernevig}\ \emph
  {et~al.}(2021{\natexlab{c}})\citenamefont {Bernevig}, \citenamefont {Song},
  \citenamefont {Regnault},\ and\ \citenamefont {Lian}}]{TBG1}%
  \BibitemOpen
  \bibfield  {author} {\bibinfo {author} {\bibfnamefont {B.~Andrei}\
  \bibnamefont {Bernevig}}, \bibinfo {author} {\bibfnamefont {Zhi-Da}\
  \bibnamefont {Song}}, \bibinfo {author} {\bibfnamefont {Nicolas}\
  \bibnamefont {Regnault}}, \ and\ \bibinfo {author} {\bibfnamefont {Biao}\
  \bibnamefont {Lian}},\ }\bibfield  {title} {\enquote {\bibinfo {title}
  {Twisted bilayer graphene. {I}. {Matrix} elements, approximations,
  perturbation theory, and a $k\cdot p$ two-band model},}\ }\href {\doibase
  10.1103/PhysRevB.103.205411} {\bibfield  {journal} {\bibinfo  {journal}
  {Physical Review B}\ }\textbf {\bibinfo {volume} {103}},\ \bibinfo {pages}
  {205411} (\bibinfo {year} {2021}{\natexlab{c}})},\ \bibinfo {note}
  {publisher: American Physical Society}\BibitemShut {NoStop}%
\bibitem [{\citenamefont {Song}\ \emph {et~al.}(2021)\citenamefont {Song},
  \citenamefont {Lian}, \citenamefont {Regnault},\ and\ \citenamefont
  {Bernevig}}]{Song-TBG2}%
  \BibitemOpen
  \bibfield  {author} {\bibinfo {author} {\bibfnamefont {Zhi-Da}\ \bibnamefont
  {Song}}, \bibinfo {author} {\bibfnamefont {Biao}\ \bibnamefont {Lian}},
  \bibinfo {author} {\bibfnamefont {Nicolas}\ \bibnamefont {Regnault}}, \ and\
  \bibinfo {author} {\bibfnamefont {B.~Andrei}\ \bibnamefont {Bernevig}},\
  }\bibfield  {title} {\enquote {\bibinfo {title} {Twisted bilayer graphene.
  ii. stable symmetry anomaly},}\ }\href {\doibase 10.1103/PhysRevB.103.205412}
  {\bibfield  {journal} {\bibinfo  {journal} {Phys. Rev. B}\ }\textbf {\bibinfo
  {volume} {103}},\ \bibinfo {pages} {205412} (\bibinfo {year}
  {2021})}\BibitemShut {NoStop}%
\bibitem [{\citenamefont {Xie}\ \emph {et~al.}(2021)\citenamefont {Xie},
  \citenamefont {Cowsik}, \citenamefont {Song}, \citenamefont {Lian},
  \citenamefont {Bernevig},\ and\ \citenamefont {Regnault}}]{TBG6}%
  \BibitemOpen
  \bibfield  {author} {\bibinfo {author} {\bibfnamefont {Fang}\ \bibnamefont
  {Xie}}, \bibinfo {author} {\bibfnamefont {Aditya}\ \bibnamefont {Cowsik}},
  \bibinfo {author} {\bibfnamefont {Zhi-Da}\ \bibnamefont {Song}}, \bibinfo
  {author} {\bibfnamefont {Biao}\ \bibnamefont {Lian}}, \bibinfo {author}
  {\bibfnamefont {B.~Andrei}\ \bibnamefont {Bernevig}}, \ and\ \bibinfo
  {author} {\bibfnamefont {Nicolas}\ \bibnamefont {Regnault}},\ }\bibfield
  {title} {\enquote {\bibinfo {title} {Twisted bilayer graphene. {VI}. {An}
  exact diagonalization study at nonzero integer filling},}\ }\href {\doibase
  10.1103/PhysRevB.103.205416} {\bibfield  {journal} {\bibinfo  {journal}
  {Physical Review B}\ }\textbf {\bibinfo {volume} {103}},\ \bibinfo {pages}
  {205416} (\bibinfo {year} {2021})},\ \bibinfo {note} {publisher: American
  Physical Society}\BibitemShut {NoStop}%
\bibitem [{\citenamefont {Zhang}\ \emph {et~al.}(2021)\citenamefont {Zhang},
  \citenamefont {Pan}, \citenamefont {Zhang}, \citenamefont {Kang},\ and\
  \citenamefont {Meng}}]{zhang2021momentum}%
  \BibitemOpen
  \bibfield  {author} {\bibinfo {author} {\bibfnamefont {Xu}~\bibnamefont
  {Zhang}}, \bibinfo {author} {\bibfnamefont {Gaopei}\ \bibnamefont {Pan}},
  \bibinfo {author} {\bibfnamefont {Yi}~\bibnamefont {Zhang}}, \bibinfo
  {author} {\bibfnamefont {Jian}\ \bibnamefont {Kang}}, \ and\ \bibinfo
  {author} {\bibfnamefont {Zi~Yang}\ \bibnamefont {Meng}},\ }\bibfield  {title}
  {\enquote {\bibinfo {title} {Momentum space quantum monte carlo on twisted
  bilayer graphene},}\ }\href {\doibase 10.1088/0256-307x/38/7/077305}
  {\bibfield  {journal} {\bibinfo  {journal} {Chinese Physics Letters}\
  }\textbf {\bibinfo {volume} {38}},\ \bibinfo {pages} {077305} (\bibinfo
  {year} {2021})}\BibitemShut {NoStop}%
\bibitem [{\citenamefont {Cha}\ \emph {et~al.}(2021)\citenamefont {Cha},
  \citenamefont {Patel},\ and\ \citenamefont {Kim}}]{cha2021strange}%
  \BibitemOpen
  \bibfield  {author} {\bibinfo {author} {\bibfnamefont {Peter}\ \bibnamefont
  {Cha}}, \bibinfo {author} {\bibfnamefont {Aavishkar~A}\ \bibnamefont
  {Patel}}, \ and\ \bibinfo {author} {\bibfnamefont {Eun-Ah}\ \bibnamefont
  {Kim}},\ }\bibfield  {title} {\enquote {\bibinfo {title} {Strange metals from
  melting correlated insulators in twisted bilayer graphene},}\ }\href
  {https://arxiv.org/abs/2105.08069} {\bibfield  {journal} {\bibinfo  {journal}
  {arXiv preprint arXiv:2105.08069}\ } (\bibinfo {year} {2021})}\BibitemShut
  {NoStop}%
\bibitem [{\citenamefont {Chichinadze}\ \emph {et~al.}(2020)\citenamefont
  {Chichinadze}, \citenamefont {Classen},\ and\ \citenamefont
  {Chubukov}}]{Chichinadze2020Nematic}%
  \BibitemOpen
  \bibfield  {author} {\bibinfo {author} {\bibfnamefont {Dmitry~V.}\
  \bibnamefont {Chichinadze}}, \bibinfo {author} {\bibfnamefont {Laura}\
  \bibnamefont {Classen}}, \ and\ \bibinfo {author} {\bibfnamefont {Andrey~V.}\
  \bibnamefont {Chubukov}},\ }\bibfield  {title} {\enquote {\bibinfo {title}
  {Nematic superconductivity in twisted bilayer graphene},}\ }\href {\doibase
  10.1103/PhysRevB.101.224513} {\bibfield  {journal} {\bibinfo  {journal}
  {Phys. Rev. B}\ }\textbf {\bibinfo {volume} {101}},\ \bibinfo {pages}
  {224513} (\bibinfo {year} {2020})}\BibitemShut {NoStop}%
\bibitem [{\citenamefont {Sheffer}\ and\ \citenamefont
  {Stern}(2021)}]{sheffer2021chiral}%
  \BibitemOpen
  \bibfield  {author} {\bibinfo {author} {\bibfnamefont {Yarden}\ \bibnamefont
  {Sheffer}}\ and\ \bibinfo {author} {\bibfnamefont {Ady}\ \bibnamefont
  {Stern}},\ }\bibfield  {title} {\enquote {\bibinfo {title} {Chiral
  magic-angle twisted bilayer graphene in a magnetic field: Landau level
  correspondence, exact wavefunctions and fractional chern insulators},}\
  }\href@noop {} {\bibfield  {journal} {\bibinfo  {journal} {arXiv preprint
  arXiv:2106.10650}\ } (\bibinfo {year} {2021})}\BibitemShut {NoStop}%
\bibitem [{\citenamefont {Kang}\ \emph {et~al.}(2021)\citenamefont {Kang},
  \citenamefont {Bernevig},\ and\ \citenamefont {Vafek}}]{kang2021cascades}%
  \BibitemOpen
  \bibfield  {author} {\bibinfo {author} {\bibfnamefont {Jian}\ \bibnamefont
  {Kang}}, \bibinfo {author} {\bibfnamefont {B~Andrei}\ \bibnamefont
  {Bernevig}}, \ and\ \bibinfo {author} {\bibfnamefont {Oskar}\ \bibnamefont
  {Vafek}},\ }\bibfield  {title} {\enquote {\bibinfo {title} {Cascades between
  light and heavy fermions in the normal state of magic angle twisted bilayer
  graphene},}\ }\href {https://arxiv.org/abs/2104.01145} {\bibfield  {journal}
  {\bibinfo  {journal} {arXiv preprint arXiv:2104.01145}\ } (\bibinfo {year}
  {2021})}\BibitemShut {NoStop}%
\bibitem [{\citenamefont {Hofmann}\ \emph {et~al.}(2021)\citenamefont
  {Hofmann}, \citenamefont {Khalaf}, \citenamefont {Vishwanath}, \citenamefont
  {Berg},\ and\ \citenamefont {Lee}}]{hofmann2021fermionic}%
  \BibitemOpen
  \bibfield  {author} {\bibinfo {author} {\bibfnamefont {Johannes~S}\
  \bibnamefont {Hofmann}}, \bibinfo {author} {\bibfnamefont {Eslam}\
  \bibnamefont {Khalaf}}, \bibinfo {author} {\bibfnamefont {Ashvin}\
  \bibnamefont {Vishwanath}}, \bibinfo {author} {\bibfnamefont {Erez}\
  \bibnamefont {Berg}}, \ and\ \bibinfo {author} {\bibfnamefont {Jong~Yeon}\
  \bibnamefont {Lee}},\ }\bibfield  {title} {\enquote {\bibinfo {title}
  {Fermionic monte carlo study of a realistic model of twisted bilayer
  graphene},}\ }\href {https://arxiv.org/abs/2105.12112} {\bibfield  {journal}
  {\bibinfo  {journal} {arXiv preprint arXiv:2105.12112}\ } (\bibinfo {year}
  {2021})}\BibitemShut {NoStop}%
\bibitem [{\citenamefont {Calder\'on}\ and\ \citenamefont
  {Bascones}(2020)}]{Calder2020Interactions}%
  \BibitemOpen
  \bibfield  {author} {\bibinfo {author} {\bibfnamefont {M.~J.}\ \bibnamefont
  {Calder\'on}}\ and\ \bibinfo {author} {\bibfnamefont {E.}~\bibnamefont
  {Bascones}},\ }\bibfield  {title} {\enquote {\bibinfo {title} {Interactions
  in the 8-orbital model for twisted bilayer graphene},}\ }\href {\doibase
  10.1103/PhysRevB.102.155149} {\bibfield  {journal} {\bibinfo  {journal}
  {Phys. Rev. B}\ }\textbf {\bibinfo {volume} {102}},\ \bibinfo {pages}
  {155149} (\bibinfo {year} {2020})}\BibitemShut {NoStop}%
\bibitem [{\citenamefont {Thomson}\ and\ \citenamefont
  {Alicea}(2021)}]{Thomson2021Recovery}%
  \BibitemOpen
  \bibfield  {author} {\bibinfo {author} {\bibfnamefont {Alex}\ \bibnamefont
  {Thomson}}\ and\ \bibinfo {author} {\bibfnamefont {Jason}\ \bibnamefont
  {Alicea}},\ }\bibfield  {title} {\enquote {\bibinfo {title} {Recovery of
  massless dirac fermions at charge neutrality in strongly interacting twisted
  bilayer graphene with disorder},}\ }\href {\doibase
  10.1103/PhysRevB.103.125138} {\bibfield  {journal} {\bibinfo  {journal}
  {Phys. Rev. B}\ }\textbf {\bibinfo {volume} {103}},\ \bibinfo {pages}
  {125138} (\bibinfo {year} {2021})}\BibitemShut {NoStop}%
\bibitem [{\citenamefont {Xie}\ and\ \citenamefont
  {MacDonald}(2020{\natexlab{b}})}]{Xie2020TBG}%
  \BibitemOpen
  \bibfield  {author} {\bibinfo {author} {\bibfnamefont {Ming}\ \bibnamefont
  {Xie}}\ and\ \bibinfo {author} {\bibfnamefont {A.~H.}\ \bibnamefont
  {MacDonald}},\ }\bibfield  {title} {\enquote {\bibinfo {title} {Nature of the
  correlated insulator states in twisted bilayer graphene},}\ }\href {\doibase
  10.1103/PhysRevLett.124.097601} {\bibfield  {journal} {\bibinfo  {journal}
  {Phys. Rev. Lett.}\ }\textbf {\bibinfo {volume} {124}},\ \bibinfo {pages}
  {097601} (\bibinfo {year} {2020}{\natexlab{b}})}\BibitemShut {NoStop}%
\bibitem [{SM()}]{SM}%
  \BibitemOpen
  \href@noop {} {\ }\bibinfo {note} {See supplementary material, which includes
  the references
  \cite{bistritzer_moire_2011,song_all_2019,tarnopolsky_origin_2019,Song-TBG2,Biao-TBG3,elcoro2020magnetic,xu2020high,Wannier90-1,Wannier90-2,Wannier90-3,winkler_spin-orbit_2003,wang2020chiral,bultinck_ground_2020,kang_strong_2019,cualuguaru2021general,Biao-TBG4,TBG5,vafek2021lattice,cea_band_2020,zhang_HF_2020,kang2021cascades,xie2019spectroscopic,wong_cascade_2020,ledwith2020,GMP1986,Sondhi2013Fractional,Roy2014BandGeometry,sheffer2021chiral,repellin_FCI_2020,abouelkomsan2020}}\BibitemShut
  {NoStop}%
\bibitem [{\citenamefont {Bradlyn}\ \emph {et~al.}(2017)\citenamefont
  {Bradlyn}, \citenamefont {Elcoro}, \citenamefont {Cano}, \citenamefont
  {Vergniory}, \citenamefont {Wang}, \citenamefont {Felser}, \citenamefont
  {Aroyo},\ and\ \citenamefont {Bernevig}}]{bradlyn_topological_2017}%
  \BibitemOpen
  \bibfield  {author} {\bibinfo {author} {\bibfnamefont {Barry}\ \bibnamefont
  {Bradlyn}}, \bibinfo {author} {\bibfnamefont {L.}~\bibnamefont {Elcoro}},
  \bibinfo {author} {\bibfnamefont {Jennifer}\ \bibnamefont {Cano}}, \bibinfo
  {author} {\bibfnamefont {M.~G.}\ \bibnamefont {Vergniory}}, \bibinfo {author}
  {\bibfnamefont {Zhijun}\ \bibnamefont {Wang}}, \bibinfo {author}
  {\bibfnamefont {C.}~\bibnamefont {Felser}}, \bibinfo {author} {\bibfnamefont
  {M.~I.}\ \bibnamefont {Aroyo}}, \ and\ \bibinfo {author} {\bibfnamefont
  {B.~Andrei}\ \bibnamefont {Bernevig}},\ }\bibfield  {title} {\enquote
  {\bibinfo {title} {Topological quantum chemistry},}\ }\href {\doibase
  10.1038/nature23268} {\bibfield  {journal} {\bibinfo  {journal} {Nature}\
  }\textbf {\bibinfo {volume} {547}},\ \bibinfo {pages} {298--305} (\bibinfo
  {year} {2017})}\BibitemShut {NoStop}%
\bibitem [{\citenamefont {Po}\ \emph {et~al.}(2018{\natexlab{b}})\citenamefont
  {Po}, \citenamefont {Watanabe},\ and\ \citenamefont
  {Vishwanath}}]{po_fragile_2018}%
  \BibitemOpen
  \bibfield  {author} {\bibinfo {author} {\bibfnamefont {Hoi~Chun}\
  \bibnamefont {Po}}, \bibinfo {author} {\bibfnamefont {Haruki}\ \bibnamefont
  {Watanabe}}, \ and\ \bibinfo {author} {\bibfnamefont {Ashvin}\ \bibnamefont
  {Vishwanath}},\ }\bibfield  {title} {\enquote {\bibinfo {title} {Fragile
  {Topology} and {Wannier} {Obstructions}},}\ }\href {\doibase
  10.1103/PhysRevLett.121.126402} {\bibfield  {journal} {\bibinfo  {journal}
  {Physical Review Letters}\ }\textbf {\bibinfo {volume} {121}},\ \bibinfo
  {pages} {126402} (\bibinfo {year} {2018}{\natexlab{b}})}\BibitemShut
  {NoStop}%
\bibitem [{\citenamefont {Ahn}\ \emph {et~al.}(2019)\citenamefont {Ahn},
  \citenamefont {Park},\ and\ \citenamefont {Yang}}]{ahn_failure_2019}%
  \BibitemOpen
  \bibfield  {author} {\bibinfo {author} {\bibfnamefont {Junyeong}\
  \bibnamefont {Ahn}}, \bibinfo {author} {\bibfnamefont {Sungjoon}\
  \bibnamefont {Park}}, \ and\ \bibinfo {author} {\bibfnamefont {Bohm-Jung}\
  \bibnamefont {Yang}},\ }\bibfield  {title} {\enquote {\bibinfo {title}
  {Failure of {Nielsen}-{Ninomiya} {Theorem} and {Fragile} {Topology} in
  {Two}-{Dimensional} {Systems} with {Space}-{Time} {Inversion} {Symmetry}:
  {Application} to {Twisted} {Bilayer} {Graphene} at {Magic} {Angle}},}\ }\href
  {\doibase 10.1103/PhysRevX.9.021013} {\bibfield  {journal} {\bibinfo
  {journal} {Physical Review X}\ }\textbf {\bibinfo {volume} {9}},\ \bibinfo
  {pages} {021013} (\bibinfo {year} {2019})}\BibitemShut {NoStop}%
\bibitem [{\citenamefont {Bouhon}\ \emph {et~al.}(2019)\citenamefont {Bouhon},
  \citenamefont {Black-Schaffer},\ and\ \citenamefont {Slager}}]{Slager2019WL}%
  \BibitemOpen
  \bibfield  {author} {\bibinfo {author} {\bibfnamefont {Adrien}\ \bibnamefont
  {Bouhon}}, \bibinfo {author} {\bibfnamefont {Annica~M.}\ \bibnamefont
  {Black-Schaffer}}, \ and\ \bibinfo {author} {\bibfnamefont {Robert-Jan}\
  \bibnamefont {Slager}},\ }\bibfield  {title} {\enquote {\bibinfo {title}
  {Wilson loop approach to fragile topology of split elementary band
  representations and topological crystalline insulators with time-reversal
  symmetry},}\ }\href {\doibase 10.1103/PhysRevB.100.195135} {\bibfield
  {journal} {\bibinfo  {journal} {Phys. Rev. B}\ }\textbf {\bibinfo {volume}
  {100}},\ \bibinfo {pages} {195135} (\bibinfo {year} {2019})}\BibitemShut
  {NoStop}%
\bibitem [{\citenamefont {Cano}\ \emph {et~al.}(2018)\citenamefont {Cano},
  \citenamefont {Bradlyn}, \citenamefont {Wang}, \citenamefont {Elcoro},
  \citenamefont {Vergniory}, \citenamefont {Felser}, \citenamefont {Aroyo},\
  and\ \citenamefont {Bernevig}}]{cano_fragile_2018}%
  \BibitemOpen
  \bibfield  {author} {\bibinfo {author} {\bibfnamefont {Jennifer}\
  \bibnamefont {Cano}}, \bibinfo {author} {\bibfnamefont {Barry}\ \bibnamefont
  {Bradlyn}}, \bibinfo {author} {\bibfnamefont {Zhijun}\ \bibnamefont {Wang}},
  \bibinfo {author} {\bibfnamefont {L.}~\bibnamefont {Elcoro}}, \bibinfo
  {author} {\bibfnamefont {M.~G.}\ \bibnamefont {Vergniory}}, \bibinfo {author}
  {\bibfnamefont {C.}~\bibnamefont {Felser}}, \bibinfo {author} {\bibfnamefont
  {M.~I.}\ \bibnamefont {Aroyo}}, \ and\ \bibinfo {author} {\bibfnamefont
  {B.~Andrei}\ \bibnamefont {Bernevig}},\ }\bibfield  {title} {\enquote
  {\bibinfo {title} {Topology of disconnected elementary band
  representations},}\ }\href {\doibase 10.1103/PhysRevLett.120.266401}
  {\bibfield  {journal} {\bibinfo  {journal} {Phys. Rev. Lett.}\ }\textbf
  {\bibinfo {volume} {120}},\ \bibinfo {pages} {266401} (\bibinfo {year}
  {2018})}\BibitemShut {NoStop}%
\bibitem [{\citenamefont {Marzari}\ and\ \citenamefont
  {Vanderbilt}(1997)}]{Wannier90-1}%
  \BibitemOpen
  \bibfield  {author} {\bibinfo {author} {\bibfnamefont {Nicola}\ \bibnamefont
  {Marzari}}\ and\ \bibinfo {author} {\bibfnamefont {David}\ \bibnamefont
  {Vanderbilt}},\ }\bibfield  {title} {\enquote {\bibinfo {title} {Maximally
  localized generalized wannier functions for composite energy bands},}\ }\href
  {\doibase 10.1103/PhysRevB.56.12847} {\bibfield  {journal} {\bibinfo
  {journal} {Phys. Rev. B}\ }\textbf {\bibinfo {volume} {56}},\ \bibinfo
  {pages} {12847--12865} (\bibinfo {year} {1997})}\BibitemShut {NoStop}%
\bibitem [{\citenamefont {Souza}\ \emph {et~al.}(2001)\citenamefont {Souza},
  \citenamefont {Marzari},\ and\ \citenamefont {Vanderbilt}}]{Wannier90-2}%
  \BibitemOpen
  \bibfield  {author} {\bibinfo {author} {\bibfnamefont {Ivo}\ \bibnamefont
  {Souza}}, \bibinfo {author} {\bibfnamefont {Nicola}\ \bibnamefont {Marzari}},
  \ and\ \bibinfo {author} {\bibfnamefont {David}\ \bibnamefont {Vanderbilt}},\
  }\bibfield  {title} {\enquote {\bibinfo {title} {Maximally localized wannier
  functions for entangled energy bands},}\ }\href {\doibase
  10.1103/PhysRevB.65.035109} {\bibfield  {journal} {\bibinfo  {journal} {Phys.
  Rev. B}\ }\textbf {\bibinfo {volume} {65}},\ \bibinfo {pages} {035109}
  (\bibinfo {year} {2001})}\BibitemShut {NoStop}%
\bibitem [{\citenamefont {Pizzi}\ \emph {et~al.}(2020)\citenamefont {Pizzi},
  \citenamefont {Vitale}, \citenamefont {Arita}, \citenamefont {Blügel},
  \citenamefont {Freimuth}, \citenamefont {G{\'{e}}ranton}, \citenamefont
  {Gibertini}, \citenamefont {Gresch}, \citenamefont {Johnson}, \citenamefont
  {Koretsune}, \citenamefont {Iba{\~{n}}ez-Azpiroz}, \citenamefont {Lee},
  \citenamefont {Lihm}, \citenamefont {Marchand}, \citenamefont {Marrazzo},
  \citenamefont {Mokrousov}, \citenamefont {Mustafa}, \citenamefont {Nohara},
  \citenamefont {Nomura}, \citenamefont {Paulatto}, \citenamefont
  {Ponc{\'{e}}}, \citenamefont {Ponweiser}, \citenamefont {Qiao}, \citenamefont
  {Thöle}, \citenamefont {Tsirkin}, \citenamefont {Wierzbowska}, \citenamefont
  {Marzari}, \citenamefont {Vanderbilt}, \citenamefont {Souza}, \citenamefont
  {Mostofi},\ and\ \citenamefont {Yates}}]{Wannier90-3}%
  \BibitemOpen
  \bibfield  {author} {\bibinfo {author} {\bibfnamefont {Giovanni}\
  \bibnamefont {Pizzi}}, \bibinfo {author} {\bibfnamefont {Valerio}\
  \bibnamefont {Vitale}}, \bibinfo {author} {\bibfnamefont {Ryotaro}\
  \bibnamefont {Arita}}, \bibinfo {author} {\bibfnamefont {Stefan}\
  \bibnamefont {Blügel}}, \bibinfo {author} {\bibfnamefont {Frank}\
  \bibnamefont {Freimuth}}, \bibinfo {author} {\bibfnamefont {Guillaume}\
  \bibnamefont {G{\'{e}}ranton}}, \bibinfo {author} {\bibfnamefont {Marco}\
  \bibnamefont {Gibertini}}, \bibinfo {author} {\bibfnamefont {Dominik}\
  \bibnamefont {Gresch}}, \bibinfo {author} {\bibfnamefont {Charles}\
  \bibnamefont {Johnson}}, \bibinfo {author} {\bibfnamefont {Takashi}\
  \bibnamefont {Koretsune}}, \bibinfo {author} {\bibfnamefont {Julen}\
  \bibnamefont {Iba{\~{n}}ez-Azpiroz}}, \bibinfo {author} {\bibfnamefont
  {Hyungjun}\ \bibnamefont {Lee}}, \bibinfo {author} {\bibfnamefont {Jae-Mo}\
  \bibnamefont {Lihm}}, \bibinfo {author} {\bibfnamefont {Daniel}\ \bibnamefont
  {Marchand}}, \bibinfo {author} {\bibfnamefont {Antimo}\ \bibnamefont
  {Marrazzo}}, \bibinfo {author} {\bibfnamefont {Yuriy}\ \bibnamefont
  {Mokrousov}}, \bibinfo {author} {\bibfnamefont {Jamal~I}\ \bibnamefont
  {Mustafa}}, \bibinfo {author} {\bibfnamefont {Yoshiro}\ \bibnamefont
  {Nohara}}, \bibinfo {author} {\bibfnamefont {Yusuke}\ \bibnamefont {Nomura}},
  \bibinfo {author} {\bibfnamefont {Lorenzo}\ \bibnamefont {Paulatto}},
  \bibinfo {author} {\bibfnamefont {Samuel}\ \bibnamefont {Ponc{\'{e}}}},
  \bibinfo {author} {\bibfnamefont {Thomas}\ \bibnamefont {Ponweiser}},
  \bibinfo {author} {\bibfnamefont {Junfeng}\ \bibnamefont {Qiao}}, \bibinfo
  {author} {\bibfnamefont {Florian}\ \bibnamefont {Thöle}}, \bibinfo {author}
  {\bibfnamefont {Stepan~S}\ \bibnamefont {Tsirkin}}, \bibinfo {author}
  {\bibfnamefont {Ma{\l}gorzata}\ \bibnamefont {Wierzbowska}}, \bibinfo
  {author} {\bibfnamefont {Nicola}\ \bibnamefont {Marzari}}, \bibinfo {author}
  {\bibfnamefont {David}\ \bibnamefont {Vanderbilt}}, \bibinfo {author}
  {\bibfnamefont {Ivo}\ \bibnamefont {Souza}}, \bibinfo {author} {\bibfnamefont
  {Arash~A}\ \bibnamefont {Mostofi}}, \ and\ \bibinfo {author} {\bibfnamefont
  {Jonathan~R}\ \bibnamefont {Yates}},\ }\bibfield  {title} {\enquote {\bibinfo
  {title} {Wannier90 as a community code: new features and applications},}\
  }\href {\doibase 10.1088/1361-648x/ab51ff} {\bibfield  {journal} {\bibinfo
  {journal} {Journal of Physics: Condensed Matter}\ }\textbf {\bibinfo {volume}
  {32}},\ \bibinfo {pages} {165902} (\bibinfo {year} {2020})}\BibitemShut
  {NoStop}%
\bibitem [{\citenamefont {Ramires}\ and\ \citenamefont
  {Lado}(2021)}]{Ramires2021Emulating}%
  \BibitemOpen
  \bibfield  {author} {\bibinfo {author} {\bibfnamefont {Aline}\ \bibnamefont
  {Ramires}}\ and\ \bibinfo {author} {\bibfnamefont {Jose~L.}\ \bibnamefont
  {Lado}},\ }\bibfield  {title} {\enquote {\bibinfo {title} {Emulating heavy
  fermions in twisted trilayer graphene},}\ }\href {\doibase
  10.1103/PhysRevLett.127.026401} {\bibfield  {journal} {\bibinfo  {journal}
  {Phys. Rev. Lett.}\ }\textbf {\bibinfo {volume} {127}},\ \bibinfo {pages}
  {026401} (\bibinfo {year} {2021})}\BibitemShut {NoStop}%
\bibitem [{\citenamefont {Si}\ and\ \citenamefont
  {Steglich}(2010)}]{si2010heavy-review}%
  \BibitemOpen
  \bibfield  {author} {\bibinfo {author} {\bibfnamefont {Qimiao}\ \bibnamefont
  {Si}}\ and\ \bibinfo {author} {\bibfnamefont {Frank}\ \bibnamefont
  {Steglich}},\ }\bibfield  {title} {\enquote {\bibinfo {title} {Heavy fermions
  and quantum phase transitions},}\ }\href {\doibase 10.1126/science.1191195}
  {\bibfield  {journal} {\bibinfo  {journal} {Science}\ }\textbf {\bibinfo
  {volume} {329}},\ \bibinfo {pages} {1161--1166} (\bibinfo {year}
  {2010})}\BibitemShut {NoStop}%
\bibitem [{\citenamefont {Gegenwart}\ \emph {et~al.}(2008)\citenamefont
  {Gegenwart}, \citenamefont {Si},\ and\ \citenamefont
  {Steglich}}]{gegenwart2008quantum}%
  \BibitemOpen
  \bibfield  {author} {\bibinfo {author} {\bibfnamefont {Philipp}\ \bibnamefont
  {Gegenwart}}, \bibinfo {author} {\bibfnamefont {Qimiao}\ \bibnamefont {Si}},
  \ and\ \bibinfo {author} {\bibfnamefont {Frank}\ \bibnamefont {Steglich}},\
  }\bibfield  {title} {\enquote {\bibinfo {title} {Quantum criticality in
  heavy-fermion metals},}\ }\href@noop {} {\bibfield  {journal} {\bibinfo
  {journal} {nature physics}\ }\textbf {\bibinfo {volume} {4}},\ \bibinfo
  {pages} {186--197} (\bibinfo {year} {2008})}\BibitemShut {NoStop}%
\bibitem [{\citenamefont {Coleman}(1984)}]{Coleman1984}%
  \BibitemOpen
  \bibfield  {author} {\bibinfo {author} {\bibfnamefont {Piers}\ \bibnamefont
  {Coleman}},\ }\bibfield  {title} {\enquote {\bibinfo {title} {New approach to
  the mixed-valence problem},}\ }\href {\doibase 10.1103/PhysRevB.29.3035}
  {\bibfield  {journal} {\bibinfo  {journal} {Phys. Rev. B}\ }\textbf {\bibinfo
  {volume} {29}},\ \bibinfo {pages} {3035--3044} (\bibinfo {year}
  {1984})}\BibitemShut {NoStop}%
\bibitem [{\citenamefont {Dzero}\ \emph {et~al.}(2010)\citenamefont {Dzero},
  \citenamefont {Sun}, \citenamefont {Galitski},\ and\ \citenamefont
  {Coleman}}]{Coleman2010Topological}%
  \BibitemOpen
  \bibfield  {author} {\bibinfo {author} {\bibfnamefont {Maxim}\ \bibnamefont
  {Dzero}}, \bibinfo {author} {\bibfnamefont {Kai}\ \bibnamefont {Sun}},
  \bibinfo {author} {\bibfnamefont {Victor}\ \bibnamefont {Galitski}}, \ and\
  \bibinfo {author} {\bibfnamefont {Piers}\ \bibnamefont {Coleman}},\
  }\bibfield  {title} {\enquote {\bibinfo {title} {Topological kondo
  insulators},}\ }\href {\doibase 10.1103/PhysRevLett.104.106408} {\bibfield
  {journal} {\bibinfo  {journal} {Phys. Rev. Lett.}\ }\textbf {\bibinfo
  {volume} {104}},\ \bibinfo {pages} {106408} (\bibinfo {year}
  {2010})}\BibitemShut {NoStop}%
\bibitem [{\citenamefont {Tsvelik}\ and\ \citenamefont
  {Wiegmann}(1983)}]{tsvelick1983exact}%
  \BibitemOpen
  \bibfield  {author} {\bibinfo {author} {\bibfnamefont {AM}~\bibnamefont
  {Tsvelik}}\ and\ \bibinfo {author} {\bibfnamefont {PB}~\bibnamefont
  {Wiegmann}},\ }\bibfield  {title} {\enquote {\bibinfo {title} {Exact results
  in the theory of magnetic alloys},}\ }\href {\doibase
  10.1080/00018738300101581} {\bibfield  {journal} {\bibinfo  {journal}
  {Advances in Physics}\ }\textbf {\bibinfo {volume} {32}},\ \bibinfo {pages}
  {453--713} (\bibinfo {year} {1983})}\BibitemShut {NoStop}%
\bibitem [{\citenamefont {Werner}\ \emph {et~al.}(2006)\citenamefont {Werner},
  \citenamefont {Comanac}, \citenamefont {de' Medici}, \citenamefont {Troyer},\
  and\ \citenamefont {Millis}}]{Millis2006}%
  \BibitemOpen
  \bibfield  {author} {\bibinfo {author} {\bibfnamefont {Philipp}\ \bibnamefont
  {Werner}}, \bibinfo {author} {\bibfnamefont {Armin}\ \bibnamefont {Comanac}},
  \bibinfo {author} {\bibfnamefont {Luca}\ \bibnamefont {de' Medici}}, \bibinfo
  {author} {\bibfnamefont {Matthias}\ \bibnamefont {Troyer}}, \ and\ \bibinfo
  {author} {\bibfnamefont {Andrew~J.}\ \bibnamefont {Millis}},\ }\bibfield
  {title} {\enquote {\bibinfo {title} {Continuous-time solver for quantum
  impurity models},}\ }\href {\doibase 10.1103/PhysRevLett.97.076405}
  {\bibfield  {journal} {\bibinfo  {journal} {Phys. Rev. Lett.}\ }\textbf
  {\bibinfo {volume} {97}},\ \bibinfo {pages} {076405} (\bibinfo {year}
  {2006})}\BibitemShut {NoStop}%
\bibitem [{\citenamefont {Lu}\ \emph {et~al.}(2013)\citenamefont {Lu},
  \citenamefont {Zhao}, \citenamefont {Weng}, \citenamefont {Fang},\ and\
  \citenamefont {Dai}}]{Dai2013Correlated}%
  \BibitemOpen
  \bibfield  {author} {\bibinfo {author} {\bibfnamefont {Feng}\ \bibnamefont
  {Lu}}, \bibinfo {author} {\bibfnamefont {JianZhou}\ \bibnamefont {Zhao}},
  \bibinfo {author} {\bibfnamefont {Hongming}\ \bibnamefont {Weng}}, \bibinfo
  {author} {\bibfnamefont {Zhong}\ \bibnamefont {Fang}}, \ and\ \bibinfo
  {author} {\bibfnamefont {Xi}~\bibnamefont {Dai}},\ }\bibfield  {title}
  {\enquote {\bibinfo {title} {Correlated topological insulators with mixed
  valence},}\ }\href {\doibase 10.1103/PhysRevLett.110.096401} {\bibfield
  {journal} {\bibinfo  {journal} {Phys. Rev. Lett.}\ }\textbf {\bibinfo
  {volume} {110}},\ \bibinfo {pages} {096401} (\bibinfo {year}
  {2013})}\BibitemShut {NoStop}%
\bibitem [{\citenamefont {Weng}\ \emph {et~al.}(2014)\citenamefont {Weng},
  \citenamefont {Zhao}, \citenamefont {Wang}, \citenamefont {Fang},\ and\
  \citenamefont {Dai}}]{Dai2014Topological}%
  \BibitemOpen
  \bibfield  {author} {\bibinfo {author} {\bibfnamefont {Hongming}\
  \bibnamefont {Weng}}, \bibinfo {author} {\bibfnamefont {Jianzhou}\
  \bibnamefont {Zhao}}, \bibinfo {author} {\bibfnamefont {Zhijun}\ \bibnamefont
  {Wang}}, \bibinfo {author} {\bibfnamefont {Zhong}\ \bibnamefont {Fang}}, \
  and\ \bibinfo {author} {\bibfnamefont {Xi}~\bibnamefont {Dai}},\ }\bibfield
  {title} {\enquote {\bibinfo {title} {Topological crystalline kondo insulator
  in mixed valence ytterbium borides},}\ }\href {\doibase
  10.1103/PhysRevLett.112.016403} {\bibfield  {journal} {\bibinfo  {journal}
  {Phys. Rev. Lett.}\ }\textbf {\bibinfo {volume} {112}},\ \bibinfo {pages}
  {016403} (\bibinfo {year} {2014})}\BibitemShut {NoStop}%
\bibitem [{\citenamefont {Kotliar}\ \emph {et~al.}(2006)\citenamefont
  {Kotliar}, \citenamefont {Savrasov}, \citenamefont {Haule}, \citenamefont
  {Oudovenko}, \citenamefont {Parcollet},\ and\ \citenamefont
  {Marianetti}}]{Kotliar2006Electronic}%
  \BibitemOpen
  \bibfield  {author} {\bibinfo {author} {\bibfnamefont {G.}~\bibnamefont
  {Kotliar}}, \bibinfo {author} {\bibfnamefont {S.~Y.}\ \bibnamefont
  {Savrasov}}, \bibinfo {author} {\bibfnamefont {K.}~\bibnamefont {Haule}},
  \bibinfo {author} {\bibfnamefont {V.~S.}\ \bibnamefont {Oudovenko}}, \bibinfo
  {author} {\bibfnamefont {O.}~\bibnamefont {Parcollet}}, \ and\ \bibinfo
  {author} {\bibfnamefont {C.~A.}\ \bibnamefont {Marianetti}},\ }\bibfield
  {title} {\enquote {\bibinfo {title} {Electronic structure calculations with
  dynamical mean-field theory},}\ }\href {\doibase 10.1103/RevModPhys.78.865}
  {\bibfield  {journal} {\bibinfo  {journal} {Rev. Mod. Phys.}\ }\textbf
  {\bibinfo {volume} {78}},\ \bibinfo {pages} {865--951} (\bibinfo {year}
  {2006})}\BibitemShut {NoStop}%
\bibitem [{\citenamefont {Emery}\ and\ \citenamefont
  {Kivelson}(1992)}]{Kivelson1992Mapping}%
  \BibitemOpen
  \bibfield  {author} {\bibinfo {author} {\bibfnamefont {V.~J.}\ \bibnamefont
  {Emery}}\ and\ \bibinfo {author} {\bibfnamefont {S.}~\bibnamefont
  {Kivelson}},\ }\bibfield  {title} {\enquote {\bibinfo {title} {Mapping of the
  two-channel kondo problem to a resonant-level model},}\ }\href {\doibase
  10.1103/PhysRevB.46.10812} {\bibfield  {journal} {\bibinfo  {journal} {Phys.
  Rev. B}\ }\textbf {\bibinfo {volume} {46}},\ \bibinfo {pages} {10812--10817}
  (\bibinfo {year} {1992})}\BibitemShut {NoStop}%
\bibitem [{\citenamefont {Freyer}\ \emph {et~al.}(2018)\citenamefont {Freyer},
  \citenamefont {Attig}, \citenamefont {Lee}, \citenamefont {Paramekanti},
  \citenamefont {Trebst},\ and\ \citenamefont {Kim}}]{Kim2018Two-Stage}%
  \BibitemOpen
  \bibfield  {author} {\bibinfo {author} {\bibfnamefont {Frederic}\
  \bibnamefont {Freyer}}, \bibinfo {author} {\bibfnamefont {Jan}\ \bibnamefont
  {Attig}}, \bibinfo {author} {\bibfnamefont {SungBin}\ \bibnamefont {Lee}},
  \bibinfo {author} {\bibfnamefont {Arun}\ \bibnamefont {Paramekanti}},
  \bibinfo {author} {\bibfnamefont {Simon}\ \bibnamefont {Trebst}}, \ and\
  \bibinfo {author} {\bibfnamefont {Yong~Baek}\ \bibnamefont {Kim}},\
  }\bibfield  {title} {\enquote {\bibinfo {title} {Two-stage multipolar
  ordering in $\mathrm{Pr}{T}_{2}{\mathrm{al}}_{20}$ kondo materials},}\ }\href
  {\doibase 10.1103/PhysRevB.97.115111} {\bibfield  {journal} {\bibinfo
  {journal} {Phys. Rev. B}\ }\textbf {\bibinfo {volume} {97}},\ \bibinfo
  {pages} {115111} (\bibinfo {year} {2018})}\BibitemShut {NoStop}%
\bibitem [{\citenamefont {Furusaki}\ and\ \citenamefont
  {Nagaosa}(1994)}]{Nagaosa1994Kondo}%
  \BibitemOpen
  \bibfield  {author} {\bibinfo {author} {\bibfnamefont {Akira}\ \bibnamefont
  {Furusaki}}\ and\ \bibinfo {author} {\bibfnamefont {Naoto}\ \bibnamefont
  {Nagaosa}},\ }\bibfield  {title} {\enquote {\bibinfo {title} {Kondo effect in
  a tomonaga-luttinger liquid},}\ }\href {\doibase 10.1103/PhysRevLett.72.892}
  {\bibfield  {journal} {\bibinfo  {journal} {Phys. Rev. Lett.}\ }\textbf
  {\bibinfo {volume} {72}},\ \bibinfo {pages} {892--895} (\bibinfo {year}
  {1994})}\BibitemShut {NoStop}%
\bibitem [{\citenamefont {Cassanello}\ and\ \citenamefont
  {Fradkin}(1996)}]{Fradkin1996Kondo}%
  \BibitemOpen
  \bibfield  {author} {\bibinfo {author} {\bibfnamefont {Carlos~R.}\
  \bibnamefont {Cassanello}}\ and\ \bibinfo {author} {\bibfnamefont {Eduardo}\
  \bibnamefont {Fradkin}},\ }\bibfield  {title} {\enquote {\bibinfo {title}
  {Kondo effect in flux phases},}\ }\href {\doibase 10.1103/PhysRevB.53.15079}
  {\bibfield  {journal} {\bibinfo  {journal} {Phys. Rev. B}\ }\textbf {\bibinfo
  {volume} {53}},\ \bibinfo {pages} {15079--15094} (\bibinfo {year}
  {1996})}\BibitemShut {NoStop}%
\bibitem [{\citenamefont {Martin}\ \emph {et~al.}(1997)\citenamefont {Martin},
  \citenamefont {Wan},\ and\ \citenamefont {Phillips}}]{Philip1997Kondo}%
  \BibitemOpen
  \bibfield  {author} {\bibinfo {author} {\bibfnamefont {Ivar}\ \bibnamefont
  {Martin}}, \bibinfo {author} {\bibfnamefont {Yi}~\bibnamefont {Wan}}, \ and\
  \bibinfo {author} {\bibfnamefont {Philip}\ \bibnamefont {Phillips}},\
  }\bibfield  {title} {\enquote {\bibinfo {title} {Size dependence in the
  disordered kondo problem},}\ }\href {\doibase 10.1103/PhysRevLett.78.114}
  {\bibfield  {journal} {\bibinfo  {journal} {Phys. Rev. Lett.}\ }\textbf
  {\bibinfo {volume} {78}},\ \bibinfo {pages} {114--117} (\bibinfo {year}
  {1997})}\BibitemShut {NoStop}%
\bibitem [{\citenamefont {Chakravarty}\ and\ \citenamefont
  {Rudnick}(1995)}]{Sudip1995Kondo}%
  \BibitemOpen
  \bibfield  {author} {\bibinfo {author} {\bibfnamefont {Sudip}\ \bibnamefont
  {Chakravarty}}\ and\ \bibinfo {author} {\bibfnamefont {Joseph}\ \bibnamefont
  {Rudnick}},\ }\bibfield  {title} {\enquote {\bibinfo {title} {Dissipative
  dynamics of a two-state system, the kondo problem, and the inverse-square
  ising model},}\ }\href {\doibase 10.1103/PhysRevLett.75.501} {\bibfield
  {journal} {\bibinfo  {journal} {Phys. Rev. Lett.}\ }\textbf {\bibinfo
  {volume} {75}},\ \bibinfo {pages} {501--504} (\bibinfo {year}
  {1995})}\BibitemShut {NoStop}%
\bibitem [{\citenamefont {Elcoro}\ \emph {et~al.}(2020)\citenamefont {Elcoro},
  \citenamefont {Wieder}, \citenamefont {Song}, \citenamefont {Xu},
  \citenamefont {Bradlyn},\ and\ \citenamefont
  {Bernevig}}]{elcoro2020magnetic}%
  \BibitemOpen
  \bibfield  {author} {\bibinfo {author} {\bibfnamefont {Luis}\ \bibnamefont
  {Elcoro}}, \bibinfo {author} {\bibfnamefont {Benjamin~J}\ \bibnamefont
  {Wieder}}, \bibinfo {author} {\bibfnamefont {Zhida}\ \bibnamefont {Song}},
  \bibinfo {author} {\bibfnamefont {Yuanfeng}\ \bibnamefont {Xu}}, \bibinfo
  {author} {\bibfnamefont {Barry}\ \bibnamefont {Bradlyn}}, \ and\ \bibinfo
  {author} {\bibfnamefont {B~Andrei}\ \bibnamefont {Bernevig}},\ }\bibfield
  {title} {\enquote {\bibinfo {title} {Magnetic topological quantum
  chemistry},}\ }\href {https://arxiv.org/abs/2010.00598} {\bibfield  {journal}
  {\bibinfo  {journal} {arXiv preprint arXiv:2010.00598}\ } (\bibinfo {year}
  {2020})}\BibitemShut {NoStop}%
\bibitem [{\citenamefont {Xu}\ \emph {et~al.}(2020)\citenamefont {Xu},
  \citenamefont {Elcoro}, \citenamefont {Song}, \citenamefont {Wieder},
  \citenamefont {Vergniory}, \citenamefont {Regnault}, \citenamefont {Chen},
  \citenamefont {Felser},\ and\ \citenamefont {Bernevig}}]{xu2020high}%
  \BibitemOpen
  \bibfield  {author} {\bibinfo {author} {\bibfnamefont {Yuanfeng}\
  \bibnamefont {Xu}}, \bibinfo {author} {\bibfnamefont {Luis}\ \bibnamefont
  {Elcoro}}, \bibinfo {author} {\bibfnamefont {Zhi-Da}\ \bibnamefont {Song}},
  \bibinfo {author} {\bibfnamefont {Benjamin~J}\ \bibnamefont {Wieder}},
  \bibinfo {author} {\bibfnamefont {MG}~\bibnamefont {Vergniory}}, \bibinfo
  {author} {\bibfnamefont {Nicolas}\ \bibnamefont {Regnault}}, \bibinfo
  {author} {\bibfnamefont {Yulin}\ \bibnamefont {Chen}}, \bibinfo {author}
  {\bibfnamefont {Claudia}\ \bibnamefont {Felser}}, \ and\ \bibinfo {author}
  {\bibfnamefont {B~Andrei}\ \bibnamefont {Bernevig}},\ }\bibfield  {title}
  {\enquote {\bibinfo {title} {High-throughput calculations of magnetic
  topological materials},}\ }\href
  {https://www.nature.com/articles/s41586-020-2837-0} {\bibfield  {journal}
  {\bibinfo  {journal} {Nature}\ }\textbf {\bibinfo {volume} {586}},\ \bibinfo
  {pages} {702--707} (\bibinfo {year} {2020})}\BibitemShut {NoStop}%
\bibitem [{\citenamefont {Winkler}(2003)}]{winkler_spin-orbit_2003}%
  \BibitemOpen
  \bibfield  {author} {\bibinfo {author} {\bibfnamefont {Roland}\ \bibnamefont
  {Winkler}},\ }\href@noop {} {\emph {\bibinfo {title} {Spin-orbit {Coupling}
  {Effects} in {Two}-{Dimensional} {Electron} and {Hole} {Systems}}}}\
  (\bibinfo  {publisher} {Springer Science \& Business Media},\ \bibinfo {year}
  {2003})\BibitemShut {NoStop}%
\bibitem [{\citenamefont {C{\u{a}}lug{\u{a}}ru}\ \emph
  {et~al.}(2021)\citenamefont {C{\u{a}}lug{\u{a}}ru}, \citenamefont {Chew},
  \citenamefont {Elcoro}, \citenamefont {Regnault}, \citenamefont {Song},\ and\
  \citenamefont {Bernevig}}]{cualuguaru2021general}%
  \BibitemOpen
  \bibfield  {author} {\bibinfo {author} {\bibfnamefont {Dumitru}\ \bibnamefont
  {C{\u{a}}lug{\u{a}}ru}}, \bibinfo {author} {\bibfnamefont {Aaron}\
  \bibnamefont {Chew}}, \bibinfo {author} {\bibfnamefont {Luis}\ \bibnamefont
  {Elcoro}}, \bibinfo {author} {\bibfnamefont {Nicolas}\ \bibnamefont
  {Regnault}}, \bibinfo {author} {\bibfnamefont {Zhi-Da}\ \bibnamefont {Song}},
  \ and\ \bibinfo {author} {\bibfnamefont {B~Andrei}\ \bibnamefont
  {Bernevig}},\ }\bibfield  {title} {\enquote {\bibinfo {title} {General
  construction and topological classification of all magnetic and non-magnetic
  flat bands},}\ }\href {https://arxiv.org/abs/2106.05272} {\bibfield
  {journal} {\bibinfo  {journal} {arXiv preprint arXiv:2106.05272}\ } (\bibinfo
  {year} {2021})}\BibitemShut {NoStop}%
\bibitem [{\citenamefont {Girvin}\ \emph {et~al.}(1986)\citenamefont {Girvin},
  \citenamefont {MacDonald},\ and\ \citenamefont {Platzman}}]{GMP1986}%
  \BibitemOpen
  \bibfield  {author} {\bibinfo {author} {\bibfnamefont {S.~M.}\ \bibnamefont
  {Girvin}}, \bibinfo {author} {\bibfnamefont {A.~H.}\ \bibnamefont
  {MacDonald}}, \ and\ \bibinfo {author} {\bibfnamefont {P.~M.}\ \bibnamefont
  {Platzman}},\ }\bibfield  {title} {\enquote {\bibinfo {title} {Magneto-roton
  theory of collective excitations in the fractional quantum hall effect},}\
  }\href {\doibase 10.1103/PhysRevB.33.2481} {\bibfield  {journal} {\bibinfo
  {journal} {Phys. Rev. B}\ }\textbf {\bibinfo {volume} {33}},\ \bibinfo
  {pages} {2481--2494} (\bibinfo {year} {1986})}\BibitemShut {NoStop}%
\bibitem [{\citenamefont {Parameswaran}\ \emph {et~al.}(2013)\citenamefont
  {Parameswaran}, \citenamefont {Roy},\ and\ \citenamefont
  {Sondhi}}]{Sondhi2013Fractional}%
  \BibitemOpen
  \bibfield  {author} {\bibinfo {author} {\bibfnamefont {Siddharth~A.}\
  \bibnamefont {Parameswaran}}, \bibinfo {author} {\bibfnamefont {Rahul}\
  \bibnamefont {Roy}}, \ and\ \bibinfo {author} {\bibfnamefont {Shivaji~L.}\
  \bibnamefont {Sondhi}},\ }\bibfield  {title} {\enquote {\bibinfo {title}
  {Fractional quantum hall physics in topological flat bands},}\ }\href
  {\doibase https://doi.org/10.1016/j.crhy.2013.04.003} {\bibfield  {journal}
  {\bibinfo  {journal} {Comptes Rendus Physique}\ }\textbf {\bibinfo {volume}
  {14}},\ \bibinfo {pages} {816--839} (\bibinfo {year} {2013})},\ \bibinfo
  {note} {topological insulators / Isolants topologiques}\BibitemShut {NoStop}%
\bibitem [{\citenamefont {Roy}(2014)}]{Roy2014BandGeometry}%
  \BibitemOpen
  \bibfield  {author} {\bibinfo {author} {\bibfnamefont {Rahul}\ \bibnamefont
  {Roy}},\ }\bibfield  {title} {\enquote {\bibinfo {title} {Band geometry of
  fractional topological insulators},}\ }\href {\doibase
  10.1103/PhysRevB.90.165139} {\bibfield  {journal} {\bibinfo  {journal} {Phys.
  Rev. B}\ }\textbf {\bibinfo {volume} {90}},\ \bibinfo {pages} {165139}
  (\bibinfo {year} {2014})}\BibitemShut {NoStop}%
\end{thebibliography}%

\clearpage

\onecolumngrid
\begin{center}
\textbf{Supplementary Materials}
\end{center}

\stepcounter{SMmark}
\stepcounter{page}

\renewcommand{\thefigure}{S\arabic{figure}}

\renewcommand{\thetable}{S\arabic{table}}

\renewcommand{\thesection}{S\arabic{section}}

\renewcommand{\theequation}{S\arabic{equation}}



\tableofcontents

\clearpage

\section{Bistritzer-MacDonald model} \label{sec:BM-model}
Here we briefly review the Bistritzer-MacDonald (BM) model \cite{bistritzer_moire_2011} of magic-angle twisted bilayer graphene (MATBG). 
Readers may refer to the supplementary materials of Ref.~\cite{song_all_2019} for more details.

\subsection{Basis and single-particle Hamiltonian}
\begin{figure}
\centering
\includegraphics[width=0.6\linewidth]{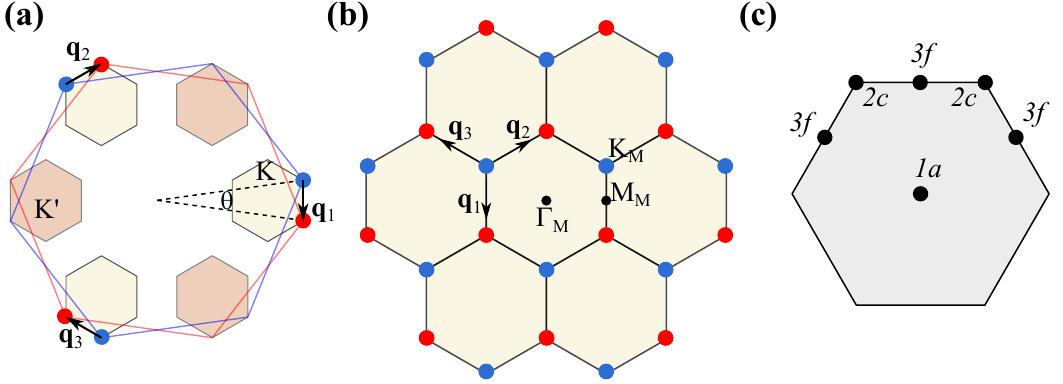}
\caption{The moir\'e BZ and the $\QQ$ lattice. (a) The blue and red hexagons represent the BZ's of the top and bottom graphene layers, respectively. The blue and red dots represent the Diract points (K momentum) from the two layers. The vectors $\qq_1,\qq_2,\qq_3$ connecting the Dirac points of the two layers span the moir\'e BZ in the valley K (shaded with light yellow). Similarly, the Dirac points at the K$^\prime$ momentum form another moir\'e BZ in the valley K$^\prime$ (shaded with light orange). (b) We denote the blue and red $\QQ$ lattices as $\mathcal{Q}_+$ and $\mathcal{Q}_{-}$ respectively. In the BM model, the states in the valley K (K$^\prime$) of the top layer and in the valley K$^\prime$ (K) of the bottom layer contribute to the lattice $\mathcal{Q}_+$ ($\mathcal{Q}_-$).
(c) The unit cell in real space, where the maximal Wyckoff positions $1a$, $2c$, $3f$ are marked by the black dots. 
}
\label{fig:QQ}
\end{figure}

As shown in \cref{fig:QQ}(a), the low energy states around the Dirac points at the K momenta of the two graphene layers form the moir\'e Brillouin zone (BZ) in the valley K. Similarly, the low energy states around the Dirac points at the K$^\prime$ momenta of the two layers form another moir\'e BZ in the valley K$^\prime$.
Thus in the BM model there are two independent sets of basis from the two valleys.
We use the index $\eta$ ($=+$ for K and $-$ for K$^\prime$) to label the two graphene valleys.
We denote the basis as $c_{\kk,\QQ,\alpha,\eta,s}$, where where $\kk$ is a momentum in the moir\'e Brillouin zone (BZ), $\QQ$ takes values in the lattice shown in \cref{fig:QQ}(b), $\alpha=1,2$ represents the graphene sublattice, $\eta=\pm$ represents the graphene valley, and $s=\up\down$ is the spin index. 
There are two types of Q lattices: the blue lattice $\mathcal{Q}_+=\{ \qq_2 + n_1 \bb_{M1} + n_2 \bb_{M2}\ |\ n_{1,2}\in \mathbb{Z} \}$ and the red lattice $\mathcal{Q}_- = \{- \qq_2 + n_1 \bb_{M1} + n_2 \bb_{M2}\ |\ n_{1,2}\in \mathbb{Z} \}$, where $\bb_{M1},\bb_{M2}$ are Moir\'e reciprocal lattice basis
\begin{equation}
    \bb_{M1} = \qq_2 - \qq_1 ,\qquad 
    \bb_{M_2} = \qq_3 - \qq_1\ . 
\end{equation}
For $\QQ \in \mathcal{Q}_+$, the basis is defined as 
\begin{equation} \label{eq:BM-basis1}
(\QQ\in \mathcal{Q}_+)\qquad
c_{\kk,\QQ,\alpha,\eta,s}^\dagger = 
\begin{cases}\displaystyle
\frac1{\sqrt{N_{\rm tot}}} \sum_{\RR \in \mrm{top}} e^{i(\KK_+ + \kk-\QQ) \cdot (\RR + \tt_\alpha) }  c_{\RR,\alpha,s}^\dagger \qquad & \text{if}\; \eta=+ \\
\displaystyle
\frac1{\sqrt{N_{\rm tot}}} \sum_{\RR' \in \mrm{bottom}} e^{i(-\KK_- + \kk-\QQ) \cdot (\RR' + \tt_\alpha') }  c_{\RR',\alpha,s}^\dagger \qquad & \text{if}\; \eta=-
\end{cases}\ ,
\end{equation}
and for $\QQ \in \mathcal{Q}_-$, the basis is defined as 
\begin{equation}\label{eq:BM-basis2}
(\QQ\in \mathcal{Q}_-)\qquad
c_{\kk,\QQ,\alpha,\eta,s}^\dagger = 
\begin{cases}\displaystyle
\frac1{\sqrt{N_{\rm tot}}} \sum_{\RR' \in \mrm{bottom}} e^{i(\KK_- + \kk-\QQ) \cdot (\RR'+ \tt_\alpha') }  c_{\RR',\alpha,s}^\dagger \qquad & \text{if}\; \eta=+ \\
\displaystyle
\frac1{\sqrt{N_{\rm tot}}} \sum_{\RR \in \mrm{top}} e^{i(-\KK_+ + \kk-\QQ) \cdot (\RR + \tt_\alpha) }  c_{\RR,\alpha,s}^\dagger \qquad & \text{if}\; \eta=-
\end{cases}\ . 
\end{equation}
Here $N_{\rm tot}$ is the number of graphene unit cells in each layer, $\RR$ and $\RR'$ indexes graphene lattices in the top and bottom layers, respectively, $\tt_\alpha$ and $\tt_\alpha'$ are the sublattice vectors of the two layers, respectively, $\KK_+$  and $\KK_-$ are the K momentum of the top and bottom layers, respectively, and $c_{\RR,\alpha,s}$ ($c_{\RR',\alpha,s}$) is the fermion operator with spin $s$ at the atom site $\RR+\tt_\alpha$ ($\RR'+\tt_\alpha'$).
The $\mathcal{Q}_{+}$ ($\mathcal{Q}_{-}$) lattice is defined in such a way that $\eta \KK_+ +\kk - \QQ$ ($\eta \KK_- +\kk - \QQ$) with $\QQ\in \mathcal{Q}_+$ ($\QQ\in \mathcal{Q}_-$) is the Dirac point position $\eta \KK_+$ ($\eta \KK_-$) when $\kk$ equals to the high symmetry point $\eta K_M$ ($\eta K_M'$) of the moir\'e BZ, as sown in \cref{fig:QQ}(b).

The BM model is given by 
\begin{equation}
\hH_{\rm BM} = \sum_{\eta s} \sum_{\kk \in {\rm MBZ}} \sum_{\alpha \alpha'} \sum_{\QQ,\QQ'} h_{\QQ\alpha,\QQ'\alpha'}^{(\eta)}(\kk) c_{\kk,\QQ,\alpha,\eta,s}^\dagger c_{\kk,\QQ',\alpha',\eta,s}
\end{equation}
where the single particle Hamiltonian reads 
\begin{equation}\label{eq:BM-Hamiltonian}
h_{\QQ\alpha,\QQ'\alpha'}^{(+)}(\kk) = v_F (\kk-\QQ)\cdot \boldsymbol{\sigma} \delta_{\QQ,\QQ'} 
+ \sum_{j=1}^3 [T_j]_{\alpha\alpha'} \delta_{\QQ,\QQ'\pm \qq_j},\qquad 
h_{\QQ\alpha,\QQ'\alpha'}^{(-)}(\kk) = h_{-\QQ\alpha,-\QQ'\alpha'}^{(+)*}(-\kk), 
\end{equation}
\begin{equation}
T_j = w_0 \sigma_0 + w_1 \sigma_x \cos \frac{2\pi(j-1)}3 + w_1 \sigma_y \sin \frac{2\pi(j-1)}3 \ .
\end{equation}
Here $\qq_{j}$ are shown in \cref{fig:QQ}, and $w_0$ and $w_1$ are the interlayer couplings in the AA-stacking and AB-stacking regions, respectively. 
The length of $\qq_j$ is determined by the twist angle $\theta$, \ie $|\qq_j|=k_\theta=2|\KK|\sin\frac{\theta}2$.
In this work, we adopt the parameters $v_F = 5.944\mrm{eV\cdot\mathring{A}}$, $|\KK|=1.703\mrm{\mathring{A}^{-1}}$, $w_1=110\mrm{meV}$, $\theta=1.05^\circ$. 
The relation $h_{\QQ\alpha,\QQ'\alpha'}^{(-)}(\kk) = h_{-\QQ\alpha,-\QQ'\alpha'}^{(+)*}(-\kk)$ is due to the time-reversal symmetry that transform the two valleys to each other. 
The single-particle Hamiltonian (upon to a unitary transformation \cite{song_all_2019}) is periodic with respect to the reciprocal lattice vectors $\bb_{M1} = \qq_2 -\qq_1 = k_\theta(\frac{\sqrt{3}}{2},\frac{3}2)$, $\bb_{M2} = \qq_3 -\qq_1 = k_\theta(-\frac{\sqrt{3}}{2},\frac{3}2)$.
The moir\'e BZ and high symmetry momenta are defined in \cref{fig:QQ}(b). 
The real space unit cell is generated by $\aa_{M1} = \frac{2\pi}{3k_\theta}(\sqrt{3},1)$, $\aa_{M2} = \frac{2\pi}{3k_\theta}(-\sqrt{3},1)$, which satisfy $\aa_{M i}\cdot\bb_{Mj} = 2\pi \delta_{ij}$. 
Area of the moir\'e unit cell is given by $\Omega_0 = \aa_{M1}\times\aa_{M2} = \frac{8\pi^2}{3\sqrt3 k_\theta^2}$. 
Maximal Wyckoff positions in the unit cell are shown in \cref{fig:QQ}(c). 
The $1a$ and $2c$ positions correspond to the AA-stacking and AB-stacking regions, respectively. 

\begin{figure}
\centering
\includegraphics[width=0.9\linewidth]{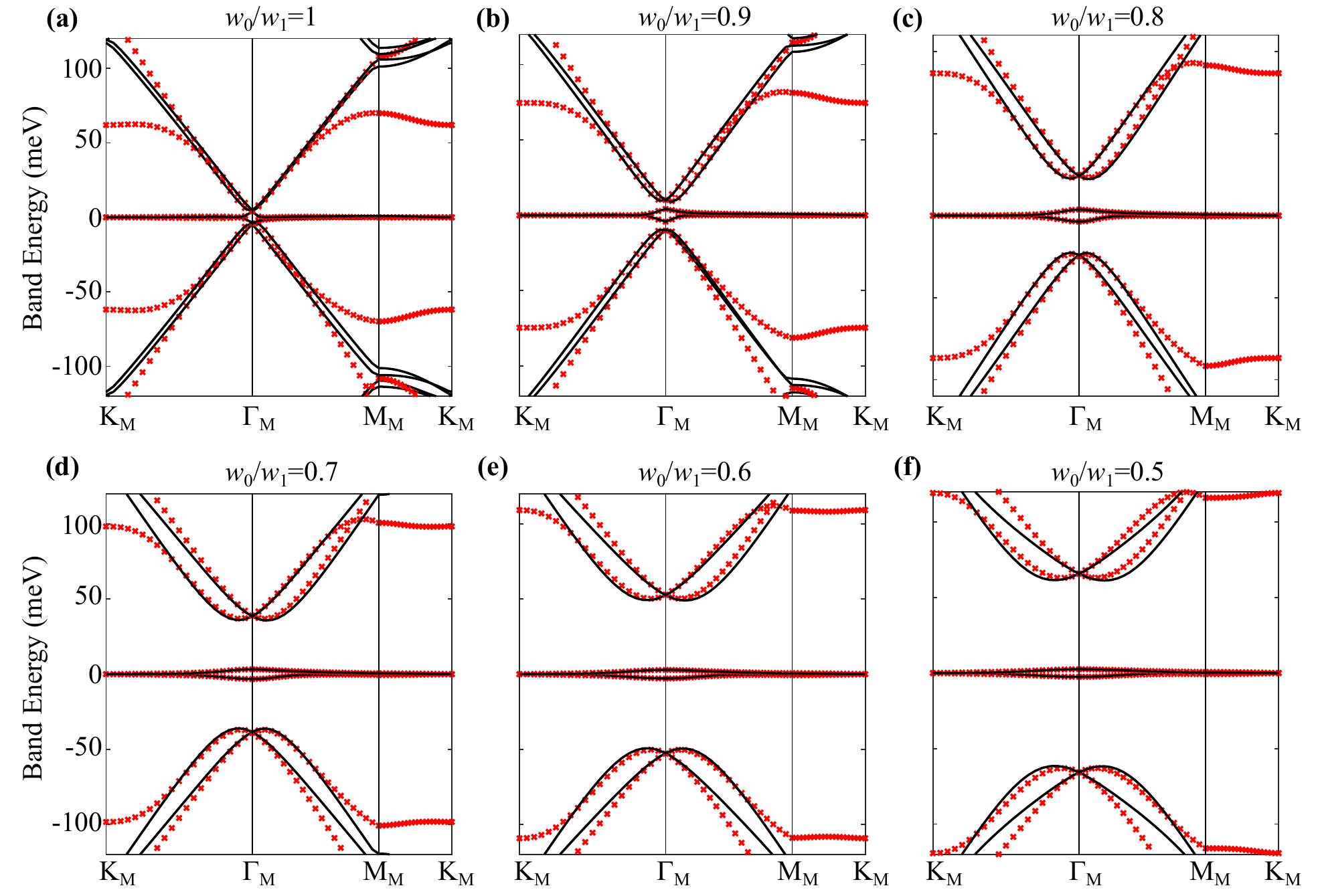}
\caption{The band structure in the valley $\eta=+$ of MATBG. In (a)-(f) the band structures with the inter-layer couplings $w_0/w_1$=1, 0.9, 0.8, 0.7, 0.6, 0.5 are plotted. The red crossings are energy eigenvalues of the BM model, and the black lines are bands of the effective model. The other parameters of the BM model are given by  $v_F = 5.944\mrm{eV\cdot\mathring{A}}$, $|\KK|=1.703\mrm{\mathring{A}^{-1}}$, $w_1=110\mrm{meV}$, $\theta=1.05^\circ$. 
The parameters of the effective model for different $w_0/w_1$ are given in \cref{tab:H0parameters}.
We have set $v_\star''=B=C=0$ for simplicity. }
\label{fig:bands_free}
\end{figure}

We plot the band structures in the valley $\eta=+$ for different $w_0$ in \cref{fig:bands_free}. 

The fact that the band structure is labeled by $\kk$ (in the moir\'e BZ) implies that $\kk$ labels the eigenvalues of translation operators of the moir\'e lattice.  
We hence {\it define} the translation operator $T_\RR$ as 
\begin{equation} \label{eq:moire-translation-momentum-basis}
    T_{\RR} c_{\kk,\QQ,\alpha,\eta,s}^\dagger T_{\RR}^{-1}
= e^{-i\kk\cdot\RR} c_{\kk,\QQ,\alpha,\eta,s}^\dagger\ ,
\end{equation}
where $\RR = n_1 \aa_{M1} + n_2 \aa_{M2}$, with $n_{1,2}\in \mathbb{Z}$, is a moir\'e lattice vector. 
One should not confuse $T_\RR$ with the time-reversal symmetry ($T$) defined in \cref{sec:BM-sym}. 
We now verify \cref{eq:moire-translation-momentum-basis} at the commensurate twist angles, where $\aa_{M1}$ and $\aa_{M2}$ are integer linear combinations of the microscopic graphene lattice vectors $\aa_1,\aa_2$. 
$T_{\RR}$ can be defined as translation acting on the atomic orbitals: $\RR'\to \RR'+\RR$. 
Then there is 
\begin{align}
T_{\RR} c_{\kk,\QQ,\alpha,\eta,s}^\dagger T_{\RR}^{-1} = \frac1{\sqrt{N_{\rm tot}}} 
    \sum_{\RR' \in l} e^{i(\eta \KK_l + \kk -\QQ)\cdot (\RR'+\tt_\alpha)} c_{\RR'+\RR,\alpha,s}^\dagger 
= \frac1{\sqrt{N_{\rm tot}}} 
\sum_{\RR' \in l} e^{-i(\eta\KK_{l}+\kk-\QQ)\cdot\RR} e^{i(\eta \KK_l + \kk -\QQ)\cdot (\RR'+\tt_\alpha)} c_{\RR',\alpha,s}^\dagger \ ,
\end{align}
where $l=\eta$ ($-\eta$) for $\QQ \in \mathcal{Q}_+$ ($\mathcal{Q}_-$). 
For commensurate angle, $\eta\KK_{l}-\QQ$ is a moir\'e reciprocal lattice and hence $e^{-i(\eta\KK_{l}+\kk-\QQ)\cdot\RR} = e^{-i\kk\cdot\RR}$. 
Then \cref{eq:moire-translation-momentum-basis} is justified at commensurate angles.
We emphasize that \cref{eq:moire-translation-momentum-basis} is also well-defined even at non-commensurate angles.

For later convenience, here we define our convention for Fourier transformation.
Let $N$, $\Omega_0$, $\Omega_{\rm tot} = N\Omega_0$ be the number of moir\'e unit cells, area of each moir\'e unit cell, and the total area of the sample. 
Then, for $\kk,\kk'$ in the moir\'e BZ and $\QQ,\QQ'\in \mcl{Q}_{l\eta}$ with $l\eta=+$ or $-$, integral over $\rr$ leads to the momentum conservation 
\begin{equation}\label{eq:Fourier-1}
  \text{for}\quad \QQ,\QQ'\in \mcl{Q}_{l\eta}\quad \int d^2\rr e^{(\kk-\kk'-\QQ+\QQ')\cdot\rr} = \Omega_{\rm tot} \delta_{\kk\kk'}\delta_{\QQ,\QQ'}\ .
\end{equation}
The completeness relation of the plane waves is 
\begin{equation}\label{eq:Fourier-2}
  \frac1{\Omega_{\rm tot}} \sum_{\kk} \sum_{\QQ \in \mcl{Q}_{l\eta}} e^{i(\kk-\QQ)\cdot\rr} = \delta(\rr),\qquad \text{for}\; l\eta=\pm. 
\end{equation}

We introduce a continuous real space basis for the BM model 
\begin{equation} \label{eq:continuous-basis}
c_{l,\alpha,\eta,s}^\dagger (\rr) = \frac1{\sqrt{\Omega_{\rm tot}}} \sum_{\kk\in {\rm MBZ}} \sum_{\QQ \in \mathcal{Q}_{l\eta}} e^{-i(\kk -\QQ)\cdot\rr } c_{\kk,\QQ,\alpha,\eta,s}^\dagger\ ,
\end{equation}
where $\Omega_{\rm tot} = N \Omega_0$ is the total area of the sample in consideration, $N$ is the number of Moir\'e unit cells, $\Omega_0$ is the area of the Moir\'e unit cell, $l=\pm$ is the layer index.
We can understand the {\it continuous} variable $\rr$ as the coarse-grained graphene lattice vector.
The graphene sublattice ($\alpha$) and valley ($\eta$) are now regarded as inner degrees of freedom. 
At each $\rr$ $c_{l,\alpha,\eta,s}^\dagger (\rr)$ has sixteen components from two layers, two Graphene sublattice, two valleys, and two spins.  
Unlike the usual real space basis, which just changes its position under a translation operation, $c_{l,\alpha,\eta,s}^\dagger (\rr)$ also gains a phase factor under translation. 
According to \cref{eq:moire-translation-momentum-basis}, there is 
\begin{align} 
T_{\RR} c_{l,\alpha,\eta,s}^\dagger (\rr) T_{\RR}^{-1}
= \frac1{\sqrt{\Omega_{\rm tot}}} \sum_{\kk\in {\rm MBZ}} \sum_{\QQ \in \mathcal{Q}_{l\eta}}
    e^{-i(\kk -\QQ)\cdot\rr } T_\RR c_{\kk,\QQ,\alpha,\eta,s}^\dagger T_{\RR}^{-1} 
= \frac1{\sqrt{\Omega_{\rm tot}}} \sum_{\kk\in {\rm MBZ}} \sum_{\QQ \in \mathcal{Q}_{l\eta}}
e^{-i(\kk -\QQ)\cdot\rr } e^{-i\kk\cdot\RR} c_{\kk,\QQ,\alpha,\eta,s}^\dagger \ .
\end{align}
For convenience, we define the momenta
\begin{equation}
    \Delta \KK_l = \begin{cases}
        \qq_2, \qquad & l=+\\
        -\qq_3,\qquad & l=-
    \end{cases}\ .
\end{equation}
Notice that $\eta\DKK_l - \QQ$ ($\QQ\in\mathcal{Q}_{\eta l}$) is a moir\'e reciprocal lattice and hence $e^{-i(\eta\DKK_l-\QQ)\cdot\RR}=1$. 
We hence have 
\begin{equation}\label{eq:continuous-basis-translation}
T_{\RR} c_{l,\alpha,\eta,s}^\dagger (\rr) T_{\RR}^{-1}
= \frac1{\sqrt{\Omega_{\rm tot}}} \sum_{\kk\in {\rm MBZ}} \sum_{\QQ \in \mathcal{Q}_{l\eta}}
    e^{-i\eta\DKK_l\cdot\RR} e^{-i(\kk -\QQ)\cdot(\rr+\RR) } c_{\kk,\QQ,\alpha,\eta,s}^\dagger 
= e^{-i\eta\DKK_l\cdot\RR} c_{l,\alpha,\eta,s}^\dagger (\rr+\RR)\ . 
\end{equation}
One may remove the factor $e^{i\eta\DKK_l\cdot\RR}$ by redefining the basis as $c_{l,\alpha,\eta,s}^\dagger (\rr) \to e^{i\eta\DKK_l\cdot\rr} c_{l,\alpha,\eta,s}^\dagger (\rr)$.
However, this redefining will complicate the representation of crystalline symmetries (\cref{sec:BM-sym}).
In this work we will stick to the definition \cref{eq:continuous-basis} of the real space basis.

\subsection{Discrete symmetries of the BM model} \label{sec:BM-sym}
The single-particle Hamiltonian in each valley has the symmetry of the magnetic space group $P6'2'2$ (\# 177.151 in the BNS setting), which is generated by $C_{2z}T$, $C_{3z}$, $C_{2x}$ and translation symmetries. 
Since the two valleys are related by time-reversal symmetry $T$, the total system also has the $C_{2z}$ symmetry (product of $C_{2z}T$ and $T$).
The full crystalline symmetries of the two valleys form the space group $P622$ (\#177), which is generated by $C_{6z}$, $C_{2x}$, and translation symmetries. 
We write the symmetry action on the fermion operators as 
\begin{equation}
\hg c_{\kk,\QQ,\alpha,\eta,s}^\dagger \hg^{-1} = \sum_{\QQ' \alpha'\eta'} c_{g\kk,\QQ',\alpha',\eta',s}^\dagger D_{\QQ'\alpha'\eta',\QQ\alpha\eta} (g) \ . 
\end{equation}
The $D$ matrices of the time-reversal symmetry and the single-valley symmetries are given by  
\begin{equation} \label{eq:D-T-C3z}
D_{\QQ'\alpha'\eta',\QQ\alpha\eta} (T) = \delta_{\QQ',-\QQ} [\sigma_0]_{\alpha'\alpha} [\tau_x]_{\eta'\eta},\qquad 
D_{\QQ'\alpha'\eta',\QQ\alpha\eta} (C_{3z}) = \delta_{\QQ',C_{3z}\QQ} [e^{i\frac{2\pi}3 \sigma_z \tau_z}]_{\alpha'\eta',\alpha\eta} 
\end{equation}
\begin{equation} \label{eq:D-C2x-C2zT}
D_{\QQ'\alpha'\eta',\QQ\alpha\eta} (C_{2x}) = \delta_{\QQ',C_{2x}\QQ} [\sigma_x]_{\alpha'\alpha} [\tau_0]_{\eta'\eta},\qquad 
D_{\QQ'\alpha'\eta',\QQ\alpha\eta} (C_{2z}T) = \delta_{\QQ',\QQ} [\sigma_x]_{\alpha'\alpha} [\tau_0]_{\eta'\eta},
\end{equation}
where $\sigma_{x,y,z}$ ($\sigma_0$), $\tau_{x,y,z}$ ($\tau_0$) are Pauli (identity) matrices in the sublattice and valley spaces, and $\zeta_{\QQ}=\pm1$ for $\QQ \in \mathcal{Q}_{\pm}$.

Besides the crystalline symmetries, the BM model also has a unitary particle-hole symmetry $P$ that anti-commutes with the single-particle Hamiltonian. 
$P$ transforms $\kk$ to $-\kk$ and the corresponding $D$ matrix reads 
\begin{equation}
D_{\QQ'\alpha'\eta',\QQ\alpha\eta} (P) = \zeta_{\QQ} \delta_{\QQ',-\QQ} [\sigma_0]_{\alpha'\alpha} [\tau_z]_{\eta'\eta}\ .
\end{equation}
(Unlike the $P$ symmetry defined in Refs.~\cite{Biao-TBG3,Biao-TBG4,TBG5}, where $P$ is valley-independent, here $P$ is chosen have opposite signs in the two valleys. The two definitions are equivalent because the Hamiltonian respects the valley-U(1) and one can redefine the particle-hole symmetry as $P\to \tau_z P$.)
In the first-quantized formalism the algebra between $P$ and other symmetries are given by
\begin{equation} \label{eq:P-algebra}
P^2 = -1,\qquad 
[P,T]=0,\qquad [P,C_{3z}]=0,\qquad \{P,C_{2x}\}=0,\qquad    [P,C_{2z}T]=0\ . 
\end{equation}
The particle-hole symmetry will be broken if the $\theta$-dependence of the single-layer Hamiltonian or quadratic terms in $\kk$ of the single-layer Hamiltonian are taken into account.

If, furthermore, $w_0=0$, the model acquires an effective chiral symmetry $C$ that anti-commute with the single-particle Hamiltonian \cite{tarnopolsky_origin_2019}. 
$C$ is referred to as the first chiral symmetry in Refs.~\cite{Song-TBG2,Biao-TBG3}. 
$C$ leaves $\kk$ invariant and its $D$ matrix reads 
\begin{equation}
D_{\QQ'\alpha'\eta',\QQ\alpha\eta} (C) = \delta_{\QQ',\QQ} [\sigma_z]_{\alpha'\alpha} [\tau_0]_{\eta'\eta}\ .
\end{equation}
In the first-quantized formalism the algebra between $C$ and other symmetries are given by
\begin{equation} \label{eq:C-first-algebra}
C^2 =1,\qquad 
[C,T]=0,\qquad    [C,C_{3z}]=0,\qquad \{C,C_{2x}\}=0,\qquad \{C,C_{2z}T\}=0,\qquad 
[C,P]=0 \ .  
\end{equation}

In the end we derive the symmetry actions on the continuous real space basis. 
For unitary or anti-unitary operators, there are 
\begin{equation}
\hg c_{l,\alpha,\eta,s}^\dagger(\rr) \hg^{-1} = \sum_{l' \alpha'\eta'} c_{l',\alpha',\eta',s}^\dagger(g\rr) D_{l'\alpha'\eta',l \alpha \eta} (g) \ .
\end{equation}
Notice that $P\rr = -\rr$. 
($P$ is an inversion-like operation that interchanges the two layers but leave the sublattice invariant.)
Using the $D$ matrices in momentum space and the definition of $c_{l,\alpha,\eta,s}(\rr)$ (\cref{eq:continuous-basis}), we can derive the $D$ matrices for the real space continuous basis 
\begin{equation}\label{eq:D-real-T-C3z}
D_{l'\alpha'\eta',l\alpha\eta} (T) = [\varrho_0]_{l'l} [\sigma_0]_{\alpha'\alpha} [\tau_x]_{\eta'\eta},\qquad 
D_{l'\alpha'\eta',l\alpha\eta} (C_{3z}) = [\varrho_0]_{l'l} [e^{i\frac{2\pi}3 \sigma_z \tau_z}]_{\alpha'\eta',\alpha\eta} \ ,
\end{equation}
\begin{equation}\label{eq:D-real-C2x-C2zT}
D_{l'\alpha'\eta',l\alpha\eta} (C_{2x}) = [\varrho_x]_{l'l} [\sigma_x]_{\alpha'\alpha} [\tau_0]_{\eta'\eta},\qquad 
D_{l'\alpha'\eta',l\alpha\eta} (C_{2z}T) = [\varrho_0]_{l'l} [\sigma_x]_{\alpha'\alpha} [\tau_0]_{\eta'\eta}\ ,
\end{equation}
\begin{equation} \label{eq:D-real-P-C}
D_{l'\alpha'\eta',l\alpha\eta} (P) = [-i\varrho_y]_{l',l} [\sigma_0]_{\alpha'\alpha} [\tau_0]_{\eta'\eta},\qquad 
D_{l'\alpha'\eta',l\alpha\eta} (C) = [\varrho_0]_{l',l} [\sigma_z]_{\alpha'\alpha} [\tau_0]_{\eta'\eta}\ . 
\end{equation}
Here $\sigma_{x,y,z}$ ($\sigma_0$), $\tau_{x,y,z}$ ($\tau_0$), $\varrho_{x,y,z}$ ($\varrho_0$) are Pauli (identity) matrices in the space of sublattice, valley, and layer, respectively.
We emphasize that, unlike the momentum space basis where $D(P)$ is proportional to $\tau_z$, the real space basis has $D$ proportional to $\tau_0$. 
Here we take the particle-hole symmetry as an example to show how the $D$ matrices are derived. 
By definition, there is 
\begin{align}
& P c_{l,\alpha,\eta,s}^\dagger(\rr) P^{-1}
= \frac1{\sqrt{\Omega_{\rm tot}}} \sum_{\kk \in {\rm MBZ}} \sum_{\QQ \in \mathcal{Q}_{l\eta}} 
    e^{-i(\kk-\QQ)\cdot\rr} P c_{\kk,\QQ,\alpha,\eta,s}^\dagger P^{-1} \nono\\
=& \frac1{\sqrt{\Omega_{\rm tot}}} \sum_{\kk \in {\rm MBZ}} \sum_{\QQ \in \mathcal{Q}_{l\eta}}
e^{-i(\kk-\QQ)\cdot\rr}  c_{-\kk,-\QQ,\alpha,\eta,s}^\dagger \ \eta \zeta_{\QQ} 
\end{align}
We can replace $-\QQ$ by $\QQ'$ with $\QQ' \in \mathcal{Q}_{-l\eta}$.
Relabeling $\QQ'$ and $\kk$ as $\QQ$ and $-\kk$, respectively, we obtain 
\begin{align}
& P c_{l,\alpha,\eta,s}^\dagger(\rr) P^{-1}
= \frac1{\sqrt{\Omega_{\rm tot}}} \sum_{\kk \in {\rm MBZ}} \sum_{\QQ \in \mathcal{Q}_{-l\eta}}
e^{-i( \kk - \QQ)\cdot (-\rr)}  c_{\kk,\QQ,\alpha,\eta,s}^\dagger  \eta \zeta_{-\QQ} 
\end{align}
Notice that, according to \cref{eq:BM-basis1,eq:BM-basis2}, for $\QQ \in \mathcal{Q}_{-l\eta}$, the state $c_{\kk,\QQ,\alpha,\eta,s}^\dagger$ is in the layer $-l$, and the factor $\eta \zeta_{-\QQ}$ equals to $l$.
We hence obtain 
\begin{equation}
P c_{l,\alpha,\eta,s}^\dagger(\rr) P^{-1} =  l c_{-l,\alpha,\eta,s}^\dagger(-\rr) \ , 
\end{equation}
consistent with \cref{eq:D-real-P-C}. 

For later convenience, we also derive the $D$ matrices of $C_{2z}P$ for the momentum space basis and real space basis as 
\begin{equation}
D_{\QQ'\alpha'\eta',\QQ\alpha\eta} (C_{2z}P) = -i\zeta_{\QQ} \delta_{\QQ',\QQ} [\sigma_x]_{\alpha'\alpha} [\tau_y]_{\eta'\eta},\qquad 
D_{l'\alpha'\eta',l\alpha\eta} (C_{2z}P) = [-i\varrho_y]_{l',l} [\sigma_x]_{\alpha'\alpha} [\tau_x]_{\eta'\eta} \ ,
\label{eq:D-C2P}
\end{equation}
respectively. 

\section{The single-particle Hamiltonian of the topological heavy fermion model}

\subsection{Maximal localized Wannier functions for the local orbitals}\label{sec:Wannier}


We first summarize the representation analysis of the Wannier obstruction given in Ref.~\cite{song_all_2019}.
As proved in Ref.~\cite{song_all_2019}, due to the particle-hole symmetry, the middle two bands in each valley must form the irreducible co-representations (dubbed as irreps) 
\begin{equation} \label{eq:irreps-top-band}
    \Gamma_1 \oplus \Gamma_2;\ M_1 \oplus M_2;\ K_2K_3
\end{equation}
of the magnetic space group $P6'2'2$ as long as they are gapped from higher and lower energy bands. 
One can find the definitions of the irreps in \cref{tab:irreps-MSG} or using the \href{https://www.cryst.ehu.es/cgi-bin/cryst/programs/corepresentations.pl}{Corepresentations} toolkit on the Bilbao Crystallographic Server \cite{elcoro2020magnetic,xu2020high}.
In \cref{tab:EBR-magSG} we tabulate all the elementary band representations (EBRs), \ie the minimal local orbitals, and their corresponding irreps in momentum space.
One can also find the definition of EBRs using the \href{https://www.cryst.ehu.es/cgi-bin/cryst/programs/mbandrep.pl}{MBANDREP} toolkit on the Bilbao Crystallographic Server \cite{elcoro2020magnetic,xu2020high}.
There are only six allowed two-band band representations, \ie $2 [A_1]_a \up G$, $2 [A_2]_a \up G$, $ [A_1\oplus A_2]_a \up G$, $[E]_a \up G$, $[A_1]_c \up G$, and $[A_2]_c\up G$, while none of them has the irreps in \cref{eq:irreps-top-band}. 
(The notation $[\rho]_w \up G$ represents the EBR of the magnetic space group $G$ induced from the representation $\rho$ at the Wyckoff possible $w$.)
Thus the middle two bands are not consistent with any local orbitals and must be topological. 

\begin{table*}
\begin{tabular}{lrrr|lrr|lrr}
\hline 
& $\Gamma_{1}$ & $\Gamma_{2}$ & $\Gamma_{3}$ &  & $M_1$ & $M_2$ &  & $K_1$ & $K_2K_3$ \\
\hline 
$E$ & 1 & 1 & 2 &        $E$ & 1 & 1 &        $E$ & 1 & 2\\
$2C_{3}$ & 1 & 1 & -1 &  $C_{2}^\prime$ & 1 & -1 & $C_{3}$ & 1 & -1 \\
$3C_{2}^\prime$ & 1 & -1 & 0 &     &   &   &       $C_{3}^{-1}$ & 1 & -1 \\
\hline
\end{tabular}
\caption{\label{tab:irreps-MSG} Character table of irreps at high symmetry momenta in magnetic space group $P6^\prime2^\prime2$ (\#177.151 in BNS settings). For the little group of $\Gamma$, $E$, $2C_3$, and $3C_2^\prime$ represent the conjugation classes generated from identity, $C_{3z}$, and $C_{2x}$, respectively. Symbols for conjugate class at $M$ and $K$ are defined in similar ways. 
}
\end{table*}

\begin{table*}
\begin{centering}
\begin{tabular}{|c|c|c|c|c|c|c|c|c|}
\hline 
Wyckoff pos. & \multicolumn{3}{c|}{$1a$ $\left(000\right)$} & \multicolumn{3}{c|}{$2c$ $\left(\frac{1}{3}\frac{2}{3}0\right)$, $\left(\frac{2}{3}\frac{1}{3}0\right)$} & \multicolumn{2}{c|}{$3f$ $(\frac1200)$, $(0\frac120)$, $(\frac12\frac120)$}\tabularnewline
\hline 
Site sym. & \multicolumn{3}{c|}{$6^{\prime}22^{\prime}$, $32$} & \multicolumn{3}{c|}{$32$, $32$} & \multicolumn{2}{c|}{$2^\prime2^\prime2$, $2$}\tabularnewline
\hline 
EBR & $[A_{1}]_a \up G $ & $[A_{2}]_a \up G$ & $[E]_a \up G$ & $[A_1]_c \up G$ & $[A_{2}]_c \up G$ & $[E]_c \up G$ & $[A]_f \up G$ & $[B]_f \up G$ \tabularnewline
\hline
Orbitals & $s$ & $p_z$ & $p_x, p_y$ & $s$ & $p_z$ & $p_{x}, p_y$ & $s$ & $p_z$ \\
\hline
$\Gamma\left(000\right)$ & $\Gamma_{1}$ & $\Gamma_{2}$ & $\Gamma_{3}$ & $2\Gamma_{1}$ & $2\Gamma_{2}$ & $2\Gamma_{3}$ & {$\Gamma_1\oplus\Gamma_3$} & $\Gamma_2\oplus \Gamma_3$ \tabularnewline
\hline 
$K\left(\frac{1}{3}\frac{1}{3}0\right)$ & $K_{1}$ & $K_{1}$ & $K_{2}K_{3}$ & $K_{2}K_{3}$ & $K_{2}K_{3}$ & $2K_{1}\oplus K_{2}K_{3}$ & {$K_1\oplus K_2K_3$} & $K_1\oplus K_2K_3$ \tabularnewline
\hline 
$M\left(\frac{1}{2}00\right)$ & $M_{1}$  & $M_{2}$  & $M_{1}\oplus M_{2}$  & $2M_{1}$  & $2M_{2}$  & $2M_{1}\oplus 2M_{2}$ & {$2M_1\oplus M_2$}  & $M_1\oplus 2M_2$ \tabularnewline
\hline
\end{tabular}
\par\end{centering}
\caption{\label{tab:EBR-magSG} EBRs of the magnetic
space group $P6^{\prime}2^{\prime}2$ (\#177.151 in the BNS setting).
In the first row are the symbols and coordinates of Wyckoff positions. 
In the second row are the corresponding site symmetry groups (magnetic point groups) and their unitary subgroups (point groups). 
In the third row are the names of EBRs. 
The notation $[\rho]_w \up G$ represents the EBR of the magnetic space group $G$ induced from the irrep $\rho$ at the Wyckoff possible $w$.
The orbitals that form the representation $[\rho]_w$ are given in the fourth row.
The fifth row to seventh row give the irreps at high symmetry momenta of each EBR.}
\end{table*}

\begin{figure}
\includegraphics[width=0.9\linewidth]{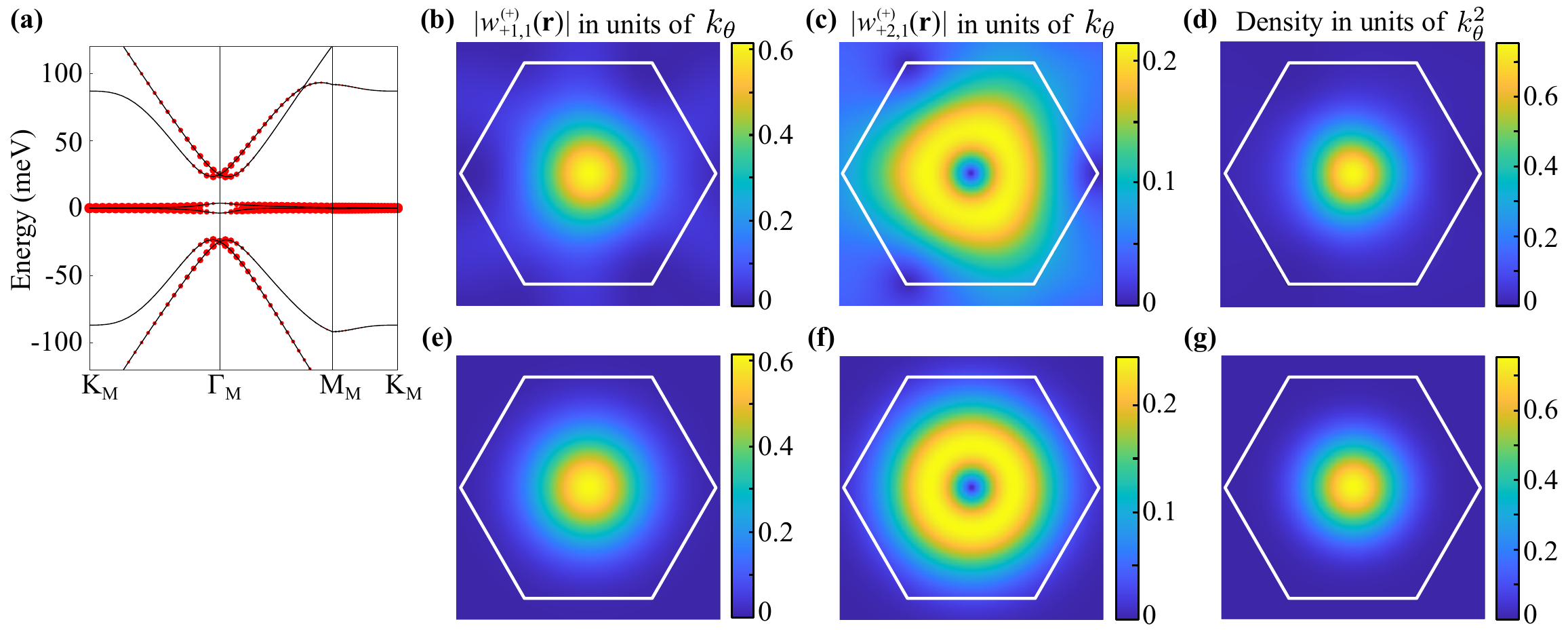}
\caption{Wannier functions for $w_0/w_1=0.8$. 
In (a) the sizes of the red dots represent the overlaps between the trial Wannier functions and the Bloch states, \ie $\sum_{\alpha} |A_{n,\alpha}(\kk)|^2 $, for the bands $\ee_{\kk,n}$ with $n=\pm1,\pm2,\pm3$.
(b) and (c) are absolute values of the two components $w_{+\beta,1}(\rr)$ ($\beta=1,2$) of the constructed maximally localized Wannier functions.
The hexagons represent the moir\'e unit cell. 
(d) The density of Wannier functions \ie $\sum_{l\beta} |w^{(\eta)}_{l\beta,\alpha}(\rr)|^2$, which does not depend on $\eta,\alpha$. 
(e), (f), (g) are $|w_{+1,1}(\rr)|$, $|w_{+2,1}(\rr)|$, and the density of the approximate Wannier functions of the Gaussian form (\cref{eq:WF-form1,eq:WF-form2}), respectively. 
The used parameters for the BM model are  $v_F = 5.944\mrm{eV\cdot\mathring{A}}$, $|\DKK|=1.703\mrm{\mathring{A}^{-1}}$, $w_1=110\mrm{meV}$, $\theta=1.05^\circ$. 
\label{fig:Wannier}
}
\end{figure}

We now resolve the Wannier obstruction by adding higher energy bands in MATBG.
Both the first ``passive'' bands above the middle two bands and the first ``passive'' bands below the middle two bands form the degenerate irrep $\Gamma_3$ at the $\Gamma_M$ point. 
If we hybridize the middle two bands with these higher energy bands, we can extract two topologically trivial bands that form the band representation (in terms of irreps)
\begin{equation}
    [E]_{a} \up G: \qquad \Gamma_3;\ M_1 \oplus M_2;\ K_2K_3
\end{equation} 
where $[E]_{a}$ is formed by a $p_x$-like orbital and a $p_y$-like orbital at the $1a$ position (triangular lattice).
In the remaining of this subsection we will discuss an explicit construction of the Wannier functions forming the representation $[E]_{a}$. 
It has large overlaps (\cref{fig:Wannier}) with the lowest two bands for $\kk$ away from $\Gamma_M$ --- in most area of the moir\'e BZ excluding a small neighborhood of $\Gamma_M$, the lowest two bands of MATBG are almost completely spanned by the constructed Wannier functions. 
While at (or around) $\Gamma_M$, the Wannier functions have zero (or small) overlap with the lowest two bands of MATBG. 
Another set of basis, which form the representation $\Gamma_1\oplus \Gamma_2 \oplus \Gamma_3$ at $\Gamma_M$, will be introduced around $\Gamma_M$. 
In \cref{sec:conduction-basis,sec:conduction-bands,sec:coupling-Hamiltonian,sec:H-single-summary} we will write the single-particle Hamiltonian on the basis of the constructed Wannier functions and the remaining low-energy bands, which together form the representation $\Gamma_1 \oplus \Gamma_2 \oplus 2\Gamma_3$ at the $\Gamma_M$ point. 

For simplicity in this subsection we will use the first-quantized formalism. 
We define the basis 
\begin{equation}
    \ket{\kk,\QQ,\alpha,\eta,s} = c_{\kk,\QQ,\alpha,\eta,s}^\dagger \ket{0}\ ,
\end{equation}
\begin{equation}
|\rr, l, \alpha,  \eta, s\rangle =  c_{l \alpha \eta s}^\dagger (\rr) \ket{0} 
=\frac{1}{\sqrt{\Omega_{\rm tot}}} \sum_{\kk \in{\rm MBZ}}\sum_{\QQ \in \mathcal{Q}_{l\eta}}
    e^{-i( \kk-\QQ)\cdot\rr}
    \ket{\kk,\QQ,\alpha,\eta,s}\ .
\end{equation}
We write the trial Wannier functions as 
\begin{equation}
\ket{W'_{\RR, \alpha=1, \eta, s}} = \frac1{\lambda_0\sqrt{2\pi}} \sum_l \int d^2 \rr\ e^{ i \frac{\pi}4 l\eta  - i \eta\DKK_l\cdot\RR -(\rr-\RR)^2/(2\lambda_0^2)} \ket{\rr,l,1,\eta,s}  
\end{equation}
\begin{equation}
\ket{W'_{\RR, \alpha=2, \eta, s}} = \frac1{\lambda_0\sqrt{2\pi}} \sum_l \int d^2\rr\ e^{ -i \frac{\pi}4 l \eta - i \eta\DKK_l\cdot\RR -(\rr-\RR)^2/(2\lambda_0^2)}  \ket{\rr,l,2,\eta,s}  \ ,
\end{equation}
where $\lambda_0$ represents the size of the Gaussian function and $\RR$ takes value in the triangular lattice sites, such that they form the representations
\begin{align}
& D^f(T) = \sigma_0 \tau_x,\quad
D^f(C_{3z}) = e^{i\frac{2\pi}3\sigma_z\tau_z},\quad 
D^f(C_{2x}) = \sigma_x\tau_0,\quad 
D^f(C_{2z}T) = \sigma_x\tau_0, \nono\\
& 
D^f(P) = i \sigma_z\tau_z, \quad
D^f(C_{2z}P) = -i\sigma_y\tau_y\ ,
\label{eq:Df}
\end{align}
same as $p_x\pm ip_y$ at the $1a$ Wyckoff position (per spin per valley). 
Here $\sigma_{0,x,y,z}$, $\tau_{0,x,y,z}$ are the identity or Pauli matrices in for the orbital ($\alpha=1,2$) and valley ($\eta=\pm$) degrees of freedom, respectively. 
Since the Wannier functions transform diagonally (\ie $D^f(P)$ in \cref{eq:Df}) under the $P$ symmetry and $P$ flips the layer (\cref{eq:D-real-P-C}), each Wannier function should be a linear combination of states from the two layers. 
The phase factors $e^{\pm i \frac{\pi}4 l \eta}$ in the Wannier functions are introduced to fix the gauge of the representation matrices $D^f$'s. 
The phase factor $e^{-i\eta\DKK_l\cdot\RR}$ is required by the translation symmetry (\cref{eq:continuous-basis-translation}).
One can verify that such defined trial Wannier functions satisfy 
\begin{equation} \label{eq:Wannier-translation}
    T_{\Delta\RR} \ket{W'_{\RR, \alpha, \eta, s}} = \ket{W'_{\RR+\Delta\RR, \alpha, \eta, s}}\ ,
\end{equation}
under a translation of the moir\'e lattice $\Delta\RR$. 
We emphasize that, for given $D^f(C_{3z})$, the sign of $D^f(P)$ is {\it not} a gauge choice.
Since $[C_{3z},P]=0$, $P$ and $C_{3z}$ can be diagonalized at the same time and the $P$-eigenvalues in each $C_{3z}$ eigen-space are determined. 
For the trial Wannier functions, the $P$ eigenvalue of the $C_{3z}$ eigenstates with the eigenvalue $e^{i\frac{2\pi}3}$ ($e^{-i\frac{2\pi}3}$) is $i$ ($-i$). 

We can express the trial Wannier functions on the plane-wave basis as 
\begin{align}
\ket{W'_{\RR, \alpha=1, \eta, s}} =& \frac1{\lambda_0\sqrt{2\pi \Omega_{\rm tot} }} \sum_{l=\pm} \sum_{\kk\in{\rm MBZ}} \sum_{\QQ\in \mathcal{Q}_{l\eta}} 
    \int d^2\rr\ 
    e^{ i \frac{\pi}4 l\eta - i\eta\DKK_l\cdot\RR -(\rr-\RR)^2/(2\lambda_0^2) - i(\kk-\QQ)\cdot \rr } \ket{\kk,\QQ,1,\eta,s}  \nono\\
=& \sqrt{\frac{2\pi\lambda_0^2 }{ \Omega_{\rm tot} }} \sum_{l=\pm} \sum_{\kk\in{\rm MBZ}} \sum_{\QQ\in \mathcal{Q}_{l\eta}} 
    e^{ i \frac{\pi}4 l\eta - i\kk\cdot \RR - \frac12 \lambda_0^2 (\kk-\QQ)^2 } \ket{\kk,\QQ,1,\eta,s}  
\end{align}
\begin{equation}
\ket{W'_{\RR, \alpha=2, \eta, s}} = \sqrt{\frac{2\pi\lambda_0^2 }{ \Omega_{\rm tot} }} \sum_{l=\pm} \sum_{\kk\in{\rm MBZ}} \sum_{\QQ\in \mathcal{Q}_{l\eta}} 
    e^{ - i \frac{\pi}4 l\eta - i\kk \cdot \RR - \frac12 \lambda_0^2 (\kk-\QQ)^2 } \ket{\kk,\QQ,2,\eta,s}  \ .
\end{equation}
We denote the energy eigenstates of the continuous BM model as $\ket{\psi_{\kk,n,\eta,s}}$.
The overlap matrix between the energy eigenstates and the trial Wannier functions are 
\begin{equation}
A_{n,1}^{(\eta)}(\kk) = \inn{  \psi_{\kk,n,\eta,s} | W'_{0, 1, \eta, s}   } 
= \sqrt{\frac{2\pi\lambda_0^2 }{ \Omega_{\rm tot} }} \sum_{l=\pm} \sum_{\QQ\in \mathcal{Q}_{l\eta}} 
e^{ i \frac{\pi}4 l\eta  - \frac12 \lambda_0^2 (\kk-\QQ)^2 } \inn{ \psi_{\kk,n,\eta,s} |\kk,\QQ,1,\eta,s} \ ,
\end{equation}
\begin{equation}
A_{n,2}^{(\eta)}(\kk) = \inn{  \psi_{\kk,n,\eta,s} | W'_{0, 2, \eta, s}   } 
= \sqrt{\frac{2\pi\lambda_0^2 }{ \Omega_{\rm tot} }} \sum_{l=\pm} \sum_{\QQ\in \mathcal{Q}_{l\eta}} 
e^{ -i \frac{\pi}4 l\eta - \frac12 \lambda_0^2 (\kk-\QQ)^2 } \inn{ \psi_{\kk,n,\eta,s} |\kk,\QQ,2,\eta,s} \ .
\end{equation}
We plot the $ \sum_{\alpha} |A_{n,\alpha}(\kk)|^2 $ for each band $\ee_{\kk,n}$ ($n=\pm1,\pm2,\pm3$) in \cref{fig:Wannier}(a). 
One can see that, as expected, at the $\Gamma_M$ point, the overlap between the trial Wannier functions and the $\Gamma_1$, $\Gamma_2$ states in the middle two bands are zero. 

We feed the overlap matrix $A_{n,\alpha}^{(\eta)}(\kk)$ ($n=\pm1,\pm2,\pm3$, $\alpha=1,2$, $\eta=\pm$) into the machinery of Wannier90 \cite{Wannier90-1,Wannier90-2,Wannier90-3} to construct the maximally localized Wannier functions.
We have set $\lambda_0 = 0.1a_M$ in practical calculations, with $a_M$ being the moir\'e lattice constant. 
We use a $18\times18$ momentum mesh and the energy window $\rm [-80meV,80meV]$ for the disentanglement and Wannierization procedures.
Wannier90 returns $\ket{W_{\RR,\alpha,\eta,s}}$ on the plane-wave basis $\ket{\kk,\QQ,\beta,\eta,s}$ as
\begin{equation} \label{eq:Wannier-from-Wannier90}
\ket{W_{\RR, \alpha, \eta, s}} = \frac1{\sqrt{N}} \sum_{l=\pm} \sum_{\kk \in{\rm MBZ}} \sum_{\beta} \sum_{\QQ\in \mathcal{Q}_{l\eta}}   \ket{\kk,\QQ,\beta,\eta,s} e^{-i\kk\cdot\RR}\td{v}^{(\eta)}_{\QQ \beta,\alpha} (\kk)
\end{equation}
Here $N$ is the number of Moir\'e unit cells. 
$\td{v}^{(\eta)}_{\QQ\beta,\alpha}(\kk)$ returned by Wannier90 is originally stored on the basis of the Bloch wave functions of the energy bands. 
Only the lowest six energy bands are involved in the Wannier functions because we have chosen the energy window $\rm [-80meV,80meV]$ for the Wannierization, which only covers the lowest six bands (\cref{fig:Wannier}(a)). 
We then transform $\td{v}^{(\eta)}_{\QQ\beta,\alpha}(\kk)$ onto the basis $\ket{\kk,\QQ,\beta,\eta,s}$. 
Since $\ket{\rr,l,\beta,\eta,s}$ gains a phase factor $e^{-i\eta\DKK_l\cdot\RR}$ under the translation $\RR$ (\cref{eq:continuous-basis-translation}), we can write $\inn{\rr, l,\beta, \eta, s| W_{\RR \alpha \eta s}}$ as 
\begin{equation}\label{eq:WR-wr}
    \inn{\rr, l,\beta, \eta, s| W_{\RR \alpha \eta s}}
= \inn{\rr, l,\beta, \eta, s|T_{\RR} |W_{0 \alpha \eta s}} 
= e^{-i\eta\DKK_l\cdot\RR} \inn{\rr-\RR, l,\beta, \eta, s | W_{0 \alpha \eta s}}
= e^{-i\eta\DKK_l\cdot\RR} w_{l\beta,\alpha}^{(\eta)}(\rr-\RR) 
\end{equation}
such that $w_{l\beta,\alpha}^{(\eta)}(\rr-\RR) = \inn{\rr-\RR, l,\beta, \eta, s | W_{0 \alpha \eta s}}$ is a function of $\rr-\RR$. 
The real space shape of the Wannier states can be calculated as 
\begin{equation}  \label{eq:wannier-bloch}
w_{l\beta,\alpha}^{(\eta)}(\rr-\RR) =e^{i\eta\DKK_l\cdot\RR} \inn{\rr, l,\beta, \eta, s| W_{\RR \alpha \eta s}} = 
    \frac{1}{\sqrt{N \Omega_{\rm tot}}} \sum_{\kk\in{\rm MBZ}} \sum_{\QQ \in \mathcal{Q}_{l\eta} } e^{  i(\kk-\QQ)\cdot(\rr-\RR)  }  
    \td{v}^{(\eta)}_{\QQ \beta,\alpha} (\kk)\ .
\end{equation}
Making use of \cref{eq:Fourier-1,eq:Fourier-2}, we obtain the inverse transformation
\begin{equation} \label{eq:bloch-wannier-local}
\td{v}^{(\eta)}_{\QQ\beta, \alpha}(\kk) = \frac1{\sqrt{\Omega_0}} \int d^2\rr\ w_{l_{\QQ,\eta} \beta ,\alpha}^{(\eta)} (\rr-\RR) e^{-i(\kk-\QQ)\cdot(\rr-\RR)}\ , 
\end{equation}
where $l_{\QQ,\eta}=\zeta_{\QQ} \eta$ is the layer the basis $ \ket{\kk,\QQ,\beta,\eta,s}$ belonging to and $\Omega_0 = \Omega_{\rm tot}/N$ is the area of the moir\'e unit cell.

We now determine the independent components of the Wannier function.
Due to the spin-SU(2) symmetry, the real space wave functions $w_{l\beta,\alpha}^{(\eta)}(\rr)$ do not depend on spin. 
Due to the time-reversal symmetry, the $w_{l\beta,\alpha}^{(\eta)}(\rr)$ in the two valleys are just complex conjugation of each other, \ie 
\begin{equation} \label{eq:wannier-constraint0}
    w_{l\beta,\alpha}^{(\eta)}(\rr) = w_{l\beta,\alpha}^{(-\eta)*}(\rr)\ . 
\end{equation}
Due to the $P$ symmetry ($\rr \to -\rr$), which has $D(P)=-i\varrho_y \sigma_0 \tau_0$ for the continuous real space basis (\cref{eq:D-real-P-C}) and $D^f(P)=i\sigma_z \tau_z$ for the Wannier functions (\cref{eq:Df}), the Wannier functions must satisfy the constraint 
$ \sum_{l'\alpha'\eta'} D_{l \alpha \eta, l' \alpha' \eta}(P) w_{l'\alpha'\eta',\alpha}^{(\eta)}(\rr) = \sum_{\alpha'} w_{l\alpha\eta,\alpha'
}^{(\eta)}(-\rr) D^f_{\eta \alpha',\eta \alpha}(P) $, \ie 
\begin{equation} \label{eq:wannier-constraint1}
w_{-l\beta,\alpha}^{(\eta)}(-\rr) =  i l \eta (-1)^{\alpha} w_{l\beta,\alpha}^{(\eta)}(\rr)
\end{equation}
Due to the $C_{2z}T$ symmetry, which has $D(C_{2z}T)=\varrho_0 \sigma_x \tau_0$ for the continuous basis (\cref{eq:D-real-C2x-C2zT}) and $D^f(C_{2z}T)=\sigma_x \tau_0$ for the Wannier functions (\cref{eq:Df}), the Wannier functions must satisfy the constraint 
\begin{equation} \label{eq:wannier-constraint2}
    w_{l\beta,\alpha}^{(\eta)}(\rr) = w_{l \ovl{\beta},\ovl{\alpha}}^{(\eta)*}(-\rr)
\end{equation}
where $\ovl{\alpha}=2,1$ for $\alpha=1,2$. 
The above two conditions together imply a constraint (given by $PC_{2z}T$) that holds at each point $\rr$ 
\begin{equation} \label{eq:wannier-constraint3}
w_{l\beta,\alpha}^{(\eta)}(\rr) = -i l \eta (-1)^{\alpha} w_{-l\ovl{\beta},\ovl{\alpha}}^{(\eta)*}(\rr)\ . 
\end{equation}
Due to the above three independent constraints, only two of the sixteen components in the real space functions $w_{l\beta,\alpha}^{(\eta)}(\rr)$ are independent. 
(Notice that each constraint reduce to half the number independent components.)
Without loss of generality, we will focus on the two independent components  $w_{+\beta,1}^{(+)}(\rr)$ ($\beta=1,2$). 
We plot the absolute values of the two components $w_{+\beta,1}^{(+)}(\rr)$ ($\beta=1,2$) in \cref{fig:Wannier}(b), (c), respectively. 
Due to \cref{eq:wannier-constraint3,eq:wannier-constraint0}, the densities of the Wannier functions, \ie $\sum_{l\beta} |w_{l\beta,\alpha}^{(\eta)}(\rr)|^2$, do not depend on $\eta$ and $\alpha$. 
We plot the density of the Wannier functions in \cref{fig:Wannier}(d).
From the density plot we can see that the Wannier functions are very well localized. 
To characterize the localization, we also calculate the square root of the spread of the Wannier functions, \ie
\begin{equation}\label{eq:lambda-def}
    \lambda = \sqrt{ \sum_{l \beta} \int d^2\rr  |w_{l\beta,\alpha}^{(\eta)}(\rr)|^2 \rr^2 }\ .
\end{equation}
$\lambda$ does not depend on $\alpha$ and $\eta$ according to  \cref{eq:wannier-constraint3,eq:wannier-constraint0}. 
As tabulated in \cref{tab:Wannier}, $\lambda$ ranges from $0.2965a_M$ to $0.5330a_M$ for $w_0/w_1$ changing from 1.0 to 0.5, where $a_M$ is the lattice constant of moir\'e unit cell. 
The magnitude of the hopping between the constructed Wannier functions at nearest neighbor sites, \ie $t_0 = |\inn{W_{\RR,\alpha,\eta,s}|H_{\rm BM}|W_{\RR',-\alpha,\eta,s}}|$ for $\RR\RR'$ being nearest neighbor pairs, is also extremely small.
($\inn{W_{\RR,\alpha,\eta,s}|H_{\rm BM}|W_{\RR',\alpha,\eta,s}}=0$ due to the crystalline and $P$ symmetries.)
As shown in \cref{tab:H0parameters}, the $t_0$ ranges from 0.01512meV to 0.3027meV for $w_0/w_1$ changing from 1.0 to 0.5. 
In the rest of this work we will omit the nearest neighbor hoppings.

\begin{table}[t]
\centering
\begin{tabular}{|c |c|c|c|c|c|}
\hline 
$w_0/w_1$ &  $\alpha_1$ & $\alpha_2$ & $\lambda_1$ ($a_M$) & $\lambda_2$ ($a_M$) & $\lambda$ ($a_M$) \\
\hline
1.0 & 0.7632 & 0.6461 & 0.1552 & 0.1731 & 0.2965 \\
\hline
0.9 & 0.7882 & 0.6155 & 0.1671 & 0.1850 & 0.3029 \\
\hline
0.8 & 0.8193 & 0.5734 & 0.1791 & 0.1910 & 0.3375 \\
\hline 
0.7 & 0.8531 & 0.5217 & 0.1970 & 0.1970 & 0.3792 \\
\hline
0.6 & 0.8878 & 0.4603 & 0.2089 & 0.2029 & 0.4380\\
\hline 
0.5 & 0.9208 & 0.3901 & 0.2268 & 0.2089 & 0.5330\\
\hline  
\end{tabular}
\caption{Parameters of Wannier functions for different $w_0/w_1$. 
$\alpha_{1,2}$ and $\lambda_{1,2}$ are parameters for the analytical form (\cref{eq:WF-form1,eq:WF-form2}) of the Wannier functions and they are fitted from the numerical Wannier functions.
$\lambda$ is the square root of the spread of the numerical Wannier functions. 
The used parameters for the BM model are  $v_F = 5.944\mrm{eV\cdot\mathring{A}}$, $|\DKK|=1.703\mrm{\mathring{A}^{-1}}$, $w_1=110\mrm{meV}$, $\theta=1.05^\circ$. 
\label{tab:Wannier}}
\end{table}

We find the Wannier functions shown in \cref{fig:Wannier}(b), (c) can be approximated by Gaussian functions of the form
\begin{equation} \label{eq:Wannier-independent-component}
w_{+1,1}^{(+)}(\rr) = \frac{\alpha_1}{\sqrt2} \frac{1}{\sqrt{\pi\lambda_1^2}} e^{i \frac{\pi}4 -\rr^2/(2\lambda_1^2)},\quad 
w_{+2,1}^{(+)}(\rr) = -\frac{\alpha_2}{\sqrt2} \frac{x+i y}{\lambda_2^2 \sqrt{\pi}} e^{i \frac{\pi}4  -\rr^2/(2\lambda_2^2)}\ .
\end{equation}
Even though the Wannier states transform as $p_{x}\pm i p_y$ orbitals, which form the representation $[E]_a$ at the $1a$ position (\cref{tab:irreps-MSG,tab:EBR-magSG}), we see that the component $w_{+1,1}^{(+)}(\rr)$ is $s$-orbital-like and the component $w_{+2,1}^{(+)}(\rr)$ is ($p_x+ip_y$)-orbital-like. 
The reason is that the inner degrees of freedom, \ie layer, sublattice, and valley, transform non-trivially under the crystalline symmetries. 
For example, the layer, sublattice, and valley degrees of freedom transform under $C_{3z}$ as $D(C_{3z}) = \varrho_0 e^{i\frac{2\pi}3 \sigma_z \tau_z}$ (\cref{eq:D-real-T-C3z}), in order for the Wannier functions $w_{+\beta,1}^{(+)}(\rr)$ to have the $C_{3z}$ eigenvalue $e^{i\frac{2\pi}3}$ (\cref{eq:Df}), there must be $  e^{i\frac{2\pi}3 [\sigma_z]_{\beta\beta}}w_{+\beta,1}^{(+)}(C_{3z}^{-1}\rr) =e^{i\frac{2\pi}3} w_{+\beta,1}^{(+)}(\rr)$.
The two components satisfy $w_{+1,1}^{(+)}(C_{3z}\rr) = w_{+1,1}^{(+)}(\rr)$ and $w_{+2,1}^{(+)}(C_{3z}\rr) = e^{i\frac{2\pi}3} w_{+2,1}^{(+)}(\rr)$, respectively. 
Thus they can be realized by the polynomials $1$ (constant) and $x+iy$, respectively. 
The phase factor $e^{i\frac{\pi}4}$ is required by the symmetry $C_{2x}PC_{2z}T=C_{2y}PT$, whose action action on the continuous basis and representation on the Wannier states are $D(C_{2y}PT) = \varrho_z\sigma_0 \tau_0$ (\cref{eq:D-real-C2x-C2zT,eq:D-real-P-C}) and $D^f (C_{2y}PT) = -i\sigma_z\tau_z$ (\cref{eq:Df}), respectively. 
Given $w_{+1,1}^{(+)}(\rr)\propto e^{-\rr^2/(2\lambda_1^2)}$, $w_{+1,1}^{(+)}(\rr)\propto (x+iy) e^{-\rr^2/(2\lambda_1^2)}$, the phases of the two components are constrained by the $C_{2y}PT$ symmetry as
\begin{equation} 
    w_{+\beta,1}^{(+)*}(x,-y) = -i w_{+\beta,1}^{(+)}(x,y) \qquad \Rightarrow \qquad 
    w_{+\beta,1}^{(+)}(\rr) = \pm e^{i\frac{\pi}4}|w_{+\beta,1}^{(+)}(\rr)|\ . 
\end{equation}
However, the symmetry representation cannot fix the relative sign of the two components: 
Suppose $w_{+\beta,1}^{(+)}(\rr)$ forms the correct representation of $C_{2y}PT$, then $(-1)^{\beta-1} w_{+\beta,1}^{(+)}(\rr)$ also forms the correct representation. 
Thus there are two symmetry-allowed choices for the analytical expressions of $w_{+\beta,1}^{(+)}(\rr)$.
We have computed the overlaps between the two choices and the numerical Wannier functions and pick the choice (\cref{eq:Wannier-independent-component}) that gives the larger overlap. 
$\lambda_1$ can be fitted using the condition $|w_{+1,1}^{(+)}(\lambda_1,0)| = e^{-\frac12} |w_{+1,1}^{(+)}(0)| $, and $\lambda_2$ can be fitted by the maximum of $|w_{+2,1}^{(+)}(x_{\max},0)|$, where $x_{\max} = \sqrt2 \lambda_2$. 
$\alpha_1$ and $\alpha_2$ are real coefficients satisfying $\alpha_1^2 + \alpha_2^2=1$. 
For $w_0/w_1=0.8$, we find $\lambda_1=0.1791a_M$, $\lambda_2=0.1850a_M$, $\alpha_1 = 0.7882$, $\alpha_2 = 0.6155$. 
These parameters for other $w_0/w_1$ values are tabulated in \cref{tab:Wannier}. 
Using the constraints \cref{eq:wannier-constraint0,eq:wannier-constraint1,eq:wannier-constraint2}, we obtain the other components as 
\begin{equation} \label{eq:WF-form1}
w_{l1,1}^{(\eta)}(\rr) = \frac{\alpha_1}{\sqrt2} \frac{1}{\sqrt{\pi\lambda_1^2}} e^{i \frac{\pi}4 l\eta  -\rr^2/(2\lambda_1^2)},\quad 
w_{l2,1}^{(\eta)}(\rr) = -l \frac{\alpha_2}{\sqrt2} \frac{x+i\eta y}{\lambda_2^2 \sqrt{\pi}} e^{i \frac{\pi}4 l\eta -\rr^2/(2\lambda_2^2)}, 
\end{equation}
\begin{equation} \label{eq:WF-form2}
w_{l1,2}^{(\eta)}(\rr) = l\frac{\alpha_2}{\sqrt2} \frac{x-i\eta y}{\lambda_2^2 \sqrt{\pi}} e^{-i \frac{\pi}4 l\eta - \rr^2/(2\lambda_2^2)},\qquad 
w_{l2,2}^{(\eta)}(\rr) = \frac{\alpha_1}{\sqrt2} \frac{1}{\sqrt{\pi\lambda_1^2}} e^{-i \frac{\pi}4 l\eta - \rr^2/(2\lambda_1^2)}\ .
\end{equation}
One can verify that the Wannier functions form the representations in \cref{eq:Df}.

We introduce the creation operators for the Wannier states as 
\begin{equation}
f_{\RR \alpha\eta s}^\dagger 
= \sum_{l\beta} \int d^2\rr\ 
    \inn{ \rr, l, \beta, s,\eta,s | W_{\RR, \alpha,\eta, s} }
    c_{l,\beta,\eta,s}^\dagger (\rr)
= \sum_{l\beta} e^{-i\eta\DKK_l\cdot\RR} \int d^2\rr\ w_{l \beta, \alpha}^{(\eta)}(\rr-\RR) c_{l,\beta,\eta,s}^\dagger (\rr)\ . 
\end{equation}
Under the translation $T_{\Delta\RR}$ the operator $f_{\RR \alpha\eta s}^\dagger $ transforms as 
\begin{equation}
    T_{\Delta\RR} f_{\RR \alpha\eta s}^\dagger T_{\Delta\RR}^{-1} = f_{\RR+\Delta\RR \alpha\eta s}^\dagger\ ,
\end{equation}
which follows \cref{eq:Wannier-translation} directly.
Correspondingly, the momentum space operators can be defined as 
\begin{equation} \label{eq:f-Fourier}
f_{\kk \alpha\eta s}^\dagger = \frac1{\sqrt{N}} \sum_{\RR} e^{i\kk\cdot\RR} f_{\RR \alpha\eta s}^\dagger
= \sum_{\QQ \beta} c_{\kk,\QQ,\beta,\eta,s}^\dagger \td{v}^{(\eta)}_{\QQ \beta, \alpha}(\kk) \ ,
\end{equation}
with $\td{v}^{(\eta)}_{\QQ \beta, \alpha}(\kk)$ from \cref{eq:bloch-wannier-local}. 
Since $f_{\RR \alpha\eta s}^\dagger $ creates the Wannier states, they follows the same representation (\cref{eq:Df}) under symmetry operations
\begin{equation}
\hg f_{\kk \alpha\eta s}^\dagger \hg^{-1} = \sum_{\alpha'\eta'} f_{g\kk \alpha'\eta' s}^\dagger D^f_{\alpha'\eta',\alpha\eta}(g), \qquad
\hg f_{\RR \alpha\eta s}^\dagger \hg^{-1} = \sum_{\alpha'\eta'} f_{g\RR \alpha'\eta' s}^\dagger D^f_{\alpha'\eta',\alpha\eta}(g),
\end{equation}
with $D^f(g)$ given by \cref{eq:Df}. 

Hereafter, we refer to $f_{\RR,\alpha,\eta,s}^\dagger$ as {\it local orbitals}. 
They transform as $p_x \pm i p_y$ orbitals, which form the 2D irreducible representation $[E]_a$ at the $1a$ position (\cref{tab:irreps-MSG,tab:EBR-magSG}), under the crystalline symmetries. 

\subsection{The topological conduction bands} \label{sec:conduction-basis}

As discussed in the last subsection, in the BM model the six low energy states per valley per spin form the representation $\Gamma_1\oplus \Gamma_2 \oplus 2\Gamma_3$ around the $\Gamma_M$ point (\cref{fig:bands_free}), where $\Gamma_1\oplus  \Gamma_2$ are contributed by the middle two bands and $2\Gamma_3$ are contributed by the higher energy bands. 
The Wannier functions (in each valley-spin sector) constructed in the last subsection can span the middle two bands for momenta away from $\Gamma_M$, where the overlap with the flat bands is finite (\cref{fig:Wannier}(a)).
At the $\Gamma_M$ point, the Wannier functions span a $\Gamma_3$ representation. 
Thus, in order to reproduce the correct band structure around the $\Gamma_M$ point, we add four additional degree of freedom that form the representation $\Gamma_1 \oplus \Gamma_2 \oplus \Gamma_3$ at $\Gamma_M$. 
We denote the four states as $c_{\kk,a,\eta,s}^\dagger$ ($a=1,2,3,4$). 
Formally they can be written as 
\begin{equation} \label{eq:c-conduction-def}
c_{\kk,a,\eta,s}^\dagger = \sum_{\QQ\beta} \td{u}^{(\eta)}_{\QQ\beta,a}(\kk) c_{\kk,\QQ,\beta,\eta,s}^\dagger\ ,
\end{equation}
where $\td{u}^{(\eta)}_{\QQ\beta,a}(\kk)$ is to be determined. 
As shown in next subsection, $c_{\kk,a,\eta,s}^\dagger $ has a huge kinetic energy for large $\kk$.
Therefore, in this work we will only consider $c_{\kk,a,\eta,s}^\dagger $ within the cutoff $|\kk|< \Lambda_c$.  
Now we determine the coefficients $\td{u}^{(\eta)}_{\QQ\beta,a}(\kk)$.
We define the projector to the six lowest bands in the two valleys as 
\begin{equation} \label{eq:projector-P}
    P_{\QQ\alpha,\QQ'\beta}^{(\eta)} (\kk) = \sum_{n=\pm1,\pm2,\pm3} u_{\QQ\alpha,n}^{(\eta)}(\kk) u_{\QQ'\beta,n}^{(\eta)*}(\kk), 
\end{equation}
where $u_{\QQ\alpha,n}^{(\eta)}(\kk)$ is the eigenstate of the BM Hamiltonian (\cref{eq:BM-Hamiltonian}) for the $|n|$-th band above ($n>0$) or below ($n<0$) the charge neutrality point. 
We then define the projector to the two Wannier functions of the local orbitals as 
\begin{equation} \label{eq:projector-Q}
    Q_{\QQ\alpha,\QQ'\beta}^{(\eta)} (\kk) = \sum_{\alpha=1,2} \td{v}_{\QQ\alpha,\alpha}^{(\eta)}(\kk) \td{v}_{\QQ'\beta,\alpha}^{(\eta)*}(\kk)\ .
\end{equation}
Since the two Wannier functions are constructed from the lowest six bands, there must be $P^{(\eta)}(\kk) Q^{(\eta)}(\kk) P^{(\eta)}(\kk) = Q^{(\eta)}(\kk)$. 
This is guaranteed by the definition of our Wannier functions: The Bloch states of the Wannier functions $\td{v}_{\QQ\alpha,\alpha}^{(\eta)}(\kk)$ are linear combinations of $u_{\QQ\alpha,n}^{(\eta)}(\kk)$ ($n=\pm1,\pm2,\pm3$), as explained after \cref{eq:Wannier-from-Wannier90}. 
The projector to the four low-energy states in the remaining bands is given by $P^{(\eta)}(\kk) - Q^{(\eta)}(\kk)$. 
The states $\td{u}^{(\eta)}_{\QQ\beta,a}(\kk)$ can be obtained as eigenstates of $P^{(\eta)}(\kk) - Q^{(\eta)}(\kk)$ with the eigenvalue 1. 

We emphasize that even though we have formally introduced $\td{u}^{(\eta)}_{\QQ\beta,a}(\kk)$, in practical calculations, \eg \cref{sec:conduction-bands}, we will always approximate $\td{u}^{(\eta)}_{\QQ\beta,a}(\kk)$ by $\td{u}^{(\eta)}_{\QQ\beta,a}(0)$ for low energy states, as the $c$-electrons are only relevant for a small $\kk$ region around $\Gamma_M$. 

There is a gauge degree of freedom of $\td{u}^{(\eta)}_{\QQ\beta,a}(\kk)$. 
We fix the gauge by requiring the representation matrices under the symmetry transformations
\begin{equation}
\hg c_{\kk, a\eta s}^\dagger \hg^{-1} = \sum_{a'\eta'} c_{g\kk, a'\eta' s}^\dagger D^c_{a'\eta',a\eta}(g)
\end{equation}
to be
\begin{align} \label{eq:Dc-crystalline}
& D^c(T) = \sigma_0\tau_x \oplus \sigma_0 \tau_x,\quad 
D^{c}(C_{3z}) = e^{i\frac{2\pi}3 \sigma_z \tau_z} \oplus \sigma_0\tau_0,\quad
D^{c}(C_{2x}) = \sigma_x\tau_0 \oplus \sigma_x\tau_0,\quad 
D^{c}(C_{2z}T) = \sigma_x\tau_0 \oplus \sigma_x\tau_0\ . \nono\\
\end{align}
such that the  $a=1,2$ and $a=3,4$ states form the $\Gamma_3$ and $\Gamma_1\oplus \Gamma_2$ representations at $\kk=0$, respectively.
For example, the $C_{3z}$ representation matrices of the $\Gamma_3$ and $\Gamma_1\oplus \Gamma_2$ states are given by $e^{i\frac{2\pi}3 \sigma_z \tau_z} $ and $\sigma_0\tau_0$, respectively. 
Since the four states form three different representations at $\Gamma_M$, the states $\td{u}^{(\eta)}_{\QQ\beta,a}(\kk)$ can be {\it uniquely} determined at $\kk=0$. 
However, for $\kk\neq 0$, there is no easy method to uniquely fix the gauge of the four states.
Nevertheless, in practical calculations, we only need $\td{u}^{(\eta)}_{\QQ\beta,a}(0)$ for the k$\cdot$p expansion. 
Using the gauge fixed by \cref{eq:Dc-crystalline} and the algebra between $P$ and the other symmetries (\cref{eq:P-algebra})
\begin{equation}
    P^2=-1,\qquad [P,T]=[P,C_{3z}]=[P,C_{2z}T]=0,\qquad \{P,C_{2x}\}=0\ ,
\end{equation}
we can further determine the representation of $P$.
To satisfy $P^2=-1$ and $[P,C_{2z}T]=0$, the representation of $P$ in either the $\Gamma_3$ or $\Gamma_1\oplus \Gamma_2$ space must have the form $\pm i \sigma_z\tau_{0,z}$.
Representations involving $\tau_{x,y}$ are not considered because $P$ preserves the valley index. 
The considered representations, $\pm i \sigma_z\tau_{0,z}$, already satisfy $\{P,C_{2x}\}=0$.  
To commute with $T$, the only allowed form is $\pm i\sigma_z \tau_z$.
Numerically, we find the representation of $P$ (and $C_{2z}P$ for later use) as
\begin{equation} \label{eq:Dc-P}
D^c(P) = (-i\sigma_z\tau_z) \oplus (-i\sigma_z\tau_z),\quad
D^c(C_{2z}P) = (i\sigma_y\tau_y) \oplus (i\sigma_y \tau_y)\ . 
\end{equation}
As explained after \cref{eq:Df}, for fixed $D^c(C_{3z})$, the sign of $D^c(P)$ is {\it not} a gauge choice.
One may notice that the first two-by-two block of $D^c(P)$, which spans the $\Gamma_3$ irrep, has an opposite sign with $D^f(P)$ of the local $f$-orbitals (\cref{eq:Df}), which also span a $\Gamma_3$ irrep. 
Now we explain that they have to have opposite signs.
We denote the $\Gamma_3$ irreps formed by the positive energy bands ($n=2,3$) and the negative energy bands ($n=-2,-3$) of the BM model as $\Gamma_{3+}$ and $\Gamma_{3-}$, respectively. 
$\Gamma_{3+}$ and $\Gamma_{3-}$ transform into each other under $P$.
Thus, in the basis of $\Gamma_{3+}$ and $\Gamma_{3-}$, in each eigen-subspace of $C_{3z}$, the $P$ operator is off-diagonal.
To be explicit, we can write the representation matrices of $C_{3z}$ and $P$
\begin{equation}
    \begin{pmatrix}
        e^{i\frac{2\pi}3}\sigma_0 & 0 \\
        0 & e^{-i\frac{2\pi}3}\sigma_0
    \end{pmatrix},\qquad 
    \begin{pmatrix}
        X & 0 \\
        0 & X'
    \end{pmatrix}\ ,
\end{equation}
respectively.
The basis is ordered as: the first component of $\Gamma_{3+}$, the first component of $\Gamma_{3-}$, the second component of $\Gamma_{3+}$, the second component of $\Gamma_{3-}$, such that the first (second) two bases have the $C_{3z}$ eigenvalue $e^{i\frac{2\pi}3}$ ($e^{-i\frac{2\pi}3}$).
$X$ and $X'$ are off-diagonal such that $P$ interchanges the two $\Gamma_3$ representations. 
After we diagonalize the $P$ operator, the two $P$-eigenvalues in each eigen-subspace of $C_{3z}$ in each valley must sum to zero. 
Hence the sign of $D^c(P)$, which spans a diagonalized $\Gamma_3$, must be opposite to the sign of $D^f(P)$, which spans the other $\Gamma_3$. 

Hereafter, we refer to $c_{\kk,a,\eta,s}^\dagger$ ($a=1,2,3,4$) as {\it conduction bands}.

\subsection{Single-particle Hamiltonian of the topological conduction bands} \label{sec:conduction-bands}

In this subsection we use the first-quantized formalism for simplicity. 
We can divide the single-particle Hamiltonian in the low energy space into four parts: 
\begin{equation}
    H^{(f,\eta)}(\kk) = Q^{(\eta)}(\kk) h^{(\eta)}(\kk) Q^{(\eta)}(\kk) , \qquad
    H^{(c,\eta)}(\kk) = (P^{(\eta)}(\kk)-Q^{(\eta)}(\kk) ) h^{(\eta)}(\kk) (P^{(\eta)}(\kk)-Q^{(\eta)}(\kk)) \ , 
\end{equation}
\begin{equation}
    H^{(fc,\eta)}(\kk) = Q^{(\eta)}(\kk) h^{(\eta)}(\kk) (P^{(\eta)}(\kk)-Q^{(\eta)}(\kk)) ,\qquad 
    H^{(cf,\eta)}(\kk) = (P^{(\eta)}(\kk)-Q^{(\eta)}(\kk)) h^{(\eta)}(\kk) Q^{(\eta)}(\kk)\ ,
\end{equation}
where $h^{(\eta)}(\kk)$ is the BM model (\cref{eq:BM-Hamiltonian}), $P^{(\eta)}(\kk)$ (\cref{eq:projector-P}) and $Q^{(\eta)}(\kk)$ (\cref{eq:projector-Q}) are projectors to the lowest six bands (per spin valley) and the Wannier functions, respectively. 
The first term is the single-particle Hamiltonian of the local orbitals, which almost vanishes ($\sim0.15$meV for $w_0/w_1=0.8$) as discussed in \cref{sec:Wannier} and shown in \cref{tab:H0parameters}. 
The second term is the Hamiltonian of the conduction bands. 
The third and fourth terms are the couplings between the local orbitals and the conduction bands.

We consider the k$\cdot$p expansion of the projected BM model $H^{(c,\eta)}(\kk)$ around $\kk=0$  ($a,a'=1\cdots 4$)
\begin{equation} \label{eq:Hc-perturbation}
H^{(c,\eta)}_{aa'}(\kk) = \bra{\td{u}_{a}^{(\eta,)}(0)}  h^{(\eta)}(\kk) \ket{\td{u}_{a'}^{(\eta)}(0)}\ .
\end{equation}
We emphasize that the usual k$\cdot$p expansion \cite{winkler_spin-orbit_2003} is made on the basis $\td{u}^{(\eta)}_{\QQ\beta,a}(0)$ rather than $\td{u}^{(\eta)}_{\QQ\beta,a}(\kk)$.
This basis choice has two advantages: (i) It is exact if we keep an infinite number of bands ($a=1\cdots \infty$), (ii) It has simple $\kk$-dependence. 
We find that in our problem keeping four conduction bands is a good enough approximation. 
Hereafter, we alway choose the Bloch states at $\kk=0$ as the basis whenever a k$\cdot$p expansion is needed. 
Since $h^{(\eta)}(\kk)$ has only zeroth order and linear order terms in $\kk$, $H^{(c,\eta)}(\kk) $ also only has constant terms and linear terms in $\kk$. 
As proved in the following paragraphs, to linear order terms of $\kk$ the effective Hamiltonian under the gauge \cref{eq:Dc-crystalline,eq:Dc-P} has the form
\begin{equation} \label{eq:Hc-full}
H^{(c,+)}(\kk) = \begin{pmatrix}
0  &  v_\star(k_x \sigma_0 + ik_y\sigma_z)    \\
v_\star(k_x \sigma_0 - ik_y\sigma_z)  & M  \sigma_x 
\end{pmatrix}\ , 
\end{equation}
\begin{equation}
H^{(c,-)}(\kk) = H^{(c,+)*}(-\kk)\ ,
\end{equation}
where the $C_{3z}$ eigenvalues of the four basis in the valley $\eta=+$ are $e^{i\frac{2\pi}3}$, $e^{-i\frac{2\pi}3}$, 1, 1, respectively. 
The parameters for different $w_0/w_1$ obtained from \cref{eq:Hc-perturbation} are tabulated in \cref{tab:H0parameters}. 
Physical meaning of $M$ is the splitting between the $\Gamma_1$ and $\Gamma_2$ states. 

\begin{table}[t]
\centering
\begin{tabular}{|c|c|c |c|c|c|a| }
\hline 
$w_0/w_1$ & $t_0$ & $\gamma$ & $M$ & 
 $v_\star$ & $v_\star'$  & $v_\star''$  \\
\hline
1.0&  0.01512 &  4.086 & 4.727 &  
    -3.829 & 1.492 & -0.01237  \\ 
\hline
0.9 & 0.1099 & -10.21 & 4.194 & 
    -4.089 & 1.579 & -0.01680  \\
\hline
0.8 & 0.1497 & -24.75 & 3.697  & 
    -4.303 & 1.623 & -0.03320 \\
\hline
0.7 & 0.2276 & -39.11 & 3.248 & 
    -4.483 & 1.624 & -0.04012 \\
\hline
0.6 & 0.2789 & -52.91 &  2.854 & 
    -4.637 & 1.580 & -0.04009  \\
\hline
0.5 & 0.3027 & -65.78 & 2.518 & 
    -4.774 & 1.483 & -0.03509 \\
\hline
Units & \multicolumn{3}{c|}{meV} &  \multicolumn{3}{c|}{ $\rm eV\cdot\mathring{A}$ }  \\
\hline  
\end{tabular}
\caption{Parameters of the single-particle Hamiltonian of the topological heavy fermion model for different $w_0/w_1$. 
Since $v_\star''$ is extremely small compared to other parameters, in all the single-particle and many-body calculations we have set $v_\star''=0$. 
Parameters of the BM model used to obtain this table are  $v_F = 5.944\mrm{eV\cdot\mathring{A}}$, $|\KK|=1.703\mrm{\mathring{A}^{-1}}$, $w_1=110\mrm{meV}$, $\theta=1.05^\circ$. 
}
\label{tab:H0parameters}
\end{table}

We now derive the form of $H^{(c,\eta)}$ in the valley $\eta=+$ to linear order of $\kk$. 
The effective Hamiltonian in the other valley can be obtained through the time-reversal symmetry, \ie $H^{(c,-)}(\kk) = H^{(c,+)*}(-\kk) $. 
For simplicity, in the following we introduce an additional set of Pauli (identity) matrices $\zeta_{0,x,y,z}$ to distinguish the $\Gamma_3$ space ($a=1,2$) and the $\Gamma_1\oplus \Gamma_2$ space ($a=3,4$). 
The $a=1,2$ and $a=3,4$ span the subspaces of $\zeta_z=1$ and $\zeta_z=-1$, respectively.
Due to \cref{eq:Dc-crystalline,eq:Dc-P}, the symmetry operators in the valley $\eta=+$ are given by 
\begin{equation} \label{eq:single-sym-dirac}
C_{3z} = e^{i\frac{2\pi}3 \sigma_z} \oplus \sigma_0,\qquad 
C_{2x} = \zeta_0 \sigma_x,\qquad 
C_{2z}T = \zeta_0 \sigma_x K,\qquad 
P = - i\zeta_0 \sigma_z\ ,
\end{equation}
where $\zeta_0 \sigma_x \equiv \sigma_x \oplus \sigma_x$, $\zeta_0 \sigma_z \equiv \sigma_z \oplus \sigma_z$. 
We first look at the subspace $\Gamma_3$.
According to $C_{2z}T$, only $\sigma_{0,x,y}$ terms are allowed in the Hamiltonian. 
According to $P$, $\sigma_0$ is odd in $k$ while $\sigma_{x,y}$ are even (constant). 
$\sigma_0$ is invariant under $C_{3z}$, hence the coefficient before $\sigma_0$ must be invariant under $C_{3z}$.
The lowest odd order polynomial satisfying this condition is in third order of $k$.
Thus the $\sigma_0$ term vanishes to linear order of $k$.
Since $\sigma_{x,y}$ are not invariant under $C_{3z}$, their (constant) coefficients must vanish to linear order of $\kk$.
(One may show that the lowest order coefficients of $\sigma_{x,y}$ terms are quadratic in $\kk$.)

We then look at the subspace $\Gamma_1\oplus \Gamma_2$. 
According to $C_{2z}T$, only $\sigma_{0,x,y}$ terms are allowed. 
According to $P$, $\sigma_0$ is odd in $k$ while $\sigma_{x,y}$ are even (constant) in $k$.
$\sigma_0$ is invariant under $C_{3z}$, hence the coefficient before $\sigma_0$ must be invariant under $C_{3z}$.
The lowest odd order polynomial satisfying this condition is in third order of $k$.
Thus the $\sigma_0$ term vanishes to linear order of $k$.
In the subspace $\Gamma_1\oplus \Gamma_2$, both $\sigma_{x,y}$ are invariant under $C_{3z}$, hence they are allowed by $C_{3z}$.
However $\sigma_y$ is forbidden by the $C_{2x}$ symmetry.
Thus the Hamiltonian in the $\Gamma_1\oplus \Gamma_2$ subspace must have the form $M\sigma_x$. 

Thirdly we look at the coupling between $\Gamma_3$ and $\Gamma_1\oplus \Gamma_2$. 
We only need to consider the off-diagonal terms $\zeta_{x,y}\sigma_{0,x,y,z}$. 
According to $C_{2z}T$, only $\zeta_x\sigma_{0,x,y}$ and $\zeta_y\sigma_z$ terms are allowed.
According to $P$, the coefficients of $\zeta_x\sigma_{x,y}$ are even (constant) in $k$ while $\zeta_x \sigma_0$, $\zeta_y \sigma_z$ are odd in $k$.
Since  $\zeta_x \sigma_{x,y}$ are not invariant under $C_{3z}$, their constant coefficients must vanish.
We then study the odd order terms $\zeta_x \sigma_0$, $\zeta_y \sigma_z$.
Under $C_{3z}$ they transform as $C_{3z} \zeta_x \sigma_0 C_{3z}^{-1} = \zeta_x \sigma_0 \cos\frac{2\pi}3 - \zeta_y \sigma_z \sin \frac{2\pi}3$,
$C_{3z} \zeta_y\sigma_z \sigma_0 C_{3z}^{-1} = \zeta_y \sigma_z \cos\frac{2\pi}3 + \zeta_x \sigma_0 \sin \frac{2\pi}3$.
Thus the $\zeta_{x}\sigma_0, \zeta_y\sigma_z$ terms can be chosen as $v_\star k_x \zeta_x \sigma_0 - v_\star k_y \zeta_y\sigma_z$, which is also symmetric under $C_{2x}$. 

In summary, the effective Hamiltonian $H^{(c,\eta)}$ has the form in \cref{eq:Hc-full}.

In Ref. \cite{Song-TBG2} the authors of the present work have shown that the single-particle Hamiltonian in each valley has a symmetry anomaly of the $C_{2z}T$ and the particle-hole $P$ symmetries.
The anomaly is reflected as unavoidable $4n+2$ ($n\in \mathbb{N}$) Dirac cones at the Fermi-level. 
Since the local orbitals are topologically trivial, the anomaly must be reflected in the band structure of the conduction bands, meaning that, if we turn off the coupling between local orbitals and conduction bands, the conduction bands (given by \cref{eq:Hc-full}) must carry the symmetry anomaly by itself and hence be gapless. 
The spectrum of \cref{eq:Hc-full} is given by 
\begin{equation}
    \pm \frac{M}2 \pm \sqrt{\frac{M^2}4  + v_\star^2\kk^2} 
\end{equation}
which has a quadratic touching at $\kk=0$. 
To relate the quadratic touching to the symmetry anomaly, we consider adding a $C_{3z}$-breaking but $C_{2z}T$- and $P$-preserving term $D$ into the Hamiltonian 
\begin{equation} 
\begin{pmatrix}
D \sigma_x  &  v_\star(k_x \sigma_0 + ik_y\sigma_z)    \\
v_\star(k_x \sigma_0 - ik_y\sigma_z)  & M  \sigma_x 
\end{pmatrix}\ . 
\end{equation}
The quadratic touching is now split into two linear Dirac points at $k_x =\pm \frac{1}{v_\star} \sqrt{DM}$, $k_y=0$ ($k_x =0$, $k_y=\pm \frac1{v_\star} \sqrt{DM}$) if $DM>0$ ($DM<0$). 
The two Dirac points have the same chirality because they are related by $P$ \cite{Song-TBG2}. 
Thus the quadratic touching is topologically equivalent to two Dirac points with the same chirality. 
According to the theorem proven in Ref. \cite{Song-TBG2}, the presence of $4n+2$ (here $n=0$) Dirac points is an equivalent condition to the symmetry anomaly. 

\subsection{Coupling between topological conduction bands and the local orbitals}\label{sec:coupling-Hamiltonian}

We now consider the coupling between the conduction band states and the local orbitals 
{\small
\begin{equation} 
\hH_{cf} = \sum_{\eta s} \sum_{a \alpha} \sum_{|\kk|<\Lambda_c} \sum_{\RR} \inn{\kk a\eta s| H |W_{\RR\alpha\eta s}} c_{\kk a\eta s}^\dagger  f_{\RR \alpha\eta s}  + h.c. =\sum_{\eta s} \sum_{a \alpha} \sum_{|\kk|<\Lambda_c} \sum_{\RR} \inn{ \kk a\eta s | H| W_{0 \alpha\eta s} } e^{-i\kk\cdot\RR} c_{\kk a\eta s}^\dagger  f_{\RR \alpha\eta s} + h.c.,
\end{equation}
}
where $\ket{W_{\RR\alpha\eta s}}$ ($\alpha=1,2$) and $\ket{\kk a\eta s}=c_{\kk,a,\eta,s}^\dagger\ket{0}$ ($a=1,2,3,4$) are the single particle states of the local orbitals and the conduction bands, respectively, $H$ is the BM Hamiltonian (\cref{eq:BM-Hamiltonian}), and $\Lambda_c$ is the cutoff for the conduction bands. 
Since the Wannier functions of the local orbitals are localized, the integral $\inn{ \kk a\eta s | H| W_{0 \alpha\eta s} }$ must decay exponentially with $\kk$.
However, the particular form of the decay is complicated due to two reasons. 
First, as discussed in \cref{sec:Wannier}, the Wannier functions have two independent components (\cref{eq:WF-form1,eq:WF-form2}). 
Thus the overlap between the conduction band states and the two components will decay differently.
Second, each conduction band state is not a simple plane-wave but a linear combination of plane-waves ($\QQ$-vectors) with different wave-vectors (\cref{eq:c-conduction-def}), hence the overlap for different plane-wave components will decay differently. 
However, the role played by the decay is simply a truncation of the couplings between the local orbitals and high energy states.
We claim that the particular form of the truncation is not relevant in the low energy physics. 
Thus, for simplicity here we assume that the overlap decays as $e^{-|\kk|\lambda^2/2}$ with $\lambda$ being the square root of the spread of the Wannier function (\cref{eq:lambda-def}). 
We hence parameterize the integral as 
\begin{equation} \label{eq:Hcf-def}
    \inn{ \kk a\eta s | H | W_{0 \alpha\eta s} } 
= \frac1{\sqrt{N}} e^{-|\kk|^2\lambda^2/2} H^{(cf,\eta)}_{a\alpha}(\kk)\ . 
\end{equation}
Applying the Fourier transformation (\cref{eq:f-Fourier}) of the local orbitals, we can rewrite the coupling Hamiltonian as 
\begin{equation}
\hH_{cf} = \sum_{\eta s} \sum_{a \alpha} \sum_{|\kk|<\Lambda_c} e^{-|\kk|^2\lambda^2/2} H^{(cf,\eta)}_{a \alpha}(\kk) c_{\kk a\eta s}^\dagger f_{\kk \alpha \eta s} + h.c. 
\end{equation}
For small $\kk$, the factor $ e^{-|\kk|^2\lambda^2/2}$ is negligible and we can obtain $H^{(cf,\eta)}_{a \alpha}(\kk)$ as the k$\cdot$p Hamiltonian 
\begin{equation} \label{eq:Hcf-calculation}
H^{(cf,\eta)}_{a \alpha}(\kk) = \bra{\td{u}_{a}^{(\eta)} (0)} h^{(\eta)}(\kk) \ket{ \td{v}_{\alpha}^{(\eta)}(0) } ,
\end{equation}
where $h^{(\eta)}(\kk)$ is the BM model (\cref{eq:BM-Hamiltonian}), $\ket{\td{v}_{\alpha}^{(\eta)}(0)}$ are the Bloch state of the Wannier functions (\cref{eq:Wannier-from-Wannier90}), and $\ket{\td{u}_{a}^{(\eta)}(0)}$ are the Bloch states of the conduction bands (\cref{eq:c-conduction-def}). 
As explained after \cref{eq:Hc-perturbation}, we always choose $\ket{\td{v}_{\alpha}^{(\eta)}(0)}$ and  $\ket{\td{u}_{a}^{(\eta)}(0)}$ rather than $\ket{\td{v}_{\alpha}^{(\eta)}(\kk)}$ and  $\ket{\td{u}_{a}^{(\eta)}(\kk)}$ as the basis for the k$\cdot$p expansion.
Since $h^{(\eta)}(\kk)$ only has zeroth and linear order terms in $\kk$, $H^{(cf,\eta)}_{a \alpha}(\kk) $ also only has constant and linear terms. 
As proved in the next paragraph, we find the coupling Hamiltonian has the form
\begin{equation}
H^{(cf,\eta)} = \begin{pmatrix}
 \gamma \sigma_0 + v_\star'( \eta k_x \sigma_x + k_y \sigma_y) \\
 v_\star''( \eta k_x \sigma_x - k_y \sigma_y)
\end{pmatrix}\ .
\end{equation}
The first two-by-two block is the coupling between the local orbitals (forming $\Gamma_3$) and the $\Gamma_3$ states of the conduction bands and hence allows a $\gamma \sigma_0$ term, the second two-by-two block the coupling between the local orbitals and the $\Gamma_1 \oplus \Gamma_2$ states of the conduction bands and vanishes at $\kk=0$. 
The $\Gamma_3$ representation carries angular momenta $L=\pm1$ and the $\Gamma_1 \oplus \Gamma_2$ representation carries the angular momentum $L=0$.
The above equation respects the angular momentum conservation up to $\Delta L=0$ mod 3 because the system only has $C_{3z}$ symmetry. 
The parameters obtained from \cref{eq:Hcf-calculation} are tabulated in \cref{tab:H0parameters}. 
We find that $v_\star''$ is very small compared to other velocities. 
We hence will omit this term. 

Here we give one more justification for the choice of the damping factor $e^{-|\kk|^2 \lambda^2/2}$.
First, around $\kk=0$, the damping factor $e^{-|\kk|^2 \lambda^2/2} \sim 1$ and works well because the $H^{(cf,\eta)}(\kk)$ matrix is by construction (\cref{eq:Hcf-calculation}) accurate at $\kk=0$. 
Second, at large $\kk$, the actual damping factor may be complicated due to the reasons given before \cref{eq:Hcf-def}, but the $f-c$ coupling is weak compared to the energy separation of $f$- and $c$-bands.
Thus, the actual form of damping factor is not important. 
Our approximation should have at least captured the main feature of the damping behavior - it assumes the Wannier function as a single exponential function and the conduction bands as monochromatic plane waves, which are reasonable due to the localized and delocalized natures of $f$- and $c$-electrons.
As will be discussed in \cref{sec:H-single-summary} and shown in \cref{fig:bands_free}, the choice

Now we use the symmetry principle to derive the form of $H^{(cf,+)}(\kk)$.
$H^{(cf,-)}(\kk)$ is given by $H^{(cf,+)*}(\kk)$ due to the time-reversal symmetry. 
We only keep zeroth and linear order terms in $\kk$.
The symmetry operators on the single-particle states of the conduction bands ($c$) are given in \cref{eq:single-sym-dirac}.
According to \cref{eq:Df}, the symmetry operators on the single-particle states of the local orbitals ($f$) are 
\begin{equation} \label{eq:single-sym}
C_{3z} = e^{i\frac{2\pi}3 \sigma_z},\qquad 
C_{2x} = \sigma_x ,\qquad 
C_{2z}T = \sigma_x K,\qquad 
P = i\sigma_z\ . 
\end{equation}
We first look at the coupling between the $\Gamma_3$ states ($a=1,2$) and the local orbitals. 
According to $C_{2z}T$, only $\sigma_{0,x,y}$ and $i\sigma_z$ are allowed. 
According to $P$, the Hamiltonian satisfy $(-i\sigma_z) H^{(cf,\eta)}(-\kk) (i\sigma_z)^\dagger = - H^{(cf,\eta)}(\kk)$, where $-i\sigma_z$ and $i\sigma_z$ are the representation matrix of $P$ formed by the conduction bands and local orbitals, respectively. 
Thus the $\sigma_{x,y}$ terms must be odd in $k$ and $\sigma_{0}$, $i\sigma_z$ terms must be even in $k$. 
The even order terms are constants to first order of $k$, \ie $\gamma \sigma_{0} + i \gamma'\sigma_{z}$. 
$\gamma$ respects both $C_{3z}$ and $C_{2x}$ symmetries.
$\gamma'$ respects $C_{3z}$ but breaks $C_{2x}$, hence the $\gamma'$ term is forbidden.
The two odd order terms $\sigma_{x,y}$ transform under $C_{3z}$ as $C_{3z} \sigma_x C_{3z}^{-1} = \sigma_x \cos\frac{2\pi}3 + \sigma_y \sin\frac{2\pi}3$, $C_{3z} \sigma_y C_{3z}^{-1} = \sigma_y \cos\frac{2\pi}3 - \sigma_x \sin\frac{2\pi}3$, thus the odd terms can be chosen as $v_\star' (k_x \sigma_x + k_y \sigma_y)$, which is also symmetric under $C_{2x}$. 
Thus the coupling between the $\Gamma_3$ states in the QDP and the local orbitals takes the form $\gamma \sigma_0 + v_\star' (k_x \sigma_x + k_y \sigma_y)$. 
We then consider couplings between the $\Gamma_1\oplus \Gamma_2$ states ($a=3,4$) in the conduction bands and the local orbitals. 
According to $C_{2z}T$, only $\sigma_{0,x,y}$ and $i\sigma_z$ are allowed. 
According to $P$, $\sigma_{x,y}$ terms must be odd in $k$ and $\sigma_{0}$, $i\sigma_z$ terms must be even in $k$. 
The even order terms are constants to first order of $k$, however they break the $C_{3z}$ symmetry as they couple states with angular momentum 0 to states with angular momenta $\pm1$. 
The two odd order terms $\sigma_{x,y}$ transform under $C_{3z}$ as $C_{3z} \sigma_x C_{3z}^{-1} = \sigma_x \cos\frac{2\pi}3 - \sigma_y \sin\frac{2\pi}3$, $C_{3z} \sigma_y C_{3z}^{-1} = \sigma_y \cos\frac{2\pi}3 + \sigma_x \sin\frac{2\pi}3$, thus the odd terms can be chosen as $v_\star'' (k_x \sigma_x - k_y \sigma_y)$, which is also symmetric under $C_{2x}$.

\subsection{Summary of the single-particle Hamiltonian} \label{sec:H-single-summary}

According to the discussions in \cref{sec:conduction-bands,sec:coupling-Hamiltonian}, we can summarize the total single-particle Hamiltonian as 
\begin{align} \label{eq:H0-def}
\hH_0 =& \sum_{\eta s} \sum_{aa'} \sum_{|\kk|<\Lambda_c} (H^{(c,\eta)}_{a,a'}(\kk) - \mu \delta_{aa'}) c_{\kk a\eta s}^\dagger c_{\kk a'\eta s}  
- \mu \sum_{\eta s} \sum_{\RR} f_{\RR \alpha \eta s}^\dagger f_{\RR \alpha \eta s} \nono\\
& + \frac1{\sqrt{N}}\sum_{\eta s} \sum_{|\kk|<\Lambda_c}\sum_{\RR} \sum_{\alpha a} 
    \pare{ e^{-|\kk|^2\lambda^2/2 - i\kk\cdot\RR} H^{(cf,\eta)}_{a \alpha}(\kk) c_{\kk a\eta s}^\dagger f_{\RR \alpha \eta s}  + h.c. }\ .
\end{align}
$\mu$ is the chemical potential, $\Lambda_c$ is the cutoff of the momentum of the conduction bands, $H^{(c,\eta)}_{a,a'}(\kk)$ and $H^{(cf)}_{a \alpha}(\kk)$ are given by 
\begin{equation}  \label{eq:Hc-summary}
H^{(c,\eta)}(\kk) = \begin{pmatrix}
0_{2\times 2}   &  v_\star(\eta k_x \sigma_0 + ik_y\sigma_z)   \\
v_\star(\eta k_x \sigma_0 - ik_y\sigma_z) & 
M \sigma_x 
\end{pmatrix},
\end{equation}
\begin{equation}  \label{eq:Hcf-summary}
        H^{(cf,\eta)} = \begin{pmatrix}
         \gamma \sigma_0 + v_\star'( \eta k_x \sigma_x + k_y \sigma_y) \\
         0_{2\times 2}
        \end{pmatrix}\ ,
\end{equation}
respectively. 
The parameters in the single-particle Hamiltonian are given in \cref{tab:H0parameters}. 

In principle, the cutoff $\Lambda_c$ should be a small quantity compared to the Moir\'e BZ. 
However, because the states at large $\kk$'s have very high energies, we can formally extend $\Lambda_c$ to infinity without changing the low energy physics.
A theoretical advantage of $\Lambda_c\to \infty$ is that the yielded band structure will become periodic over the moir\'e BZ. 
We label the momentum for the conduction bands as $\kk+\GG$ with $\kk$ being in the moir\'e BZ and $\GG$ moir\'e reciprocal lattice. 
Using a larger $\Lambda_c$ implies a larger lattice of $\GG$. 
In \cref{fig:bands_free} we have plotted the band structures of \cref{eq:H0-def} using the cutoff $\Lambda_c=5\sqrt{3} k_\theta$. 

As shown in Ref.~\cite{Biao-TBG3}, the {\it anti-unitary} single-particle symmetry $PC_{2z}T$ (anti-commuting with single-particle Hamiltonian) implies a {\it unitary} many-body charge-conjugation symmetry $\PH_c$ that commutes with the Hamiltonian, \ie 
\begin{equation} \label{eq:charge-conjugation-def}
    \PH_c f_{\RR \alpha\eta s}^\dagger \PH_c^{-1} = \sum_{\alpha'\eta'} D^f_{\alpha'\eta',\alpha\eta}(PC_{2z}T) f_{\RR\alpha'\eta' s},\qquad
    \PH_c c_{\kk a\eta s}^\dagger \PH_c^{-1} = \sum_{a'\eta'} D^c_{a'\eta',a\eta}(PC_{2z}T) c_{-\kk a'\eta' s}\ ,
\end{equation}
where 
\begin{equation}
    D^{f}(C_{2z}TP) = -\sigma_y \tau_z,\qquad 
    D^{f}(C_{2z}TP) = (\sigma_y \tau_z) \oplus (\sigma_y \tau_z)\ . 
\end{equation}
One can verify that $\PH_c$ commutes with $\hH_0$ in \cref{eq:H0-def}. 

\label{sec:H0-symmetry}

\subsection{The (first) chiral limit and the chiral U(4) symmetry} \label{sec:chiral-U4}

\begin{figure}
\includegraphics[width=1\linewidth]{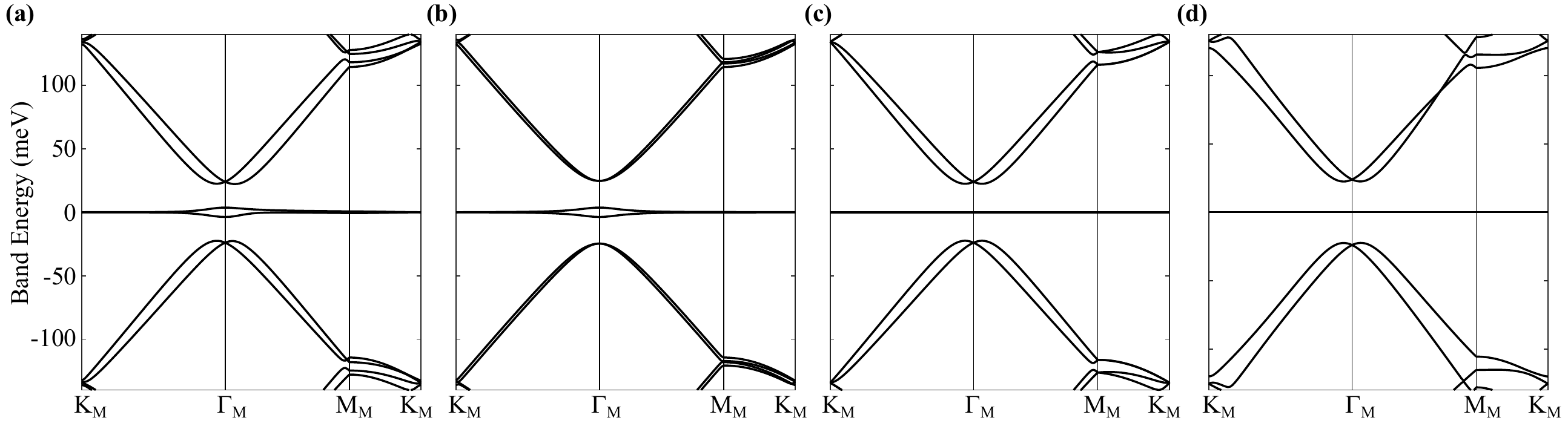}
\caption{Band structures in the first, third chiral or flat limits. 
(a) Band structure of the original topological heavy fermion model (\cref{eq:H0-def}) at $w_0/w_1=0.8$. (b) Band structure in the first chiral symmetry, where $v_\star'$ is imposed to zero. 
(c) Band structure in the third chiral limit (flat limit), where $M$ is imposed to zero. 
(d) Band structure in the flat limit in the absence of the third chiral symmetry, where $M$ is imposed to zero and a chiral breaking term $A$ is taken into account. (See \cref{sec:flat-beyond-chiral}). 
The other parameters are given in \cref{tab:H0parameters}.
\label{fig:bands_chiral}
}
\end{figure}

We find that when $v_\star'=0$ the Hamiltonian (\cref{eq:H0-def}) {\it anti-commute} with the {\it unitary} chiral operator 
\begin{equation} \label{eq:first-chiral}
C f_{\kk \alpha\eta s}^\dagger C^{-1} =\sum_{\alpha'\eta'} f_{\kk \alpha'\eta' s}^\dagger D^f_{\alpha'\eta', a \eta} (C),\qquad 
C c_{\kk a\eta s}^\dagger C^{-1} = \sum_{a'\eta'} c_{\kk a'\eta' s}^\dagger D^c_{a'\eta', a \eta} (C)
\end{equation}
with 
\begin{equation}
D^f(C) = \sigma_z\tau_0,\qquad 
D^c(C) = (-\sigma_z \tau_0) \oplus (\sigma_z \tau_0)\ . 
\end{equation}
According to \cref{eq:Df,eq:Dc-crystalline,eq:Dc-P}, we find that $C$ satisfies the algebra at the single-particle level
\begin{equation}
C^2=1,\qquad [C,T]=0,\qquad [C,C_{3z}]=0,\qquad \{C,C_{2x}\}=0,\qquad \{C,C_{2z}T\}=0,\qquad 
[C,P]=0\ ,
\end{equation}
which are same as the first chiral symmetry discussed in \cref{sec:BM-sym} (\cref{eq:C-first-algebra}) and in Refs. \cite{Song-TBG2,Biao-TBG3,tarnopolsky_origin_2019,wang2020chiral}. 
Therefore, we identify $C$ as the first chiral symmetry. 
The band structure after imposing this chiral symmetry is shown in \cref{fig:bands_chiral}(b).

Refs. \cite{Biao-TBG3,bultinck_ground_2020,kang_strong_2019} have shown that the presence of chiral symmetry implies a so-called chiral non-flat U(4) symmetry of the projected interaction Hamiltonian.
(A similar U(4) symmetry was also found in Ref.~\cite{kang_strong_2019}.)
Here we find that $C$ indeed implies a U(4) symmetry of the single-particle Hamiltonian by using the symmetry $C_{2z}PC$. 
In next paragraph we will show the chiral implied U(4) is consistent with the chiral non-flat U(4) discussed in Refs. \cite{Biao-TBG3,bultinck_ground_2020,kang_strong_2019}. 
In \cref{sec:interaction-symmetry} we will show that the chiral U(4) symmetry is also an approximate symmetry of the interaction Hamiltonian.
Since the Hamiltonian anti-commutes with $P$ and $C$, it must commute with the product $PC$ as well as $C_{2z}PC$.
As discussed in \cref{sec:BM-sym}, $C$ is local in real space, $P$ transforms $\rr$ to $-\rr$, hence $C_{2z}PC$ is also local in real space.
Its representation matrices are 
\begin{equation}
D^f(C_{2z}PC) = \sigma_x \tau_y, \qquad 
D^c(C_{2z}PC) = (\sigma_x\tau_y) \oplus (-\sigma_x\tau_y) . 
\end{equation}
We introduce the continuous symmetry generator
\begin{equation}
\UC_{y0} = \frac12 \sum_{\kk \in \mrm{MBZ}} \sum_{\alpha\alpha'\eta\eta's}   [\sigma_x \tau_y]_{\alpha\eta,\alpha'\eta'} f_{\kk \alpha\eta s}^\dagger f_{\kk \alpha'\eta's} 
+ \frac12 \sum_{|\kk|<\Lambda_c} \sum_{aa'\eta\eta's}   [(\sigma_x \tau_y)\oplus (-\sigma_x \tau_y)]_{a\eta,a'\eta'} c_{\kk a\eta s}^\dagger c_{\kk a'\eta's}\ ,
\end{equation}
and it commutes with $H_0$.
Then we find that the Hamiltonian (\cref{eq:H0-def}) has the continuous symmetry
\begin{equation}
    [e^{-i\theta \UC_{y0} } , \hat{H}_0] = 0\ . 
\end{equation}
$\UC_{y0}$ together with the U(2)$\times$U(2) symmetry form the U(4) symmetry group.
The generators of the U(4) group can be explicitly written as 
\begin{equation} \label{eq:chiralU4}
\UC_{\mu\nu} = \frac12 \sum_{\eta\eta'}\sum_{ss'} \pare{
    \sum_{\kk \in \mrm{MBZ}} \sum_{\alpha\alpha'}  \Theta^{(\mu\nu,f)}_{ \alpha \eta s, \alpha' \eta'  s'} f_{\kk \alpha  \eta s}^\dagger f_{\kk  \alpha'  \eta' s'} + 
    \sum_{|\kk|<\Lambda_c}  \sum_{a a'}  \Theta^{(\mu\nu,c)}_{a \eta s, a' \eta'  s'} c_{\kk a \eta  s}^\dagger c_{\kk a' \eta' s'} 
}\ , 
\end{equation}
where $\mu,\nu=0,x,y,z$ and
\begin{equation} \label{eq:chiralU4-mat1}
\Theta^{(0\nu,f)} = \sigma_0 \tau_0 \spin_{\nu}, \qquad
\Theta^{(0\nu,c)} = ( \sigma_0 \tau_0 \spin_{\nu}) \oplus (\sigma_0 \tau_0 \spin_{\nu} ) \ , 
\end{equation}
\begin{equation}\label{eq:chiralU4-mat2}
\Theta^{(x\nu,f)} = \sigma_x \tau_x \spin_{\nu}, \qquad
\Theta^{(x\nu,c)} = ( \sigma_x \tau_x \spin_{\nu}) \oplus ( - \sigma_x \tau_x \spin_{\nu} ) \ , 
\end{equation}
\begin{equation}\label{eq:chiralU4-mat3}
\Theta^{(y\nu,f)} = \sigma_x \tau_y \spin_{\nu}, \qquad
\Theta^{(y\nu,c)} = ( \sigma_x \tau_y \spin_{\nu}) \oplus ( - \sigma_x \tau_y \spin_{\nu} ) \ , 
\end{equation}
\begin{equation}\label{eq:chiralU4-mat4}
\Theta^{(z\nu,f)} = \sigma_0 \tau_z \spin_{\nu}, \qquad
\Theta^{(z\nu,c)} = ( \sigma_0 \tau_z \spin_{\nu}) \oplus ( \sigma_0 \tau_z \spin_{\nu} ) \ . 
\end{equation}
Here $\sigma_{\mu}$, $\tau_{\mu}$, $\spin_{\mu}$ are Pauli matrices for the orbital, valley, and spin degree of freedom, respectively. 
The first and second eight-by-eight blocks in $\Theta^{(\mu\nu,c)}$ are for the $\Gamma_3$ orbitals ($a=1,2$) and the $\Gamma_1\oplus \Gamma_2$ ($a=3,4$), respectively. 

In the end we explicitly show that the U(4) symmetry in this chiral limit is consistent with the (first) chiral non-flat U(4) symmetry in Refs.~\cite{Biao-TBG3,kang_strong_2019,bultinck_ground_2020}. 
In the Appendix B2 of  Ref.~\cite{Biao-TBG3}, the sewing matrices of $T$, $C_{2z}$, and $P$ have been fixed as $\kk$-independent matrices 
\begin{equation} \label{eq:gauge-TBG3}
B(T) = \sigma_0 \tau_x,\qquad 
B(C_{2z}) = \sigma_0 \tau_x, \qquad 
B(P) = -i\sigma_y \tau_z\ . 
\end{equation}
The generators of the (first) chiral non-flat U(4) symmetry in Ref.~\cite{Biao-TBG3} in this gauge are 
\begin{equation} \label{eq:chiralU4-TBG3}
\sigma_0\tau_0 \spin_\nu,\qquad 
\sigma_0\tau_x \spin_\nu,\qquad 
\sigma_0\tau_y \spin_\nu,\qquad 
\sigma_0\tau_z \spin_\nu,\qquad 
(\nu=0,x,y,z)\ ,
\end{equation}
where $\sigma_{\mu}$, $\tau_{\mu}$, abd  $\spin_\mu$ are Pauli matrices for the band, valley, and spin degrees of freedom, respectively. 
As shown in \cref{sec:Wannier} and \cref{fig:Wannier}, in most area of the moir\'e BZ, the middle two bands are mainly contributed by the local orbitals. 
Thus we expect that, for $\kk\neq 0$ where the overlap between local orbitals and the middle two bands is finite, the U(4) generators on the local orbitals (after certain gauge transformation) should be same as those in \cref{eq:chiralU4-TBG3}.
We apply the gauge transformation $U=e^{i\frac{\pi}4\sigma_0\tau_z} e^{-i\frac{\pi}4 \sigma_x\tau_z }$ to the local orbitals such that the transformed representation matrices $U D^f(g) U^\dagger$ become same as \cref{eq:gauge-TBG3}.
(See \cref{eq:Df} for definition of $D^f(g)$'s.)
The U(4) generators for the local orbitals in \cref{eq:chiralU4-mat1,eq:chiralU4-mat2,eq:chiralU4-mat3,eq:chiralU4-mat4} transform to
\begin{equation}
\sigma_0 \tau_0 \spin_\nu,\qquad \sigma_0 \tau_y \spin_\nu, \qquad 
\sigma_0 \tau_x \spin_\nu,\qquad \sigma_0 \tau_z \spin_\nu,\qquad (\nu=0,x,y,z)\ . 
\end{equation}
which are indeed same as \cref{eq:chiralU4-TBG3}. 
For $\kk=0$, our local orbital basis (forming the representation $\Gamma_3$) is different from the basis in Ref.~\cite{Biao-TBG3} (forming the representation $\Gamma_1\oplus \Gamma_2$), thus the above  comparison is only valid for $\kk\neq 0$. 

\subsection{The flat limit and the flat U(4) symmetry}\label{sec:flat-U4}

Refs. \cite{Biao-TBG3,bultinck_ground_2020} discussed a flat limit where the kinetic energy of the flat middle two bands are discarded.
A so-called non-chiral flat U(4) symmetry emerges in this limit \cite{Biao-TBG3,bultinck_ground_2020}. 
We find that in the topological heavy fermion model we can achieve an exact flat band if we set $M=0$, \ie the $\Gamma_1-\Gamma_2$ splitting is set to zero. 
In the limit $M=0$, the Hamiltonian (\cref{eq:H0-def}) {\it anti-commutes} with the {\it unitary} chiral operator 
\begin{equation}\label{eq:third-chiral}
S f_{\kk \alpha\eta s}^\dagger S^{-1} =\sum_{\alpha'\eta'} f_{\kk \alpha'\eta' s}^\dagger D^f_{\alpha'\eta', a \eta} (S),\qquad 
S c_{\kk a\eta s}^\dagger S^{-1} = \sum_{a'\eta'} c_{\kk a'\eta' s}^\dagger D^c_{a'\eta', a \eta} (S)
\end{equation}
with 
\begin{equation}
    D^{f}(S) = \sigma_0\tau_0,\qquad 
    D^{c}(S) =  (-\sigma_0\tau_0) \oplus (\sigma_0\tau_0)\ .
\end{equation}
Since $S$ is different from the first and second chiral symmetries discussed in Ref.~\cite{Song-TBG2,Biao-TBG3}, we refer to $S$ as the third chiral symmetry.  
The local orbitals and the $a=3,4$ conduction band basis have the chiral eigenvalue 1, whereas the $a=1,2$ conduction band basis has the chiral eigenvalue $-1$.
Since only hoppings between opposite chiral eigenvalues are allowed (such that the Hamiltonian anti-commutes with $S$), the rank of the single-particle Hamiltonian at each $\kk$ (in a valley-spin sector) is at most 4.
Therefore, in this limit there must be two exact flat bands at the zero energy. 
Correspondingly, we also refer to this chiral limit as the flat limit.
The band structure in this limit is given in \cref{fig:bands_chiral}(c).

Similar to the first chiral limit, the third chiral limit also implies a U(4) symmetry of the single-particle Hamiltonian. 
In \cref{sec:interaction-symmetry} we will show that the flat U(4) symmetry is also an approximate symmetry of the interaction Hamiltonian.
Since the Hamiltonian anti-commutes with $P$ and $S$, it must commute with the product $PS$ as well as $C_{2z}PS$.
$S$ is local in real space, while $P$ transforms $\rr$ to $-\rr$, hence $C_{2z}PS$ is also local in real space.
Its representation matrices are 
\begin{equation}
D^f(C_{2z}PS) = -i\sigma_y \tau_y, \qquad 
D^c(C_{2z}PS) = (-i\sigma_y\tau_y) \oplus (i\sigma_y\tau_y)\ . 
\end{equation}
We introduce the continuous symmetry generator
\begin{equation}
\UF_{y0} = \frac12 \sum_{\kk\in \mrm{MBZ}} \sum_{\alpha\alpha'\eta\eta's}   [\sigma_y \tau_y]_{\alpha\eta,\alpha'\eta'} f_{\kk \alpha\eta s}^\dagger f_{\kk \alpha'\eta's} 
+ \frac12 \sum_{|\kk|<\Lambda_c} \sum_{aa'\eta\eta's}   [(\sigma_y \tau_y)\oplus (-\sigma_y \tau_y)]_{a\eta,a'\eta'} c_{\kk a\eta s}^\dagger c_{\kk a'\eta's}\ ,
\end{equation}
and it commutes with $\hH_0$.
Then we find that the Hamiltonian (\cref{eq:H0-def}) has the continuous symmetry
\begin{equation}
    [e^{-i\theta \UF_{y0} } , \hat{H}_0] = 0\ . 
\end{equation}
Similar to the case in the first chiral symmetry, the above rotation together with the U(2)$\times$U(2) symmetry lead to a U(4) symmetry group.
The generators of the U(4) group can be explicitly written as 
\begin{equation} \label{eq:flatU4}
\UF_{\mu\nu} = \frac12 \sum_{\eta\eta'}\sum_{ss'} \pare{
    \sum_{\kk \in \mrm{MBZ}} \sum_{\alpha\alpha'}  \Sigma^{(\mu\nu,f)}_{ \alpha \eta s, \alpha' \eta'  s'} f_{\kk \alpha  \eta s}^\dagger f_{\kk  \alpha'  \eta' s'} + 
    \sum_{|\kk|< \Lambda_c} \sum_{a a'}  \Sigma^{(\mu\nu,c)}_{a \eta s, a' \eta'  s'} c_{\kk a \eta  s}^\dagger c_{\kk a' \eta' s'} 
}\ , 
\end{equation}
where $\mu,\nu=0,x,y,z$ and
\begin{equation}\label{eq:flatU4-mat1}
\Sigma^{(0\nu,f)} = \sigma_0 \tau_0 \spin_{\nu}, \qquad
\Sigma^{(0\nu,c)} = ( \sigma_0 \tau_0 \spin_{\nu}) \oplus (\sigma_0 \tau_0 \spin_{\nu} ) \ , 
\end{equation}
\begin{equation}\label{eq:flatU4-mat2}
\Sigma^{(x\nu,f)} = \sigma_y \tau_x \spin_{\nu}, \qquad
\Sigma^{(x\nu,c)} = ( \sigma_y \tau_x \spin_{\nu}) \oplus ( - \sigma_y \tau_x \spin_{\nu} ) \ , 
\end{equation}
\begin{equation}\label{eq:flatU4-mat3}
\Sigma^{(y\nu,f)} = \sigma_y \tau_y \spin_{\nu}, \qquad
\Sigma^{(y\nu,c)} = ( \sigma_y \tau_y \spin_{\nu}) \oplus ( - \sigma_y \tau_y \spin_{\nu} ) \ , 
\end{equation}
\begin{equation}\label{eq:flatU4-mat4}
\Sigma^{(z\nu,f)} = \sigma_0 \tau_z \spin_{\nu}, \qquad
\Sigma^{(z\nu,c)} = ( \sigma_0 \tau_z \spin_{\nu}) \oplus ( \sigma_0 \tau_z \spin_{\nu} ) \ . 
\end{equation}
Here $\sigma_{\mu}$, $\tau_{\mu}$, $\spin_{\mu}$ are Pauli matrices for the orbital, valley, and spin degree of freedom, respectively. 
The first and second eight-by-eight blocks in $\Sigma^{(\mu\nu,c)}$ are for the orbitals $a=1,2$ and $a=3,4$, respectively. 

Now we explicitly show that the U(4) symmetry in the third chiral limit (flat limit) of our heavy fermion model is consistent with the non-chiral flat U(4) symmetry in Ref.~\cite{Biao-TBG3,bultinck_ground_2020}, which commutes with the interaction Hamiltonian but anti-commutes with the single-particle BM Hamiltonian.  
It is worth mentioning that our U(4) rotations, unlike the ones in Ref.~\cite{Biao-TBG3},  commutes with both the interaction Hamiltonian (approximately) and the single-particle Hamiltonian (if $M=0$). 
In the Appendix B2 of  Ref.~\cite{Biao-TBG3}, the sewing matrices of $T$, $C_{2z}$, and $P$ have been fixed as $\kk$-independent matrices shown in \cref{eq:gauge-TBG3}. 
The generators of the non-chiral flat U(4) symmetry in this gauge are 
\begin{equation} \label{eq:flatU4-TBG3}
\sigma_0\tau_0 \spin_\nu,\qquad 
\sigma_y\tau_x \spin_\nu,\qquad 
\sigma_y\tau_y \spin_\nu,\qquad 
\sigma_0\tau_z \spin_\nu,\qquad 
(\nu=0,x,y,z)\ ,
\end{equation}
where $\sigma_{\mu}$, $\tau_{\mu}$, abd  $\spin_\mu$ are Pauli matrices for the band, valley, and spin degrees of freedom, respectively. 
As shown in \cref{sec:Wannier} and \cref{fig:Wannier}, in most area of the moir\'e BZ, the middle two bands are mainly contributed by the local orbitals. 
Thus we expect that, for $\kk\neq 0$ where the overlap between local orbitals and the middle two bands is finite, the U(4) generators on the local orbitals (after certain gauge transformation) should be same as those in \cref{eq:flatU4-TBG3}.
We apply the gauge transformation $U=e^{i\frac{\pi}4\sigma_0\tau_z} e^{-i\frac{\pi}4 \sigma_x\tau_z }$ to the local orbitals such that the transformed representation matrices $U D^f(g) U^\dagger$ become same as \cref{eq:gauge-TBG3}.
(See \cref{eq:Df} for definition of $D^f(g)$'s.)
The U(4) generators for the local orbitals in \cref{eq:flatU4-mat1,eq:flatU4-mat2,eq:flatU4-mat3,eq:flatU4-mat4} transform to
\begin{equation}
\sigma_0 \tau_0 \spin_\nu,\qquad -\sigma_y \tau_y \spin_\nu, \qquad 
\sigma_y \tau_x \spin_\nu,\qquad \sigma_0 \tau_z \spin_\nu,\qquad (\nu=0,x,y,z)\ . 
\end{equation}
which generate the same U(4) rotations as \cref{eq:flatU4-TBG3}. 
For $\kk=0$, our local orbital basis (forming the representation $\Gamma_3$) is different from the basis in Ref.~\cite{Biao-TBG3} (forming the representation $\Gamma_1\oplus \Gamma_2$), thus the above  comparison is only valid for $\kk\neq 0$.

As will be revealed by the numerical calculations in \cref{sec:GS-nu=0}, our flat-U(4) symmetry is a better approximation than our (first) chiral-U(4) symmetry.
A heuristic argument is the following. 
The flat-U(4) breaking parameter $M$ only affects low energy single-particle states of $H^{(c,\eta)}$ (\cref{eq:Hc-summary}) because the $M$ term is just a weak perturbation for high energy states, while the chiral-U(4) breaking parameter $v_\star'$ mainly affects high energy single-particle states because it enters the Hamiltonian in the form of $v_\star'k_x \sigma_0 \pm iv_\star'k_y \sigma_z$. 
Roughly speaking, in the presence of $v_\star'$ and $M$, low energy components of the single-particle Hamiltonian do not commute with the flat-U(4) rotation, while high energy components do not commute with the chiral-U(4) rotation.
Therefore, the energy change after a chiral-U(4) rotation will be larger than that after a flat-U(4) rotation, implying that the flat-U(4) symmetry is better approximation of the Hamiltonian. 

In the end of this section we split the flat-U(4) generators to the fundamental representations of the flat-U(4) group. 
In Ref.~\cite{Biao-TBG3}, the fundamental basis is obtained by diagonalizing the Pauli matrix $\sigma_y$ in the band basis that is fixed by the sewing matrices in \cref{eq:gauge-TBG3}. 
As discussed below \cref{eq:flatU4-TBG3}, for $\kk\neq0$, the $f$-orbital basis is related to the basis of Ref.~\cite{Biao-TBG3} by the rotation $U=e^{i\frac{\pi}4\sigma_0\tau_z} e^{-i\frac{\pi}4\sigma_x\tau_z}$. 
Thus now the fundamental U(4) representations should be obtained by diagonalizing $U^\dagger \sigma_y U = -\sigma_z\tau_z$. 
In the following paragraph we prove that the eigenvalue $\xi$ of $\sigma_z\tau_z$ indeed labels fundamental U(4) representations. 
It is worth mentioning that the choice of fundamental U(4) representations is not unique.
For example, since all the U(4) rotations in \cref{eq:flatU4-mat1,eq:flatU4-mat2,eq:flatU4-mat3,eq:flatU4-mat4} commute with $\sigma_y$, we can also use eigenvalues of $\sigma_y$ in our heavy fermion basis to label fundamental representations. 
We choose $\xi$ because (i) it has the physical meaning of Chern number (\cref{sec:Chern}) \cite{Biao-TBG3} and (ii) it simplifies the exchange interaction (\cref{sec:interaction-exchange}). 

First, the flat-U(4) operators are already block-diagonal: They do not mix the eight $f$-orbitals, eight $\Gamma_3$ basis of $c$-bands, and eight $\Gamma_1\oplus\Gamma_2$ basis of $c$-bands with each other. 
We need to further split each block into two fundamental representations of the flat-U(4) group.
We notice that the flat-U(4) generators either do not change the $\alpha,\eta$ ($a,\eta$) indices of the $f$-electrons ($c$-electrons) or flip them at the same time. 
Therefore, the index $\xi=(-1)^{\alpha-1}\eta$ ($\xi=(-1)^{a-1}\eta$) of $f$-electrons ($c$-electrons) is unchanged under the flat-U(4) rotations. 
(In general, $\xi$ is not conserved by $\hH_0$. For example, the $M$ and $v_\star'$ terms in \cref{eq:Hc-summary,eq:Hcf-summary} flip the $\xi$ index.) 
Consequently, the four components with the same $\xi$ in each block form a fundamental representation of the flat-U(4) group. 
For later convenience, we introduce the U(4) generator operators for fundamental representations separately
\begin{align} \label{eq:flatU4-xi-f}
\UF_{\mu\nu}^{(f,\xi)} = & \frac12 \sum_{\eta\eta'}\sum_{ss'} 
    \sum_{\kk \in \mrm{MBZ}} \sum_{\alpha\alpha'} 
    \delta_{\xi,(-1)^{\alpha-1}\eta}
    \delta_{\xi,(-1)^{\alpha'-1}\eta'}
    \Sigma^{(\mu\nu,f)}_{ \alpha \eta s, \alpha' \eta'  s'} f_{\kk \alpha  \eta s}^\dagger f_{\kk  \alpha'  \eta' s'} \ ,
\end{align}
\begin{align} \label{eq:flatU4-xi-cp}
\UF_{\mu\nu}^{(c\prime,\xi)} = & \frac12 \sum_{\eta\eta'}\sum_{ss'} 
    \sum_{|\kk| < \Lambda_c} \sum_{aa'=1,2} 
    \delta_{\xi,(-1)^{a-1}\eta}
    \delta_{\xi,(-1)^{a'-1}\eta'}
    \Sigma^{(\mu\nu,c)}_{ a \eta s, a' \eta'  s'} c_{\kk a  \eta s}^\dagger c_{\kk  a'  \eta' s'} \ ,
\end{align}
\begin{align} \label{eq:flatU4-xi-cpp}
\UF_{\mu\nu}^{(c\prime\prime,\xi)} = & \frac12 \sum_{\eta\eta'}\sum_{ss'} 
    \sum_{|\kk| < \Lambda_c} \sum_{aa'=3,4} 
    \delta_{\xi,(-1)^{a-1}\eta}
    \delta_{\xi,(-1)^{a'-1}\eta'}
    \Sigma^{(\mu\nu,c)}_{ a \eta s, a' \eta'  s'} c_{\kk a  \eta s}^\dagger c_{\kk  a'  \eta' s'} \ ,
\end{align}
respectively. 
The total flat-U(4) generators are given by $\UF_{\mu\nu} =  \sum_{\xi} \pare{ \UF_{\mu\nu}^{(f,\xi)} + \UF_{\mu\nu}^{(c\prime,\xi)} + \UF_{\mu\nu}^{(c\prime\prime,\xi)} }$. 

\subsection{Flat bands beyond the third chiral symmetry}\label{sec:flat-beyond-chiral}

Consider a system consisting of sublattices $A$ and $B$, if only hoppings between $A$- and $B$-sublattices are allowed, then the system has a chiral symmetry, eigenvalues of which are $1$ and $-1$ for the two sublattices, respectively.
Then the Hamiltonian matrix $H(\kk)$ satisfies 
\begin{equation}
     H_{\alpha,\alpha'}(\kk)  = H_{\beta,\beta'}(\kk)  = 0\ , 
\end{equation}
where $\alpha\in A$, $\beta\in B$. 
We use $L_A$ and $L_B$ to represent the number of orbitals (per unit cell) in the $A$- and $B$-sublattices, respectively.
Without loss of generality, in the following we always label the sublattice with more (or equal number) orbitals as $A$ such that $L_A\ge L_B$.
Because the dimension of the Hamiltonian matrix $H(\kk)$ is $L_A+L_B$ and the rank of the Hamiltonian is at most $2L_B$, which are contributed by $L_B$ nonzero columns $H_{\alpha,\beta}(\kk)$ and $L_B$ nonzero rows $H_{\beta,\alpha}(\kk)$ ($\alpha\in A$, $\beta\in B$), $H(\kk)$ will have at least $L_A-L_B$ zero-energy flat bands if $L_A> L_B$. 
A recent study on flat bands based on the so-called $S$-matrix \cite{cualuguaru2021general} pointed out that, in many cases, breaking the chiral symmetry does not necessarily destroy the flatness of flat bands.
A general condition for the presence of flat bands and their topological classifications are derived in Ref. \cite{cualuguaru2021general}.
Here we only focus on a simple situation, where intra-sublattice hoppings in the smaller sublattice ($B$) are taken into account.  
Adding such hoppings do break the chiral symmetry but do not change the rank of $H(\kk)$ --- with $H_{\beta,\beta'}(\kk)$ being nonzero, $H(\kk)$ still has at most $L_B$ nonzero columns and $L_B$ nonzero rows. 
Therefore, the flatness is robust against the hoppings within the smaller sublattice. 

The above discussion explains the flat bands when we take into account certain chiral breaking terms in the heavy fermion Hamiltonian. 
Following the symmetry argument in \cref{sec:conduction-bands}, we can add the quadratic term $A(k_x^2-k_y^2) \sigma_x - 2 \eta A k_x k_y \sigma_y$ to the $a=1,2$ block of \cref{eq:Hc-summary}.  
This term commutes with $S$ (\cref{eq:third-chiral}) and hence breaks the third chiral symmetry. 
However, it does not affect the flatness of the flat bands as long as $M=0$, as shown in \cref{fig:bands_chiral}(d), where $A=-30 \mrm{meV}/k_\theta^2$, $M=0$. 
Now we explain the flatness in the presence of $A$ using the $S$-matrix theory \cite{cualuguaru2021general}. 
In \cref{sec:flat-U4} we have shown that the eigenvalues of $S$ are $-1$, $+1$, and $+1$ for the $\Gamma_3 $  conduction band basis ($a=1,2$), $\Gamma_1\oplus \Gamma_2$ conduction band basis ($a=3,4$), and the local orbitals, respectively. 
Before we add the $A$ term, only hoppings between bases with opposite $S$ eigenvalues are present.
We thus can label the $\Gamma_3$ conduction band basis ($a=1,2$) and the local orbitals as the $A$-sublattice, and the $\Gamma_1\oplus \Gamma_2$ conduction band basis ($a=3,4$) as the $B$-sublattice. 
There is $L_A-L_B=2$ and hence there will be two flat bands.
Since the $A$ term is nothing but the hopping within the smaller ($B$) sublattice, according to the discussion in the last paragraph, the flatness of flat bands are stable against $A$.

\subsection{The \texorpdfstring{U(4)$\times$U(4)}{U(4)xU(4)} symmetry in chiral flat limit} \label{sec:U4xU4}

As discussed in Refs.~\cite{Biao-TBG3,bultinck_ground_2020}, the two-band projected interaction Hamiltonian will have a U(4)$\times$U(4) group when both the chiral and flat limits are achieved. 
It should be emphasized that neither chiral-U(4) nor flat-U(4) is a factor U(4) group of the total U(4)$\times$U(4) group. 
In the heavy fermion basis, when both limits are achieved ($M=v_\star'=0$), the $\xi$ index defined in \cref{sec:flat-U4}, which equals $(-1)^{\alpha-1}\eta$ and $(-1)^{a-1}\eta$ for $f$- and $c$-electrons, respectively, and labels the fundamental flat-U(4) representations, is conserved in $\hH_0$. 
Moreover, every bilinear term in $\hH_0$ only involves a single $\xi$ flavor, implying that we can write $\hH_0$ as $\hH_0 = \sum_{\xi} \hH_0^{(\xi)} $ where $\hH_0^{(\xi)}$ only contains fermion operators of the $\xi$ flavor. 
Then, not only the total flat-U(4) generators $\UF_{\mu\nu}$, but also the generators acting in each $\xi$ flavor, \ie $\UF_{\mu\nu}^{(\xi)} = \UF_{\mu\nu}^{(f,\xi)} + \UF_{\mu\nu}^{(c\prime,\xi)} + \UF_{\mu\nu}^{(c\prime\prime,\xi)}$ (\cref{eq:flatU4-xi-f,eq:flatU4-xi-cp,eq:flatU4-xi-cpp}), commute with $\hH_0$ because there is $[\UF_{\mu\nu}^{(\xi)}, \hH_0^{(\xi)}]=0$ for the two $\xi$'s separately.
A generic rotation that is commuting with $\hH_0$ can be parameterized as
\begin{equation}
    \exp\pare{ -i\sum_{\mu\nu\xi} \theta_{\mu\nu}^{(\xi)} \UF_{\mu\nu}^{(\xi)} }\ ,
\end{equation}
where $\theta_{\mu\nu}^{(+)}$ can be different with $\theta_{\mu\nu}^{(-)}$. Therefore, when $M=v_\star'=0$, $\hH_0$ has a U(4)$\times$U(4) symmetry group with the two factor U(4)'s generated by $\UF_{\mu\nu}^{(+)}$ and $\UF_{\mu\nu}^{(-)}$, respectively. 

\begin{table}[t]
\centering
\begin{tabular}{|c|c|c|c|c|c|c|}
\hline
 &  $\hH_0$ & $\hH_0(M=0)$ &  $\hH_0(v_\star'=0)$ & $\hH_0(M=v_\star'=0)$ 
 & $\hH_U + \hH_V + \hH_W + \hH_J + \hH_{\td{J}}$ & $\hH_K$ \\
\hline
Local Symmetry & U(2)$\times$U(2) & flat-U(4) & chiral-U(4) & U(4)$\times$U(4) & U(4)$\times$U(4) & chiral-U(4) \\
\hline
\end{tabular}
\caption{Summary of the local symmetries of the single-particle and interaction Hamiltonians. The single-particle Hamiltonian $\hH_0$ is summarized in \cref{sec:H-single-summary}, and interaction Hamiltonians $\hH_X$ ($X=U,V,W,J,\td{J},K$) are summarized in \cref{sec:summary-interaction}. }
\label{tab:U4-summary}
\end{table}

Now we prove that, in general, when the single-particle or interaction Hamiltonian has both the chiral-U(4) and flat-U(4) symmetry, it must at least have the SU(4)$\times$SU(4) symmetry. 
Since $\delta_{\xi,(-1)^{a-1}\eta} \delta_{\xi,(-1)^{a'-1}\eta'}$, $\sigma_y \tau_x$, $\sigma_x \tau_y $ enter the coefficient matrices of $\UF^{(\xi)}_{x\nu}$, $\UF_{x\nu}$, and $\UC_{y\nu}$, respectively, and 
\begin{equation}
     \delta_{\xi,(-1)^{a-1}\eta} \delta_{\xi,(-1)^{a'-1}\eta'} [\sigma_y \tau_x]_{\alpha \eta, \alpha'\eta'}
     = \frac12 [\sigma_y \tau_x + \xi \sigma_x \tau_y ]
\end{equation} 
we have 
\begin{equation}
    \UF^{(\xi)}_{x\nu} = \frac12 \UF_{x\nu} + \frac{\xi}2 \UC_{y\nu} \ .
\end{equation}
Similarly, we have 
\begin{equation}
    \UF^{(\xi)}_{y\nu} = \frac12 \UF_{y\nu} + \frac{\xi}2 \UC_{x\nu} \ .
\end{equation}
Therefore, when the Hamiltonian commutes with $\UF_{\mu\nu}$ and $\UC_{\mu\nu}$ separately, it must also commute with $\UF^{(\xi)}_{x\nu}$ and $\UF^{(\xi)}_{y\nu}$ for $\xi=\pm$ separately, which generate a SU(4)$\times$SU(4) group. 
To see if the Hamiltonian has a higher U(4)$\times$U(4) symmetry, one has to explicitly check if the Hamiltonian also commute $\UF^{(\xi)}_{00}$ for $\xi=\pm$ separately. 
Since the Hamiltonian must respect charge-U(1), \ie $\UF^{(+)}_{00} + \UF^{(-)}_{00}$, one only need to check if $\UF^{(+)}_{00} - \UF^{(-)}_{00}$, or equivalently the index $\xi$, is conserved in the Hamiltonian. 
In summary, the Hamiltonian will at least have a SU(4)$\times$SU(4) symmetry if it respects both flat-U(4) and chiral-U(4) symmetries.
Given the charge-U(1) is respected, the SU(4)$\times$SU(4) group will be promoted to U(4)$\times$U(4) if the Hamiltonian further commutes with $\UF^{(+)}_{00} - \UF^{(-)}_{00}$. 

Using the criterion derived in the above paragraph, we find that most terms in the interaction Hamiltonian also has the U(4)$\times$U(4) symmetry (\cref{sec:summary-interaction}). 
We summarize the local symmetries of all the single-particle and interaction Hamiltonians in \cref{tab:U4-summary}. 

\subsection{Fubini-Study metric and Berry's curvature of the flat bands in the chiral flat limit}
\label{sec:band-geometry}

It has been shown that the flat (Chern) bands of the BM model in the chiral limit are analytically solvable \cite{tarnopolsky_origin_2019}.
Because the Chern band solutions are holomorphic functions of $k_x + ik_y$ (or $k_x - ik_y$), there is a simple relation between the Berry's curvature $\Omega(\kk)$ and the Fubini-Study metric $g_{ij}(\kk)$, \ie, $ g_{ij}(\kk)= \frac12 |\Omega(\kk)| \delta_{ij}$ \cite{ledwith2020}, which further leads to $4\det g(\kk) = |\Omega(\kk)|^2$, also known as the ideal droplet condition. 
The ideal droplet condition together with the flatness of Berry's curvature, \ie $\Omega(\kk) = const.$, lead to the GMP algebra of density operators \cite{GMP1986,Sondhi2013Fractional,Roy2014BandGeometry} and hence can give rise to fractional Chern insulator phases in spin and valley polarized MATBG \cite{sheffer2021chiral,repellin_FCI_2020,abouelkomsan2020,ledwith2020}. 

In this subsection we will show that in the chiral flat limit ($v_\star'=M=0$) of our heavy fermion model the Chern band states are also analytically solvable and are holomorphic functions of $k_x + ik_y$ or $k_x - ik_y$ (up to a normalization constant). 
Thus the ideal droplet condition of the flat band will be automatically satisfied. 
(It is worth noting that this chiral limit is theoretically achieved by imposing $v_\star'=M=0$.
It shares the same symmetries as the BM model in the chiral limit, where $w_0/w_1=0$. 
But the other parameters, \eg $M$, $v_\star$, are not required to be obtained at $w_0/w_1=0$.)
According to \cref{sec:flat-U4,sec:chiral-U4}, when both (first) chiral and flat limits are achieved the single-particle Hamiltonian must commute with $C\cdot S$, representation of which is given by 
\begin{equation}
    D^f (C\cdot S) = \sigma_z \tau_0,\qquad 
    D^c(C\cdot S) = \sigma_z \tau_0 \oplus \sigma_z \tau_0 \ .  
\end{equation}
Therefore, we can block-diagonalize the Hamiltonian using the eigenvalues ($\zeta$) of $C\cdot S$.
The total Hamiltonian in momentum space can be rewritten as 
\begin{align} \label{eq:H0-chiral-flat}
\hH_0 =& \sum_{\eta s} \sum_{aa'=1,3} \sum_{|\kk|<\Lambda_c} H^{(c,\eta,+)}_{a,a'}(\kk) c_{\kk a\eta s}^\dagger c_{\kk a'\eta s}  
 + \sum_{\eta s} \sum_{|\kk|<\Lambda_c} \sum_{a=1,3} 
    \pare{ e^{-|\kk|^2\lambda^2/2 } H^{(cf,\eta,+)}_{a 1}(\kk) c_{\kk a\eta s}^\dagger f_{\kk 1 \eta s}  + h.c. } \nono\\
+ & \sum_{\eta s} \sum_{aa'=2,4} \sum_{|\kk|<\Lambda_c} H^{(c,\eta,-)}_{a,a'}(\kk) c_{\kk a\eta s}^\dagger c_{\kk a'\eta s}  
 + \sum_{\eta s} \sum_{|\kk|<\Lambda_c} \sum_{a=2,4} 
    \pare{ e^{-|\kk|^2\lambda^2/2 } H^{(cf,\eta,-)}_{a 2}(\kk) c_{\kk a\eta s}^\dagger f_{\kk 2 \eta s}  + h.c. }
\end{align}
where the two-by-two matrix $H^{(c,\eta,\zeta)}$ and two  $H^{(cf,\eta,\zeta)}$ (two-by-one) are given by 
\begin{equation} 
H^{(c,\eta,\zeta)} = v_\star\begin{pmatrix}
    0 & \eta k_x + i \zeta k_y \\
    \eta k_x - i \zeta k_y & 0
    \end{pmatrix},\qquad 
H^{(cf,\eta,\zeta)} = \begin{pmatrix}
    \gamma \\ 0
    \end{pmatrix}\ . 
\end{equation}
The Hamiltonian in each $\zeta$ sector still anti-commutes with $S$ and hence must have flat bands (\cref{sec:flat-U4}).  
As explained in \cref{sec:H-single-summary}, for the band structure to be periodic in MBZ, we can formally extend the cutoff of $c$-bands, $\Lambda_c$, to infinity. 
Here we just assume $\Lambda_c$ is much larger than the size of MBZ. 
One must be aware that, for a moire reciprocal lattice $\GG$, there is $c_{\kk+\GG,a\eta s} \neq c_{\kk, a\eta s}$ and $f_{\kk+\GG,\alpha\eta s} = f_{\kk, \alpha\eta s}$ because the former is a continuous field while the second is a lattice. 
Then the flat band solution is in the sector $(\zeta, \eta, s)$ can be written as 
\begin{equation}
d_{\kk, \zeta=+, \eta, s}^\dagger = \frac1{\sqrt{\mcl{N}_\kk}} 
     f_{\kk,1,\eta,s}^\dagger 
     + \frac1{\sqrt{\mcl{N}_\kk}} \sum_{\substack{\GG \\ |\kk+\GG|<\Lambda_c}} \frac{\gamma/v_\star}{\eta (k_x+G_x) + i (k_y+G_y) } c_{\kk+\GG,3,\eta,s}^\dagger\ ,
\end{equation}
\begin{equation}
d_{\kk, \zeta=-, \eta, s}^\dagger = \frac1{\sqrt{\mcl{N}_\kk}} 
     f_{\kk,2,\eta,s}^\dagger 
     + \frac1{\sqrt{\mcl{N}_\kk}} \sum_{\substack{\GG \\ |\kk+\GG|<\Lambda_c}} \frac{\gamma/v_\star}{\eta (k_x+G_x) - i (k_y+G_y) } c_{\kk+\GG,4,\eta,s}^\dagger\ ,
\end{equation}
where 
\begin{equation}
    \mcl{N}_\kk = 1 + \sum_{\substack{\GG \\ |\kk+\GG|<\Lambda_c}} \frac{\gamma^2/v_\star^2}{|\kk+\GG|^2}\ . 
\end{equation}
The states $d_{\kk, \zeta, \eta, s}^\dagger$ are also eigenstate of the (first) chiral operator $C$  with eigenvalues equal to $\zeta$.
Their Chern numbers are computed to be $\zeta\cdot\eta$. 

In order to calculate the quantum metric tensor, we rewrite the flat band wavefunction in the first-quantized formalism.
For simplicity, here we only focus on the $\zeta=+$, $\eta=+$, $\spin=\up$ sector and will omit the indices $\zeta,\eta,\spin$.  
There is 
\begin{equation}
    \ket{ u_{\kk} } = \frac1{\sqrt{ \mcl{N}_\kk }} \ket{\td{u}_\kk} ,\qquad 
    \ket{\td{u}_\kk} = \ket{f_1,\kk} + \sum_{\substack{\GG \\ |\kk+\GG|<\Lambda_c}} \frac{\gamma/v_\star}{ (k_x+G_x) + i (k_y+G_y) }  \ket{c_3,\kk+\GG}\ ,
\end{equation}
where $\ket{f_1,\kk}$ and $\ket{c_3,\kk+\GG}$ are $f$- ($\alpha=1$) and $c$-band basis ($a=3$), respectively, and $\ket{\td{u}_\kk}$ is a holomorphic function of $k_x+ik_y$. 
The quantum metric can be written as a sum of four terms
\begin{align}
\mathfrak{g}_{ij} (\kk) =& \bra{\partial_{k_i} u_\kk} (1 - \ket{u_\kk}\bra{u_\kk}) \ket{\partial_{k_j} u_\kk} \nono\\
=& \frac1{ { \mcl{N}_\kk }} \bra{\partial_{k_i} \td{u}_\kk} (1 - \ket{u_\kk}\bra{u_\kk}) \ket{\partial_{k_j} \td{u}_\kk}
 + \frac1{ \sqrt{ \mcl{N}_\kk }} \pare{ \partial_{k_i} \frac1{ \sqrt{ \mcl{N}_\kk }} } \bra{ \td{u}_\kk} (1 - \ket{u_\kk}\bra{u_\kk}) \ket{\partial_{k_j} \td{u}_\kk} \nono\\
+ & \frac1{ \sqrt{ \mcl{N}_\kk }} \pare{ \partial_{k_j} \frac1{ \sqrt{ \mcl{N}_\kk }} } \bra{\partial_{k_i} \td{u}_\kk} (1 - \ket{u_\kk}\bra{u_\kk}) \ket{ \td{u}_\kk} 
+ \pare{ \partial_{k_i} \frac1{ \sqrt{ \mcl{N}_\kk }} } \pare{ \partial_{k_j} \frac1{ \sqrt{ \mcl{N}_\kk }} } \bra{ \td{u}_\kk} (1 - \ket{u_\kk}\bra{u_\kk}) \ket{ \td{u}_\kk} \ .
\end{align}
Because $(1 - \ket{u_\kk}\bra{u_\kk}) \ket{\td{u}_\kk} =0$, the last three terms in the above equation vanish. 
Hence we have 
\begin{equation}
    \mathfrak{g}_{ij} (\kk) = \frac1{ { \mcl{N}_\kk }} \bra{\partial_{k_i} \td{u}(\kk)} (1 - \ket{u(\kk)}\bra{u(\kk)}) \ket{\partial_{k_j} \td{u}(\kk)}\ .
\end{equation} 
Since $\td{u}_\kk$ is holomorphic in $k=k_x + ik_y$, there are $ \partial_{k_x}\td{u}_\kk = \partial_{k}\td{u}_\kk $, $ \partial_{k_y}\td{u}_\kk = i\partial_{k}\td{u}_\kk$, and the quantum metric must have the form
\begin{equation}
    \mathfrak{g}_{ij} (\kk) = \begin{pmatrix}
     1 & i \\ -i & 1   
    \end{pmatrix} f(\kk)
\end{equation}
where $f(\kk) =\bra{\partial_k u(\kk)} (1 - \ket{u(\kk)}\bra{u(\kk)}) \ket{\partial_k u(\kk)} $ is a real function of $\kk$. 
The Fubini-Study metric and Berry's curvature are given by $g_{ij} (\kk)= \mrm{Re}[\mathfrak{g}_{ij}(\kk)] = f(\kk) \delta_{ij}$ and $\Omega(\kk) =2 \mrm{Im}[\mathfrak{g}_{xy}(\kk)] = 2f(\kk)$, respectively, thus the ideal droplet condition $ g_{ij}(\kk)= \frac12 |\Omega(\kk)| \delta_{ij}$ is satisfied.

\begin{figure}[t]
\includegraphics[width=0.8\linewidth]{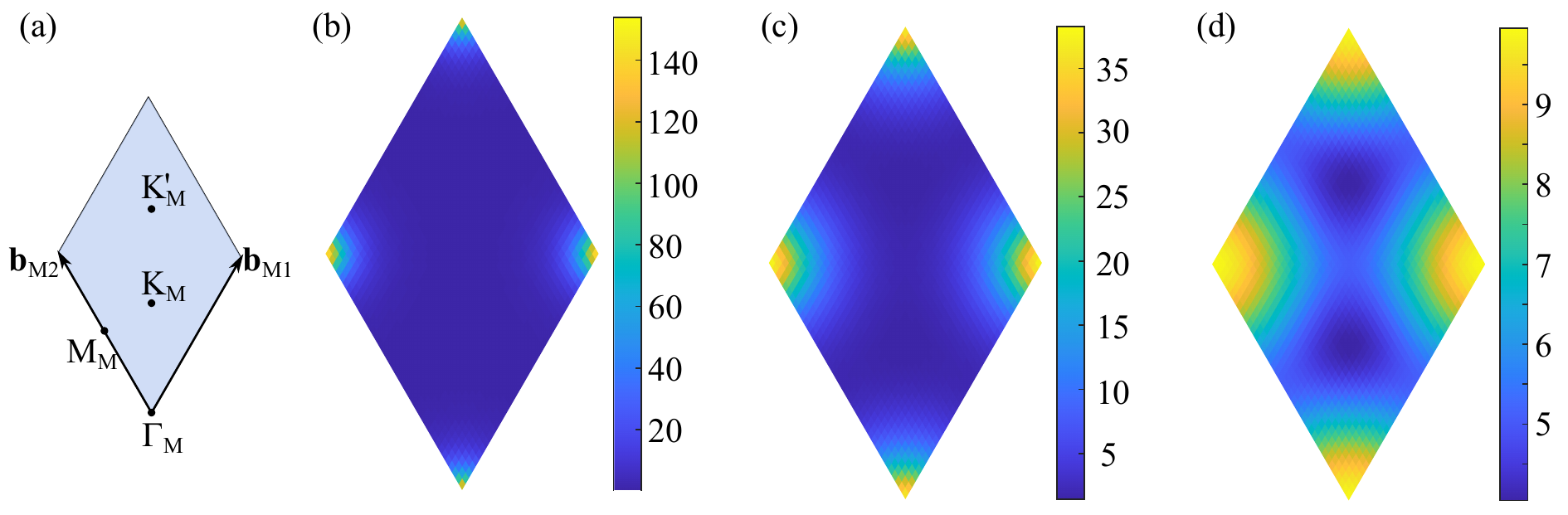}
\caption{Berry's curvatures of the flat Chern bands in the (first) chiral flat limit. (a) is an illustration of the moir\'e Brillouin zone and high symmetry momenta. (b)-(d) are the Berry's curvatures in the (first) chiral flat limit, where $\gamma$ equals to -24.75meV, -50meV, -100.8meV, respectively, and $v_\star'=M=0$, $v_\star = 1.623\mrm{eV\cdot\mathring{A}}$. The values of the Berry's curvature is normalized in the convention where the area of moir\'e Brillouin zone is $4\pi^2$.}
\label{fig:curvature}
\end{figure}

We then investigate the flatness of their Berry's curvatures. 
We first consider the parameters $v_\star=1.623\mrm{eV\cdot\mathring{A}}$, $\gamma=-24.75$meV obtained at $w_0/w_1=0.8$ (\cref{tab:H0parameters}).
The corresponding Berry's curvature is shown in \cref{fig:curvature}(b). 
We can see that it is not flat at all. 
Its deviation from perfect flatness can be estimated by the number
\begin{equation}
    \delta C = \sqrt{  \int \frac{d^2\kk} {(2\pi)^2} \pare{ \frac{\Omega(\kk)}{2\pi} - C }^2  }\ ,
\end{equation}
which is computed to be 2.723 for $\gamma = -24.75$meV.
The actual chiral limit of the BM model, \ie $w_0/w_1=0$, leads to a larger gap ($100.8$meV) between the flat bands the lowest passive bands.
That means in the actual chiral limit $\gamma$ should be $\gamma=-100.8$meV.
For comparison, in \cref{fig:curvature}(c) and (d) we present the distributions of Berry's curvatures given by $\gamma=-50$meV and $\gamma=-100.8$meV, respectively. 
It is clear that a larger $|\gamma|$ gives a flatter Berry's curvature. 
The derivations from perfect flatness are 1.1585 and 0.2570 for $\gamma=-50$meV and $\gamma=-100.8$meV, respectively. 

In summary, because of the holomorphic or anti-holomorphic property of the flat band wavefunction the ideal droplet condition is automatically achieved in the chiral flat limit of our heavy fermion model, where $v_\star'$ and $M$ are artificially imposed to zero.
The relative flatness of the Berry's curvature can also be reproduced by using parameters fitting the actual BM model bands in the chiral limit. 
These lead to the GMP algebra of density operators.
Given that the Hamiltonian (\cref{eq:H0-chiral-flat}) only involves $k_x+ik_y$ or $k_x-ik_y$ and is periodic in real space, the Jastrow type wavefunctions for fractional Chern insulators of spin and valley polarized MATBG \cite{ledwith2020} should also be applicable to our model. 

\section{The interaction Hamiltonian and its symmetries}\label{sec:interaction-Hamiltonian}

\subsection{The projected density operator and Coulomb interaction}\label{sec:Coulomb}

We can project the original basis onto the basis of the topological heavy fermion model as 
\begin{equation} \label{eq:original-basis-on-low-energy-basis}
c_{\kk,\QQ,\beta,\eta,s}^\dagger \approx \sum_{\alpha} \td{v}^{(\eta)*}_{\QQ\beta, \alpha}(\kk) f_{\kk\alpha \eta s}^\dagger + \sum_{a} \td{u}^{(\eta)*}_{\QQ\beta, a}(\kk) c_{\kk a \eta s}^\dagger
\end{equation}
Correspondingly, the projected real space basis (\cref{eq:continuous-basis}) can be written as 
\begin{equation} \label{eq:original-basis-on-low-energy-basis-real}
c_{l \beta \eta s}^\dagger(\rr) \approx \sum_{\RR \alpha} e^{i\eta\DKK_l\cdot\RR} w^{(\eta)*}_{l\beta, \alpha}(\rr-\RR) f_{\RR \alpha\eta s}^\dagger + \frac1{\sqrt{\Omega_{\rm tot}}} \sum_{|\kk|<\Lambda_c} \sum_{a} \sum_{\QQ\in \mathcal{Q}_{l \eta }} e^{-i(\kk-\QQ)\cdot\rr} \td{u}^{(\eta)*}_{\QQ\beta,a}(\kk) c_{\kk, a\eta s}^\dagger\ ,
\end{equation}
where $\Lambda_c$ is the cutoff for the conduction band basis. 
We emphasize that even though we formally introduce $\td{u}^{(\eta)*}_{\QQ\beta, a}(\kk)$, in practical calculations we will only use $\td{u}^{(\eta)*}_{\QQ\beta, a}(0)$, as we did in the k$\cdot$p expansion in \cref{sec:conduction-bands}. 
The phase factor $e^{i\eta\DKK_l\cdot\RR}$ comes from the phase shift gained by the real space basis under translation, as explained around \cref{eq:WR-wr}. 
We now can write the density operator as 
{\small
\begin{align}
& \hat{\rho} (\rr) = \sum_{\beta l \eta s} c_{l \beta \eta s}^\dagger(\rr) c_{l \beta \eta s}(\rr) \nono\\
=& \sum_{\beta l\eta s}\bigg[ \sum_{\substack{\RR \alpha \\ \RR' \alpha'} }  
    e^{i\eta\DKK_l\cdot(\RR-\RR')}
    w^{(\eta)*}_{l \beta, \alpha}(\rr-\RR)  w^{(\eta)}_{l\beta, \alpha'}(\rr-\RR') f_{\RR \alpha\eta s}^\dagger f_{\RR' \alpha' \eta s} 
    + \frac1{\Omega_{\rm tot}}\sum_{\substack{|\kk|,|\kk'|<\Lambda_c \\a a'\\ \QQ\QQ'\in \mathcal{Q}_{l\eta} }}  e^{-i(\kk-\QQ-\kk'+\QQ')\cdot\rr} 
    \td{u}^{(\eta)*}_{\QQ\beta,a}(\kk) \td{u}^{(\eta)}_{\QQ'\beta,a'}(\kk') c_{\kk a\eta s}^\dagger c_{\kk' a' \eta s} \nono\\
+& \frac1{\sqrt{\Omega_{\rm tot}}} \sum_{\RR \alpha a} \sum_{\substack{|\kk|<\Lambda_c \\ \QQ\in \mathcal{Q}_{l\eta} }}  \pare{
    w^{(\eta)*}_{l \beta, \alpha}(\rr-\RR) \td{u}^{(\eta)}_{\QQ\beta,a}(\kk) e^{i\eta\DKK_l\cdot\RR + i(\kk-\QQ)\cdot\rr}
    f_{\RR \alpha\eta s}^\dagger c_{ \kk a\eta s}
    + w^{(\eta)}_{l\beta, \alpha}(\rr-\RR) \td{u}^{(\eta)*}_{\QQ\beta,a}(\kk) e^{-i \eta\DKK_l\cdot\RR -i(\kk-\QQ)\cdot\rr}
    c^\dagger_{ \kk a\eta s} f_{\RR \alpha\eta s}  }  \bigg ]
\end{align}}
Since $w_{l\beta,\alpha}^{(\eta)}(\rr-\RR)$ is a sharp Gaussian function centered at $\RR$, we can omit the overlap between Wannier functions at different sites, \ie $w^{(\eta)*}_{l\beta, \alpha}(\rr-\RR)  w^{(\eta)}_{l\beta, \alpha'}(\rr-\RR')$ for $\RR\neq\RR'$. 
From \cref{eq:WF-form1,eq:WF-form2} we find $\sum_{l\beta} w^{(\eta)*}_{l\beta, \alpha}(\rr-\RR)  w^{(\eta)}_{l\beta, \alpha'}(\rr-\RR)$ is zero if $\alpha\neq \alpha'$.
This property is guaranteed by the $PC_{2z}T$ symmetry, which implies the constraint \cref{eq:wannier-constraint3} and hence 
\begin{align} \label{eq:no-Hund}
   & \sum_{l\beta} w^{(\eta)*}_{l\beta, \alpha}(\rr-\RR)  w^{(\eta)}_{l\beta, \ovl{\alpha}}(\rr-\RR)
= \sum_{l\beta} i l \eta (-1)^{\alpha}  w^{(\eta)*}_{-l\ovl{\beta}, \ovl{\alpha}}(\rr-\RR)  
    (-i) l\eta (-1)^{\ovl{\alpha}} w^{(\eta)}_{-l\ovl{\beta}, {\alpha}}(\rr-\RR) \nono\\
=& - \sum_{l\beta}   w^{(\eta)*}_{-l\ovl{\beta}, \ovl{\alpha}}(\rr-\RR) w^{(\eta)}_{-l\ovl{\beta}, {\alpha}}(\rr-\RR)
= - \sum_{l\beta} w^{(\eta)*}_{l\beta, \alpha}(\rr-\RR)  w^{(\eta)}_{l\beta, \ovl{\alpha}}(\rr-\RR)\ = 0 \ . 
\end{align}
We can understand this property as a consequence of the (single-particle) anti-unitary particle-hole symmetry $\PH = PC_{2z}T$.
According to the representations in \cref{eq:Df}, $D^f(P)=i\sigma_z \tau_z$, $D^f(C_{2z}T)=\sigma_x \tau_0$, there is $D^f(\PH) = -\sigma_y \tau_z$. 
Since $\PH^2=-1$ and $\ket{W_{\RR2 \eta s}} = -i \eta \PH \ket{W_{\RR 1 \eta s}}$ there must be $\bra{W_{\RR2 \eta s}} O \ket{W_{\RR1 \eta s}}= 0$ for any hermitian operator that commutes with $\PH$ (Kramers theorem). 
\cref{eq:no-Hund} can be obtained if we choose $O$ as the density operator at $\rr$. 
According to the constraints \cref{eq:wannier-constraint0,eq:wannier-constraint3}, the density of the Wannier functions
\begin{equation} \label{eq:nf(r)}
    n_f(\rr) = \sum_{l \beta} |w_{l\beta,\alpha}^{(\eta)} (\rr)|^2
\end{equation}
does not depend on $\alpha$ and $\eta$. 
Therefore, we can simplify the density operator as 
{\small
\begin{align} \label{eq:rho}
&\hat{\rho} (\rr) 
= \sum_{\eta s}\bigg[ \sum_{\RR \alpha  }  
    n_f(\rr-\RR) f_{\RR \alpha\eta s}^\dagger f_{\RR \alpha \eta s} 
    + \frac1{\Omega_{\rm tot}} \sum_{l\beta aa'} \sum_{\substack{|\kk|,|\kk'|<\Lambda_c }} \sum_{ \QQ\QQ'\in \mathcal{Q}_{l\eta} } e^{-i(\kk-\QQ-\kk'+\QQ')\cdot\rr} 
    \td{u}^{(\eta)*}_{\QQ\beta,a}(\kk) \td{u}^{(\eta)}_{\QQ'\beta,a'}(\kk') c_{\kk a\eta s}^\dagger c_{\kk' a' \eta s} \nono\\
& + \frac1{\sqrt{\Omega_{\rm tot}}} \sum_{\substack{l\beta\\\RR \alpha a}}  \sum_{\substack{|\kk|<\Lambda_c \\ \QQ\in \mathcal{Q}_{l\eta} }} \pare{
    w^{(\eta)*}_{l \beta, \alpha}(\rr-\RR) \td{u}^{(\eta)}_{\QQ\beta,a}(\kk) e^{i\eta\DKK_l\cdot\RR + i( \kk-\QQ)\cdot\rr}
    f_{\RR \alpha\eta s}^\dagger c_{ \kk a\eta s}
    + w^{(\eta)}_{l\beta, \alpha}(\rr-\RR) \td{u}^{(\eta)*}_{\QQ\beta,a}(\kk) e^{-i\eta\DKK_l\cdot\RR  -i(\kk-\QQ)\cdot\rr}
    c^\dagger_{ \kk a\eta s} f_{\RR \alpha\eta s}  }  \bigg ]\ . 
\end{align}}


We now consider the interaction Hamiltonian of the form
\begin{equation} \label{eq:HI-rho-rho}
    \hat{H}_I = \frac12 \int d^2\rr_1 d^2\rr_2 V(\rr_1-\rr_2) :\hrho(\rr_1): :\hrho(\rr_2): 
\end{equation}
where $V$ is the double-gate-screened Coulomb interaction (will be defined in next paragraph) and $:\hrho: = \hrho - \bra{G_0} \hrho \ket{G_0}$ is the normal ordered density operator with respect to a state $\ket{G_0}$ at charge neutrality point. 
We assume $\ket{G_0}$ is such a state that 
\begin{equation}
\bra{G_0} c_{\kk,\QQ,\alpha,\eta,s}^\dagger c_{\kk',\QQ',\alpha',\eta',s'} \ket{G_0}
= \frac12 \delta_{\kk\kk'} \delta_{\QQ\QQ'} \delta_{\alpha\alpha'} \delta_{\eta\eta'} \delta_{ss'}\ . 
\end{equation}
One can verify 
\begin{equation}
\bra{G_0} f_{\RR\alpha\eta s}^\dagger f_{\RR'\alpha'\eta' s'} \ket{G_0} 
= \frac12 \delta_{\RR\RR'} \delta_{\alpha\alpha'} \delta_{\eta\eta'} \delta_{ss'} ,\qquad 
\bra{G_0} c_{\kk a\eta s}^\dagger c_{\kk' a' \eta' s'} \ket{G_0}
= \frac12 \delta_{\kk\kk'} \delta_{aa'} \delta_{\eta\eta'} \delta_{ss'}\ ,
\end{equation}
\begin{equation}
\bra{G_0} f_{\RR\alpha\eta s}^\dagger  c_{\kk' a' \eta' s'} \ket{G_0} 
    = \bra{G_0} c_{\kk a\eta s}^\dagger f_{\RR'\alpha'\eta' s'} \ket{G_0} = 0\ . 
\end{equation}
Therefore, we can write the normal ordered density operator as 
{\small
\begin{align} \label{eq:rho-normal}
&:\hat{\rho} (\rr):
= \sum_{\eta s}\bigg[ \sum_{\RR \alpha  }  
    n_f(\rr-\RR) :f_{\RR \alpha\eta s}^\dagger f_{\RR \alpha \eta s} :
    + \frac1{\Omega_{\rm tot}} \sum_{l\beta aa'} \sum_{\substack{|\kk|, |\kk'|< \Lambda_c \\ \QQ\QQ'\in \mathcal{Q}_{l\eta}  }}  e^{-i(\kk-\QQ-\kk'+\QQ')\cdot\rr} 
    \td{u}^{(\eta)*}_{\QQ\beta,a}(\kk) \td{u}^{(\eta)}_{\QQ'\beta,a'}(\kk') 
    :c_{\kk a\eta s}^\dagger c_{\kk' a' \eta s}: \nono\\
& + \frac1{\sqrt{\Omega_{\rm tot}}} \sum_{ \substack{ l\beta \\ \RR \alpha a} } \sum_{\substack{ |\kk|<\Lambda_c \\ \QQ\in \mathcal{Q}_{l\eta} }}  \pare{
    w^{(\eta)*}_{l \beta, \alpha}(\rr-\RR) \td{u}^{(\eta)}_{\QQ\beta,a}(\kk) e^{i\eta\DKK_l\cdot\RR + i(\kk-\QQ)\cdot\rr}
    f_{\RR \alpha\eta s}^\dagger c_{ \kk a\eta s}
    + w^{(\eta)}_{l\beta, \alpha}(\rr-\RR) \td{u}^{(\eta)*}_{\QQ'\beta,a'}(\kk) e^{-i\eta\DKK_l\cdot\RR-i(\kk-\QQ)\cdot\rr}
    c^\dagger_{ \kk a\eta s} f_{\RR \alpha\eta s}  }  \bigg ]\ . 
\end{align}}
where 
\begin{equation} \label{eq:f-c-normal}
:f_{\RR \alpha\eta s}^\dagger f_{\RR \alpha \eta s}:  = f_{\RR \alpha\eta s}^\dagger f_{\RR \alpha \eta s} - \frac12, \qquad 
:c_{\kk a\eta s}^\dagger c_{\kk' a' \eta s}: = c_{\kk a\eta s}^\dagger c_{\kk' a' \eta s} - \frac12 \delta_{aa'}\ ,
\end{equation}
\begin{equation}
:c^\dagger_{\kk \eta s } f_{\RR' \alpha'\eta's'}: = c^\dagger_{\kk \eta s } f_{\RR' \alpha'\eta's'},\qquad 
:f_{\RR' \alpha'\eta's'} c^\dagger_{\kk \eta s }: = f_{\RR' \alpha'\eta's'} c^\dagger_{\kk \eta s }\ . 
\end{equation}

The double-gate-screened Coulomb interaction is given by 
\begin{equation}
    V(\rr) = U_{\xi} \sum_{n=-\infty}^{\infty} \frac{(-1)^n}{ \sqrt{(\rr/\xi )^2 + n^2} }
\end{equation}
with $\xi$ being distance between the two gates, $U_{\xi} = e^2/(4\pi \epsilon \xi)$, and $\epsilon\approx 6$ is the dielectric constant. 
For $\xi=10$nm there is $U_{\xi}=24$meV. 
For convenience we introduce the Fourier transformation of $V(\rr)$ as 
\begin{equation}\label{eq:Vr-Vq}
V(\rr) = \int \frac{d^2\qq}{(2\pi)^2}V(\qq) e^{-i\qq\cdot\rr},\qquad 
V(\qq) = {\pi \xi^2 U_{\xi}}\cdot \frac{\tanh (\xi|\qq|/2)}{\xi|\qq|/2}\ . 
\end{equation}

For simplicity, we denote the four terms in \cref{eq:rho-normal} as $:\rho_{ff}:$, $:\rho_{cc}:$, $:\rho_{fc}:$, $:\rho_{cf}:$, respectively. 
Their explicit forms are given by 
\begin{equation}
:\rho_{ff}: = \sum_{\eta s} \sum_{\RR \alpha  }  
    n_f(\rr-\RR) :f_{\RR \alpha\eta s}^\dagger f_{\RR \alpha \eta s}: \ , 
\end{equation}
\begin{equation}
:\rho_{cc}: = \frac1{\Omega_{\rm tot}} \sum_{\eta s}  \sum_{l\beta a a'} \sum_{\substack{|\kk|,|\kk'|<\Lambda_c \\ \QQ\QQ'\in \mathcal{Q}_{l\eta}  }}  e^{-i(\kk-\QQ-\kk'+\QQ')\cdot\rr} 
\td{u}^{(\eta)*}_{\QQ\beta,a}(\kk) \td{u}^{(\eta)}_{\QQ'\beta,a'}(\kk') 
:c_{\kk a\eta s}^\dagger c_{\kk' a' \eta s}:\ , 
\end{equation}
\begin{equation}
:\rho_{fc}: = \frac1{\sqrt{\Omega_{\rm tot}}} \sum_{\eta s} \sum_{\substack{l\beta\\ \RR \alpha a}} \sum_{\substack{|\kk|<\Lambda_c \\ \QQ\in \mathcal{Q}_{l\eta} }} 
    w^{(\eta)*}_{l \beta, \alpha}(\rr-\RR) \td{u}^{(\eta)}_{\QQ\beta,a}(\kk) e^{i\eta\DKK_l\cdot\RR + i( \kk-\QQ)\cdot\rr}
    f_{\RR \alpha\eta s}^\dagger c_{ \kk a\eta s},\qquad 
:\rho_{cf}: = (:\rho_{fc}:)^\dagger\ .
\end{equation}
There are four diagonal terms in the interaction:
\begin{enumerate}
\item The term $:\rho_{ff}(\rr_1):V(\rr_1-\rr_2):\rho_{ff}(\rr_2):$ is a   density-density interaction of the local $f$-orbitals. As will be shown in \cref{sec:interaction-density-density}, the on-site repulsion ($U_1$) is the largest energy scale of the problem. 
\item The term $:\rho_{cc}(\rr_1):V(\rr_1-\rr_2):\rho_{cc}(\rr_2):$ is the density-density interaction of the conduction bands. 
\item The term $:\rho_{fc}(\rr_1):V(\rr_1-\rr_2):\rho_{fc}(\rr_2):$ creates two particles in the local orbitals and two holes in the conduction bands. Exciting particles or holes in the local orbitals must overcome a large repulsion energy of the $f$-electrons. Thus this term is a high energy process. We will omit this term in the Hartree-Fock calculation. 
\item The term $:\rho_{cf}(\rr_1):V(\rr_1-\rr_2):\rho_{cf}(\rr_2):$ creates two particles in the conduction bands  and two holes in the local orbitals. 
\end{enumerate}
There are also $\begin{pmatrix}
4\\2
\end{pmatrix}=6$ off-diagonal terms in the interaction:
\begin{enumerate}
\setcounter{enumi}{4}
\item The term $:\rho_{ff}(\rr_1):V(\rr_1-\rr_2):\rho_{cc}(\rr_2): + (\leftrightarrow)$ is the density-density interaction between the local orbitals and the conduction bands. This term will effectively shift the energy of conduction bands if $:\rho_{ff}:\neq 0$. 
\item The term $:\rho_{ff}(\rr_1):V(\rr_1-\rr_2):\rho_{fc}(\rr_2): + (\leftrightarrow)$ creates two particles plus one hole in the local orbitals and a hole in the conduction bands. 
As shown in \cref{sec:interaction-density-hybridization}, this term is guaranteed to be  vanishing by the symmetries. 
\item The term $:\rho_{ff}(\rr_1):V(\rr_1-\rr_2):\rho_{cf}(\rr_2): + (\leftrightarrow)$ creates two holes plus a particle in the local orbitals and a particle in the conduction bands. 
As the 6th term, this term is guaranteed to be vanishing by the symmetries.
\item The term $:\rho_{cc}(\rr_1):V(\rr_1-\rr_2):\rho_{fc}(\rr_2): + (\leftrightarrow)$ creates a particle in the local orbitals and two holes plus one particle in the conduction bands.  Exciting particles or holes in the local orbitals must overcome a large repulsion energy of the $f$-electrons. Thus this term is a high energy process. This term is also weak because the hybridization, \ie $\inn{f^\dagger c}$, is small, as discussed in \cref{sec:CI-phases}. We will omit this term in the Hartree-Fock calculation. 
\item The term $:\rho_{cc}(\rr_1):V(\rr_1-\rr_2):\rho_{cf}(\rr_2): + (\leftrightarrow)$ creates a hole in the local orbitals and two particles plus one hole in the conduction bands.  Exciting particles or holes in the local orbitals must overcome a large repulsion energy of the $f$-electrons. Thus this term is a high energy process. This term is also weak because the hybridization, \ie $\inn{c^\dagger f}$, is small, as discussed in \cref{sec:CI-phases}. We will omit this term in the Hartree-Fock calculation. 
\item The term $:\rho_{fc}(\rr_1):V(\rr_1-\rr_2):\rho_{cf}(\rr_2): + (\leftrightarrow)$ is an exchange interaction between local orbitals and conduction bands.
\end{enumerate}
We will derive their explicit forms and calculate their interaction strengths in \cref{sec:interaction-density-density,sec:interaction-density-hybridization,sec:interaction-exchange,sec:interaction-double-hybridization}.

\subsection{Density-density terms (the first and fifth terms)} \label{sec:interaction-density-density}

\subsubsection{The density-density terms of local orbitals (the first term)}

We can write the density-density interaction of the local orbitals, \ie the first term discussed in \cref{sec:Coulomb}, as 
{\small
\begin{align} \label{eq:HI1}
\hH_{U} =  & \frac12 \int d^2\rr_1 d^2\rr_2 V(\rr_1-\rr_2) :\hrho_{ff}(\rr_1): :\hrho_{ff}(\rr_2): =
\frac12 \sum_{\RR\RR'} \sum_{\alpha\eta s} \sum_{\alpha'\eta's'} 
    U(\RR-\RR'):f_{\RR \alpha \eta s}^\dagger f_{\RR \alpha\eta s}:
    :f_{\RR' \alpha' \eta' s'}^\dagger f_{\RR' \alpha'\eta' s'}: \ ,
\end{align}
}
where 
\begin{equation}
U(\RR) = \int d^2 \rr_1 d^2\rr_2\ V(\rr_1-\rr_2-\RR) n_f(\rr_1) n_f(\rr_2)\ .
\end{equation}

We will not apply the above equation for practical calculation of $U(\RR)$ because $V(\rr)$ is divergent when $\rr$ approaches zero. 
Instead we calculate $U(\RR)$ in momentum space.
According to the transformation (\cref{eq:wannier-bloch})
\begin{equation}   
w_{l\beta,\alpha}^{(\eta)}(\rr) = 
    \frac{1}{\sqrt{N \Omega_{\rm tot}}} \sum_\kk \sum_{\QQ \in \mathcal{Q}_{l\eta} } e^{i(\eta\DKK_l + \kk-\QQ)\cdot\rr }  
    \td{v}^{(\eta)}_{\QQ \beta,\alpha} (\kk)\ ,
\end{equation}
where $\Omega_{\rm tot} = N \Omega_0$, 
we can express the density function (\cref{eq:nf(r)}) as 
\begin{equation}
n_f(\rr) = \frac{1}{N^2\Omega_0} \sum_{\kk \kk'} \sum_{l\beta} \sum_{\QQ,\QQ'\in \mathcal{Q}_{l\eta}} e^{i(\kk-\QQ-\kk'+\QQ')\cdot\rr} \td{v}^{(\eta)*}_{\QQ'\beta,\alpha}(\kk')  \td{v}^{(\eta)}_{\QQ\beta,\alpha}(\kk)\ .
\end{equation}
We define the Fourier transformation of the density function as 
\begin{equation}
n_f(\rr) = \frac{1}{N\Omega_0} \sum_{\qq,\GG} n_f(\qq+\GG) e^{-i(\qq+\GG)\cdot\rr}, \qquad
n_f( \qq + \GG ) = \frac{1}{N} \sum_{\kk} \sum_{\QQ\beta}
    \td{v}^{(\eta)*}_{\QQ-\GG \beta,\alpha}(\kk+\qq)
    \td{v}^{(\eta)}_{\QQ\beta,\alpha}(\kk)\ . 
\end{equation}
The function $U(\RR)$ can be calculated as 
\begin{align}
U(\RR) =& \frac{1}{N^2\Omega_0^2} \int d^2 \rr_1 d^2\rr_2\ V(\rr_1-\rr_2-\RR) 
    \sum_{\qq\GG} \sum_{\qq'\GG'} e^{-i(\qq+\GG)\cdot\rr_1 - i(\qq'+\GG')\cdot\rr_2} n_f(\qq+\GG) n_f(\qq'+\GG') \nono\\
=& \frac{1}{N^2\Omega_0^2} \int d^2\rr_2\ 
\sum_{\qq\GG} \sum_{\qq'\GG'} V(\qq+\GG) e^{- i(\qq+\GG+\qq'+\GG')\cdot\rr_2 -i(\qq+\GG)\cdot\RR} n_f(\qq+\GG) n_f(\qq'+\GG') \nono\\
=& \frac{1}{N\Omega_0} \sum_{\qq\GG}  V(\qq+\GG)  
    e^{-i\qq\cdot\RR} n_f(\qq+\GG) n_f(-\qq-\GG)\ . 
\end{align}
For $w_0/w_1=0.8$, we find that the onsite, nearest neighbor, next-nearest neighbor, and third nearest neighbor interactions are 
\begin{equation}
U(0) = 57.95\mrm{meV},\qquad 
U(\aa_{M1}) = 2.157\mrm{meV},\qquad 
U(\aa_{M1} - \aa_{M2} ) = 0.1144\mrm{meV},\qquad 
U(2\aa_{M1}) = 0.04145\mrm{meV}\ ,
\end{equation}
respectively. 
We can see that the interaction decays quickly as $\RR$ and the onsite term ($U_1=U(0)$) is much larger than the other terms.
In fact, we find that $U_1$ is the largest energy scale in this problem. 
Thus the onsite interaction will dominate the correlation physics.
Nevertheless, the $U(\RR)$ ($\RR\neq 0$) terms will contribute to a chemical potential term (for the local orbitals) in the $U_1\to \infty$ limit.
Suppose $U_1$ has frozen the charge fluctuation of local orbitals, \ie $\forall \RR$, $\sum_{\alpha\eta s}\inn{f_{\RR\alpha\eta s}^\dagger f_{\RR \alpha\eta s} }= N_f$ is constant, then all terms in \cref{eq:HI1} with $\RR\neq\RR'$ can be decoupled in the Hartree channel as 
\begin{align} \label{eq:chemical-U-local-orbitals}
 & \frac12 \sum_{\RR\neq \RR'} \sum_{\alpha\eta s} \sum_{\alpha'\eta's'} 
    U(\RR-\RR'):f_{\RR \alpha \eta s}^\dagger f_{\RR \alpha\eta s}:
    :f_{\RR' \alpha' \eta' s'}^\dagger f_{\RR' \alpha'\eta' s'}:\nono\\
=& \frac12 \sum_{\RR\neq \RR'} \sum_{\alpha\eta s} \sum_{\alpha'\eta's'} 
    U(\RR-\RR') \pare{ 2\inn{:f_{\RR \alpha \eta s}^\dagger f_{\RR \alpha\eta s}:}
    :f_{\RR' \alpha' \eta' s'}^\dagger f_{\RR' \alpha'\eta' s'}:
    - \inn{:f_{\RR \alpha \eta s}^\dagger f_{\RR \alpha\eta s}:} \inn{:f_{\RR' \alpha' \eta' s'}^\dagger f_{\RR' \alpha'\eta' s'}:} 
    } \nono\\
=&   \pare{ (N_f-4) \sum_{\RR'\neq 0 } U(\RR') }
    \sum_{\RR} \sum_{\alpha\eta s} f_{\RR \alpha\eta s}^\dagger f_{\RR \alpha\eta s} + const.  
\end{align}
In order to capture the physics of chemical potential shift in an as simple as possible model, we consider the approximate Hamiltonian of the form 
{\small
\begin{equation}\label{eq:HU-explicit}
\hH_{U}  = \frac{U_1}{2} \sum_{\RR} \sum_{\alpha\eta s} \sum_{\alpha'\eta's'} 
    :f_{\RR \alpha \eta s}^\dagger f_{\RR \alpha\eta s}:
    :f_{\RR \alpha' \eta' s'}^\dagger f_{\RR \alpha'\eta' s'}: 
+ \frac{U_2}{2} \sum_{\inn{\RR\RR'}} \sum_{\alpha\eta s} \sum_{\alpha'\eta's'} 
:f_{\RR \alpha \eta s}^\dagger f_{\RR \alpha\eta s}:
:f_{\RR' \alpha' \eta' s'}^\dagger f_{\RR' \alpha'\eta' s'}: \ ,
\end{equation}}
where $\inn{\RR\RR'}$ only sums over nearest neighbor pairs on the triangular lattice.
Rather than naively choosing $U_2 = U(\aa_{M1})$, we choose $U_2$ as such a value that it gives the same chemical potential shift as \cref{eq:chemical-U-local-orbitals}.
In other words, we absorb all the chemical potential shift effects coming from the $U(\RR)$ ($\RR\neq 0$) terms into the nearest neighbor repulsion.
$U_2$ can be calculated as 
\begin{equation}
U_{2} = \frac1{6} \sum_{\RR\neq 0} U(\RR) 
= \frac16 \pare{  - U_1 +  \frac1{\Omega_0} \sum_{\GG} V(\GG)  n_f(\GG) n_f(-\GG)  }\ .
\end{equation}
For $w_0/w_1=0.8$ we find $U_2 = 2.329$meV, which is only slightly larger than $U(\aa_{M1})$.

\begin{table}[t]
\begin{tabular}{|c|c|c|c|c|c|c|}
\hline 
$w_0/w_1$ & $U_1$ & $U_2$ & $W_1=W_2$ &  $W_{3}=W_4$ & $J$ & $K$  \\
\hline
1.0 &  69.35 & 1.914 &  43.46 & 52.90 & 14.48 & 5.122\\
\hline 
0.9 &  64.56 & 2.006 &  43.84 & 51.37 & 15.40 & 4.958 \\
\hline
0.8 & 57.95 & 2.329 & 44.03 & 50.20 & 16.38 & 4.887\\
\hline
0.7 & 51.72 & 2.656 & 44.05 & 49.33 & 18.27 & 4.897\\
\hline
0.6 & 45.02 & 3.103 & 44.03 & 48.73 & 20.75 & 4.975\\
\hline 
0.5 & 37.61 & 3.709 & 44.06 & 48.35 & 23.75 & 5.111\\
\hline 
\end{tabular}
\caption{Parameters of the interaction Hamiltonian.
Parameters of the BM model used to obtain this table are  $v_F = 5.944\mrm{eV\cdot\mathring{A}}$, $|\DKK|=1.703\mrm{\mathring{A}^{-1}}$, $w_1=110\mrm{meV}$, $\theta=1.05^\circ$.
Parameters of the double-gate-screened interaction used to obtain this table are $U_{\xi}=24 {\rm meV}$, $\xi=10{\rm nm}$.  
}
\label{tab:HIparameters}
\end{table}

\subsubsection{The density-density terms between local orbitals and conduction bands (the fifth term)}
Second we consider the interaction between local orbital density and conduction electron density
{\small
\begin{align}
\hH_{W} 
=& \frac1{\Omega_{\rm tot}} \int d^2\rr_1 d^2\rr_2 \sum_{\RR \alpha\eta_1 s_1} 
    :f_{\RR \alpha\eta_1 s_1}^\dagger f_{\RR \alpha\eta_1 s_1}: n_f(\rr_1-\RR) V(\rr_1-\rr_2) \sum_{|\kk|, |\kk'|<\Lambda_c }  \sum_{\eta_2s_2 aa' } \sum_{l \beta} \sum_{\QQ,\QQ'\in \mathcal{Q}_{l\eta_2} } \nono\\
 & \times \td{u}^{(\eta_2)*}_{\QQ\beta,a}(\kk) \td{u}^{(\eta_2)}_{\QQ'\beta,a'}(\kk') e^{-i(\kk-\QQ-\kk'+\QQ')\cdot\rr_2} 
    :c_{\kk a\eta_2 s_2}^\dagger  c_{\kk' a' \eta_2 s_2}: 
\end{align}}%
Applying the transformation (\cref{eq:Vr-Vq})
\begin{equation}
    \int d^2\rr_2  V(\rr_1-\rr_2)  e^{-i(\kk-\QQ-\kk'+\QQ')\cdot\rr_2} = V(\kk-\QQ-\kk'+\QQ') e^{-i(\kk-\QQ-\kk'+\QQ')\cdot\rr_1}
\end{equation}
and the variable substitution $\rr_1\to \rr_1 + \RR$, we can rewrite $\hH_W$ as  
{\small
\begin{align}
\hH_W =& \frac1{{\Omega_{\rm tot}}} \int d^2\rr_1 \sum_{\RR \alpha\eta_1 s_1} 
:f_{\RR \alpha\eta_1 s_1}^\dagger f_{\RR \alpha\eta_1 s_1}: n_f(\rr_1) 
    \sum_{|\kk|, |\kk'|<\Lambda_c}  \sum_{\eta_2s_2 aa' l } \sum_{\QQ,\QQ'\in \mathcal{Q}_{l\eta_2} }  
    e^{-i(\kk-\QQ-\kk'+\QQ')\cdot(\rr_1+\RR)}  \nono\\ 
& \times  V(\kk-\QQ-\kk'+\QQ') \td{u}^{(\eta_2)*}_{\QQ\beta,a}(\kk) \td{u}^{(\eta_2)}_{\QQ'\beta,a'}(\kk')
    :c_{\kk a\eta_2 s_2}^\dagger  c_{\kk' a' \eta_2 s_2}:
\end{align}}
We introduce the integral
\begin{align}
X_{aa'}^{(\eta_2)} (\kk,\kk')
= \frac1{\Omega_{0}} \int d^2\rr\  n_f(\rr) \sum_{l\beta} \sum_{\QQ,\QQ'\in \mathcal{Q}_{l\eta_2}} V(\kk-\QQ-\kk'+\QQ')  e^{-i(\kk-\QQ-\kk'+\QQ')\cdot\rr}  \td{u}^{(\eta_2)*}_{\QQ\beta,a}(\kk) \td{u}^{(\eta_2)}_{\QQ'\beta,a'}(\kk')
\end{align}
such that the interaction can be written as ($e^{i(\QQ-\QQ')\cdot\RR}=1$)
\begin{equation}
\hH_{W} = \frac1{N} \sum_{\RR \alpha\eta_1 s_1} \sum_{|\kk|, |\kk'|<\Lambda_c}  \sum_{\eta_2s_2 aa'} 
    X_{aa'}^{(\eta_2)} (\kk,\kk') e^{-i(\kk-\kk')\cdot\RR} 
    :f_{\RR \alpha\eta_1 s_1}^\dagger f_{\RR \alpha\eta_1 s_1}: 
    :c_{\kk a\eta_s s_2}^\dagger  c_{\kk' a' \eta_s s_2}:
\end{equation}
The conduction bands have low energy states around $\kk\approx 0$.
For the low energy states we can approximate the integral as $X_{aa'}^{(\eta_2)} (\kk,\kk') \approx X_{aa'}^{(\eta_2)} (0,0)$
\begin{align}
X_{aa'}^{(\eta_2)} (0,0)
=& \frac1{\Omega_{0}} \int d^2\rr_1 n_f(\rr) \sum_{l\beta} \sum_{\QQ,\QQ'\in \mathcal{Q}_{l\eta_2}} V(\QQ-\QQ') e^{i(\QQ-\QQ')\cdot\rr}  \td{u}^{(\eta_2)*}_{\QQ\beta,a}(0) \td{u}^{(\eta_2)}_{\QQ'\beta,a'}(0) \nono\\
=& \frac1{\Omega_0 } \sum_{l\beta} \sum_{\QQ\GG} n_f(\GG) V(\GG)   \td{u}^{(\eta_2)*}_{\QQ\beta,a}(0) \td{u}^{(\eta_2)}_{\QQ-\GG\beta,a'}(0)\ . 
\end{align}
Here $\QQ$ indexes all vectors in $\mcl{Q}_+ \oplus \mcl{Q}_-$ and $\GG$ indexes reciprocal lattices. 
Under the crystalline symmetries  the function $ n_f(\GG) V(\GG)$ forms an identity representation, thus the summation over $\QQ,\GG,\beta$ is non-vanishing only if $\td{u}^{(\eta_2)*}_{\QQ\beta,a}(0) \td{u}^{(\eta_2)}_{\QQ-\GG\beta,a'}(0)$ also form an identity representation of the valley-preserving symmetries. 
To be concrete, because $n_f(g\GG) V(g\GG) = n_f(\GG) V(\GG)$ for any crystalline symmetry $g$ that preserves the valley, we can rewrite the $X$ matrix as 
\begin{align}
X_{aa'}^{(\eta_2)} (0,0) =\frac1{\Omega_0 } \sum_{l\beta} \sum_{\QQ\GG} n_f(g\GG) V(g\GG)   \td{u}^{(\eta_2)*}_{\QQ\beta,a}(0) \td{u}^{(\eta_2)}_{\QQ-\GG\beta,a'}(0)\ ,
\end{align}  
Relabeling $\QQ \to g^{-1}\QQ$, $\GG \to g^{-1}\GG$, we obtain 
\begin{align} \label{eq:X00-tmp1}
X_{aa'}^{(\eta_2)} (0,0) =&  \frac1{\Omega_0 } \sum_{l\beta} \sum_{\QQ\GG} n_f(\GG) V(\GG)   \td{u}^{(\eta_2)*}_{g^{-1}\QQ\beta,a}(0) \td{u}^{(\eta_2)}_{g^{-1}\QQ-g^{-1}\GG\beta,a'}(0)\ .
\end{align}
For simplicity, we denote the symmetry operator matrices in \cref{eq:D-T-C3z,eq:D-C2x-C2zT} as $D_{\QQ \alpha\eta, \QQ'\alpha'\eta'}(g) = \delta_{\QQ,g\QQ'} \delta_{\eta\eta'} \ovl{D}_{\alpha,\alpha'}^{(\eta)}(g)$ for $g$ that preserve the valley. 
Since $ \ovl{D}_{\alpha,\alpha'}^{(\eta)}(g)$'s are unitary, \ie $ \sum_{\alpha} \ovl{D}_{\alpha,\beta}^{(\eta)*}(g) \ovl{D}_{\alpha,\beta'}^{(\eta)}(g) = \delta_{\beta\beta'}$, there are 
\begin{align}
  & \sum_{\beta \QQ} \td{u}^{(\eta_2)*}_{g^{-1}\QQ\beta,a}(0) \td{u}^{(\eta_2)}_{g^{-1}\QQ-g^{-1}\GG\beta,a'}(0)
= \sum_{\beta \beta' \QQ} \sum_{\alpha} \ovl{D}_{\alpha,\beta}^{(\eta_2)*}(g)  \td{u}^{(\eta_2)*}_{g^{-1}\QQ\beta,a}(0) \ovl{D}_{\alpha,\beta'}^{(\eta_2)}\td{u}^{(\eta_2)}_{g^{-1}\QQ-g^{-1}\GG\beta',a'}(0) 
\end{align}
Since the vectors $\td{u}^{(\eta_2)}$ form irreducible representations given by \cref{eq:Dc-crystalline}, \ie 
\begin{equation}
 \sum_{\beta} \ovl{D}_{\alpha,\beta'}^{(\eta_2)}(g)\td{u}^{(\eta_2)}_{g^{-1}\QQ \beta',a}(0)
 = \sum_{\beta \QQ'} {D}_{\QQ\alpha\eta_2,\QQ'\beta'\eta_2} \td{u}^{(\eta_2)}_{\QQ' \beta',a}(0) 
 = \sum_{b} \td{u}^{(\eta_2)}_{\QQ \alpha,a'}(0) D_{b\eta_2, a\eta_2}^{(c)}(g) 
\end{equation}
there is 
\begin{align}
  & \sum_{\beta \QQ} \td{u}^{(\eta_2)*}_{g^{-1}\QQ\beta,a}(0) \td{u}^{(\eta_2)}_{g^{-1}\QQ-g^{-1}\GG\beta,a'}(0)
= \sum_{bb'}\sum_{\alpha \QQ} \td{u}^{(\eta_2)*}_{\QQ\alpha,b}(0) D_{b\eta_2, a\eta_2}^{(c)*}(g)  \td{u}^{(\eta_2)}_{\QQ-\GG\alpha,b'}(0)
D_{b'\eta_2, a'\eta_2}^{(c)}(g) 
\end{align}
Substituting this relation into \cref{eq:X00-tmp1}, we obtain
\begin{align}
X_{aa'}^{(\eta_2)} (0,0)
=& \sum_{bb'}  D_{b\eta_2,a\eta_2}^{(c)*}(g)  X_{bb'}^{(\eta_2)} (0,0) D_{b'\eta_2,a'\eta_2}^{(c)}(g)\ .
\end{align}
In other words, $X$ commutes with all $D^{(c)}(g)$ for crystalline symmetries $g$ that preserve the valley.
These symmetries are generated by $g=C_{3z}$, $C_{2x}$, $C_{2z}T$.  
Since $D^{(c)}(g)$ consists of three different irreducible representations of the crystalline symmetry, \ie $\Gamma_3\oplus \Gamma_1\oplus\Gamma_2$, $X$ must take the following form in the $\Gamma_3\oplus\Gamma_1\oplus \Gamma_2$ basis according to the Schur's lemma 
\begin{equation}
    X^{(\eta)} = \begin{pmatrix}
        W_1^{(\eta)} & 0 & 0 & 0\\
        0 & W_1^{(\eta)} & 0 & 0\\
        0 & 0 & W_3^{(\eta)} & W_3^{\prime(\eta)} \\
        0 & 0 & W_3^{\prime(\eta)} & W_3^{(\eta)} 
    \end{pmatrix}
\end{equation}
The off diagonal term $W_3^{\prime(\eta)}$ is allowed by the crystalline symmetries because the third and fourth components are mixed states of the $\Gamma_1\oplus \Gamma_2$ basis.
To be concrete, in the $\Gamma_1\oplus\Gamma_2$ subspace in each valley $\eta$, $C_{3z}$, $C_{2z}$, and $C_{2z}T$ are represented by $\sigma_0$, $\sigma_x$, $\sigma_xK$ (with $K$ being complex conjugation), respectively, and hence $W_3^{(\eta)}\sigma_0 + W_3^{\prime(\eta)}\sigma_x$ commutes with all of them. 
However, $W_3^{\prime(\eta)}$ is forbidden by the particle-hole symmetry $P$. 
Due to $[X^{(\eta)},D^{(c,\eta)}(P)]=0$ and $D^{(c,\eta)}_{aa'}(P) = D^{(c)}_{a\eta,a'\eta}(g) = \eta[(-i\sigma_z) \oplus (-i\sigma_z)]_{aa'}$,  $W_3^{\prime(\eta)}$ must vanish and hence $X^{(\eta)}$ is diagonal. 
Since $n_f(\rr)$ is a real function, there must be $n_f(\QQ-\QQ') = n_f^*(\QQ'-\QQ)$ and hence $X^{(\eta)}_{aa'}(0,0)$ is a hermitian matrix, implying that $  X^{(\eta)}_{aa}(0,0)$ are all real numbers. 
Due to the time-reversal symmetry, $X^{(\eta)}_{aa'}(0,0)$ does not depend on the $\eta$ index. 
We hence write the $X$ matrix as $X^{(\eta)}_{aa'}(0,0) = W_a \delta_{aa'}$, where $W_1=W_2$, $W_3=W_4$. 
For $w_0/w_1=0.8$ we find $W_1= 44.03$meV and $W_3 = 50.20$meV. 
Values of $W_a$ at other $w_0/w_1$ are tabulated in \cref{tab:HIparameters}. 


In summary, the density-density interaction between the local orbitals and the conduction bands can be written as 
\begin{equation} \label{eq:HW-explicit}
\hH_{W} = \frac1{N} \sum_{\RR \alpha\eta_1 s_1} \sum_{|\kk|, |\kk'|<\Lambda_c} \sum_{\eta_2s_2 a} 
    W_a  e^{-i(\kk-\kk')\cdot\RR} 
    :f_{\RR \alpha\eta_1 s_1}^\dagger f_{\RR \alpha\eta_1 s_1}: 
    :c_{\kk a\eta_2 s_2}^\dagger  c_{\kk' a \eta_2 s_2}:
\end{equation}

\subsubsection{Density-density interaction of the conduction bands (the second term)}

We third consider the density-density interaction of the conduction bands
\begin{align}
\hH_{V} =& \frac1{2\Omega_{\rm tot}^2} 
    \sum_{\beta_1 l_1 \eta_1 s_1} \sum_{\beta_2 l_2 \eta_2 s_2} \sum_{a_1 a_1' a_2 a_2'}
    \sum_{\substack{|\kk_1|,|\kk_1'| < \Lambda_c \\ \QQ_1\QQ_1'\in \mathcal{Q}_{l_1\eta_1} }} 
    \sum_{\substack{|\kk_2|,|\kk_2'|<\Lambda_c \\ \QQ_2\QQ_2'\in \mathcal{Q}_{l_2\eta_2} }} 
    \int d^2\rr_1 d^2\rr_2 V(\rr_1-\rr_2) \nono\\
&\times
    e^{-i(\kk_1-\QQ_1-\kk_1'+\QQ_1')\cdot\rr_1} 
    \td{u}^{(\eta_1)*}_{\QQ_1\beta_1,a_1}(\kk_1) \td{u}^{(\eta_1)}_{\QQ'_1\beta_1,a_1'}(\kk_1') 
    e^{-i(\kk_2'-\QQ_2'-\kk_2+\QQ_2)\cdot\rr_2} 
    \td{u}^{(\eta_2)*}_{\QQ_2'\beta_2,a_2'}(\kk_2') \td{u}^{(\eta_2)}_{\QQ_2\beta_2,a_2}(\kk_2) \nono\\
&\times :c_{\kk_1 a_1\eta_1 s_1}^\dagger c_{\kk_1' a_1' \eta_1 s_1}:
    :c_{\kk_2' a_2' \eta_2 s_2}^\dagger c_{\kk_2 a_2 \eta_2 s_2}:\ .
\end{align}
After integration over $\rr_1$, there is 
\begin{align}
\hH_{V} =& \frac1{2\Omega_{\rm tot}^2} 
    \sum_{\beta_1 l_1 \eta_1 s_1} \sum_{\beta_2 l_2 \eta_2 s_2}\sum_{a_1 a_1' a_2 a_2'}
    \sum_{\substack{|\kk_1|,|\kk_1'| < \Lambda_c \\ \QQ_1\QQ_1'\in \mathcal{Q}_{l_1\eta_1} }} 
    \sum_{\substack{|\kk_2|,|\kk_2'|<\Lambda_c \\ \QQ_2\QQ_2'\in \mathcal{Q}_{l_2\eta_2} }} 
    \int d^2\rr_2 V(\kk_1'-\QQ_1'-\kk_1+\QQ_1) \nono\\
&\times
e^{-i(\kk_2'-\QQ_2'-\kk_2+\QQ_2 -\kk_1' + \QQ_1' +\kk_1-\QQ_1 )\cdot\rr_2} 
    \td{u}^{(\eta_1)*}_{\QQ_1\beta_1,a_1}(\kk_1) \td{u}^{(\eta_1)}_{\QQ'_1\beta_1,a_1'}(\kk_1') 
    \td{u}^{(\eta_2)*}_{\QQ_2'\beta_2,a_2'}(\kk_2') \td{u}^{(\eta_2)}_{\QQ_2\beta_2,a_2}(\kk_2) \nono\\
&\times :c_{\kk_1 a_1\eta_1 s_1}^\dagger c_{\kk_1' a_1' \eta_1 s_1}:
    :c_{\kk_2' a_2' \eta_2 s_2}^\dagger c_{\kk_2 a_2 \eta_2 s_2}:\ .
\end{align}
Integration over $\rr_2$ leads to the momentum conservation $\kk_1'-\QQ_1' - \kk_1 + \QQ_1 = \kk_2'-\QQ_2' - \kk_2 + \QQ_2$.
Since $-\QQ_1' + \QQ_1$ and $-\QQ_2'+\QQ_2$ are reciprocal lattices and $\kk_1'$, $\kk_1$, $\kk_2$, $\kk_2'$ are small momenta around $\Gamma_M$, there must be $\kk_1'-\kk_1 = \kk_2'-\kk_2$ and $-\QQ_1' + \QQ_1 = -\QQ_2'+\QQ_2$ separately. 
We introduce the momentum $\qq = \kk_1' - \kk_1 = \kk_2'-\kk_2$, the reciprocal lattice $\GG = -\QQ_1' + \QQ_1 = -\QQ_2'+\QQ_2$ and rewrite the interaction as 
{\small
\begin{align}
\hH_{V} =& \frac1{2\Omega_{0} N}  
    \sum_{\beta_1 l_1 \eta_1 s_1} \sum_{\beta_2 l_2 \eta_2 s_2} \sum_{a_1 a_1' a_2 a_2'}
    \sum_{\substack{ |\kk_1| < \Lambda_c \\ \QQ_1\in \mathcal{Q}_{l_1\eta_1} }} 
    \sum_{\substack{ |\kk_2| < \Lambda_c \\ \QQ_2\in \mathcal{Q}_{l_2\eta_2} }} 
    \sum_{\GG} \sum_{\substack{\qq \\ |\kk_1+\qq|, |\kk_2+\qq| <\Lambda_c}}
     V(\qq + \GG) \nono\\
&\times
    \td{u}^{(\eta_1)*}_{\QQ_1\beta_1,a_1}(\kk_1) \td{u}^{(\eta_1)}_{\QQ_1-\GG\beta_1,a_1'}(\kk_1+\qq) 
    \td{u}^{(\eta_2)*}_{\QQ_2-\GG\beta_2,a_2'}(\kk_2+\qq) \td{u}^{(\eta_2)}_{\QQ_2\beta_2,a_2}(\kk_2) :c_{\kk_1 a_1\eta_1 s_1}^\dagger c_{\kk_1+\qq a_1' \eta_1 s_1}:
    :c_{\kk_2+\qq a_2' \eta_2 s_2}^\dagger c_{\kk_2 a_2 \eta_2 s_2}:\ .
\end{align}}%
We introduce the matrix 
\begin{equation}
X_{\eta_1 a_1 a_1', \eta_2 a_2 a_2'}(\kk_1,\kk_2; \qq) = \frac1{\Omega_0} \sum_{\GG} V(\qq+\GG) 
    \inn{ \td{u}^{(\eta_1)}_{a_1}(\kk_1) | \td{u}^{(\eta_1)}_{a_1'}(\kk_1+\qq+\GG)   }
    \inn{ \td{u}^{(\eta_2)}_{a_2'}(\kk_2+\qq+\GG) | \td{u}^{(\eta_2)}_{a_2}(\kk_2)   }
\end{equation}
such that the interaction can be simplified to 
\begin{align}
\hH_{V} =& \frac1{2N} \sum_{\eta_1 s_1 a_1 a_1'} \sum_{\eta_2 s_2 a_2 a_2'} \sum_{|\kk_1|,|\kk_2|<\Lambda_c} \sum_{\substack{\qq\\ |\kk_1+\qq|, |\kk_2+\qq|<\Lambda_c } } 
    X_{\eta_1 a_1 a_1', \eta_2 a_2 a_2'}(\kk_1,\kk_2; \qq) \nono\\
    & \qquad \times
    :c_{\kk_1 a_1\eta_1 s_1}^\dagger c_{\kk_1+\qq a_1' \eta_1 s_1}:
    :c_{\kk_2+\qq a_2' \eta_2 s_2}^\dagger c_{\kk_2 a_2 \eta_2 s_2}:\ .
\end{align}
For low energy states around $\kk=0$, we can approximate the $X$ matrix by replacing $\td{u}^{(\eta)}_{a}(\kk)$ with $\td{u}^{(\eta)}_{a}(0)$
\begin{equation}
    X_{\eta_1 a_1 a_1', \eta_2 a_2 a_2'}(\kk_1,\kk_2; \qq) \approx X_{\eta_1 a_1 a_1', \eta_2 a_2 a_2'}(0,0;\qq)  =  \frac1{\Omega_0} \sum_{\GG} V(\qq+\GG) 
    \inn{ \td{u}^{(\eta_1)}_{a_1}(0) | \td{u}^{(\eta_1)}_{a_1'}(\GG)   }
    \inn{ \td{u}^{(\eta_2)}_{a_2'}(\GG) | \td{u}^{(\eta_2)}_{a_2}(0)   } \ . 
\end{equation}
We find that the summation over $\GG$ is dominated by the $\GG=0$ component. 
For example, the diagonal element $X_{\eta_1 1 1, \eta_2 1 1}(0,0;0)$ is found to be 49.22meV, and the $\GG=0$ component is $\frac1{\Omega_0}V(0) = 48.33$meV. 
Therefore, it is reasonable to keep only the $\GG=0$ component. 
Then we can write $\hH_{V}$  as 
\begin{equation} \label{eq:HV-explicit}
\hH_{V} = \frac{1}{2\Omega_0 N} \sum_{\eta_1 s_1 a_1} \sum_{\eta_2 s_2 a_2} 
    \sum_{|\kk_1|,|\kk_2|<\Lambda_c} \sum_{\substack{\qq\\ |\kk_1+\qq|, |\kk_2+\qq|<\Lambda_c } } 
    V(\qq) 
    :c_{\kk_1 a_1\eta_1 s_1}^\dagger c_{\kk_1+\qq a_1 \eta_1 s_1}:
    :c_{\kk_2+\qq a_2 \eta_2 s_2}^\dagger c_{\kk_2 a_2 \eta_2 s_2}:\ .
\end{equation}

\subsection{Exchange terms (the tenth term)} \label{sec:interaction-exchange}

\subsubsection{Constraints of the coupling matrix}

Now we consider the $:\hrho_{fc}: V :\hrho_{cf}: + (\leftrightarrow)$ interaction discussed in \cref{sec:Coulomb} (the tenth term)
{\small
\begin{align}
\hH_{J} =& \frac1{2\Omega_{\rm tot}} \sum_{\substack{\beta_1l_1\eta_1 s_1\\\beta_2 l_2 \eta_2 s_2 } } 
    \sum_{\substack{\RR_1 \alpha_1 \\ \RR_2 \alpha_2}} 
    \sum_{a_1 a_2} \sum_{\substack{ |\kk_1|<\Lambda_c \\ \QQ_1\in \mathcal{Q}_{l_1\eta_1}}}  \sum_{\substack{|\kk_2| < \Lambda_c \\ \QQ_2 \in \mathcal{Q}_{l_2\eta_2}}}  
    \int d^2\rr_1 d^2 \rr_2 \ V(\rr_1 -\rr_2) \nono\\
& \times w^{(\eta_1)*}_{l_1 \beta_1, \alpha_1}(\rr_1-\RR_1) 
    w^{(\eta_2)}_{l_2 \beta_2, \alpha_2}(\rr_2-\RR_2) 
    \td{u}^{(\eta_1)}_{\QQ_1\beta_1,a_1}(\kk_1) e^{i\eta_1\DKK_{l_1} \cdot\RR_1 + i(\kk_1-\QQ_1)\cdot\rr_1} 
    \td{u}^{(\eta_2)*}_{\QQ_2\beta_2,a_2}(\kk_2) e^{-i\eta_2\DKK_{l_2} \cdot\RR_2 -i(\kk_2-\QQ_2)\cdot\rr_2} 
\nono\\
& \times \pare{
    f_{\RR_1 \alpha_1\eta_1 s_1}^\dagger c_{ \kk_1 a_1\eta_1 s_1} c_{\kk_2 a_2 \eta_2 s_2}^\dagger f_{\RR_2 \alpha_2 \eta_2 s_2} 
    + c_{\kk_2 a_2 \eta_2 s_2}^\dagger f_{\RR_2 \alpha_2 \eta_2 s_2} f_{\RR_1 \alpha_1\eta_1 s_1}^\dagger c_{ \kk_1 a_1\eta_1 s_1}    } 
\end{align}}
Now we argue that, since the Wannier functions are well localized, the $\RR_1=\RR_2$ terms will be much stronger than the $\RR_1\neq \RR_2$ terms. 
In \cref{sec:interaction-density-density} we find that the terms $\int d^2\rr_1 d^2 \rr_2 \ V(\rr_1 -\rr_2) n_f(\rr_1-\RR_1) n_f(\rr_2 -\RR_2)$ with $\RR_1=\RR_2$ are about 30 times the terms with $\RR_1,\RR_2$ being nearest neighbors. 
Now $n_f(\rr-\RR)$ is replaced by $w(\rr-\RR)$.
Since $w(\rr-\RR)$ ($\sim \sqrt{n(\rr-\RR)}$) has a similar profile as $n_f(\rr-\RR)$, the $\RR_1= \RR_2$ terms will still be much stronger than the $\RR_1\neq \RR_2$ terms. 
(We have also numerically verified this argument. We find that, for $w_0/w_1=0.8$, the largest coupling for $|\RR_1-\RR_2|=0$ and $|\RR_1-\RR_2|=a_M$ are about 16meV and 1meV, respectively.)
Thus we can simplify this term as 
{\small
\begin{align}
\hH_{J} =& \frac1{2\Omega_{\rm tot}} \sum_{\substack{\beta_1l_1\eta_1 s_1\\\beta_2 l_2 \eta_2 s_2 } } 
    \sum_{\RR \alpha_1 \alpha_2} 
    \sum_{a_1 a_2} \sum_{\substack{ |\kk_1|<\Lambda_c \\ \QQ_1\in \mathcal{Q}_{l_1\eta_1}}}  \sum_{\substack{|\kk_2| < \Lambda_c \\ \QQ_2 \in \mathcal{Q}_{l_2\eta_2}}}  
    \int d^2\rr_1 d^2 \rr_2 \ V(\rr_1 -\rr_2) \nono\\
& \times w^{(\eta_1)*}_{l_1\beta_1 , \alpha_1}(\rr_1-\RR) 
    w^{(\eta_2)}_{l_2\beta_2 , \alpha_2}(\rr_2-\RR) 
    e^{i(\eta_1\DKK_{l_1} - \eta_2\DKK_{l_2})\cdot\RR} 
    \td{u}^{(\eta_1)}_{\QQ_1\beta_1,a_1}(\kk_1) e^{i(\kk_1-\QQ_1)\cdot\rr_1} 
    \td{u}^{(\eta_2)*}_{\QQ_2\beta_2,a_2}(\kk_2) e^{-i(\kk_2-\QQ_2)\cdot\rr_2} 
\nono\\
& \times \pare{
    f_{\RR \alpha_1\eta_1 s_1}^\dagger c_{ \kk_1 a_1\eta_1 s_1} c_{\kk_2 a_2 \eta_2 s_2}^\dagger f_{\RR \alpha_2 \eta_2 s_2} 
    + c_{\kk_2 a_2 \eta_2 s_2}^\dagger f_{\RR \alpha_2 \eta_2 s_2} f_{\RR \alpha_1\eta_1 s_1}^\dagger c_{ \kk_1 a_1\eta_1 s_1}    } 
\end{align}}
Substituting the Fourier transformation of the interaction into the above equation, we obtain 
{\small
\begin{align}
\hH_{J} =& \frac1{2\Omega_{\rm tot}} \sum_{\substack{\beta_1l_1\eta_1 s_1\\\beta_2 l_2 \eta_2 s_2 } } 
    \sum_{\RR \alpha_1 \alpha_2} 
    \sum_{a_1 a_2} \sum_{\substack{ |\kk_1|<\Lambda_c \\ \QQ_1\in \mathcal{Q}_{l_1\eta_1}}}  \sum_{\substack{|\kk_2| < \Lambda_c \\ \QQ_2 \in \mathcal{Q}_{l_2\eta_2}}}  
    \int d^2\rr_1 d^2 \rr_2 \int \frac{d^2\qq}{(2\pi)^2} \ V(\qq) e^{-i\qq(\rr_1-\rr_2)}\nono\\
& \times w^{(\eta_1)*}_{l_1\beta_1 , \alpha_1}(\rr_1-\RR) 
    w^{(\eta_2)}_{l_2\beta_2, \alpha_2}(\rr_2-\RR) 
    e^{i(\eta_1\DKK_{l_1} - \eta_2\DKK_{l_2})\cdot\RR} 
    \td{u}^{(\eta_1)}_{\QQ_1\beta_1,a_1}(\kk_1) e^{i(\kk_1-\QQ_1)\cdot\rr_1} 
    \td{u}^{(\eta_2)*}_{\QQ_2\beta_2,a_2}(\kk_2) e^{-i(\kk_2-\QQ_2)\cdot\rr_2} 
\nono\\
& \times \pare{
    f_{\RR \alpha_1\eta_1 s_1}^\dagger c_{ \kk_1 a_1\eta_1 s_1} c_{\kk_2 a_2 \eta_2 s_2}^\dagger f_{\RR \alpha_2 \eta_2 s_2} 
    + c_{\kk_2 a_2 \eta_2 s_2}^\dagger f_{\RR \alpha_2 \eta_2 s_2} f_{\RR \alpha_1\eta_1 s_1}^\dagger c_{ \kk_1 a_1\eta_1 s_1}    } 
\end{align}}
According to \cref{eq:bloch-wannier-local}, we have 
\begin{equation}
    \int d^2 \rr_2  w^{(\eta_2)}_{l_2\beta_2 , \alpha_2}(\rr_2-\RR) e^{-i(\kk_2-\QQ_2-\qq)\cdot \rr_2} = \sqrt{\Omega_0}\ e^{-i(\kk_2-\QQ_2-\qq)\cdot\RR} \td{v}^{(\eta_2)}_{\QQ_2\beta_2,\alpha_2}(\kk_2-\qq) \ . 
\end{equation}
\begin{equation}
    \int d^2 \rr_1 w^{(\eta_1)*}_{l_1\beta_1 , \alpha_1}(\rr_1-\RR) e^{i(\kk_1-\QQ_1-\qq)\cdot \rr_1} = \sqrt{\Omega_0}\ e^{i(\kk_1-\QQ_1-\qq)\cdot\RR} \td{v}^{(\eta_1)*}_{\QQ_1\beta_1,\alpha_1}(\kk_1-\qq)\ . 
\end{equation}
Therefore, we find 
{\small
\begin{align}
\hH_{J} =& \frac1{2N} \sum_{\substack{\beta_1l_1\eta_1 s_1\\\beta_2 l_2 \eta_2 s_2 } } 
\sum_{\RR \alpha_1 \alpha_2} 
    \sum_{a_1 a_2} \sum_{\substack{ |\kk_1|<\Lambda_c \\ \QQ_1\in \mathcal{Q}_{l_1\eta_1}}}  \sum_{\substack{|\kk_2| < \Lambda_c \\ \QQ_2 \in \mathcal{Q}_{l_2\eta_2}}}  
    \int \frac{d^2\qq}{(2\pi)^2} \ V(\qq) 
    \td{v}^{(\eta_1)*}_{\QQ_1\beta_1,\alpha_1}(\kk_1-\qq) \td{u}^{(\eta_1)}_{\QQ_1\beta_1, a_1} (\kk_1) 
    \td{u}^{(\eta_2)*}_{\QQ_2\beta_2,a_2}(\kk_2) \td{v}^{(\eta_2)}_{\QQ_1\beta_1,\alpha_2}(\kk_2-\qq) 
    \nono\\
\times  &   e^{i(\eta_1\DKK_{l_1} + \kk_1-\QQ_1 - \eta_2\DKK_{l_2} - \kk_2 + \QQ_2)\cdot\RR } \pare{
    f_{\RR \alpha_1\eta_1 s_1}^\dagger c_{ \kk_1 a_1\eta_1 s_1} c_{\kk_2 a_2 \eta_2 s_2}^\dagger f_{\RR \alpha_2 \eta_2 s_2} 
    + c_{\kk_2 a_2 \eta_2 s_2}^\dagger f_{\RR \alpha_2 \eta_2 s_2} f_{\RR \alpha_1\eta_1 s_1}^\dagger c_{ \kk_1 a_1\eta_1 s_1}    } 
\end{align}}
Since $ \eta_1\DKK_{l_1}-\QQ_1 - \eta_2\DKK_{l_2} + \QQ_2 $ is a reciprocal lattice vector of the Moire lattice, there is $e^{i(\eta_1\DKK_{l_1}-\QQ_1 - \eta_2\DKK_{l_2} + \QQ_2)\cdot\RR}=1$ and hence 
{\small
\begin{align}
\hH_{J} =& \frac1{2N} \sum_{\eta_1 s_1 \eta_2 s_2} 
\sum_{\RR \alpha_1 \alpha_2} \sum_{a_1 a_2} \sum_{|\kk_1|,|\kk_2|<\Lambda_c }  
    \int \frac{d^2\qq}{(2\pi)^2} \ V(\qq) 
    \inn{\td{v}^{(\eta_1)}_{\alpha_1}(\kk_1-\qq) | \td{u}^{(\eta_1)}_{a_1}(\kk_1) }
    \inn{ \td{u}^{(\eta_2)}_{a_2}(\kk_2) |  \td{v}^{(\eta_2)}_{\alpha_2}(\kk_2-\qq) } 
    \nono\\
& \times     e^{i( \kk_1- \kk_2 )\cdot\RR } \pare{
    f_{\RR \alpha_1\eta_1 s_1}^\dagger c_{ \kk_1 a_1\eta_1 s_1} c_{\kk_2 a_2 \eta_2 s_2}^\dagger f_{\RR \alpha_2 \eta_2 s_2} 
    + c_{\kk_2 a_2 \eta_2 s_2}^\dagger f_{\RR \alpha_2 \eta_2 s_2} f_{\RR \alpha_1\eta_1 s_1}^\dagger c_{ \kk_1 a_1\eta_1 s_1}    } 
\end{align}}
We define the matrix
\begin{equation} \label{eq:exchange-matrix}
X_{\eta_1 \alpha_1 a_1, \eta_2 \alpha_2 a_2}(\kk_1,\kk_2) = \int \frac{d^2\qq}{(2\pi)^2} \ V(\qq) 
\inn{\td{v}^{(\eta_1)}_{\alpha_1}(\kk_1-\qq) | \td{u}^{(\eta_1)}_{a_1}(\kk_1) }
\inn{ \td{u}^{(\eta_2)}_{a_2}(\kk_2) |  \td{v}^{(\eta_2)}_{\alpha_2}(\kk_2-\qq) } 
\end{equation}
such that the interaction can be written as 
\begin{align}
\hH_{J} =& \frac1{2N} \sum_{ \substack{\eta_1\alpha_1a_1 \\ {\eta_2\alpha_2 a_2} } } \sum_{\RR} \sum_{|\kk_1|,|\kk_2|<\Lambda_c }  
    X_{\eta_1\alpha_1a_1, \eta_2\alpha_2 a_2}(\kk_1,\kk_2) e^{i(\kk_1-\kk_2)\cdot\RR}   \nono\\
& \times
    \pare{
    f_{\RR \alpha_1\eta_1 s_1}^\dagger c_{ \kk_1 a_1\eta_1 s_1} c_{\kk_2 a_2 \eta_2 s_2}^\dagger f_{\RR \alpha_2 \eta_2 s_2} 
    + c_{\kk_2 a_2 \eta_2 s_2}^\dagger f_{\RR \alpha_2 \eta_2 s_2} f_{\RR \alpha_1\eta_1 s_1}^\dagger c_{ \kk_1 a_1\eta_1 s_1}    } \ . 
\end{align}
Since the low energy physics only involves conduction band around $\kk_{1,2}\approx 0$, we can approximate the interaction matrix elements by those at $\kk_1=\kk_2=0$, \ie
\begin{equation} \label{eq:exchange-matrix0}
\mJ_{\eta_1 \alpha_1 a_1, \eta_2 \alpha_2 a_2} = 
X_{\eta_1 \alpha_1 a_1, \eta_2 \alpha_2 a_2}(0,0) = \int \frac{d^2\qq}{(2\pi)^2} \ V(\qq) 
\inn{\td{v}^{(\eta_1)}_{\alpha_1}(-\qq) | \td{u}^{(\eta_1)}_{a_1}(0) }
\inn{ \td{u}^{(\eta_2)}_{a_2}(0) |  \td{v}^{(\eta_2)}_{\alpha_2}(-\qq) } 
\end{equation}
We hence approximate $\hH_{J}$ as 
\begin{align}
\hH_{J} =& \frac1{2N} \sum_{ \substack{\eta_1\alpha_1a_1 \\ {\eta_2\alpha_2 a_2} } } \sum_{\RR} \sum_{|\kk_1|,|\kk_2|<\Lambda_c }  
    \mJ_{\eta_1\alpha_1a_1, \eta_2\alpha_2 a_2} e^{i(\kk_1-\kk_2)\cdot\RR}   \nono\\
& \times
    \pare{
    f_{\RR \alpha_1\eta_1 s_1}^\dagger c_{ \kk_1 a_1\eta_1 s_1} c_{\kk_2 a_2 \eta_2 s_2}^\dagger f_{\RR \alpha_2 \eta_2 s_2} 
    + c_{\kk_2 a_2 \eta_2 s_2}^\dagger f_{\RR \alpha_2 \eta_2 s_2} f_{\RR \alpha_1\eta_1 s_1}^\dagger c_{ \kk_1 a_1\eta_1 s_1}    } \ . 
\end{align}
After rearranging the creation and annihilation operators, $\hH_{J}$ can be rewritten as (up to a constant)
\begin{align} \label{eq:HJ-generic}
\hH_{J} =& - \frac1{N} \sum_{ \substack{\eta_1\alpha_1a_1 \\ {\eta_2\alpha_2 a_2} } } \sum_{\RR} \sum_{|\kk_1|,|\kk_2|<\Lambda_c }  
    \mJ_{\eta_1\alpha_1a_1, \eta_2\alpha_2 a_2} e^{i(\kk_1-\kk_2)\cdot\RR}  
    :f_{\RR \alpha_1\eta_1 s_1}^\dagger  f_{\RR \alpha_2 \eta_2 s_2} :
    :c_{\kk_2 a_2 \eta_2 s_2}^\dagger c_{ \kk_1 a_1\eta_1 s_1} :\ ,
\end{align}
where the normal ordered operators are given in \cref{eq:f-c-normal}. 

We now study the $\mJ$ matrix (\cref{eq:exchange-matrix0}). 
One may explicitly apply all the crystalline symmetries to $\mJ$ as we did around \cref{eq:X00-tmp1} to obtain the constraints satisfied by $\mJ$. 
Here we adopt a simplified notation, we use $\ket{\td{u}^{(\eta)}_{a}(\kk)}$ to represent the vector $\td{u}^{(\eta)}_{\QQ\alpha,a}(\kk)$ and $D(g)\ket{\td{u}^{(\eta)}_{a}(\kk)}$ to represent the vector obtained by acting $D(g)$ (\cref{eq:D-C2P,eq:D-C2x-C2zT,eq:D-T-C3z}) on the vector $\ket{\td{u}^{(\eta)}_{a}(\kk)}$. 
According to the $C_{2z}P$ representations formed by $f$- and $c$-electrons (\cref{eq:Df,eq:Dc-P}):
\begin{equation}
    D^{(f)}_{\alpha\eta,\alpha'\eta'}(C_{2z}P) = -i[\sigma_y]_{\alpha \alpha'} [\tau_y]_{\eta \eta'}
\end{equation}
\begin{equation}
    D^{(c)}_{a\eta, a'\eta'}(C_{2z}P) = i[\sigma_y \oplus \sigma_y]_{a a'} [\tau_y]_{\eta \eta'}
\end{equation}
there are 
\begin{equation}
D(C_{2z}P) \ket{\td{v}^{(\eta)}_{\alpha}(\kk)} = \sum_{\beta\eta'} \ket{\td{v}^{(\eta')}_{\beta}(\kk)} D_{\beta\eta',\alpha\eta}^{(f)}(C_{2z}P) = - e^{i\pi \alpha} e^{i\frac{\pi}2 \eta} 
    \ket{\td{v}^{(-\eta)}_{\ovl{\alpha}} (\kk)}\ , 
\end{equation}
\begin{equation}
D(C_{2z}P) \ket{\td{u}^{(\eta)}_{a}(\kk)} =  
  \sum_{b\eta'} \ket{\td{u}^{(\eta)}_{a}(\kk)} D^{(c)}_{b\eta',a\eta}(C_{2z}P) = 
  e^{i\pi a} e^{i\frac{\pi}2 \eta} 
  \ket{\td{u}^{(-\eta)}_{\ovl{a}} (\kk)},
\end{equation}
where $\ovl{\alpha}=2,1$ for $\alpha=1,2$ and $\ovl{a}=2,1,4,3$ for $a=1,2,3,4$. 
Inserting $D^\dagger(C_{2z}P)D(C_{2z}P)=1$ into the inner products in \cref{eq:exchange-matrix0}, it follows that
\begin{equation}
\mJ_{\eta_1 \alpha_1 a_1, \eta_2 \alpha_2 a_2} = - e^{i\pi(\alpha_2 - a_2)} \mJ_{\eta_1 \alpha_1 a_1, -\eta_2 \ovl{\alpha}_2 \ovl{a}_2} = - e^{i\pi (a_1 - \alpha_1)} \mJ_{-\eta_1 \ovl{\alpha}_1 \ovl{a}_1, \eta_2 \alpha_2 a_2}
\end{equation}
Similarly, due to the $C_{2x}$ symmetry (\cref{eq:Df,eq:Dc-crystalline}), the $\mJ$ matrix satisfies
\begin{equation}
\mJ_{\eta_1 \alpha_1 a_1, \eta_2 \alpha_2 a_2} = \mJ_{\eta_1 \ovl{\alpha}_1 \ovl{a}_1, \eta_2 \ovl{\alpha}_2 \ovl{a}_2}\ ,
\end{equation}
due to the $C_{2y}$ ($=C_{2x}C_{2z}$) symmetry (\cref{eq:Df,eq:Dc-crystalline}), the $\mJ$ matrix satisfies
\begin{equation}
\mJ_{\eta_1 \alpha_1 a_1, \eta_2 \alpha_2 a_2} = \mJ_{-\eta_1 \alpha_1 a_1, -\eta_2 \alpha_2 a_2}\ , 
\end{equation}
and due to the $C_{3z}$ symmetry (\cref{eq:Df,eq:Dc-crystalline}), the $\mJ$ matrix satisfies
\begin{equation}
\mJ_{\eta_1 \alpha_1 a_1, \eta_2 \alpha_2 a_2} = \zeta_{\eta_1 \alpha_1}^*\zeta_{\eta_1 a_1} \zeta_{\eta_2 a_2}^*\zeta_{\eta_2 \alpha_2} \mJ_{\eta_1 \alpha_1 a_1, \eta_2 \alpha_2 a_2} \ . 
\end{equation}
Here $\zeta_{\eta \alpha} = D^{(f)}_{\alpha\eta,\alpha\eta}(C_{3z})$ and $\zeta_{\eta a} = D^{(c)}_{a\eta,a\eta}(C_{3z})$ are the $C_{3z}$ eigenvalues.

\subsubsection{Leading order coupling}

For $w_0/w_1=0.8$, we find that the largest matrix elements of $\mJ$ are $16.38$meV and the second largest matrix elements are $4.28$meV.
We will only keep the largest matrix elements for simplicity.
Due to the constraints derived above, we find the largest elements of $\mJ$ are given by 
\begin{equation} \label{eq:exchange-matrix-elements}
\mJ_{\eta 13,\eta 13} = \mJ_{\eta 24,\eta 24} = - \mJ_{\eta 13,-\eta 24} = - \mJ_{\eta 24, -\eta 13} = J\ .
\end{equation}
where $J$ is the value of the largest matrix element.
The values of $J$ for other $w_0/w_1$ are tabulated in \cref{tab:HIparameters}. 
The interaction Hamiltonian can be simplified as 
{\small
\begin{align}
\hH_{J} =& -\frac{J}{N}
\sum_{\RR \alpha \eta s_1 s_2} \sum_{|\kk_1|,|\kk_2|<\Lambda_c }   
     e^{i( \kk_1- \kk_2 )\cdot\RR } 
     \big( :f_{\RR \alpha \eta s_1}^\dagger f_{\RR \alpha \eta s_2}:  :c_{\kk_2, \alpha+2, \eta s_2}^\dagger  c_{ \kk_1, \alpha+2, \eta s_1}: 
    -  f_{\RR \alpha \eta s_1}^\dagger f_{\RR, \ovl{\alpha}, -\eta s_2}  c_{\kk_2, \ovl{\alpha}+2, -\eta s_2}^\dagger c_{ \kk_1, \alpha+2, \eta s_1}  \big)\ .
\end{align}}
We can write $\hH_{J}$ in a more compact form as 
\begin{equation}\label{eq:HJ-explicit}
H_J = - \frac{J}{2N} \sum_{\RR s_1 s_2} \sum_{\alpha\alpha'\eta\eta'} \sum_{|\kk_1|,|\kk_2|<\Lambda_c }   
     e^{i( \kk_1- \kk_2 )\cdot\RR } 
     ( \eta\eta' + (-1)^{\alpha+\alpha'} )
     :f_{\RR \alpha \eta s_1}^\dagger f_{\RR \alpha' \eta' s_2}:  :c_{\kk_2, \alpha'+2, \eta' s_2}^\dagger  c_{ \kk_1, \alpha+2, \eta s_1}:  
\end{equation}

In \cref{sec:flat-U4} we have block-diagonalized the flat-U(4) generators (\cref{eq:flatU4-xi-f,eq:flatU4-xi-cp,eq:flatU4-xi-cpp}) such that each block carries a fundamental U(4) representation.
We find that $\hH_J$ is nothing but the coupling between two fundamental flat-U(4) ``spins''.
We first define the fundamental flat-U(4) generators at each site (momentum) for $f$-electrons ($c$-electrons) as 
\begin{align} \label{eq:flatU4-xi-f-R}
\UF_{\mu\nu}^{(f,\xi)}(\RR) = & \frac12 \sum_{\eta\eta'}\sum_{ss'}  \sum_{\alpha\alpha'} 
    \delta_{\xi,(-1)^{\alpha-1}\eta}
    \delta_{\xi,(-1)^{\alpha'-1}\eta'}
    \Sigma^{(\mu\nu,f)}_{ \alpha \eta s, \alpha' \eta'  s'} f_{\RR \alpha  \eta s}^\dagger f_{\RR  \alpha'  \eta' s'} \ ,
\end{align}
\begin{align} \label{eq:flatU4-xi-cp-q}
\UF_{\mu\nu}^{(c\prime,\xi)}(\qq) = & \frac1{2N} \sum_{\eta\eta'}\sum_{ss'} 
    \sum_{\substack{|\kk| < \Lambda_c \\ |\kk+\qq|<\Lambda_c }}\sum_{aa'=1,2} 
    \delta_{\xi,(-1)^{a-1}\eta}
    \delta_{\xi,(-1)^{a'-1}\eta'}
    \Sigma^{(\mu\nu,c)}_{ a \eta s, a' \eta'  s'} c_{\kk+\qq \alpha  \eta s}^\dagger c_{\kk  \alpha'  \eta' s'} \ ,
\end{align}
\begin{align} \label{eq:flatU4-xi-cpp-q}
\UF_{\mu\nu}^{(c\prime\prime,\xi)}(\qq) = & \frac1{2N} \sum_{\eta\eta'}\sum_{ss'} 
    \sum_{\substack{|\kk| < \Lambda_c \\ |\kk+\qq|<\Lambda_c }} \sum_{aa'=3,4} 
    \delta_{\xi,(-1)^{a-1}\eta}
    \delta_{\xi,(-1)^{a'-1}\eta'}
    \Sigma^{(\mu\nu,c)}_{ a \eta s, a' \eta'  s'} c_{\kk+\qq \alpha  \eta s}^\dagger c_{\kk  \alpha'  \eta' s'} \ .
\end{align}
The $\Sigma^{(\mu\nu,f)}$ and $\Sigma^{(\mu\nu,c)}$ matrices are given in \cref{eq:flatU4-mat1,eq:flatU4-mat2,eq:flatU4-mat3,eq:flatU4-mat4}. 
\cref{eq:flatU4-xi-f,eq:flatU4-xi-cp,eq:flatU4-xi-cpp} are related to the above equations as $\UF_{\mu\nu}^{(f,\xi)}= \sum_{\RR} \UF_{\mu\nu}^{(f,\xi)}(\RR)$, 
$\UF_{\mu\nu}^{(c\prime,\xi)} = N\UF_{\mu\nu}^{(c\prime,\xi)}(\qq=0)$,
$\UF_{\mu\nu}^{(c\prime\prime,\xi)} = N\UF_{\mu\nu}^{(c\prime\prime,\xi)}(\qq=0)$. 
Now we consider the product of flat-U(4) generators $\sum_{\mu\nu\xi} e^{-i\qq\cdot\RR} \UF_{\mu\nu}^{(f,\xi)}(\RR) \UF_{\mu\nu}^{(c\prime\prime,\xi)}(\qq) $. 
Substituting the explicit forms of the $\Sigma^{(\mu\nu,f)}$ and $\Sigma^{(\mu\nu,c)}$ matrices into the product and expanding the summation over $\mu$, we obtain 
\begin{align} \label{eq:HJ-U4-expansion-tmp1}
  &  \sum_{\mu\nu\xi} e^{-i\qq\cdot\RR} \UF_{\mu\nu}^{(f,\xi)}(\RR) \UF_{\mu\nu}^{(c\prime\prime,\xi)}(\qq)\nono\\
=& \frac1{4N} \sum_{\xi} \sum_{\nu=0,x,y,z} \sum_{\eta \alpha}'
    \sum_{\eta', a=3,4}' \sum_{s_1s_2s_3s_4}\sum_{\kk}' e^{-i\qq\cdot\RR} \bigg(  
    f_{\RR,\alpha,\eta, s_1}^\dagger [\spin_\nu]_{s_1 s_2} f_{\RR,\alpha,\eta s_2}  c_{\kk+\qq, a,\eta', s_3}^\dagger [\spin_\nu]_{s_3 s_4} c_{\kk, a, \eta', s_4} \nono\\
& + \pare{ (-1)^{\alpha+a}  - (-1)^{\alpha+a}  \eta\eta' } f_{\RR,\ovl{\alpha},-\eta s_1}^\dagger [\spin_\nu]_{s_1 s_2} f_{\RR,\alpha,\eta s_2} c_{\kk+\qq,\ovl{a},-\eta',s_3}^\dagger [\spin_\nu]_{s_3 s_4} c_{\kk, a, \eta', s_4} \nono\\
& + \eta \eta' f_{\RR\alpha\eta s_1}^\dagger [\spin_\nu]_{s_1 s_2} f_{\RR\alpha\eta s_2} c_{\kk+\qq a\eta' s_3}^\dagger [\spin_\nu]_{s_3 s_4} c_{\kk a \eta' s_4} \bigg) \ ,
\end{align}
where $\sum_{\eta \alpha}'$ sums over $\eta,\alpha$ satisfying $\xi = (-1)^{\alpha-1}\eta$, $\sum_{\eta, a=3,4}'$ sums over $\eta,a$ satisfying $\xi = (-1)^{a-1}\eta$, and $\sum_{\kk}'$ sums over $\kk$ satisfying $|\kk|<\Lambda_c$, $|\kk+\qq|<\Lambda_c$. 
Due to the identity $\sum_{\nu} [\spin_\nu]_{s_1 s_2} [\spin_\nu]_{s_3 s_4} = 2\delta_{s_1 s_4} \delta_{s_2 s_3}$, there is 
\begin{align}
  &  \sum_{\mu\nu\xi} e^{-i\qq\cdot\RR} \UF_{\mu\nu}^{(f,\xi)}(\RR) \UF_{\mu\nu}^{(c\prime\prime,\xi)}(\qq)\nono\\
=& \frac1{2N} \sum_{\xi} \sum_{\eta \alpha}'
    \sum_{\eta', a=3,4}' \sum_{s_1s_2}\sum_{\kk}' e^{-i\qq\cdot\RR} \bigg(  
    f_{\RR,\alpha,\eta, s_1}^\dagger f_{\RR,\alpha,\eta s_2}  c_{\kk+\qq, a,\eta', s_2}^\dagger  c_{\kk, a, \eta', s_1} 
    + \eta \eta' f_{\RR\alpha\eta s_1}^\dagger  f_{\RR\alpha\eta s_2} c_{\kk+\qq a\eta' s_2}^\dagger c_{\kk a \eta' s_1}\nono\\
& + \pare{ (-1)^{\alpha+a}  - (-1)^{\alpha+a}  \eta\eta' } f_{\RR,\ovl{\alpha},-\eta s_1}^\dagger  f_{\RR,\alpha,\eta s_2} c_{\kk+\qq,\ovl{a},-\eta',s_2}^\dagger  c_{\kk, a, \eta', s_1}  \bigg) \ . 
\end{align}
The first line after the equal sign vanishes unless $\eta=\eta'$, which, according to $\xi = (-1)^{\alpha-1}\eta = (-1)^{a-1}\eta$, also implies $a=\alpha+2$. 
The second line after the equal sign vanishes unless $\eta=-\eta'$, which, according to $\xi = (-1)^{\alpha-1}\eta = (-1)^{a-1}\eta$, also implies $a=\ovl{\alpha}+2$. 
Therefore, the above equation can be simplified to 
\begin{align}
  &  \sum_{\mu\nu\xi} e^{-i\qq\cdot\RR} \UF_{\mu\nu}^{(f,\xi)}(\RR) \UF_{\mu\nu}^{(c\prime\prime,\xi)}(\qq)\nono\\
=& \frac1{N} \sum_{\xi} \sum_{\eta \alpha}'
     \sum_{s_1s_2}\sum_{\kk}' e^{-i\qq\cdot\RR} \bigg(  
    f_{\RR,\alpha,\eta, s_1}^\dagger f_{\RR,\alpha,\eta s_2}  c_{\kk+\qq, \alpha+2,\eta, s_2}^\dagger  c_{\kk, \alpha+2, \eta, s_1} 
& - f_{\RR,\ovl{\alpha},-\eta s_1}^\dagger  f_{\RR,\alpha,\eta s_2} c_{\kk+\qq,\alpha+2,\eta,s_2}^\dagger  c_{\kk, \ovl{\alpha}+2, -\eta, s_1}  \bigg) \ . 
\end{align}
The summation $\sum_{\xi} \sum_{\eta \alpha}'$ can be equivalently replaced by $\sum_{\eta \alpha}$ that has no restriction on $\eta \alpha$. 
Comparing it with $\hH_J$, we find $\hH_J$ equals to 
\begin{equation}\label{eq:HJ-U4}
    \hH_J = -J \sum_{\RR \qq} \sum_{\mu\nu\xi} e^{-i\qq\cdot\RR} :\UF_{\mu\nu}^{(f,\xi)}(\RR): :\UF_{\mu\nu}^{(c\prime\prime,\xi)}(\qq): \ . 
\end{equation}


\subsubsection{Other coupling terms}
\label{sec:HJ-other-terms}

For $w_0/w_1 = 0.8$, we find that the second largest and third largest matrix elements are given by 
\begin{equation} \label{eq:exchange-matrix-elements-2}
\mJ_{\eta 14,\eta 14} = \mJ_{\eta 23,\eta 23} = \mJ_{\eta 14,-\eta 23} = \mJ_{\eta 14, -\eta 13} = J' = \mathrm{4.28meV} \ , 
\end{equation}
and 
\begin{align} \label{eq:exchange-matrix-elements-3}
 & \mJ_{\eta 12,\eta 23} = \mJ_{\eta 14,\eta 21} = \mJ_{\eta 21,\eta 14} = \mJ_{\eta 23, \eta 12}  \nono\\
=&  \mJ_{\eta 12,-\eta 14} = \mJ_{\eta 14, -\eta 12} = \mJ_{\eta 21,-\eta 23} = \mJ_{\eta 23, -\eta 21} = - J'' = - \mathrm{1.17meV}\ ,
\end{align}
respectively. 
Following the calculation around \cref{eq:HJ-U4-expansion-tmp1} in last subsubsection, we find that the $J'$ term can be rewritten as a ferro-magnetic coupling btween $f$-electrons and $c$-electrons 
\begin{equation}
-J' \sum_{\RR \qq} \sum_{\mu\nu \xi} e^{-i\qq\cdot\RR} 
    :\UF_{\mu\nu}^{(f,\xi)}(\RR): :\UF_{\mu\nu}^{(c\prime\prime,-\xi)}(\qq):\ . 
\end{equation}
Different with $J$, which couples U(4) moments of $f$- and $c$-electrons with the same $\xi$ index, $J'$ couples U(4) moments $f$- and $c$-electrons with opposite $\xi$ indices. 
The $J'$ term respects the flat-U(4) symmetry because it is a product of flat-U(4) moments, but it breaks the U(4)$\times$U(4) symmetry. 

The $J''$ term can be rewritten as 
\begin{equation}
J'' \frac1{N} \sum_{\RR} \sum_{|\kk_1|, |\kk_2|<\Lambda_c}  \sum_{ss' \eta} e^{i(\kk_1-\kk_2)\cdot\RR}  
    \pare{ f_{\RR \eta 1 s}^\dagger c_{\kk_1 \eta 2 s} + f_{\RR -\eta 2 s}^\dagger c_{\kk_1 -\eta 1 s}  }
    \pare{ c_{\kk_2 \eta 3 s'}^\dagger  f_{\RR \eta 2 s' }  +   c_{\kk_2 -\eta 4 s'}^\dagger  f_{\RR -\eta 1 s' } } + h.c.
\end{equation}
$J''$ cannot be written as a product of flat-U(4) moments and it breaks the flat-U(4) symmetry. 
To see the symmetry breaking, we consider to apply the valley rotation $e^{-i\frac{\pi}2\hat{\Sigma}_{x 0} }$ (\cref{eq:flatU4}). 
Using the explicit U(4) representations defined in \cref{eq:flatU4-mat1,eq:flatU4-mat2,eq:flatU4-mat3,eq:flatU4-mat4}, we find that the $J''$ term transforms to 
\begin{equation}
\to  - J'' \frac1{N} \sum_{\RR} \sum_{|\kk_1|, |\kk_2|<\Lambda_c}  \sum_{ss' \eta} e^{i(\kk_1-\kk_2)\cdot\RR}  
    \pare{ f_{\RR \eta 1 s}^\dagger c_{\kk_1 \eta 2 s} + f_{\RR -\eta 2 s}^\dagger c_{\kk_1 -\eta 1 s}  }
    \pare{ c_{\kk_2 \eta 3 s'}^\dagger  f_{\RR \eta 2 s' }  +   c_{\kk_2 -\eta 4 s'}^\dagger  f_{\RR -\eta 1 s' } } + h.c.
\end{equation}
Thus the $J''$ term is odd under $e^{-i\frac{\pi}2\hat{\Sigma}_{x 0} }$. 
Since $J''=-1.17$meV is small compared to other energy scales, \eg $U_1=57.95$meV, $J=16.38$meV, flat-U(4) is still an approximate symmetry of the full interaction Hamiltonian. 

It was shown in Ref.~\cite{Biao-TBG3} that in the flat nonchiral limit the flat-U(4) is an exact symmetry of the projected Coulomb Hamiltonian. 
But here we find that the flat-U(4) symmetry is broken by $J''$. 
The reason for this seemingly contradiction is that the $\Gamma_3$ states, which are absent in the projected Coulomb Hamiltonian, are added in the heavy fermion model.  
Notice that the inter-valley rotation in Ref.~\cite{Biao-TBG3} is generated by the unitary operator $C_{2z}P$, while the inter-valley rotation here is generated by $C_{2z}P S$, where $S$ is the third-chiral operator defined in \cref{sec:flat-U4}.  
$S$ is needed in our model because otherwise the inter-valley rotation would not commute with the kinetic energy of the dispersive $c$-bands. 
The representations of the two operators are given by $D^f(C_{2z}P)=-i\sigma_y\tau_y$, $D^c(C_{2z}P)=(i\sigma_y\tau_y) \oplus (i\sigma_y\tau_y)$ and $D^f(C_{2z}PS)=-i\sigma_y\tau_y$, $D^c(C_{2z}PS)=(-i\sigma_y\tau_y) \oplus (i\sigma_y\tau_y)$, respectively, where the first and second blocks of $D^c$ act on $\Gamma_3$ and $\Gamma_1\oplus \Gamma_2$ states, respectively. 
Since $J$ and $J'$ terms only involve $f$-electrons and $\Gamma_1\oplus\Gamma_2$ $c$-electrons, on which $C_{2z}P$ and $C_{2z}PS$ actions are the same, the conclusion of Ref.~\cite{Biao-TBG3} applies and hence $J$, $J'$ terms respect the flat-U(4) symmetry. 
However the $J''$ term involves $\Gamma_3$ $c$-electrons, on which $C_{2z}P$ and $C_{2z}PS$ actions are different, hence the conclusion of Ref.~\cite{Biao-TBG3} does not apply to $J''$. 

\subsection{Double hybridization terms (the third and fourth terms)} \label{sec:interaction-double-hybridization}

Here we determine the double hybridization terms, \ie the third and fourth terms discussed in \cref{sec:Coulomb}. 
We first consider the $:\rho_{fc}:V:\rho_{fc}:$ term
\begin{align}
\hH_{\td{J} +} =&  \frac1{2\Omega_{\rm tot}} \int d^2\rr_1 d^2\rr_2 V(\rr_1-\rr_2) 
    \sum_{ \substack{ \beta_1 l_1\eta_1 s_1 \\ \beta_2 l_2 \eta_2 s_2}  } 
    \sum_{ \substack{ \RR_1 \alpha_1 a_1 \\ \RR_2\alpha_2 a_2} }  
    \sum_{ \substack{ |\kk_1|<\Lambda_c \\ |\kk_2|<\Lambda_c }}
    \sum_{\substack{\QQ_1 \in \mathcal{Q}_{l_1\eta_1} \\ \QQ_2\in \mathcal{Q}_{l_2\eta_2}}}
    w^{(\eta_1)*}_{l_1 \beta_1, \alpha_1}(\rr_1-\RR_1) 
    \td{u}^{(\eta_1)}_{\QQ_1\beta_1,a_1}(\kk_1) e^{i\eta_1\DKK_{l_1} \cdot\RR_1} e^{i(\kk_1-\QQ_1)\cdot\rr_1} \nono\\
& \times w^{(\eta_2)*}_{l_2 \beta_2, \alpha_2}(\rr_2-\RR_2) 
        \td{u}^{(\eta_2)}_{\QQ_2\beta_2,a_2}(\kk_2) e^{i\eta_2\DKK_{l_2} \cdot\RR_2} e^{i(\kk_2-\QQ_2)\cdot\rr_2}
        f_{\RR_1 \alpha_1 \eta_1 s_1}^\dagger c_{\kk_1 a_1 \eta_1 s_1} 
        f_{\RR_2 \alpha_2 \eta_2 s_2}^\dagger c_{\kk_2 a_2 \eta_2 s_2}  
\end{align}
As explained at the beginning of \cref{sec:interaction-exchange}, since the Wannier functions are sharp Gaussian functions and $V(\rr_1-\rr_2)$ decays quickly as $|\rr_1-\rr_2|$ grows, the largest contribution to the above equation should come from the $\RR_1=\RR_2$ terms.
Therefore, in the following we only focus on the $\RR_1=\RR_2$ term.  
Substituting the Fourier transformation of the interaction, we can rewrite the interaction as 
{\small
\begin{align}
\hH_{\td{J} +} =&  \frac1{2\Omega_{\rm tot}} \int \frac{d^2\qq}{(2\pi)^2} \int d^2\rr_1 d^2\rr_2 V(\qq) e^{-i\qq\cdot(\rr_1-\rr_2)}
    \sum_{ \substack{ \beta_1 l_1\eta_1 s_1 \\ \beta_2 l_2 \eta_2 s_2}  } 
    \sum_{ \substack{ \RR \alpha_1 \alpha_2 \\ a_1 a_2} }  
    \sum_{ \substack{ |\kk_1|<\Lambda_c \\ |\kk_2|<\Lambda_c }}
    \sum_{\substack{\QQ_1 \in \mathcal{Q}_{l_1\eta_1} \\ \QQ_2\in \mathcal{Q}_{l_2\eta_2}}}
    w^{(\eta_1)*}_{l_1 \beta_1, \alpha_1}(\rr_1-\RR) 
    \td{u}^{(\eta_1)}_{\QQ_1\beta_1,a_1}(\kk_1) e^{i( \kk_1-\QQ_1)\cdot\rr_1} \nono\\
& \times w^{(\eta_2)*}_{l_2 \beta_2, \alpha_2}(\rr_2-\RR) 
        \td{u}^{(\eta_2)}_{\QQ_2\beta_2,a_2}(\kk_2) e^{i(\kk_2-\QQ_2)\cdot\rr_2}
        e^{i(\eta_1\DKK_{l_1} + \eta_2\DKK_{l_2})\cdot\RR} 
        f_{\RR \alpha_1 \eta_1 s_1}^\dagger c_{\kk_1 a_1 \eta_1 s_1} 
        f_{\RR \alpha_2 \eta_2 s_2}^\dagger c_{\kk_2 a_2 \eta_2 s_2}  
\end{align}}
According to \cref{eq:bloch-wannier-local}, there is
\begin{equation}
    \int d^2 \rr_1  w^{(\eta_1)*}_{l_1\beta_1 , \alpha_1}(\rr_1-\RR) e^{i( \kk_1-\QQ_1-\qq)\cdot \rr_1} = \sqrt{\Omega_0}\ e^{i(\kk_1-\QQ_1-\qq)\cdot\RR} \td{v}^{(\eta_1)*}_{\QQ_1\beta_1,\alpha_1}(\kk_1-\qq)\ . 
\end{equation}
Therefore, we have 
\begin{align}
\hH_{\td{J} +} =&  \frac1{2N} \int \frac{d^2\qq}{(2\pi)^2}  V(\qq) 
    \sum_{ \substack{ \beta_1 l_1\eta_1 s_1 \\ \beta_2 l_2 \eta_2 s_2}  } 
    \sum_{ \substack{ \RR \alpha_1 \alpha_2 \\ a_1 a_2} }  
    \sum_{ \substack{ |\kk_1|<\Lambda_c \\ |\kk_2|<\Lambda_c }}
    \sum_{\substack{\QQ_1 \in \mathcal{Q}_{l_1\eta_1} \\ \QQ_2\in \mathcal{Q}_{l_2\eta_2}}}
    \td{v}^{(\eta_1)*}_{\QQ_1\beta_1, \alpha_1}(\kk_1-\qq) \td{u}^{(\eta_1)}_{\QQ_1\beta_1,a_1}(\kk_1)
    \td{v}^{(\eta_2)*}_{\QQ_2\beta_2, \alpha_2}(\kk_2+\qq) \td{u}^{(\eta_2)}_{\QQ_2\beta_2,a_2}(\kk_2)
     \nono\\
& \times e^{i(\eta_1 \DKK_{l_1}  + \kk_1-\QQ_1-\qq + \eta_2 \DKK_{l_2} + \kk_2 - \QQ_2 +\qq)\cdot\RR}
        f_{\RR \alpha_1 \eta_1 s_1}^\dagger c_{\kk_1 a_1 \eta_1 s_1} 
        f_{\RR \alpha_2 \eta_2 s_2}^\dagger c_{\kk_2 a_2 \eta_2 s_2}  \ .
\end{align}
Since $ \eta_1 \DKK_{l_1} -\QQ_1 + \eta_2 \DKK_{l_2} - \QQ_2$ is always a moir\'e reciprocal lattice, there must be $e^{i(\eta_1 \DKK_{l_1} -\QQ_1 + \eta_2 \DKK_{l_2} - \QQ_2)\cdot\RR}=1$.
We can further simplify the Hamiltonian as 
\begin{align}
\hH_{\td{J} +} =&  \frac1{2N} \int \frac{d^2\qq}{(2\pi)^2}  V(\qq) 
    \sum_{  \eta_1 s_1  \eta_2 s_2  } 
    \sum_{ \substack{ \RR \alpha_1 \alpha_2 \\ a_1 a_2} }  
    \sum_{ \substack{ |\kk_1|<\Lambda_c \\ |\kk_2|<\Lambda_c }}
    \inn{\td{v}^{(\eta_1)}_{\alpha_1}(\kk_1-\qq) | \td{u}^{(\eta_1)}_{a_1}(\kk_1)}
    \inn{\td{v}^{(\eta_2)}_{\alpha_2}(\kk_2+\qq) | \td{u}^{(\eta_2)}_{a_2}(\kk_2)}
     \nono\\
& \times e^{i(\kk_1+ \kk_2)\cdot\RR}
        f_{\RR \alpha_1 \eta_1 s_1}^\dagger c_{\kk_1 a_1 \eta_1 s_1} 
        f_{\RR \alpha_2 \eta_2 s_2}^\dagger c_{\kk_2 a_2 \eta_2 s_2}  \ .
\end{align}
We define the matrix 
\begin{equation}
X_{\eta_1\alpha_1 a_1, \eta_2\alpha_2 a_2}(\kk_1,\kk_2) = \int \frac{d^2\qq}{(2\pi)^2}  V(\qq) 
    \inn{\td{v}^{(\eta_1)}_{\alpha_1}(\kk_1-\qq) | \td{u}^{(\eta_1)}_{a_1}(\kk_1)}
    \inn{\td{v}^{(\eta_2)}_{\alpha_2}(\kk_2+\qq) | \td{u}^{(\eta_2)}_{a_2}(\kk_2)}
\end{equation}
such that the interaction can be written as 
\begin{align}
\hH_{\td{J} +} =&  \frac1{2N} 
    \sum_{  \eta_1 s_1  \eta_2 s_2  } 
    \sum_{ \substack{ \RR \alpha_1 \alpha_2 \\ a_1 a_2} }  
    \sum_{ \substack{ |\kk_1|<\Lambda_c \\ |\kk_2|<\Lambda_c }}
    X_{\eta_1\alpha_1 a_1, \eta_2\alpha_2 a_2}(\kk_1,\kk_2)
     e^{i(\kk_1+ \kk_2)\cdot\RR}
     f_{\RR \alpha_1 \eta_1 s_1}^\dagger f_{\RR \alpha_2 \eta_2 s_2}^\dagger  
      c_{\kk_2 a_2 \eta_2 s_2} c_{\kk_1 a_1 \eta_1 s_1} \ .
\end{align}

For low energy states around $\kk=0$, we can approximate $X_{\eta_1\alpha_1 a_1, \eta_2\alpha_2 a_2}(\kk_1,\kk_2)$ by $X_{\eta_1\alpha_1 a_1, \eta_2\alpha_2 a_2}(0,0)$ (dubbed as $\td{J}_{\eta_1\alpha_1 a_1, \eta_2\alpha_2 a_2}$)
\begin{equation}
\td{J}_{\eta_1\alpha_1 a_1, \eta_2\alpha_2 a_2} = \int \frac{d^2\qq}{(2\pi)^2}  V(\qq) 
    \inn{\td{v}^{(\eta_1)}_{\alpha_1}(-\qq) | \td{u}^{(\eta_1)}_{a_1}(0)}
    \inn{\td{v}^{(\eta_2)}_{\alpha_2}(\qq) | \td{u}^{(\eta_2)}_{a_2}(0)}
\end{equation}
We find that, according to the time-reversal symmetry, the matrix elements of $\td{J}$ are same as those of $\mJ$ (\cref{eq:exchange-matrix0}). 
The $T$ symmetry (\cref{eq:D-T-C3z}) and its representations (\cref{eq:Df,eq:Dc-crystalline}) imply $\td{v}^{(\eta_2)}_{\QQ\beta,\alpha_2}(\kk) =  \td{v}^{(-\eta_2)*}_{-\QQ\beta,\alpha_2}(-\kk)$,  $\td{u}^{(\eta_2)}_{\QQ\beta,a_2}(0) = \td{u}^{(-\eta_2)*}_{-\QQ\beta,a_2}(0)$ and hence
\begin{equation}
    \inn{\td{v}^{(\eta_2)}_{\alpha_2}(\qq) | \td{u}^{(\eta_2)}_{a_2}(0)}
= \inn{\td{u}^{(-\eta_2)}_{{a}_2}(0) | \td{v}^{(-\eta_2)}_{{\alpha}_2}(-\qq)}\ .
\end{equation}
Thus we have
\begin{align}
\td{J}_{\eta_1\alpha_1 a_1, \eta_2\alpha_2 a_2} = \int \frac{d^2\qq}{(2\pi)^2}  V(\qq) 
    \inn{\td{v}^{(\eta_1)}_{\alpha_1}(-\qq) | \td{u}^{(\eta_1)}_{a_1}(0)}
    \inn{\td{u}^{(-\eta_2)}_{{a}_2}(0) | \td{v}^{(-\eta_2)}_{{\alpha}_2}(-\qq)}
= \mJ_{\eta_1 \alpha_1 a_1, -\eta_2 {\alpha}_2 {a}_2}\ . 
\end{align}
Thus $\hH_{\td{J} +}$ can be written as 
\begin{align} \label{eq:HJstar-plus-generic}
\hH_{\td{J} +} =&  \frac1{2N} 
    \sum_{  \eta_1 s_1  \eta_2 s_2  } 
    \sum_{ \substack{ \RR \alpha_1 \alpha_2 \\ a_1 a_2} }  
    \sum_{ \substack{ |\kk_1|<\Lambda_c \\ |\kk_2|<\Lambda_c }}
    \mJ_{\eta_1\alpha_1 a_1, \eta_2 \ovl{\alpha}_2 \ovl{a}_2}
     e^{i(\kk_1+ \kk_2)\cdot\RR}
     f_{\RR \alpha_1 \eta_1 s_1}^\dagger f_{\RR \alpha_2 \eta_2 s_2}^\dagger  
      c_{\kk_2 a_2 \eta_2 s_2} c_{\kk_1 a_1 \eta_1 s_1} \ .
\end{align}
According to \cref{eq:exchange-matrix-elements}, the non-negligible matrix elements of $\mJ$ and $\td{\mJ}$ are given by 
\begin{equation}
\mJ_{\eta 13,\eta 13} = \mJ_{\eta 24,\eta 24} = - \mJ_{\eta 13,-\eta 24} = - \mJ_{\eta 24, -\eta 13} = J\ ,
\end{equation}
\begin{equation}
\td{\mJ}_{\eta 13,-\eta 13} = \td{\mJ}_{\eta 24,-\eta 24} = - \td{\mJ}_{\eta 13,\eta 24} = - \td{\mJ}_{\eta 24, \eta 13} = J\ .
\end{equation}
The interaction Hamiltonian $\hH_{\td{J}+}$ can be explicitly written as 
{\small
\begin{align}
\hH_{\td{J} +} =&  - \frac{J}{2N} 
    \sum_{  \eta s_1  s_2  } 
    \sum_{ \RR \alpha }  
    \sum_{  |\kk_1|, |\kk_2|<\Lambda_c } 
     e^{i(\kk_1+ \kk_2)\cdot\RR}
\pare{ f_{\RR \alpha \eta s_1}^\dagger f_{\RR \ovl{\alpha} \eta s_2}^\dagger  
      c_{\kk_2, \ovl{\alpha}+2, \eta s_2} c_{\kk_1, \alpha+2, \eta s_1} 
- f_{\RR \alpha \eta s_1}^\dagger f_{\RR \alpha, -\eta, s_2}^\dagger  
c_{\kk_2, \alpha+2, -\eta, s_2} c_{\kk_1, \alpha+2, \eta s_1}      
      }\ .
\end{align}}
We can write $\hH_{\td{J}+}$ in a more compact form as
\begin{equation} \label{eq:HJstar-plus-explicit}
\hH_{\td{J} +} =  - \frac{J}{4N} 
    \sum_{ \RR s_1 s_2 } \sum_{\alpha\alpha'\eta\eta'}  
    \sum_{  |\kk_1|, |\kk_2|<\Lambda_c } 
     e^{i(\kk_1+ \kk_2)\cdot\RR}
    ( \eta\eta' - (-1)^{\alpha+\alpha'} )
    f_{\RR \alpha \eta s_1}^\dagger f_{\RR \alpha' \eta' s_2}^\dagger
    c_{\kk_2, \alpha'+2, \eta' s_2} c_{\kk_1, \alpha+2, \eta s_1} 
\end{equation}

The fourth term in \cref{sec:Coulomb}, \ie $:\rho_{cf}:V:\rho_{cf}:$, is just the hermitian conjugation of $\hH_{\td{J}+}$.
Thus the total double hybridization interaction is given by
\begin{equation} \label{eq:HJstar-two-term}
    \hH_{\td{J}} = \hH_{\td{J} +} + \hH_{\td{J} +}^\dagger \ .   
\end{equation}

\subsection{Density-hybridization terms (the sixth, seventh, eighth, and ninth terms)} \label{sec:interaction-density-hybridization}

\subsubsection{Vanishing sixth and seventh terms}

We now study the sixth and seventh terms discussed in \cref{sec:Coulomb}. 
The sixth term ($:\hrho_{ff}:V:\hrho_{fc}: + (\leftrightarrow)$) has the form 
{\small
\begin{align}
\hH_{67} 
=& \frac1{2\sqrt{\Omega_{\rm tot}}} \int d^2\rr_1 d^2\rr_2  V(\rr_1-\rr_2) 
    \sum_{\RR \alpha\eta_1 s_1} n_f(\rr_1-\RR)  
    \sum_{|\kk|<\Lambda_c} \sum_{\eta_2s_2 \alpha' a} \sum_{l \beta} \sum_{\QQ\in \mathcal{Q}_{l\eta_2} } 
    w^{(\eta_2)*}_{l\beta,\alpha'}(\rr_2-\RR) \td{u}^{(\eta_2)}_{\QQ\beta,a}(\kk) 
    e^{i(\kk-\QQ)\cdot \rr_2} e^{i\eta_2\DKK_{l}\cdot\RR}
    \nono\\
 & \times 
    \pare{ :f_{\RR \alpha\eta_1 s_1}^\dagger f_{\RR \alpha\eta_1 s_1}: f_{\RR\alpha'\eta_2 s_2}^\dagger  c_{\kk a \eta_2 s_2}
    + f_{\RR\alpha'\eta_2 s_2}^\dagger  c_{\kk a \eta_2 s_2} :f_{\RR \alpha\eta_1 s_1}^\dagger f_{\RR \alpha\eta_1 s_1}: }\ .  
\end{align}}
We have omitted the overlap of local orbitals at different sites. 
Changing the variables $\rr_1 \to \rr_1 + \RR$, $\rr_2 \to \rr_2 + \RR$, we can rewrite the interaction as 
{\small
\begin{align}
\hH_{67} 
=& \frac1{2\sqrt{\Omega_{\rm tot}}} \int d^2\rr_1 d^2\rr_2  V(\rr_1-\rr_2) 
    \sum_{\RR \alpha\eta_1 s_1} n_f(\rr_1)  
    \sum_{|\kk|<\Lambda_c} \sum_{\eta_2s_2 \alpha' a} \sum_{l \beta} \sum_{\QQ\in \mathcal{Q}_{l\eta_2} } 
    w^{(\eta_2)*}_{l\beta,\alpha'}(\rr_2) \td{u}^{(\eta_2)}_{\QQ\beta,a}(\kk) 
    e^{i(\kk-\QQ)\cdot \rr_2} e^{i\kk\cdot\RR} 
    \nono\\
 & \times 
    \pare{ :f_{\RR \alpha\eta_1 s_1}^\dagger f_{\RR \alpha\eta_1 s_1}: f_{\RR\alpha'\eta_2 s_2}^\dagger  c_{\kk a \eta_2 s_2}
    + f_{\RR\alpha'\eta_2 s_2}^\dagger  c_{\kk a \eta_2 s_2} :f_{\RR \alpha\eta_1 s_1}^\dagger f_{\RR \alpha\eta_1 s_1}: }\ .  
\end{align}}
We have made use of the fact that $\eta\DKK_l-\QQ$ is a moir\'e reciprocal lattice and $e^{i(\eta\DKK_{l}-\QQ)\cdot\RR}=1$. 
We introduce the matrix 
\begin{align}
X_{\alpha',a}^{(\eta_2)}(\kk) =& \frac1{\sqrt{\Omega_0}} \int d^2\rr_1 d^2\rr_2  V(\rr_1-\rr_2) n_f(\rr_1) 
    \sum_{l \beta} \sum_{\QQ\in \mathcal{Q}_{l\eta_2} } 
    w^{(\eta_2)*}_{l\beta,\alpha'}(\rr_2) \td{u}^{(\eta_2)}_{\QQ\beta,a}(\kk) 
    e^{i(\kk-\QQ)\cdot \rr_2} 
\end{align}
such that the interaction can be written as 
{\small
\begin{align}
\hH_{67} =& \frac1{2\sqrt{N}} \sum_{\RR \alpha\eta_1 s_1} \sum_{\eta_2 s_2 \alpha' a} \sum_{|\kk|<\Lambda_c} X^{(\eta_2)}_{\alpha',a}(\kk) e^{i\kk \cdot\RR } 
\pare{ :f_{\RR \alpha\eta_1 s_1}^\dagger f_{\RR \alpha\eta_1 s_1}: f_{\RR\alpha'\eta_2 s_2}^\dagger  c_{\kk a \eta_2 s_2}
+ f_{\RR\alpha'\eta_2 s_2}^\dagger  c_{\kk a \eta_2 s_2} :f_{\RR \alpha\eta_1 s_1}^\dagger f_{\RR \alpha\eta_1 s_1}: }\ . 
\end{align}}
Since only the conduction states around $\kk=0$ are relevant in the low energy physics, we will approximate $X_{\alpha',a}^{(\eta_2)}(\kk)$ by $X_{\alpha',a}^{(\eta_2)}(0)$
\begin{align} \label{eq:X-tmp-density-hybridization-1}
X_{\alpha',a}^{(\eta_2)}(0) =& \frac1{\sqrt{\Omega_0}} \int d^2\rr_1 d^2\rr_2  V(\rr_1-\rr_2) n_f(\rr_1) 
    \sum_{l \beta} \sum_{\QQ\in \mathcal{Q}_{l\eta_2} } 
    w^{(\eta_2)*}_{l\beta,\alpha'}(\rr_2) \td{u}^{(\eta_2)}_{\QQ\beta,a}(0) 
    e^{-i\QQ\cdot \rr_2} \ . 
\end{align}
Now we show that $X_{\alpha',a}^{(\eta_2)}(0)$ is guaranteed to be vanishing by symmetries. 
According to the $C_{3z}$ symmetry (\cref{eq:D-T-C3z,eq:D-real-T-C3z}), there is 
\begin{equation}
    e^{i\frac{2\pi}3 \eta_2 [\sigma_z]_{\beta\beta} } w^{(\eta_2)}_{l\beta,\alpha'}(\rr_2)
= w^{(\eta_2)}_{l\beta,\alpha'}(C_{3z}\rr_2) \zeta_{\eta_2 \alpha}^f,\qquad 
e^{i\frac{2\pi}3 \eta_2 [\sigma_z]_{\beta\beta} } \td{u}^{(\eta_2)}_{\QQ\beta,a}(0)
=  \td{u}^{(\eta_2)}_{C_{3z}\QQ\beta,a}(0) \zeta^c_{\eta_2 a}
\end{equation}
with $\zeta_{\eta \alpha}^f$ and $\zeta_{\eta \alpha}^c$ being $C_{3z}$ eigenvalues given in \cref{eq:Df} and \cref{eq:Dc-crystalline}, respectively, \ie 
\begin{equation}
\zeta_{\eta_2 \alpha}^f = e^{i\frac{2\pi}3 \eta_2 [\sigma_z]_{\alpha\alpha}},\qquad
\zeta_{\eta_2 \alpha}^c = \begin{cases}
    e^{i\frac{2\pi}3 \eta_2 [\sigma_z]_{aa}},\qquad & a=1,2\\
    1,\qquad & a=3,4
\end{cases}\ .
\end{equation} 
Substituting this condition into \cref{eq:X-tmp-density-hybridization-1}, we have 
\begin{align} 
X_{\alpha',a}^{(\eta_2)}(0) =& \zeta_{\eta_2,\alpha'}^{f*} \zeta_{\eta_2,a}^c \frac1{\sqrt{\Omega_0}} \int d^2\rr_1 d^2\rr_2  V(\rr_1-\rr_2) n_f(\rr_1) 
    \sum_{l \beta} \sum_{\QQ\in \mathcal{Q}_{l\eta_2} } 
    w^{(\eta_2)*}_{l\beta,\alpha'}(C_{3z}\rr_2) \td{u}^{(\eta_2)}_{C_{3z}\QQ\beta,a}(0) 
    e^{-i\QQ\cdot \rr_2} 
\end{align}
Changing the variables as $\rr_1 \to C_{3z}^{-1} \rr_1$,  $\rr_2 \to C_{3z}^{-1} \rr_2$, $\QQ \to C_{3z}^{-1}\QQ$, we have 
{\small
\begin{align} 
X_{\alpha',a}^{(\eta_2)}(0) =& \zeta_{\eta_2,\alpha'}^{f*} \zeta_{\eta_2,a}^c \frac1{\sqrt{\Omega_0}} \int d^2\rr_1 d^2\rr_2  V(C_{3z}^{-1}\rr_1-C_{3z}^{-1}\rr_2) n_f(C_{3z}^{-1}\rr_1) 
    \sum_{l \beta} \sum_{\QQ\in \mathcal{Q}_{l\eta_2} } 
    w^{(\eta_2)*}_{l\beta,\alpha'}(\rr_2) \td{u}^{(\eta_2)}_{\QQ\beta,a}(0) 
    e^{-i\QQ\cdot \rr_2} 
\end{align}}
Making use of $V(C_{3z}\rr) = V(\rr)$, $n_{f}(C_{3z}\rr) = n_f(\rr)$, we have 
\begin{equation}
    X_{\alpha',a}^{(\eta_2)}(0)  = \zeta_{\eta_2,\alpha'}^{f*} \zeta_{\eta_2,a}^c X_{\alpha',a}^{(\eta_2)}(0) \ . 
\end{equation}
Therefore, $X_{\alpha',a}^{(\eta_2)}(0) $ is zero if $\zeta_{\eta_2,\alpha'}^{f} \neq \zeta_{\eta_2,a}^c$. 
Due to the $C_{3z}$ eigenvalues the only possible nonzero matrix elements are $X_{1,1}^{(\eta_2)}(0) $ and $X_{2,2}^{(\eta_2)}(0) $. 
The above symmetry analysis also applies for other symmetries, for example the particle-hole symmetry $P$. 
As explained in the end of \cref{sec:conduction-basis}, the $a=1,2$ states have opposite $P$ eigenvalues with the $\alpha=1,2$ states. 
Thus $X_{1,1}^{(\eta_2)}(0) $ and $X_{2,2}^{(\eta_2)}(0) $ are also guaranteed to be zero. 
Thus the matrix elements in $\hH_{67}$ vanish to zeroth order of $\kk$. 

Since the seventh term discussed in \cref{sec:Coulomb} is the hermitian conjugation of the sixth term, its matrix elements also vanish to zeroth order of $\kk$. 
The sixth and seventh interaction terms will be non-vanishing if we include higher order terms of $\kk$ in the matrix elements, \ie the $\kk$-dependence in $\td{u}^{(\eta)}_{\QQ\alpha,a}(\kk)$. 
But these matrix elements will be small for low energy conduction band states around $\kk=0$.
We claim that these $\kk$-depend matrix elements in the density-hybridization interaction are irrelevant in the low energy physics. 

\subsubsection{The eighth and ninth terms}

We now study the eighth and ninth terms discussed in \cref{sec:Coulomb}. 
They are weak because the hybridization, \ie $\inn{f^\dagger c}$ and $\inn{c^\dagger f}$, is small, as will be discussed in \cref{sec:CI-phases}.
The eighth term ($:\hrho_{cc}:V:\hrho_{fc}: + (\leftrightarrow)$) has the form 
{\small
\begin{align}
\hH_{K+} 
=& \frac1{2\Omega_{\rm tot}^{\frac32}} \int d^2\rr_1 d^2\rr_2  V(\rr_1-\rr_2) 
    \sum_{\beta_1 l_1 \eta_1 s_1} \sum_{ \substack{|\kk_1|,|\kk_1'|<\Lambda_c\\ a_1  a_1'} } \sum_{\QQ_1,\QQ_1' \in \mathcal{Q}_{l_1\eta_1}} 
    e^{-i(\kk_1-\QQ_1-\kk_1'+\QQ_1')\cdot\rr_1} 
    \td{u}^{(\eta_1)*}_{\QQ_1\beta_1,a_1}(\kk_1) \td{u}^{(\eta_1)}_{\QQ_1'\beta_1,a_1'}(\kk_1') 
    \nono\\ 
& \times \sum_{\beta_2 l_2 \eta_2 s_2} \sum_{\RR \alpha} \sum_{ \substack{|\kk_2|<\Lambda_c \\a_2} } \sum_{\QQ_2\in \mathcal{Q}_{l_2\eta_2} } 
    w^{(\eta_2)*}_{l_2\beta_2,\alpha}(\rr_2-\RR) \td{u}^{(\eta_2)}_{\QQ_2\beta_2,a}(\kk_2) 
    e^{i( \kk_2-\QQ_2)\cdot \rr_2} e^{i\eta_2\DKK_{l_2} \cdot\RR} 
    \nono\\
 & \times 
    \pare{ :c_{\kk_1 a_1 \eta_1 s_1}^\dagger c_{\kk_1' a_1' \eta_1 s_1}: f_{\RR\alpha \eta_2 s_2}^\dagger  c_{\kk_2 a_2 \eta_2 s_2}
    + f_{\RR\alpha \eta_2 s_2}^\dagger  c_{\kk_2 a_2 \eta_2 s_2} :c_{\kk_1 a_1 \eta_1 s_1}^\dagger c_{\kk_1' a_1' \eta_1 s_1}: }\ .  
\end{align}}
Applying the Fourier transformation of $V(\rr)$, we obtain 
{\small
\begin{align}
\hH_{K+} 
=& \frac1{2\Omega_{\rm tot}^{\frac32}} \int  d^2\rr_2   
    \sum_{\beta_1 l_1 \eta_1 s_1} \sum_{ \substack{|\kk_1|,|\kk_1'|<\Lambda_c\\ a_1  a_1'} } \sum_{\QQ_1,\QQ_1' \in \mathcal{Q}_{l_1\eta_1}} V(\kk_1-\QQ_1-\kk_1'+\QQ_1')
    \td{u}^{(\eta_1)*}_{\QQ_1\beta_1,a_1}(\kk_1) \td{u}^{(\eta_1)}_{\QQ_1'\beta_1,a_1'}(\kk_1') 
    \nono\\ 
& \times \sum_{\beta_2 l_2 \eta_2 s_2} \sum_{\RR \alpha}  \sum_{ \substack{|\kk_2|<\Lambda_c \\a_2} }  \sum_{\QQ_2\in \mathcal{Q}_{l_2\eta_2} } 
    w^{(\eta_2)*}_{l_2\beta_2,\alpha}(\rr_2-\RR) \td{u}^{(\eta_2)}_{\QQ_2\beta_2,a}(\kk_2) 
    e^{i( \kk_2-\QQ_2 -\kk_1 + \QQ_1 + \kk_1' -\QQ_1')\cdot \rr_2} e^{i\eta_2\DKK_{l_2} \cdot\RR} 
    \nono\\
 & \times 
    \pare{ :c_{\kk_1 a_1 \eta_1 s_1}^\dagger c_{\kk_1' a_1' \eta_1 s_1}: f_{\RR\alpha \eta_2 s_2}^\dagger  c_{\kk_2 a_2 \eta_2 s_2}
    + f_{\RR\alpha \eta_2 s_2}^\dagger  c_{\kk_2 a_2 \eta_2 s_2} :c_{\kk_1 a_1 \eta_1 s_1}^\dagger c_{\kk_1' a_1' \eta_1 s_1}: }\ .  
\end{align}}
According to \cref{eq:bloch-wannier-local}, there are 
\begin{align}
   & \int d^2 \rr_2  w^{(\eta_2)*}_{l_2\beta_2 , \alpha}(\rr_2-\RR) e^{i( \kk_2-\QQ_2-\kk_1+\QQ_1+\kk_1'-\QQ_1')\cdot \rr_2} = \nono\\
 =& \sqrt{\Omega_0}\ e^{i(\kk_2-\QQ_2- \kk_1+\QQ_1+\kk_1'-\QQ_1')\cdot\RR} \td{v}^{(\eta_2)*}_{\QQ_2-\QQ_1+\QQ_1' \beta_2,\alpha}(\kk_2-\kk_1+\kk_1')\ . 
\end{align}
Since $\eta_2 \DKK_{l_2} -\QQ_2 +\QQ_1-\QQ_1'$ is a moir\'e reciprocal lattice, it must be that $e^{i(\eta_2 \DKK_{l_2} -\QQ_2 +\QQ_1-\QQ_1') \cdot \RR} = 1$.
We hence obtain 
{\small
\begin{align}
\hH_{K+} 
=& \frac1{2N^{\frac32}\Omega_0}
    \sum_{\beta_1 l_1 \eta_1 s_1} \sum_{ \substack{|\kk_1|,|\kk_1'|<\Lambda_c\\ a_1  a_1'} } \sum_{\QQ_1,\QQ_1' \in \mathcal{Q}_{l_1\eta_1}} V(\kk_1-\QQ_1-\kk_1'+\QQ_1')
    \td{u}^{(\eta_1)*}_{\QQ_1\beta_1,a_1}(\kk_1) \td{u}^{(\eta_1)}_{\QQ_1'\beta_1,a_1'}(\kk_1') 
    \nono\\ 
& \times \sum_{\beta_2 l_2 \eta_2 s_2} \sum_{\RR \alpha} \sum_{ \substack{|\kk_2|<\Lambda_c \\a_2} }  \sum_{\QQ_2\in \mathcal{Q}_{l_2\eta_2} } 
    \td{v}^{(\eta_2)*}_{\QQ_2-\QQ_1+\QQ_1', \beta_2; \alpha}(\kk_2-\kk_1+\kk_1') \td{u}^{(\eta_2)}_{\QQ_2\beta_2,a}(\kk_2) 
    e^{i( \kk_2-\kk_1 + \kk_1')\cdot \RR}
    \nono\\
 & \times 
    \pare{ :c_{\kk_1 a_1 \eta_1 s_1}^\dagger c_{\kk_1' a_1' \eta_1 s_1}: f_{\RR\alpha \eta_2 s_2}^\dagger  c_{\kk_2 a_2 \eta_2 s_2}
    + f_{\RR\alpha \eta_2 s_2}^\dagger  c_{\kk_2 a_2 \eta_2 s_2} :c_{\kk_1 a_1 \eta_1 s_1}^\dagger c_{\kk_1' a_1' \eta_1 s_1}: }\ .  
\end{align}}
We define the matrix 
\begin{align}
X_{\eta_1 a_1 a_1', \eta_2 \alpha a_2}(\kk_1,\kk_1',\kk_2) =& \sum_{l_1 \beta_1 l_2 \beta_2} 
    \sum_{\QQ_1,\QQ_1' \in \mathcal{Q}_{l_1\eta_1}}  \sum_{\QQ_2\in \mathcal{Q}_{l_2\eta_2} }  
    V(\kk_1-\QQ_1-\kk_1'+\QQ_1') \nono\\
& \times 
    \td{u}^{(\eta_1)*}_{\QQ_1\beta_1,a_1}(\kk_1) \td{u}^{(\eta_1)}_{\QQ_1'\beta_1,a_1'}(\kk_1') 
    \td{v}^{(\eta_2)*}_{\QQ_2-\QQ_1+\QQ_1', \beta_2; \alpha}(\kk_2-\kk_1+\kk_1') \td{u}^{(\eta_2)}_{\QQ_2\beta_2,a}(\kk_2) 
\end{align}
such that the interaction can be written as 
{\small
\begin{align}
\hH_{K+} 
=& \frac1{2N^{\frac32}\Omega_0}
    \sum_{\eta_1 s_1 \eta_2 s_2} \sum_{ \substack{|\kk_1|,|\kk_1'|<\Lambda_c\\ a_1  a_1'} } 
    \sum_{\RR \alpha} \sum_{ \substack{|\kk_2|<\Lambda_c \\a_2} }   e^{i( \kk_2-\kk_1 + \kk_1')\cdot \RR}
    X_{\eta_1 a_1 a_1', \eta_2 \alpha a_2}(\kk_1,\kk_1',\kk_2)
    \nono\\
 & \times 
    \pare{ :c_{\kk_1 a_1 \eta_1 s_1}^\dagger c_{\kk_1' a_1' \eta_1 s_1}: f_{\RR\alpha \eta_2 s_2}^\dagger  c_{\kk_2 a_2 \eta_2 s_2}
    + f_{\RR\alpha \eta_2 s_2}^\dagger  c_{\kk_2 a_2 \eta_2 s_2} :c_{\kk_1 a_1 \eta_1 s_1}^\dagger c_{\kk_1' a_1' \eta_1 s_1}: }\ .  
\end{align}}
Since the conduction bands only have low energy states around $\kk=0$, thus in the following we approximate $X_{\eta_1 a_1 a_1', \eta_2 \alpha a_2}(\kk_1,\kk_1',\kk_2)$ by $X_{\eta_1 a_1 a_1', \eta_2 \alpha a_2}(0,0,0)$, dubbed as $\mK_{\eta_1 a_1 a_1', \eta_2 \alpha a_2}$
{\small 
\begin{align}
\mK_{\eta_1 a_1 a_1', \eta_2 \alpha a_2} =& \sum_{l_1 \beta_1 l_2 \beta_2} 
    \sum_{\QQ_1,\QQ_1' \in \mathcal{Q}_{l_1\eta_1}}  \sum_{\QQ_2\in \mathcal{Q}_{l_2\eta_2} }  
    V(-\QQ_1+\QQ_1') 
    \td{u}^{(\eta_1)*}_{\QQ_1\beta_1,a_1}(0) \td{u}^{(\eta_1)}_{\QQ_1'\beta_1,a_1'}(0) 
    \td{v}^{(\eta_2)*}_{\QQ_2-\QQ_1+\QQ_1', \beta_2; \alpha}(0) \td{u}^{(\eta_2)}_{\QQ_2\beta_2,a}(0) \nono\\
=&  \sum_{l_1 \beta_1 l_2 \beta_2} 
    \sum_{\QQ_1\in \mathcal{Q}_{l_1\eta_1}}  \sum_{\QQ_2\in \mathcal{Q}_{l_2\eta_2} }  \sum_{\GG} 
    V(\GG)   \td{u}^{(\eta_1)*}_{\QQ_1\beta_1,a_1}(0) \td{u}^{(\eta_1)}_{\QQ_1-\GG,\beta_1; a_1'}(0) 
    \td{v}^{(\eta_2)*}_{\QQ_2-\GG, \beta_2; \alpha}(0) \td{u}^{(\eta_2)}_{\QQ_2\beta_2,a}(0) \nono\\
=& \sum_{\GG} V(\GG) \inn{ \td{u}^{(\eta_1)}_{a_1}(0) | \td{u}^{(\eta_1)}_{a_1'}(\GG)  }
    \inn{ \td{v}^{(\eta_2)}_{\alpha}(\GG) | \td{u}^{(\eta_2)}_{a_2}(0)  }\ .
\end{align}
}
The Hamiltonian $\hH_{K+}$ can be written in a simple form 
\begin{align}\label{eq:HK-plus-generic}
\hH_{K+} 
=& \frac1{2N^{\frac32}\Omega_0}
    \sum_{\eta_1 s_1 \eta_2 s_2} \sum_{a_1 a_1' a_2} \sum_{|\kk_1|,|\kk_1'|,|\kk_2|<\Lambda_c}   \sum_{\RR \alpha}  e^{i( \kk_2-\kk_1 + \kk_1')\cdot \RR}
    \mK_{\eta_1 a_1 a_1', \eta_2 \alpha a_2}
    \nono\\
 & \times 
    \pare{ :c_{\kk_1 a_1 \eta_1 s_1}^\dagger c_{\kk_1' a_1' \eta_1 s_1}: f_{\RR\alpha \eta_2 s_2}^\dagger  c_{\kk_2 a_2 \eta_2 s_2}
    + f_{\RR\alpha \eta_2 s_2}^\dagger  c_{\kk_2 a_2 \eta_2 s_2} :c_{\kk_1 a_1 \eta_1 s_1}^\dagger c_{\kk_1' a_1' \eta_1 s_1}: }\ .  
\end{align}

We now study the $\mK$ matrix. 
According to the $C_{2z}P$ symmetry (\cref{eq:Df,eq:Dc-P}), there are 
\begin{equation}
C_{2z}P \ket{\td{v}^{(\eta)}_{\alpha}(\kk)} = - e^{i\pi \alpha} e^{i\frac{\pi}2 \eta} 
    \ket{\td{v}^{(-\eta)}_{\ovl{\alpha}} (\kk)}, \qquad 
C_{2z}P \ket{\td{u}^{(\eta)}_{a}(\kk)} =  e^{i\pi a} e^{i\frac{\pi}2 \eta} 
    \ket{\td{v}^{(-\eta)}_{\ovl{a}} (\kk)},
\end{equation}
where $\ovl{\alpha}=2,1$ for $\alpha=1,2$ and $\ovl{a}=2,1,4,3$ for $a=1,2,3,4$. 
Then if follows that the $\mK$ matrix satisfies
\begin{equation}
    \mK_{\eta_1 a_1 a_1', \eta_2 \alpha a_2} = e^{i\pi(a_1'-a_1)} \mK_{-\eta_1 \ovl{a}_1 \ovl{a}_1', \eta_2 \alpha a_2}
 = -e^{i\pi(a_2-\alpha)} \mK_{\eta_1 a_1 a_1', -\eta_2 \ovl{\alpha} \ovl{a}_2}
\end{equation}
Similarly, due to the $C_{2x}$ symmetry (\cref{eq:Df,eq:Dc-crystalline}), the $\mK$ matrix satisfies
\begin{equation}
\mK_{\eta_1 a_1 a_1', \eta_2 \alpha a_2}=\mK_{\eta_1 \ovl{a}_1 \ovl{a}_1', \eta_2 \ovl{\alpha} \ovl{a}_2}\ .
\end{equation}
Due to the $C_{2z}TP$ symmetry (\cref{eq:Df,eq:Dc-P}), the $\mK$ matrix satisfies
\begin{equation}
\mK_{\eta_1 a_1 a_1', \eta_2 \alpha a_2} = (-1)^{a_1+a_1'} \mK_{\eta_1 \ovl{a}_1' \ovl{a}_1, \eta_2 \alpha a_2}\ . 
\end{equation}
Numerically we find that there is only one independent non-negligible matrix element
\begin{equation}
\mK_{\eta_1 14, \eta_2 24} =  \mK_{\eta_1 23, \eta_2 13} = - \mK_{\eta_1 32, \eta_2 24} = - \mK_{\eta_1 41, \eta_2 13} = \eta_1\eta_2 K\ ,
\end{equation}
where $K=4.887$meV for $w_0/w_1=0.8$.
The second largest matrix element is $0.82$meV.
One can verify that the $\mK$ matrix satisfies the constraints derived above.  
Therefore, we can write the Hamiltonian $\hH_{K+}$ as 
{\small
\begin{align} \label{eq:HK-plus-explicit}
\hH_{K+} 
=& \frac{K}{N^{\frac32}\Omega_0}
\sum_{\eta_1 s_1 \eta_2 s_2} \sum_{|\kk_1|,|\kk_1'|,|\kk_2|<\Lambda_c} \sum_{ \RR \alpha } 
    e^{i(\kk_2+\kk_1'-\kk_1)\cdot\RR}
\eta_1\eta_2\bigg(c_{\kk_1 \ovl{\alpha} \eta_1 s_1}^\dagger c_{\kk_1', \alpha+2, \eta_1 s_1} f_{\RR\alpha \eta_2 s_2}^\dagger  c_{\kk_2, \alpha+2, \eta_2 s_2} \nono\\
 & \qquad  -  f_{\RR\alpha \eta_2 s_2}^\dagger  c_{\kk_2, \alpha+2, \eta_2 s_2}  c_{\kk_1, \ovl{\alpha}+2, \eta_1 s_1}^\dagger c_{\kk_1', \alpha, \eta_1 s_1} \bigg) \ . 
\end{align}}

The ninth term discussed in \cref{sec:Coulomb}, \ie $:\hrho_{cc}:V:\hrho_{cf}: + (\leftrightarrow)$, is just the hermitian conjugation of $\hH_{K+}$.
Thus the total density-hybridization Hamiltonian is given by 
\begin{equation} \label{eq:HK-two-terms}
    \hH_{K} = \hH_{K+} + \hH_{K+}^\dagger \ .
\end{equation}

\subsection{Summary of the interaction Hamiltonian and its symmetries}

\subsubsection{Summary of the interaction Hamiltonian}\label{sec:summary-interaction}

In summary, the interaction Hamiltonian is given by 
\begin{equation} \label{eq:interaction-summary}
\hH_I = \hH_{U} + \hH_{V} + \hH_{W} + \hH_{J} + \hH_{\td{J}} + \hH_{K}\ .
\end{equation}
$\hH_U$ (\cref{eq:HU-explicit}) is the density-density interaction of the local orbitals 
{\small
\begin{equation}
\hH_{U}  = \frac{U_1}{2} \sum_{\RR} \sum_{\alpha\eta s} \sum_{\alpha'\eta's'} 
    :f_{\RR \alpha \eta s}^\dagger f_{\RR \alpha\eta s}:
    :f_{\RR \alpha' \eta' s'}^\dagger f_{\RR \alpha'\eta' s'}: 
+ \frac{U_2}{2} \sum_{\inn{\RR\RR'}} \sum_{\alpha\eta s} \sum_{\alpha'\eta's'} 
:f_{\RR \alpha \eta s}^\dagger f_{\RR \alpha\eta s}:
:f_{\RR' \alpha' \eta' s'}^\dagger f_{\RR' \alpha'\eta' s'}: \ ,
\end{equation}}
where $\inn{\RR\RR'}$ indexes nearest neighbor pairs on the triangular lattice, $U_1$ is the on-site repulsion interaction, and $U_2$ is the repulsion between nearest neighbors.
$\hH_V$ (\cref{eq:HV-explicit}) is the density-density interaction of the conduction bands 
\begin{equation}
\hH_{V} = \frac{1}{2\Omega_0 N} \sum_{\eta_1 s_1 a_1} \sum_{\eta_2 s_2 a_2} \sum_{|\kk_1|, |\kk_2|<\Lambda_c} \sum_{\substack{\qq \\ |\kk_1+\qq|,|\kk_2+\qq|<\Lambda_c}} V(\qq) 
    :c_{\kk_1 a_1\eta_1 s_1}^\dagger c_{\kk_1+\qq a_1 \eta_1 s_1}:
    :c_{\kk_2+\qq a_2 \eta_2 s_2}^\dagger c_{\kk_2 a_2 \eta_2 s_2}:\ ,
\end{equation}
where $V(\qq)$ (\cref{eq:Vr-Vq}) is the Fourier transformation of the double-gate-screened Coulomb interaction.
$\hH_W$ (\cref{eq:HW-explicit}) is the density-density interaction between the local orbitals and the conduction bands
\begin{equation} 
\hH_{W} = \frac1{N} \sum_{\RR \alpha\eta_1 s_1} \sum_{|\kk|, |\kk'|<\Lambda_c}  \sum_{\eta_2s_2 a} 
    W_a  e^{-i(\kk-\kk')\cdot\RR} 
    :f_{\RR \alpha\eta_1 s_1}^\dagger f_{\RR \alpha\eta_1 s_1}: 
    :c_{\kk a\eta_2 s_2}^\dagger  c_{\kk' a \eta_2 s_2}:\ ,
\end{equation}
where $W_{1}=W_2$ and $W_3=W_4$ due to symmetries.
$\hH_J$ (\cref{eq:HJ-explicit}) is the exchange interaction between local orbitals and the conduction bands
{\small
\begin{align}
H_J = - \frac{J}{2N} \sum_{\RR s_1 s_2} \sum_{\alpha\alpha'\eta\eta'} \sum_{|\kk_1|,|\kk_2|<\Lambda_c }   
     e^{i( \kk_1- \kk_2 )\cdot\RR } 
     ( \eta\eta' + (-1)^{\alpha+\alpha'} )
     :f_{\RR \alpha \eta s_1}^\dagger f_{\RR \alpha' \eta' s_2}:  :c_{\kk_2, \alpha'+2, \eta' s_2}^\dagger  c_{ \kk_1, \alpha+2, \eta s_1}: \ .
\end{align}}
$\hH_J$ can be equivalently written as ferromagnetic coupling between the flat-U(4) momenta (\cref{eq:HJ-U4})
\begin{equation}
    \hH_J = -J \sum_{\mu\nu\xi} e^{-i\qq\cdot\RR} \UF_{\mu\nu}^{(f,\xi)}(\RR) \UF_{\mu\nu}^{(c\prime\prime,\xi)}(\qq)\ ,
\end{equation}
with the momenta given by \cref{eq:flatU4-xi-f-R,eq:flatU4-xi-cpp-q}. 
$\hH_{\td{J}} = \hH_{\td{J} + } + \hH_{\td{J} + }^\dagger  $  is the double-hybridization interaction, with $\hH_{\td{J} + }$ (\cref{eq:HJstar-plus-explicit}) given by 
{\small
\begin{align} 
\hH_{\td{J} +} = - \frac{J}{4N} 
    \sum_{ \RR s_1 s_2 } \sum_{\alpha\alpha'\eta\eta'}  
    \sum_{  |\kk_1|, |\kk_2|<\Lambda_c } 
     e^{i(\kk_1+ \kk_2)\cdot\RR}
    ( \eta\eta' - (-1)^{\alpha+\alpha'} )
    f_{\RR \alpha \eta s_1}^\dagger f_{\RR \alpha' \eta' s_2}^\dagger
    c_{\kk_2, \alpha'+2, \eta' s_2} c_{\kk_1, \alpha+2, \eta s_1} \ .
\end{align}}
Notice that the coefficients $J$ in $\hH_J$ and $\hH_{\td{J}}$ are the same coefficient because the corresponding matrix elements are related by $T$ symmetry, as explained in \cref{sec:interaction-double-hybridization}.  
$\hH_{K} = \hH_{K + } + \hH_{K + }^\dagger  $  is the density-hybridization interaction, with $\hH_{K + }$ (\cref{eq:HK-plus-explicit}) given by 
{\small
\begin{align} 
\hH_{K+} 
=& \frac{K}{N^{\frac32}\Omega_0}
\sum_{\eta_1 s_1 \eta_2 s_2} \sum_{|\kk_1|, |\kk_1'|, |\kk_2|<\Lambda_c } \sum_{ \RR \alpha } 
    e^{i(\kk_2+\kk_1'-\kk_1)\cdot\RR}
\eta_1\eta_2\bigg(c_{\kk_1 \ovl{\alpha} \eta_1 s_1}^\dagger c_{\kk_1', \alpha+2, \eta_1 s_1} f_{\RR\alpha \eta_2 s_2}^\dagger  c_{\kk_2, \alpha+2, \eta_2 s_2} \nono\\
 & \qquad  -  f_{\RR\alpha \eta_2 s_2}^\dagger  c_{\kk_2, \alpha+2, \eta_2 s_2}  c_{\kk_1, \ovl{\alpha}+2, \eta_1 s_1}^\dagger c_{\kk_1', \alpha, \eta_1 s_1} \bigg) \ . 
\end{align}}
The parameters $U_{1,2}$, $W_{1,3}$, $J$, $K$ for different $w_0/w_1$ are given in \cref{tab:HIparameters}.


\subsubsection{Symmetries of the interaction Hamiltonian}\label{sec:interaction-symmetry}

The continuous BM model (with interaction) has a U(2)$\times$U(2) symmetry when valley is a good quantum number.
Thus the effective topological heavy fermion model must also have a U(2)$\times$U(2) symmetry.
As discussed in \cref{sec:flat-U4}, when $M=0$, the presence of the $C_{2z}P S$ symmetry, where $S$ is the third chiral symmetry, will enhance the U(2)$\times$U(2) symmetry to a U(4) symmetry, referred to as the flat U(4) symmetry. 
Recall that the Coulomb interaction is given by (\cref{eq:HI-rho-rho})
\begin{equation}
    \frac12 \int d^2\rr_1 \int d^2\rr_2 V(\rr_1-\rr_2) :\hrho(\rr_1)::\hrho(\rr_2):
\end{equation}
with $:\hrho(\rr_1):$ being the normal ordered projected density operator (\cref{eq:rho-normal}).
Since $C_{2z}P$ (\cref{eq:Df,eq:Dc-P}) is a symmetry of the original model, $:\hrho(\rr):$ and hence $\hH_I$ commute with $C_{2z}P$.
(In Ref. \cite{Biao-TBG3}, it is $C_{2z}P$ that gives the U(4) symmetry, provided that remote bands are projected out and the kinetic energy of the flat bands is neglected.) 
Thus, in order for $H_I$ to have the flat-U(4) symmetry, we only need $H_I$ to commute with $S$ defined in \cref{eq:third-chiral}. 
One can easily verify that $\hH_{U}$, $\hH_{V}$, $\hH_{W}$, $\hH_{J}$, $\hH_{\td{J}}$ commute with $S$ whereas $\hH_{K}$ anti-commutes with $S$. 
Therefore, $\hH_{U}$, $\hH_{V}$, $\hH_{W}$, $\hH_{J}$, $\hH_{\td{J}}$ respect the flat-U(4) symmetry while $\hH_{K}$ breaks the flat-U(4) symmetry. 
(The neglected terms $J'$ and $J''$ discussed in \cref{sec:HJ-other-terms} commute and anti-commute with $S$, respectively. Thus $J'$ respects the flat-U(4) symmetry whereas $J''$ breaks the flat-U(4) symmetries.)
Since $K$ (and $J''$) is small compared to other interaction parameters (\cref{tab:HIparameters}), flat-U(4) is still an approximate symmetry of the interaction Hamiltonian. 
(Including the omitted small matrix elements in the interaction Hamiltonian and the $\kk$-dependence of these matrix elements may yield more terms that break the flat-U(4).
However, we claim these effects are weak in the low energy physics.)

For the same reason as the flat-U(4) symmetry, $H_I$ will have the chiral-U(4) symmetry as long as it commutes with $C$ defined in \cref{eq:first-chiral}. 
(The free part $\hH_0$ anti-commute with $C$ when $v_\star'=0$.)
One can easily verify that $\hH_{U}$, $\hH_{V}$, $\hH_{W}$, $\hH_{J}$, $\hH_{\td{J}}$, and $\hH_{K}$ all commute with $C$.
Therefore, the chiral-U(4) symmetry is an exact symmetry of the interaction Hamiltonian summarized in \cref{sec:summary-interaction}. 
Since for nonzero $w_0/w_1$, the continuous model does not really respect the first chiral symmetry, we expect that including the omitted small terms in $\hH_{I}$ will break the chiral symmetry and hence the chiral-U(4). 
Nevertheless, chiral-U(4) can be thought as an approximate symmetry of $\hH_I$.

As discussed in \cref{sec:U4xU4}, the presence of both flat-U(4) and chiral-U(4) implies an SU(4)$\times$SU(4) symmetry. 
Given that the charge-U(1) is respected, the SU(4)$\times$SU(4) group will be promoted to U(4)$\times$U(4) if the Hamiltonian further conserves the index $\xi$ ($\xi=\eta (-1)^{\alpha-1}$ for $f$-electrons and $\xi=\eta (-1)^{a-1}$ for $c$-electrons). 
Since $\hH_{U}$, $\hH_{V}$, $\hH_{W}$, $\hH_{J}$, $\hH_{\td{J}}$ all have both flat-U(4) and chiral-U(4) symmetries and $\xi$ is repsected in all of them, they all have the U(4)$\times$U(4) symmetry. 
However, as discussed in the last two paragraphs $\hH_K$ only has the chiral-U(4) symmetry.
In \cref{tab:U4-summary} we summarize the local symmetries of all the single-particle and interaction Hamiltonian terms. 

In \cref{eq:charge-conjugation-def} we defined the charge-conjugation symmetry
\begin{equation} 
    \PH_c f_{\RR \alpha\eta s}^\dagger \PH_c^{-1} = \sum_{\alpha'\eta'} D^f_{\alpha'\eta',\alpha\eta}(PC_{2z}T) f_{\RR\alpha'\eta' s},\qquad
    \PH_c c_{\kk a\eta s}^\dagger \PH_c^{-1} = \sum_{a'\eta'} D^c_{a'\eta',a\eta}(PC_{2z}T) c_{-\kk a'\eta' s}\ ,
\end{equation}
where
\begin{equation}
    D^{f}(C_{2z}TP) = -\sigma_y \tau_z,\qquad 
    D^{f}(C_{2z}TP) = (\sigma_y \tau_z) \oplus (\sigma_y \tau_z)\ ,
\end{equation}
and showed that the single-particle Hamiltonian (\cref{eq:H0-def}) respects the charge conjugation symmetry. 
One can show $\PH_c :\hrho(\rr): \PH_c^{-1}  = -:\hrho(\rr): $, which automatically implies that the Coulomb interaction of the form
\begin{equation}
    \frac12 \int d^2\rr_1 \int d^2\rr_2 V(\rr_1-\rr_2) :\hrho(\rr_1)::\hrho(\rr_2):
\end{equation}
is invariant under the charge conjugation. 

\subsection{Failure of the strong coupling picture in the first chiral limit of the BM model \texorpdfstring{($w_0=0$)}{(w0=0)} }

The first chiral limit discussed in \cref{sec:chiral-U4} is achieved by artificially enforcing $v_\star'=0$ in the single-particle Hamiltonian $\hH_0$, which is obtained from the BM model with nonzero $w_0$. 
In the actual chiral limit of the BM model, where $w_0=0$, the parameter $v_\star'$ automatically vanishes. 
In this subsection we discuss the heavy fermion model in this actual chiral limit ($w_0=0$).
We construct the maximally localized Wannier functions by the same procedure as in \cref{sec:Wannier}. 
Since for $w_0=0$ the $\Gamma_3$ states have energies about $\pm$100meV, the previously used Wannierization energy window [$-80$meV, $80$meV] now do not include the $\Gamma_3$ states and hence do not support local Wannier functions. 
We hence change the energy window to [$-120$meV, $120$meV] and use all the same other parameters (given around \cref{eq:Wannier-from-Wannier90}) for the Wannierization. 
The resulted Wannier functions still decay exponentially but are much less localized. 
The on-site repulsion $U_1$ of $f$-electrons becomes extremely small ($<1$meV) while the hybridization $\gamma$ between $f$- and $c$-electrons becomes quite strong ($\sim$100meV). 
The property $U_1\ll \gamma$ does not change if we use a larger Wannierization energy window, \eg [$-160$meV, $160$meV]. 
Therefore, in the actual first chiral limit, the strong coupling picture ($U_1>\gamma$) is no longer valid, and the quantum-dot-like behaviors observed in STMs \cite{xie2019spectroscopic,wong_cascade_2020} cannot be easily explained. 

Nevertheless, from a theoretical perspective, as explained in \cref{sec:band-geometry}, our model have the key geometric features of the BM model in the chiral limit: The flat bands are analytically solvable, and they satisfy the ideal droplet condition and have relatively flat Berry's curvature. 

\section{Correlated insulator states} \label{sec:CI-phases}

\subsection{Mean field Hamiltonian} \label{sec:mean-field}

$\hH_{\td{J}}$  will create two particles (or holes) in the local orbitals, which will cost an energy of the order $U_1$. 
Since $U_1$ is the largest energy scale of the problem, $\hH_{\td{J}}$ is an high-energy process and will be omitted in the following. 
From a mean field aspect, the energy contributed by the $\hH_{\td{J}}$ term is $\sim J |\inn{f^\dagger c}|^2$.
Since $U_1$ is much larger than the coupling between $f$- and $c$-electrons, $\inn{f^\dagger c}$ can be treated as a small quantity and its second order terms, \eg $J |\inn{f^\dagger c}|^2$ can be omitted. 
We will also omit $\hH_{K}$ for a similar reason.
From the mean field aspect, the energy contributed by $\hH_{K}$ is $\sim K \inn{f^\dagger c} \inn{c^\dagger c}$.
According to \cref{eq:HK-plus-explicit}, the involved $\inn{c^\dagger c}$ is off-diagonal in the $a$-index, \eg $\inn{c^\dagger_{\kk \ovl{\alpha} \eta s} c_{\kk,\alpha+2,\eta s}}$, which must be small because only diagonal terms are non-vanishing in the normal state. 
Thus $\sim K \inn{f^\dagger c} \inn{c^\dagger c}$ can also be treated as a second order small quantity. 
Therefore, in the following we approximate the interaction Hamiltonian as $\hH_I \approx \hH_U + \hH_W + \hH_V + \hH_J$. 

In the following we apply Hartree-Fock (HF) calculation to the remaining terms.
Let us denote the variational slater determinant state as $\ket{\Psi}$. 
$\ket{\Psi}$ is supposed to minimize the total energy $\inn{\Psi|\hH_0 + \hH_I|\Psi}$. 
To simplify the calculation, we rewrite the operator $\hH_I$ as $\hH_I = :\hH_I: + \hH_{I,MF}$, where $:\hH_I:$ is the normal ordered operator of $\hH$ with respect to $\ket{\Psi}$ (not $\ket{G_0}$) and $\hH_{I,MF}$ is {\it defined} as $\hH_I - :\hH_I:$. 
$:\hH_I:$ by definition will either annihilate $\ket{\Psi}$ or excite two particles plus two holes upon $\ket{\Psi}$. 
Then it follows that $\inn{\Psi|\hH_0 + \hH_I|\Psi} = \inn{\Psi | \hH_0 +  \hH_{I,MF} | \Psi}$. 
Therefore, one only need to minimize $\inn{\Psi | \hH_0 +  \hH_{I,MF} | \Psi}$. 

In this subsection we explicitly derive $\hH_{I,MF}$. 
For simplicity, we assume that the ground state ($\ket{\Psi}$) preserves the translation symmetry.
For later convenience, we introduce the density matrices 
\begin{equation}\label{eq:Of-def}
O^f_{\alpha\eta s, \alpha'\eta's'} = \inn{\Psi| f_{\RR\alpha\eta s}^\dagger f_{\RR \alpha'\eta's'}|\Psi}
= \frac1{N} \sum_{\kk \in {\rm MBZ}} \inn{\Psi| f_{\kk\alpha\eta s}^\dagger f_{\kk \alpha'\eta's'}|\Psi}\ ,
\end{equation}
\begin{equation}\label{eq:Oc-def}
O^c_{a\eta s, a'\eta's'} = \frac1{N} \sum_{|\kk|<\Lambda_c}  \pare{\inn{\Psi| c_{\kk a\eta s}^\dagger c_{\kk a'\eta's'}|\Psi} - \frac12 \delta_{aa'} \delta_{\eta\eta'} \delta_{ss'} } 
\end{equation}
\begin{equation}\label{eq:Ocf-def}
O^{cf}_{a\eta_2 s_2, \alpha\eta_1 s_1} = \frac1{\sqrt{N}} \sum_{|\kk|<\Lambda_c}  e^{-i\kk\cdot\RR} \inn{\Psi|c^\dagger_{\kk a\eta_2s_2} f_{\RR \alpha\eta_1 s_1} |\Psi} \ . 
\end{equation}
We have defined $O^c_{a\eta s, a'\eta's'}$ as the expectation of the conduction band density operator with respect to the charge neutrality point such that it will remain finite even we take the cutoff $\Lambda_c$ to infinity. 
The filling of local orbitals and conduction bands are given by 
\begin{equation}
    \nu_f = \Tr[O^f] -4,\qquad \nu_c = \Tr[O^c]\ .
\end{equation}
We also assume that there is no pairing in $\ket{\Psi}$. 
The total filling with respect to the charge neutrality point is given by $\nu = \nu_f + \nu_c$. 
For the normal state $\ket{G_0}$ (defined in \cref{sec:Coulomb}) at charge neutrality point, there is $\nu = \nu_f = \nu_c = 0$.

We first decouple the density-density interactions of local orbitals, \ie $\hH_U$ in \cref{eq:interaction-summary}. 
To derive the mean field Hamiltonian, we first divide the density-density interaction into constant, bilinear, and quartic terms as 
\begin{align}
\hat{H}_{U} =& 8N U_1 + 48N U_2 -( 4 U_1 + 24U_2) \sum_{\RR} \sum_{\alpha \eta s } f_{\RR\alpha\eta s}^\dagger f_{\RR \alpha\eta s} 
    + \frac{U_1}2 \sum_{\RR} \sum_{ \substack{\alpha\eta s\\ \beta\eta's'}} 
    f_{\RR\alpha\eta s}^\dagger f_{\RR \alpha\eta s}  f_{\RR\beta\eta' s'}^\dagger f_{\RR \beta\eta' s'} \nono\\
& + \frac{U_2}2 \sum_{\inn{\RR\RR'}} \sum_{ \substack{\alpha\eta s\\ \beta\eta's'}} f_{\RR\alpha\eta s}^\dagger f_{\RR \alpha\eta s}  f_{\RR'\beta\eta' s'}^\dagger f_{\RR' \beta\eta' s'}\nono\\
=& 8N U_1 + 48N U_2 -( 3.5 U_1 + 24U_2) \sum_{\RR} \sum_{\alpha \eta s } f_{\RR\alpha\eta s}^\dagger f_{\RR \alpha\eta s} 
+ \frac{U_1}2 \sum_{\RR} \sum_{ \substack{\alpha\eta s\\ \beta\eta's'}} 
f_{\RR\alpha\eta s}^\dagger  f_{\RR\beta\eta' s'}^\dagger f_{\RR \beta\eta' s'} f_{\RR \alpha\eta s}  \nono\\
& + \frac{U_2}2 \sum_{\inn{\RR\RR'}} \sum_{ \substack{\alpha\eta s\\ \beta\eta's'}} f_{\RR\alpha\eta s}^\dagger  f_{\RR'\beta\eta' s'}^\dagger f_{\RR' \beta\eta' s'} f_{\RR \alpha\eta s} 
\end{align}
The mean field Hamiltonian is obtained by projecting out the terms that create two particles and two holes upon the ground state $\ket{\Psi}$. 
According to the Wick's theorem we have 
{\small
\begin{align}
& f_{\RR\alpha\eta s}^\dagger  f_{\RR\beta\eta' s'}^\dagger f_{\RR \beta\eta' s'} f_{\RR \alpha\eta s} 
= :f_{\RR\alpha\eta s}^\dagger f_{\RR\beta\eta' s'}^\dagger f_{\RR \beta\eta' s'} f_{\RR \alpha\eta s}  :
+ \wick{ \c1 f_{\RR\alpha\eta s}^\dagger   :f_{\RR\beta\eta' s'}^\dagger f_{\RR \beta\eta' s'}: \c1 f_{\RR \alpha\eta s}}
+ \wick{ :f_{\RR\alpha\eta s}^\dagger   \c1 f_{\RR\beta\eta' s'}^\dagger \c1 f_{\RR \beta\eta' s'}  f_{\RR \alpha\eta s} : } \nono\\
+ &  \wick{ : \c1 f_{\RR\alpha\eta s}^\dagger  f_{\RR\beta\eta' s'}^\dagger \c1 f_{\RR \beta\eta' s'}  f_{\RR \alpha\eta s}  : } + \wick{ : f_{\RR\alpha\eta s}^\dagger \c1 f_{\RR\beta\eta' s'}^\dagger  f_{\RR \beta\eta' s'}  \c1 f_{\RR \alpha\eta s}  : }
+ \wick{ \c1 f_{\RR\alpha\eta s}^\dagger   \c2 f_{\RR\beta\eta' s'}^\dagger \c2 f_{\RR \beta\eta' s'}  \c1 f_{\RR \alpha\eta s} }
+ \wick{ \c1 f_{\RR\alpha\eta s}^\dagger  \c2 f_{\RR\beta\eta' s'}^\dagger \c1 f_{\RR \beta\eta' s'}  \c2 f_{\RR \alpha\eta s}  }
\end{align}}
Here $:A:$ is the normal ordered form of $A$ with respect to $\ket{\Psi}$ ({\it not} $\ket{G_0}$ defined in \cref{sec:Coulomb}).  
The first term will either annihilate $\ket{\Psi}$ or excite two particle-hole pairs on top of $\ket{\Psi}$. 
We will omit the first term and approximate the four fermion operator by bilinear terms and constant terms  
{\small
\begin{align}
& f_{\RR\alpha\eta s}^\dagger  f_{\RR\beta\eta' s'}^\dagger f_{\RR \beta\eta' s'} f_{\RR \alpha\eta s} 
\approx
\wick{ \c1 f_{\RR\alpha\eta s}^\dagger   f_{\RR\beta\eta' s'}^\dagger f_{\RR \beta\eta' s'} \c1 f_{\RR \alpha\eta s}}
+ \wick{ f_{\RR\alpha\eta s}^\dagger   \c1 f_{\RR\beta\eta' s'}^\dagger \c1 f_{\RR \beta\eta' s'}  f_{\RR \alpha\eta s} } + \wick{ \c1 f_{\RR\alpha\eta s}^\dagger  f_{\RR\beta\eta' s'}^\dagger \c1 f_{\RR \beta\eta' s'}  f_{\RR \alpha\eta s}  } \nono\\
+ &  \wick{ f_{\RR\alpha\eta s}^\dagger \c1 f_{\RR\beta\eta' s'}^\dagger  f_{\RR \beta\eta' s'}  \c1 f_{\RR \alpha\eta s}  }
- \wick{ \c1 f_{\RR\alpha\eta s}^\dagger   \c2 f_{\RR\beta\eta' s'}^\dagger \c2 f_{\RR \beta\eta' s'}  \c1 f_{\RR \alpha\eta s} }
- \wick{ \c1 f_{\RR\alpha\eta s}^\dagger  \c2 f_{\RR\beta\eta' s'}^\dagger \c1 f_{\RR \beta\eta' s'}  \c2 f_{\RR \alpha\eta s}  }\ .
\end{align}}
Similarly, there are 
{\small
\begin{align}
& f_{\RR\alpha\eta s}^\dagger  f_{\RR'\beta\eta' s'}^\dagger f_{\RR' \beta\eta' s'} f_{\RR \alpha\eta s} 
\approx
\wick{ \c1 f_{\RR\alpha\eta s}^\dagger   f_{\RR'\beta\eta' s'}^\dagger f_{\RR' \beta\eta' s'} \c1 f_{\RR \alpha\eta s}}
+ \wick{ f_{\RR\alpha\eta s}^\dagger   \c1 f_{\RR'\beta\eta' s'}^\dagger \c1 f_{\RR' \beta\eta' s'}  f_{\RR \alpha\eta s} } +   \wick{ \c1 f_{\RR\alpha\eta s}^\dagger  f_{\RR'\beta\eta' s'}^\dagger \c1 f_{\RR' \beta\eta' s'}  f_{\RR \alpha\eta s}  } \nono\\
+& \wick{ f_{\RR\alpha\eta s}^\dagger \c1 f_{\RR'\beta\eta' s'}^\dagger  f_{\RR' \beta\eta' s'}  \c1 f_{\RR \alpha\eta s}  }
- \wick{ \c1 f_{\RR\alpha\eta s}^\dagger   \c2 f_{\RR'\beta\eta' s'}^\dagger \c2 f_{\RR' \beta\eta' s'}  \c1 f_{\RR \alpha\eta s} }
- \wick{ \c1 f_{\RR\alpha\eta s}^\dagger  \c2 f_{\RR'\beta\eta' s'}^\dagger \c1 f_{\RR' \beta\eta' s'}  \c2 f_{\RR \alpha\eta s}  }\ . 
\end{align}}
Since the on-site repulsion $U_1$ is the largest energy scale and dominates the correlation physics, we {\it assume} that only on-site order parameters, \ie $\inn{f_{\RR \alpha\eta s}^\dagger f_{\RR \alpha'\eta' s'}}$, are formed. 
Then we can write the approximate on-site and off-site quartic terms as 
{\small
\begin{align}
 \sum_{ \substack{\alpha\eta s\\ \beta\eta's'}}  f_{\RR\alpha\eta s}^\dagger  f_{\RR\beta\eta' s'}^\dagger f_{\RR \beta\eta' s'} f_{\RR \alpha\eta s} 
\approx & 2 \sum_{\alpha\eta s}\Tr[O^f] f_{\RR\alpha\eta s}^\dagger f_{\RR \alpha\eta s} 
 - 2\sum_{ \substack{\alpha\eta s\\ \beta\eta's'}}  f_{\RR \beta \eta' s'}^\dagger  f_{\RR \alpha\eta s} O^f_{\alpha\eta s,\beta\eta's'} 
   - \Tr[O^f] \Tr[O^f] + \Tr[O^f O^f ] 
\end{align}}
and
{\small
\begin{align}
 \sum_{ \substack{\alpha\eta s\\ \beta\eta's'}}  f_{\RR\alpha\eta s}^\dagger  f_{\RR'\beta\eta' s'}^\dagger f_{\RR' \beta\eta' s'} f_{\RR \alpha\eta s} 
\approx & \sum_{\beta\eta' s'}\Tr[O^f] f_{\RR'\beta\eta' s'}^\dagger f_{\RR'\beta\eta' s'} 
+ \sum_{\alpha\eta s}\Tr[O^f] f_{\RR\alpha\eta s}^\dagger f_{\RR\alpha\eta s} 
  - \Tr[O^f] \Tr[O^f]\ , 
\end{align}}
respectively. 
Substituting these equations into the interaction Hamiltonian $\hH_{U}$, we obtain
\begin{equation}
\hH_U \approx \hH_{U,MF} = \ovl{H}_{U} - E_U
\end{equation}
with 
{\small
\begin{align} \label{eq:HU-MF}
\ovl{H}_{U} =  \sum_{\RR} \sum_{\alpha\eta s}  
    \pare{  U_1(\nu_f + 0.5) + 6 U_2 \nu_f  }
    f_{\RR\alpha\eta s}^\dagger f_{\RR \alpha\eta s} - U_1 \sum_{\RR} O^f_{\alpha\eta s, \beta\eta's'} f_{\RR \beta\eta's'}^\dagger f_{\RR \alpha\eta s}\ ,
\end{align}}
and 
{\small
\begin{equation}
E_U = \frac{N U_1}2  \pare{\nu_f^2 + 8\nu_f - \Tr[O^f O^f] } + 3N U_2  ( \nu_f^2 + 8\nu_f) .
\end{equation}}

Following the same logic as decoupling $\hH_{U}$, one can decouple $\hH_W$ as
{\small
\begin{align}
\hH_W \approx &
\sum_{|\kk|<\Lambda_c} \sum_{a \eta s} W_a \nu_f :c_{\kk a \eta s}^\dagger c_{\kk a\eta s}: 
+ \sum_{\RR a \alpha \eta s} W_a :f_{\RR \alpha\eta s}^\dagger f_{\RR \alpha\eta s}: \nu_{c,a} 
- N \sum_{a} W_a \nu_f \nu_{c,a} \nono\\
& - \frac1{\sqrt{N}} \sum_{\RR \alpha\eta_1 s_1} \sum_{|\kk|<\Lambda_c} \sum_{\eta_2 s_2 a} W_a \pare{ O^{cf}_{a\eta_2 s_2, \alpha\eta_1 s_1} e^{i\kk\cdot\RR} f_{\RR \alpha\eta_1 s_1}^\dagger c_{\kk a\eta_2 s_2} + h.c. } 
- N \sum_{a \alpha} \sum_{\eta_2 \eta_1 s_2 s_1} |O^{cf}_{a\eta_2 s_2, \alpha\eta_1 s_1}|^2 W_a \ ,
\end{align}}
where 
\begin{equation}
    \nu_{c,a} = \sum_{\eta s} O^c_{a\eta s, a\eta s}\ .
\end{equation}
In the first term we have omitted the Umklapp scatterings, \ie $W_a \nu_f :c_{\kk+\GG a \eta s}^\dagger c_{\kk a\eta s}: $ for $\GG\neq 0$, because the state $c_{\kk+\GG a \eta s}^\dagger$ has a huge kinetic energy and the Umklapp scattering will be weak. 
We can further organize the terms as 
\begin{equation}
    \hH_W \approx \hH_{W,MF} = \ovl{H}_W - E_W\ ,
\end{equation}
with 
\begin{align} \label{eq:HW-MF}
\ovl{H}_{W} =& \sum_{|\kk|<\Lambda_c} \sum_{a \eta s} \nu_f W_a  
     c_{\kk a\eta s}^\dagger c_{\kk a\eta s} 
+ \sum_{\RR} \sum_{\alpha\eta s} \nu_{c,a} W_a f_{\RR \alpha\eta s}^\dagger f_{\RR \alpha\eta s} \nono\\
& - \frac1{\sqrt{N}} \sum_{\RR \alpha\eta_1 s_1} \sum_{|\kk|<\Lambda_c} \sum_{\eta_2 s_2 a} W_a \pare{ O^{cf}_{a\eta_2 s_2, \alpha\eta_1 s_1} e^{i\kk\cdot\RR} f_{\RR \alpha\eta_1 s_1}^\dagger c_{\kk a\eta_2 s_2}
    + h.c. } 
\end{align}
\begin{equation}
    E_W = N \sum_{a} \pare{W_a \nu_f \nu_{c,a} + 4 W_a \nu_{c,a}}  + \sum_{a}\sum_{|\kk|<\Lambda_c} 2 W_a \nu_f  - N \sum_{a \alpha} \sum_{\eta_2 \eta_1 s_2 s_1} |O^{cf}_{a\eta_2 s_2, \alpha\eta_1 s_1}|^2 W_a \ . 
\end{equation}
We emphasize that the term $\sum_{|\kk|<\Lambda_c} 2 W_a \nu_f$ depends on the cutoff ($\Lambda_c$) of the conduction bands and diverges as $\Lambda_c\to \infty$.
However, the total energy contributed by $\hH_W$ will converge as $\Lambda_c \to \infty$ because the divergent part of $\sum_{|\kk|<\Lambda_c} 2 W_a \nu_f$ will be canceled by the divergent part of the first term of $\ovl{H}_W$. 

Then we decouple the exchange interaction $\hH_J$. 
Following the same method as for $\hH_U$, $\hH_J$ can be approximated as 
{\small
\begin{align}
\hH_J \approx & -J \sum_{\RR \alpha\eta s_1 s_2} :f_{\RR \alpha\eta s_1}^\dagger f_{\RR \alpha\eta s_2}: O^c_{\alpha+2 \eta s_2, \alpha+2 \eta s_1} 
- J \sum_{\alpha\eta s_1 s_2} \sum_{|\kk|<\Lambda_c}  
    \pare{ O^f_{\alpha\eta s_1, \alpha\eta s_2} - \frac12 \delta_{s_1 s_2} }
    :c_{\kk \alpha+2 \eta s_2}^\dagger c_{\kk\alpha+2 \eta s_1}:  \nono\\
& +N J  \sum_{ \alpha\eta s_1 s_2} \pare{ O^f_{\alpha\eta s_1, \alpha\eta s_2}- \frac12 \delta_{s_1 s_2} } O^{c}_{\alpha+2\eta s_2, \alpha+2 \eta s_1} 
+ J \sum_{\alpha\eta s_1 s_2}  \sum_{|\kk|<\Lambda_c}  
    O^f_{\alpha\eta s_1, \ovl{\alpha} -\eta s_2} c_{\kk,\ovl{\alpha}+2,-\eta,s_2}^\dagger c_{\kk,\alpha+2,\eta,s_1}  \nono\\
&+ J \sum_{\alpha\eta s_1 s_2}  \sum_{\RR}  
    f_{\RR\alpha\eta s_1}^\dagger f_{\RR \ovl{\alpha} -\eta s_2} O_{\ovl{\alpha}+2,-\eta,s_2\ ;\ \alpha+2,\eta,s_1}^c   
- N J  \sum_{\alpha \eta s_1 s_2} O^f_{\alpha\eta s_1, \ovl{\alpha} -\eta s_2} O_{\ovl{\alpha}+2,-\eta,s_2\ ;\ \alpha+2,\eta,s_1}^c \nono\\
& + \frac{J}{\sqrt{N}} \sum_{\RR} \sum_{|\kk|<\Lambda_c} \sum_{\alpha \eta s_1 s_2}\pare{ e^{i\kk\cdot\RR} O^{cf}_{\alpha+2\eta s_2,  \alpha \eta s_2} f_{\RR \alpha \eta s_1}^\dagger c_{\kk,\alpha+2, \eta, s_1} + h.c. } 
 - N J \sum_{\alpha \eta s_1 s_2}  O^{cf}_{\alpha+2\eta s_2,  \alpha \eta s_2}  O^{cf*}_{\alpha+2\eta s_1,  \alpha \eta s_1} \nono\\
& - \frac{J}{\sqrt{N}} \sum_{\RR} \sum_{|\kk|<\Lambda_c} \sum_{\alpha \eta s_1 s_2}\pare{ e^{i\kk\cdot\RR} O^{cf}_{\ovl{\alpha}+2,-\eta,s_2\ ;\   \ovl{\alpha},-\eta,s_2} f_{\RR \alpha \eta s_1}^\dagger c_{\kk,\alpha+2, \eta, s_1} + h.c. } 
+ NJ \sum_{\alpha \eta s_1 s_2}  O^{cf}_{\ovl{\alpha}+2,-\eta,s_2\ ;\  \ovl{\alpha},-\eta,s_2}  O^{cf*}_{\alpha+2\eta s_1,  \alpha \eta s_1}\ .
\end{align}}
We can organize the terms as 
\begin{equation}
    \hH_J \approx \hH_{J,MF} = \ovl{H}_J - E_J\ ,
\end{equation}
with 
{\small
\begin{align} 
\ovl{H}_J = & -J \sum_{\RR \alpha\eta s_1 s_2} f_{\RR \alpha\eta s_1}^\dagger f_{\RR \alpha\eta s_2} O^c_{\alpha+2 \eta s_2, \alpha+2 \eta s_1} 
- J \sum_{\alpha\eta s_1 s_2} \sum_{|\kk|<\Lambda_c}  
    \pare{ O^f_{\alpha\eta s_1, \alpha\eta s_2} - \frac12 \delta_{s_1 s_2} }
    c_{\kk \alpha+2 \eta s_2}^\dagger c_{\kk\alpha+2 \eta s_1}  \nono\\
& + J \sum_{\alpha\eta s_1 s_2}  \sum_{|\kk|<\Lambda_c}  
    O^f_{\alpha\eta s_1, \ovl{\alpha} -\eta s_2} c_{\kk,\ovl{\alpha}+2,-\eta,s_2}^\dagger c_{\kk,\alpha+2,\eta,s_1}  
 + J \sum_{\alpha\eta s_1 s_2}  \sum_{\RR}  
    f_{\RR\alpha\eta s_1}^\dagger f_{\RR \ovl{\alpha} -\eta s_2} O_{\ovl{\alpha}+2,-\eta,s_2\ ;\ \alpha+2,\eta,s_1}^c  \nono\\
& + \frac{J}{\sqrt{N}} \sum_{\RR} \sum_{|\kk|<\Lambda_c} \sum_{\alpha \eta s_1 s_2}\pare{ e^{i\kk\cdot\RR} O^{cf}_{\alpha+2\eta s_2,  \alpha \eta s_2} f_{\RR \alpha \eta s_1}^\dagger c_{\kk,\alpha+2, \eta, s_1} + h.c. }  \nono\\
& - \frac{J}{\sqrt{N}} \sum_{\RR} \sum_{|\kk|<\Lambda_c} \sum_{\alpha \eta s_1 s_2}\pare{ e^{i\kk\cdot\RR} O^{cf}_{\ovl{\alpha}+2,-\eta,s_2\ ;\   \ovl{\alpha},-\eta,s_2} f_{\RR \alpha \eta s_1}^\dagger c_{\kk,\alpha+2, \eta, s_1} + h.c. } \ ,
\end{align}}
and
{\small
\begin{align}
E_J =& - NJ \sum_{\alpha\eta s_1 s_2} O^f_{\alpha\eta s_1, \alpha\eta s_2}O^{c}_{\alpha+2\eta s_2, \alpha+2 \eta s_1} - J \sum_{\alpha\eta s} \sum_{|\kk|<\Lambda_c} \pare{ O^f_{\alpha\eta s, \alpha\eta s} -\frac12 } \frac12
 + N J  \sum_{\alpha \eta s_1 s_2} O^f_{\alpha\eta s_1, \ovl{\alpha} -\eta s_2} O_{\ovl{\alpha}+2,-\eta,s_2\ ;\ \alpha+2,\eta,s_1}^c \nono\\
& + N J \sum_{\alpha \eta s_1 s_2}  O^{cf}_{\alpha+2\eta s_2,  \alpha \eta s_2}  O^{cf*}_{\alpha+2\eta s_1,  \alpha \eta s_1} -  NJ \sum_{\alpha \eta s_1 s_2}  O^{cf}_{\ovl{\alpha}+2,-\eta,s_2\ ;\  \ovl{\alpha},-\eta,s_2}  O^{cf*}_{\alpha+2\eta s_1,  \alpha \eta s_1}\ . 
\end{align}}
The second term in $E_J$ diverges as $\Lambda_c \to \infty$.
However, the total energy contributed by $\hH_J$ will converge as $\Lambda_c \to \infty$ because the divergent part of the second term in $E_J$ will be canceled by the (divergent part of) second term of $\ovl{H}_J$. 
We can equivalently write $\hH_J$ in a more compact form as 
\begin{align} \label{eq:HJ-MF}
\hH_J =& - \frac{J}2 \sum_{\RR} \sum_{\alpha_1 \alpha_2 \eta_1 \eta_2 s_1 s_2} 
    ( \eta_1\eta_2 + (-1)^{\alpha_1 + \alpha_2} ) f_{\RR \alpha_1 \eta_1 s_1}^\dagger f_{\RR \alpha_2 \eta_2 s_2} O^c_{\alpha_2+2 \eta_2 s_2, \alpha_1+2 \eta_1 s_1} \nono\\
&   - \frac{J}2 \sum_{|\kk|<\Lambda_c} \sum_{\alpha_1 \alpha_2 \eta_1 \eta_2 s_1 s_2} 
    ( \eta_1\eta_2 + (-1)^{\alpha_1 + \alpha_2} ) 
    c_{\kk \alpha_2+2, \eta_2 s_2}^\dagger c_{\kk \alpha_1+2, \eta_1 s_1} 
    \pare{ O^f_{\alpha_1\eta_1s_1,\alpha_2\eta_2s_2} - \frac12 \delta_{\alpha_1\alpha_2} \delta_{\eta_1\eta_2} \delta_{s_1 s_2} } \nono\\
& + \frac{J}{2\sqrt{N}} \sum_{\RR} \sum_{|\kk|<\Lambda_c} \sum_{\alpha_1 \alpha_2 \eta_1 \eta_2 s_1 s_2} 
    e^{i\kk\cdot\RR} (\eta_1\eta_2 + (-1)^{\alpha_1+\alpha_2} ) 
    \pare{ O^{cf}_{\alpha_2+2\eta_2s_2,  \alpha_2\eta_2 s_2 } f^\dagger_{\RR \alpha_1\eta_1 s_1} c_{\kk \alpha_1+2 \eta_1 s_1} + h.c. }\ . 
\end{align}

In the end we decouple the Coulomb interaction of the conduction bands.
Usually, it is the Fock term that leads to symmetry breaking in the inner degrees of freedom.
We claim that the symmetry breaking mainly come from the on-site repulsion of the local orbitals and the interaction of conduction bands only renormalize the dispersion of the conduction bands.
Therefore, for simplicity, we will omit the Fock channel of this interaction. 
The Hartree-mean field is
\begin{equation}
    \hH_V \approx \hH_{V,MF} = \ovl{H}_V - E_V\ ,
\end{equation}
with 
\begin{equation} \label{eq:HV-MF}
    \ovl{H}_V = \frac{V(0)}{\Omega_0} \nu_c \sum_{\eta s a} \sum_{|\kk|<\Lambda_c} c^\dagger_{\kk a \eta s} c_{\kk a \eta s}  
\end{equation}
and 
\begin{equation}
E_V = \frac{V(0)}{2\Omega_0} N \nu_c^2 
    + \frac{V(0)}{\Omega_0} \sum_{|\kk|<\Lambda_c} 8 \nu_c  \ . 
\end{equation}
$\ovl{H}_V$ only shifts the energy the conduction bands.

In summary, the mean field Hamiltonian of the interaction is given by 
\begin{equation} \label{eq:mean-field}
    \hH_I \approx \hH_{I,MF} = \ovl{H}_{U} + \ovl{H}_{W} + \ovl{H}_{V} + \ovl{H}_{J} - E_U - E_W - E_V - E_J\ . 
\end{equation}

\subsection{Correlated insulator phases at the filling \texorpdfstring{$\nu=0$}{nu=0}}\label{sec:GS-nu=0}

\begin{table}[t]
\begin{tabular}{|c|c|c|c| c|c|c|c|}
\hline
Fillings & \multicolumn{3}{c|}{$\nu=0$} & \multicolumn{4}{c|}{$\nu=-1$}\\ 
\hline
Phases  & VP & K-IVC & IVC & VP & K-IVC & IVC & VP + K-IVC\\
\hline
One-shot HF energy (meV) & -698.645 & -698.702 & -694.888  
    & -666.326 & -666.359 & -664.607 & -666.360 \\
\hline 
SCF HF energy (meV) & -704.019 & -704.109  & -700.616 
    & -671.809 & -671.857 & -670.330 & -671.860 \\
\hline
\hline 
Fillings & \multicolumn{3}{c|}{$\nu=-2$} & \multicolumn{4}{c|}{$\nu=-3$}\\ 
\hline
Phases  & VP & K-IVC & IVC & VP & K-IVC &  &  \\
\hline
One-shot HF energy (meV) & -575.492 & -575.507 & -574.528 & -434.015 & -434.015 & & \\
\hline 
SCF HF energy (meV)  & -578.975 &  -579.036 &  -578.203 & -426.129 & -426.113  & & \\
\hline 
\end{tabular}
\caption{The one-shot and self-consistent HF energies of the correlated insulator phases at integer fillings. 
Since the energy difference between flat-U(4) rotation related states are small ($\lesssim 0.1$meV) there may be strong order-parameter fluctuations at finite temperatures. 
\label{tab:HFEnergy}
}
\end{table}

\begin{figure}
\centering
\includegraphics[width=0.7\linewidth]{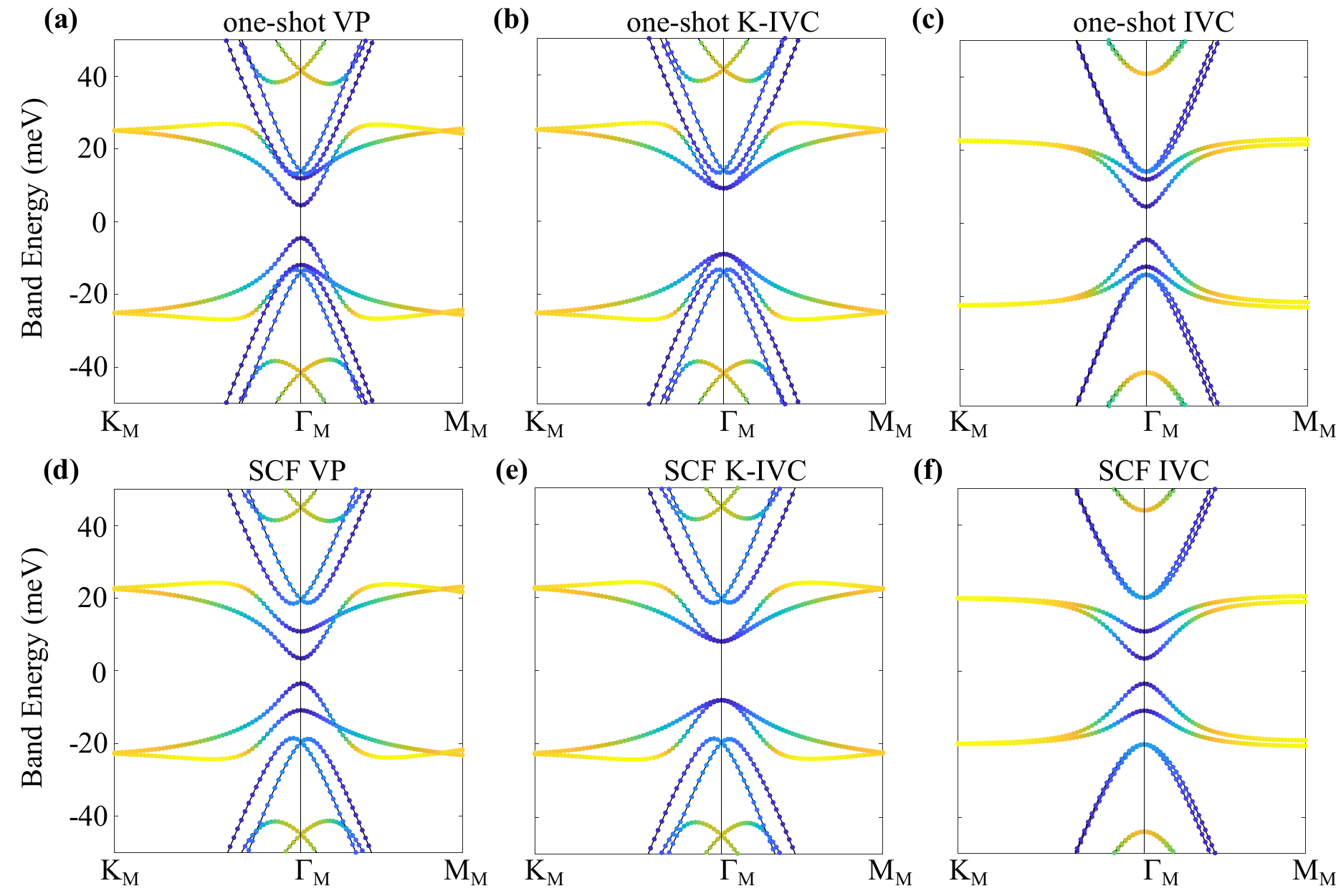}
\caption{HF band structures of correlated insulator phases at $\nu=0$.
(a), (b), (c) are the one-shot HF band structures of the VP, K-IVC, and IVC phases, respectively.
(d), (e), (f) are the self-consistent HF band structures of the VP, K-IVC, and IVC phases, respectively.  
The color represents the composition of the energy bands, where yellow corresponds to the local orbitals and blue corresponds to the conduction bands. 
We have chosen $w_0/w_1=0.8$ in the calculation. Other parameters of the single-particle and interaction hamiltonians are given in \cref{tab:H0parameters,tab:HIparameters}. 
\label{fig:HFbands_nu0} 
}
\end{figure}

\subsubsection{Numerical results}\label{sec:GS-nu=0-numeric}

We consider the parent wave function of valley polarized (VP) state as
\begin{equation}
\ket{\mrm{VP}_0^{\nu=0}} = \prod_{\RR} f_{\RR 1 + \up}^\dagger f_{\RR 1 + \down}^\dagger f_{\RR 2 + \up}^\dagger f_{\RR 2 + \down}^\dagger \ket{\mrm{FS}}\ .
\end{equation}
Here $\ket{\rm FS}$ is the Fermi sea state occupying the lower two bands (per spin per valley) of $H^{(c,\eta)}(\kk)$ (\cref{eq:Hc-summary}), \ie
\begin{equation} \label{eq:FS-def}
    \ket{\mrm{FS}} = \prod_{\eta s} \prod_{n=1,2} \prod_{|\kk|<\Lambda_c}  \pare{\sum_{a} \mathcal{U}_{a,n}^{(\eta)} c_{\kk a \eta s}^\dagger } \ket{0} \ ,
\end{equation}
where $ \mathcal{U}_{a,n}^{(\eta)}$ ($n=1,2,3,4$) is the $n$-th eigenvector of $H^{(c,\eta)}(\kk)$. 
At $\kk=0$, the second and third eigenstates are degenerate, we should choose $\mathcal{U}_{a,2}^{(\eta)}(0)$ as the vector that smoothly connects to $\mathcal{U}_{a,2}^{(\eta)}(\kk\to 0)$. 
By definition, the yielded $O^c$ is diagonal in valley and spin indices   
{\small
\begin{align}
O^{c}_{a\eta s, a'\eta's'} =& \frac1{N} \sum_{|\kk|<\Lambda_c}\pare{\inn{\mrm{FS} |  c_{\kk a \eta s}^\dagger c_{\kk' a'\eta' s'} | \mrm{FS} } - \frac12 \delta_{aa'}\delta_{\eta\eta'} \delta_{ss'}} 
= \frac1{N}\delta_{ss'} \delta_{\eta\eta'} \sum_{|\kk|<\Lambda_c} \pare{ \pare{\sum_{n=1,2} \mathcal{U}_{a,n}^{(\eta)}(\kk) \mathcal{U}^{(\eta)*}_{a',n}(\kk)}  - \frac12 \delta_{aa'}}  \ . 
\end{align}}
Given $O^{c}_{a\eta s, a'\eta's'} = \delta_{\eta\eta'}\delta_{ss'} O^c_{a\eta s, a'\eta s}$, only the diagonal elements $O^c_{a\eta s, a\eta s}$ will enter the mean field Hamiltonian in \cref{eq:HW-MF,eq:HJ-MF,eq:HV-MF}. 
Now we show that all the diagonal elements are zero.
Due to the particle-hole symmetry $P$ (\cref{eq:Dc-P}), we can always pick the gauge $(-1)^{a-1} \mathcal{U}_{a,n}^{(\eta)}(\kk) = \mathcal{U}_{a,5-n}^{(\eta)}(-\kk)$. 
Using this condition and the completeness $\sum_{n=1}^4 \mathcal{U}_{n}^{(\eta)}(\kk) \mathcal{U}^{(\eta)\dagger}_{n}(\kk) = \mathbb{I}_{4\times 4}$, one can show that
\begin{equation}
    O^{c}_{a\eta s, a'\eta s} + (-1)^{a+a'}O^{c}_{a\eta s, a'\eta s} =0\ ,
\end{equation}
which implies $O^{c}_{a\eta s, a\eta s}=0$. 
Therefore, $O^c$ does not enter the mean field Hamiltonian because it is valley-diagonal and all its diagonal elements vanish.
The density matrix $O^{cf}$ (\cref{eq:Ocf-def}) given by the parent VP state also vanishes because the state respects the particle numbers of local orbitals and conduction bands separately. 
The density matrix $O^f$ (\cref{eq:Of-def}) given by the parent VP state is nonzero. 
We will discuss it in detail in next subsection. 
Substituting these density matrices into the mean field equations, we can obtain the mean field approximation of the total Hamiltonian as 
\begin{equation}
\hH_0 + \hH_I \approx \hH_0 + \ovl{H}_{U} + \ovl{H}_{W} + \ovl{H}_{V} + \ovl{H}_{J} - E_U - E_W - E_V - E_J\ ,
\end{equation}
where $\ovl{H}_{X}$ and $E_X$ ($X=U,W,V,J$) are functions the density matrices (\cref{sec:mean-field}). 
We refer to the ground state energy of the above mean field Hamiltonian as the ``one-shot'' HF energy (\cref{tab:HFEnergy}).
The spectrum given by $\hH_0 + \sum_{X} \ovl{H}_{X}$ is referred to as the ``one-shot'' HF spectrum (\cref{fig:HFbands_nu0}(a)).
The next step wave function, denoted as $\ket{\mrm{VP}_1^{\nu=0}}$, occupies the one-shot bands to the filling of the charge neutrality point.   
One can then calculate the density matrices of $\ket{\mrm{VP}_1^{\nu=0}}$ and generate a new mean field Hamiltonian. 
The hybridization $O^{cf}$ given by $\ket{\mrm{VP}_1^{\nu=0}}$ is in general nonzero.  
Repeating this process until the ground state converges, one achieve the self-consistent HF ground state, whose energy is called the self-consistent HF energy.   
We denote the self-consistent wave function as $\ket{\mrm{VP}_{\infty}^{\nu=0}}$
The (one-shot and self-consistent) HF energies and HF spectra of the VP state are summarized in \cref{tab:HFEnergy} and \cref{fig:HFbands_nu0}, respectively. 

In the process of self-consistent iterations, the one-shot spectrum in \cref{fig:HFbands_nu0}(a) continuously changes to the self-consistent spectrum in \cref{fig:HFbands_nu0}(d).
Thus, the state $\ket{\mrm{VP}_1^{\nu=0}}$, which occupies the one-shot bands to the filling $\nu=0$, is adiabatically connected to the self-consistent state $\ket{\mrm{VP}_{\infty}^{\nu=0}}$. 
However, the parent VP state $\ket{\mrm{VP}_0^{\nu=0}}$ may not adiabatically connected to them because it occupies the local orbitals and conduction separately but not the quasi-particle bands. 
A key difference between $\ket{\mrm{VP}_1^{\nu=0}}$ and $\ket{\mrm{VP}_0^{\nu=0}}$ is that $O^{cf}$ given by $\ket{\mrm{VP}_1^{\nu=0}}$ is in general nonzero while $O^{cf}$ given by $\ket{\mrm{VP}_0^{\nu=0}}$ is zero. 
In this work, we refer to $\ket{\mrm{VP}_0^{\nu=0}}$, $\ket{\mrm{VP}_1^{\nu=0}}$, and $\ket{\mrm{VP}_\infty^{\nu=0}}$ as the parent VP, one-shot VP, and self-consistent VP states, respectively. 

The parent wave functions for the inter-valley-coherent (IVC) and Kramers inter-valley-coherent (K-IVC) states are rotated from $\ket{\mrm{VP}_0^{\nu=0}}$ by the chiral-U(4) (\cref{eq:chiralU4}) and flat-U(4) (\cref{eq:flatU4}) symmetries as 
{\small
\begin{equation}
\ket{\mrm{IVC}_0^{\nu=0}} = e^{-i\frac{\pi}2 \UC_{x0} } \ket{\mrm{VP}_0^{\nu=0}}
= \prod_{\RR} \frac14 (f_{\RR 1 + \up}^\dagger -i f_{\RR 2 - \up}^\dagger) (f_{\RR 1 + \down}^\dagger - i f_{\RR 2 - \down}^\dagger) ( - i f_{\RR 1 - \up}^\dagger + f_{\RR 2 + \up}^\dagger) ( - i f_{\RR 1 - \down}^\dagger + f_{\RR 2 + \down}^\dagger) \ket{\rm FS}\ ,  
\end{equation}}
and 
{\small
\begin{equation}
\ket{\text{K-IVC}_0^{\nu=0}} = e^{-i\frac{\pi}2 \UF_{x0} } \ket{\mrm{VP}_0^{\nu=0}}
= \prod_{\RR} \frac14 (f_{\RR 1 + \up}^\dagger + f_{\RR 2 - \up}^\dagger) (f_{\RR 1 + \down}^\dagger + f_{\RR 2 - \down}^\dagger) ( - f_{\RR 1 - \up}^\dagger + f_{\RR 2 + \up}^\dagger) ( - f_{\RR 1 - \down}^\dagger + f_{\RR 2 + \down}^\dagger) \ket{\rm FS}\ ,  
\end{equation}}
respectively.
One can verify that the parent IVC state respects the spinless time-reversal symmetry $T$ (\cref{eq:Df}) whereas the parent K-IVC state breaks $T$. 
Substituting the parent wave functions into the HF loop, we obtain the one-shot and self-consistent HF energies (\cref{tab:HFEnergy}). 
The one-shot IVC and K-IVC states, \ie $\ket{\mrm{IVC}^{\nu=0}_{1}}$ and $\ket{\text{K-IVC}^{\nu=0}_{1}}$, are obtained as Fock states occupying the corresponding one-shot HF bands to the filling $\nu=0$.
The one-shot K-IVC state has a lower energy. 

As shown in \cref{tab:HFEnergy}, the energy difference of K-IVC and VP states is only of the order 0.1meV whereas the energy difference between the IVC state and VP is of the order 3meV. 
Since K-IVC and IVC are rotated from VP by the flat-U(4) and chiral-U(4), respectively, the result suggests that the flat-U(4) symmetry is a better approximation of the considered Hamiltonian $\hH_0 + \sum_{X} \hH_{X}$ ($X=U,W,J,V$).
This observation is consistent with the argument at the end of \cref{sec:flat-U4}.

\subsubsection{Analytical analysis of the one-shot HF}
\label{sec:GS-nu=0-analytical}

The one-shot and self-consistent HF band structures of the VP, K-IVC, and IVC states at $\nu=0$ are summarized in \cref{fig:HFbands_nu0}. 
We can see that the self-consistent band structures are very close to the one-shot HF band structures.
Here we study the one-shot spectra of the VP and K-IVC states around the charge neutrality point.
We do not study the IVC state here because it has a higher energy in the mean field.
The density matrices given by $\ket{\text{VP}^{\nu=0}_0}$ are 
\begin{equation}
    O^f = \sigma_0 (\frac{\tau_z+\tau_0}2) \spin_0 ,\qquad  O^c_{a\eta s,a'\eta's'}=\delta_{\eta\eta'} \delta_{ss'} O^c_{a\eta s,a'\eta s},\qquad O^c_{a\eta s,a\eta s}=0, \qquad O^{cf}=0\ . 
\end{equation}
Substituting these density matrices into the mean field equations around \cref{eq:HU-MF,eq:HW-MF,eq:HJ-MF,eq:HV-MF}, we obtain the one-shot HF mean field and energies as 
\begin{equation}
\ovl{H}_U = - \frac{U_1}{2} \sum_{\RR} \sum_{\alpha\eta s} \eta f_{\RR \alpha\eta s}^\dagger f_{\RR \alpha\eta s},\qquad \ovl{H}_W = \ovl{H}_V = 0,\qquad 
\ovl{H}_J = - \frac{J}2 \sum_{|\kk|<\Lambda_c} \sum_{\alpha \eta s} \eta c_{\kk \alpha+2 \eta s}^\dagger c_{\kk \alpha+2 \eta s}
\end{equation}
and 
\begin{equation} \label{eq:VP-nu=0-EU}
E_U = -2N U_1,\qquad E_W=E_V=E_J=0\ ,
\end{equation}
respectively. 
The low energy band structure can be described by the k$\cdot$p expansion of $\hH_0 + \ovl{H}_U + \ovl{H}_J$
\begin{equation} \label{eq:HMF-nu=0-VP}
H_{\rm VP}^{\rm(MF)} (\kk) \approx \begin{pmatrix}
0   &  v_\star( k_x \sigma_0 \tau_z + ik_y\sigma_z\tau_0 )  & \gamma \sigma_0 \tau_0 + v_\star'( k_x \sigma_x\tau_z + k_y\sigma_y \tau_0 ) \\
v_\star( k_x \sigma_0 \tau_z - ik_y\sigma_z\tau_0 ) & M \sigma_x\tau_0 - \frac{J}2 \sigma_0\tau_z & 0 \\
\gamma \sigma_0\tau_0 + v_\star'( k_x \sigma_x\tau_z + k_y  \sigma_y\tau_0) & 0 & - \frac{U_1}2  \sigma_0\tau_z 
\end{pmatrix} \otimes \spin_0 \ .
\end{equation}
For large $\kk$ ($v_\star|\kk|\gg \frac12 U_1$), the low energy bands are contributed by the local orbitals, which have the energies $\pm U_1/2$. 
We can see this in \cref{fig:HFbands_nu0} at large $\kk$.
For $\kk=0$, the energy eigenvalues are given by 
\begin{equation} \label{eq:VP-nu=0-levels}
    \pm M \pm \frac{J}2, \qquad \pm \frac{U_1}4 \pm \sqrt{\pare{\frac{U_1}4}^2 + \gamma^2} \ .
\end{equation}
In \cref{fig:HFbands_nu0}(a), (d), the 24 energy levels closest to Fermi level from low to high are 
{\small
\begin{equation}
-\frac{U_1}4 - \sqrt{ \pare{ \frac{U_1}4 }^2 + \gamma^2 }\; (\text{4-fold}), \qquad 
\frac{U_1}4 - \sqrt{ \pare{ \frac{U_1}4 }^2 + \gamma^2 }\; (\text{4-fold}),\qquad 
-M - \frac{J}2\; (\text{2-fold}),\qquad 
M - \frac{J}2\; (\text{2-fold}), 
\end{equation}
\begin{equation}
-M + \frac{J}2\; (\text{2-fold}),\qquad 
M + \frac{J}2\; (\text{2-fold}),\qquad 
-\frac{U_1}4 + \sqrt{ \pare{ \frac{U_1}4 }^2 + \gamma^2 }\; (\text{4-fold}),\qquad 
\frac{U_1}4 + \sqrt{ \pare{ \frac{U_1}4 }^2 + \gamma^2 }\; (\text{4-fold})\ .
\end{equation}
}
Notice that in \cref{fig:HFbands_nu0}(a), (d) every single band is spin degenerate. 

The density matrices given by $\ket{\text{K-IVC}^{\nu=0}_0}$ are 
\begin{equation}
    O^f = \frac12 \sigma_0 \tau_0 \spin_0 - \frac12 \sigma_y \tau_y \spin_0 ,\qquad  
     O^c_{a\eta s,a'\eta's'}=\delta_{\eta\eta'} \delta_{ss'} O^c_{a\eta s,a'\eta s},\qquad O^c_{a\eta s,a\eta s}=0, \qquad 
     O^{cf}=0\ . 
\end{equation}
According to the mean field equations in \cref{sec:mean-field}, the one-shot HF mean field and energies are 
\begin{equation}
\ovl{H}_U = \frac{U_1}{2} \sum_{\RR} \sum_{\alpha\eta \alpha' \eta' s}  f_{\RR \alpha\eta s}^\dagger [\sigma_y]_{\alpha\alpha'} [\tau_y]_{\eta\eta'} f_{\RR \alpha'\eta' s},\qquad 
\ovl{H}_W = \ovl{H}_V = 0\ ,
\end{equation}
\begin{align}
\ovl{H}_J =& - \frac{J}2 \sum_{|\kk|<\Lambda_c} \sum_{\alpha_1 \alpha_2 \eta_1 \eta_2 s_1 s_2} 
    ( \eta_1\eta_2 + (-1)^{\alpha_1 + \alpha_2} ) 
    c_{\kk \alpha_2+2, \eta_2 s_2}^\dagger c_{\kk \alpha_1+2, \eta_1 s_1} 
    \pare{ O^f_{\alpha_1\eta_1s_1,\alpha_2\eta_2s_2} - \frac12 \delta_{\alpha_1\alpha_2} \delta_{\eta_1\eta_2} \delta_{s_1 s_2} } \nono\\
& = \frac{J}4 \sum_{|\kk|<\Lambda_c} \sum_{\alpha_1 \alpha_2 \eta_1 \eta_2 s} 
    ( \eta_1\eta_2 + (-1)^{\alpha_1 + \alpha_2} ) 
    c_{\kk \alpha_2+2, \eta_2 s}^\dagger c_{\kk \alpha_1+2, \eta_1 s} 
    [\sigma_y]_{\alpha_1\alpha_2} [\tau_y]_{\eta_1\eta_2} \nono\\
& = - \frac{J}2 \sum_{|\kk|<\Lambda_c}
\sum_{\alpha_1 \eta_1 \alpha_2 \eta_2 s} [\sigma_y]_{\alpha_1\alpha_2} [\tau_y]_{\eta_1\eta_2}  c_{\kk \alpha_2+2 \eta' s}^\dagger c_{\kk \alpha_1+2 \eta s} \ , 
\end{align}
and 
\begin{equation} \label{eq:KIVC-nu=0-EU}
E_U = -2N U_1,\qquad E_W=E_V=E_J=0\ .
\end{equation}
Notice that the $\frac12 \sigma_0 \tau_0 \spin_0$ component in $O^f$ does not contribute to the mean field Hamiltonian because they are canceled by the constant bilinear terms in \cref{eq:HU-MF,eq:HJ-MF}, respectively. 
The low energy band structure can be described by the k$\cdot$p expansion of $\hH_0 + \ovl{H}_U + \ovl{H}_J$.

\begin{equation} 
H_{\text{K-IVC}}^{\rm(MF)} (\kk) \approx \begin{pmatrix}
0  &  v_\star( k_x \sigma_0\tau_z  + ik_y\sigma_z\tau_0 )  & \gamma \sigma_0\tau_0 + v_\star'( k_x \sigma_x\tau_z + k_y \sigma_y\tau_0 ) \\
v_\star( k_x \sigma_0\tau_z - ik_y\sigma_z\tau_0 ) & M \sigma_x\tau_0 - \frac{J}2 \sigma_y\tau_y & 0 \\
\gamma \sigma_0\tau_0 + v_\star'( k_x \sigma_x\tau_z + k_y  \sigma_y\tau_0) & 0 &  \frac{U_1}2  \sigma_y \tau_y
\end{pmatrix} \otimes \spin_0 \ .
\end{equation}
For large $\kk$ ($v_\star|\kk|\gg \frac12 U_1$), the low energy bands are contributed by the local orbitals, which have the energies $\pm U_1/2$. 
We can see this in \cref{fig:HFbands_nu0} at large $\kk$.
For $\kk=0$, the energy eigenvalues are given by 
\begin{equation} \label{eq:KIVC-nu=0-levels}
    \pm \sqrt{M^2 + \frac{J^2}4}, \qquad \pm \frac{U_1}4 \pm \sqrt{\pare{\frac{U_1}4}^2 + \gamma^2} \ ,
\end{equation}
where the levels $\frac{U_1}4 \pm \sqrt{\pare{\frac{U_1}4}^2 + \gamma^2}$ are same as those in the VP state. 
In \cref{fig:HFbands_nu0}(b), (e), the 24 energy levels closest to Fermi level from low to high are 
{\small
\begin{equation}
-\frac{U_1}4 - \sqrt{ \pare{ \frac{U_1}4 }^2 + \gamma^2 }\; (\text{4-fold}), \qquad 
\frac{U_1}4 - \sqrt{ \pare{ \frac{U_1}4 }^2 + \gamma^2 }\; (\text{4-fold}),\qquad 
-\sqrt{M^2 + \frac{J^2}4 }; (\text{4-fold}),
\end{equation}
\begin{equation}
\sqrt{M^2 + \frac{J^2}4 }\; (\text{4-fold}),\qquad 
-\frac{U_1}4 + \sqrt{ \pare{ \frac{U_1}4 }^2 + \gamma^2 }\; (\text{4-fold}),\qquad 
\frac{U_1}4 + \sqrt{ \pare{ \frac{U_1}4 }^2 + \gamma^2 }\; (\text{4-fold})\ .
\end{equation}
}
Notice that in \cref{fig:HFbands_nu0}(b), (e) every single band is spin degenerate.


\subsection{General rules for the ground states}\label{sec:ground-state-rules}

When $M=0$, one can see that the energy levels at $\kk=0$ of the VP (\cref{eq:VP-nu=0-levels}) and K-IVC  (\cref{eq:KIVC-nu=0-levels}) states at $\nu=0$ become the same. 
A symmetry reason is that, when $M=0$, the flat-U(4) symmetry discussed in \cref{sec:flat-U4} is an exact symmetry of the considered Hamiltonian. 
Therefore, when $M=0$, the VP and K-IVC states must have the same energy and band structure because they are related by a flat-U(4) rotation. 
Nonzero $M$ not only makes a difference in the band structure but also lifts the energy degeneracy between the two states. 
Now we argue that nonzero $M$ stabilizes the K-IVC state. 
As shown in \cref{eq:mean-field}, the total energy can be calculated as the interaction energy, $E_U + E_W + E_J + E_V$, subtracted from the quasi-particle energy, \ie 
\begin{equation}
    -E_U - E_W - E_J - E_V + \inn{\Psi| \hH_0 + \ovl{H}_U + \ovl{H}_W + \ovl{H}_J + \ovl{H}_V |\Psi}
    = -E_U - E_W - E_J - E_V + \sum_{\kk}\sum_{n\in \mrm{occ}} \mathcal{E}_{\kk n}\ , 
\end{equation}
where $\mathcal{E}_{\kk n}$ is the HF spectrum and $n$ indexes occupied bands. 
Since the considered interaction Hamiltonian respects the flat-U(4) symmetry, the VP and K-IVC states have the same interaction energy (\cref{eq:VP-nu=0-EU,eq:KIVC-nu=0-EU}). 
Thus, the state having lower quasi-particle energy will have a lower total energy.
Because the K-IVC state opens a larger gap ($2\sqrt{M^2 + J^2/4}$) than the VP state ($2|M-J/2|$), the K-IVC state has the lower total energy.

We find that similar arguments apply to other fillings.
Here we summarize two rules for the (translationally invariant) ground states at the one-shot level. 
First, as argued at the end of \cref{sec:flat-U4} and numerically confirmed at the end of \cref{sec:GS-nu=0-numeric}, the flat-U(4) symmetry is a better approximation (of the Hamiltonian) than the chiral-U(4) symmetry. 
Heuristically, the wave function tends to be symmetric under permutation of flavor indices, \eg the four components $\{\alpha,\eta,s\}$ fixed $\xi=\eta (-1)^{\alpha-1}$ which labels the fundamental U(4) representations (\cref{sec:flat-U4}), such that the real space wave function has as many as nodes and hence the Coulomb interaction can be saved (Hund's rule). 
Thus, we summarize the first rule
\begin{enumerate}
\item \it  When possible, two $f$-electrons tend to occupy one flavor of the flat-U(4) symmetry, \eg $f_{\RR1\eta s}^\dagger f_{\RR2\eta s}^\dagger$, such that the two $f$-electrons and their flat-U(4) rotations form a maximal weight representation of the flat-U(4) group. 
\end{enumerate}
The state $f_{\RR1\eta s}^\dagger f_{\RR2\eta s}^\dagger$ and its flat-U(4) rotations form a degenerate U(4) multiplet if $M=0$.
The multiplet form the irreducible representation $[2]_4$. 
(Here $[\lambda_1,\lambda_2\cdots]_4$ is the Young tableau notation for SU(4) irreducible representations.)
Readers may refer to Ref.~\cite{Biao-TBG4} for how the U(4) irreps are used to label the ground states of MATBG. 
We then need to consider how nonzero $M$ will split the U(4) multiplet and select the ground state.
Because the considered interaction Hamiltonian respects the flat-U(4) symmetry, states in the multiplet must have the same interaction energy.
Thus, it is the single-particle energy of the conduction bands that determines the ground state - the phase opening the largest gap has the lowest energy. 
However, since there are many levels, \ie 8 from local orbitals and 16 from conduction bands, it is still not immediate to determine the gap at generic fillings.
We now {\it assume} that the $\Gamma_3$ states from the local orbitals and the $\Gamma_3$ states ($a=1,2$) from the conduction bands are all at high energies.
This is a reasonable assumption based on the BM model because previous studies \cite{Biao-TBG4,kang_strong_2019,bultinck_ground_2020} has obtained the ground states by projecting the Hamiltonian into the topological flat bands, which only has the $\Gamma_1\oplus \Gamma_2$ states ($a=3,4$). 
We have also confirmed that the lowest (closest to the gap) energy levels are indeed contributed by the $\Gamma_1\oplus\Gamma_2$ ($a=3,4$) states for all the integer fillings, in agreement with those in Refs.~\cite{TBG5,bultinck_ground_2020,vafek2021lattice,cea_band_2020,zhang_HF_2020,kang2021cascades}. 
Thus we only need to look at the gap formed by the $\Gamma_1\oplus \Gamma_2$ states ($a=3,4$). 
The effective Hamiltonian for the $\Gamma_1\oplus \Gamma_2$ states ($a=3,4$) can be obtained by restricting the $a$ index of the one-shot mean field Hamiltonian $\hH_0+\ovl{H}_U + \ovl{H}_W + \ovl{H}_J + \ovl{H}_V$ (\cref{eq:H0-def,eq:HU-MF,eq:HW-MF,eq:HJ-MF,eq:HV-MF}) into the subspace spanned by $c_{\kk a\eta s}$ ($a=3,4$).
It reads
\begin{align} 
H_{a \eta s, a' \eta' s'}^{(\Gamma_1\oplus\Gamma_2)} (\kk=0) =& \nu_f W_{3} \delta_{aa'}\delta_{\eta\eta'} \delta_{ss'}  +  M [\sigma_x]_{a-2,a'-2} \delta_{\eta\eta'} \delta_{ss'} - J \delta_{aa'} \delta_{\eta\eta'} \pare{O_{a'-2,\eta',s'\ ;\ a-2,\eta,s}^f -  \frac12  \delta_{ss'} } \nono\\
 & + J \delta_{\ovl{a},a'} \delta_{-\eta,\eta'} O^f_{\ovl{a}-2,-\eta,s' \ ;\ a-2,\eta,s}\ ,
\end{align}
where $a,a'=3,4$.
We can equivalently write it in a more compact form as 
\begin{equation}\label{eq:MF-GM1GM2}
H^{(\Gamma_1\oplus\Gamma_2)}(\kk=0)  =  
    \nu_f W_3 \sigma_0 \tau_0 \spin_0 + M \sigma_x \tau_0 \spin_0 
    - \frac{J}2 \tau_z  ( O^{fT} - \frac12 \sigma_0\tau_0\spin_0) \tau_z  
    - \frac{J}2 \sigma_z ( O^{fT} - \frac12 \sigma_0\tau_0\spin_0) \sigma_z\ .
\end{equation}
The second rule then follows 
\begin{enumerate}
\setcounter{enumi}{1}
\item \it For a set of flat-U(4) rotation related states, the state minimizing the energy of \cref{eq:MF-GM1GM2} is the ground state. 
\end{enumerate}
The onsite repulsion $U_1$, and the couplings between local orbitals and conduction bands, \ie the $\gamma$ and $v_\star$ terms (\cref{eq:Hcf-summary}), do not enter the second rule because they only affect states at higher energies contributed by the $f$-orbitals and the $\Gamma_3$ $c$-band basis but not the low energy states contributed by the $\Gamma_1\oplus \Gamma_2$ $c$-band basis, as exampled in \cref{sec:GS-nu=0-analytical}. 

In the next subsection, we will apply the two rules to the ground states at other integer fillings.

\subsection{Correlated insulator phases at the fillings \texorpdfstring{$\nu=-1,-2,-3$}{nu=-1,-2,-3}}\label{sec:GS-other-fillings}

\begin{figure}[t]
\centering
\includegraphics[width=0.9\linewidth]{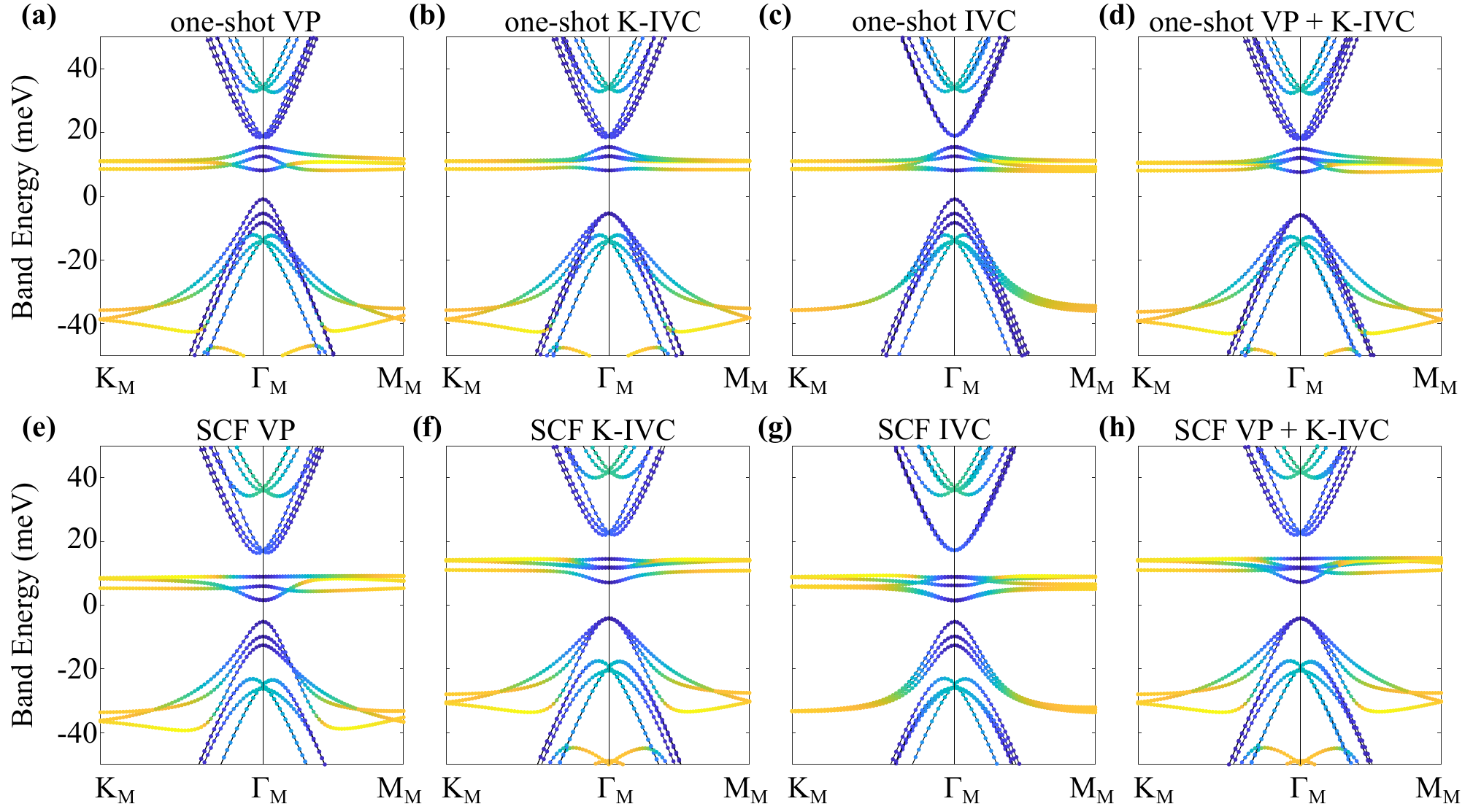}
\caption{HF band structures of correlated insulator phases at $\nu=-1$.
(a), (b), (c), (d) are the one-shot HF band structures of the VP, K-IVC, IVC, and mixed VP + K-IVC phases, respectively.
(e), (f), (g), (h) are the self-consistent HF band structures of the VP, K-IVC, IVC, and mixed VP + K-IVC phases, respectively.
The color represents the composition of the energy bands, where yellow corresponds to the local orbitals and blue corresponds to the conduction bands. 
We have chosen $w_0/w_1=0.8$ in the calculation. Other parameters of the single-particle and interaction hamiltonians are given in \cref{tab:H0parameters,tab:HIparameters}. 
\label{fig:HFbands_nu1} 
}
\end{figure}

\subsubsection{Filling \texorpdfstring{$\nu=-1$}{nu=-1}}
\label{sec:GS-nu=-1}

We consider the parent wave function of (fully) valley polarized (VP) and (partially) spin polarized state at $\nu=-1$ as 
\begin{equation}
\ket{\mrm{VP}_0^{\nu=-1}} = \prod_{\RR} f_{\RR 1 + \up}^\dagger f_{\RR 1 + \down}^\dagger f_{\RR 2 + \up}^\dagger \ket{\mrm{FS}}\ , 
\end{equation}
where $\ket{\rm FS}$ is the Fermi sea state of the conduction bands at the charge neutrality point (\cref{eq:FS-def}).
The parent wave functions for the IVC and K-IVC states are rotated from $\ket{\mrm{VP}_0^{\nu=-1}}$ by the chiral-U(4) (\cref{eq:chiralU4}) and flat-U(4) (\cref{eq:flatU4}) symmetries as 
{\small
\begin{equation}
\ket{\mrm{IVC}_0^{\nu=-1}} = e^{-i\frac{\pi}2 \UC_{x0} } \ket{\mrm{VP}_0^{\nu=-1}}
= \prod_{\RR} \frac1{2\sqrt2} (f_{\RR 1 + \up}^\dagger -i f_{\RR 2 - \up}^\dagger) (f_{\RR 1 + \down}^\dagger - i f_{\RR 2 - \down}^\dagger) ( - i f_{\RR 1 - \up}^\dagger + f_{\RR 2 + \up}^\dagger)  \ket{\rm FS}\ ,  
\end{equation}}
and 
{\small
\begin{equation}
\ket{\text{K-IVC}_0^{\nu=-1}} = e^{-i\frac{\pi}2 \UF_{x0} } \ket{\mrm{VP}_0^{\nu=-1}}
= \prod_{\RR} \frac1{2\sqrt2} (f_{\RR 1 + \up}^\dagger + f_{\RR 2 - \up}^\dagger) (f_{\RR 1 + \down}^\dagger + f_{\RR 2 - \down}^\dagger) ( - f_{\RR 1 - \up}^\dagger + f_{\RR 2 + \up}^\dagger)  \ket{\rm FS}\ ,  
\end{equation}}
respectively.
Ref.~\cite{Biao-TBG4} found that a mixed state of VP and K-IVC  has the lowest energy at $\nu=-1$. 
The parent wave function of the mixed state can be written as 
{\small
\begin{equation}
\ket{\text{VP+K-IVC}_0^{\nu=-1}} = 
\prod_{\RR} f_{\RR 1 + \down}^\dagger e^{-i\frac{\pi}2 \UF_{x0} } f_{\RR 1 + \up}^\dagger  f_{\RR 2 + \up}^\dagger e^{i\frac{\pi}2 \UF_{x0} } \ket{\mrm{FS}}
= \prod_{\RR} \frac1{2} f_{\RR 1 + \down}^\dagger (f_{\RR 1 + \up}^\dagger + f_{\RR 2 - \up}^\dagger) ( - f_{\RR 1 - \up}^\dagger + f_{\RR 2 + \up}^\dagger)  \ket{\rm FS}\ . 
\end{equation}}
According to Ref.~\cite{Biao-TBG4}, the two states  ($f_{\RR 1 + \up}^\dagger  f_{\RR 2 + \up}^\dagger$) from the same valley-spin flavor are rotated to a K-IVC state, whereas the state $f_{\RR 1 + \down}^\dagger$ remains valley and spin polarized. 
Feeding the parent states into the HF loop, we obtain the one-shot and self-consistent HF band structures (\cref{fig:HFbands_nu1}) and energies (\cref{tab:HFEnergy}). 
We find that the mixed state indeed has the lowest energy.

All the one-shot states, \ie $\ket{X^{\nu=-1}_{1}}$ ($X=$VP, IVC, K-IVC, VP+K-IVC), and self-consistent states, \ie $\ket{X^{\nu=-1}_{\infty}}$, are gapped Fock states occupying the HF bands (\cref{fig:HFbands_nu1}) up to the filling $\nu=-1$.
As will be explained in \cref{sec:Chern}, all these states have the Chern number $C=1$. 
It is worth mentioning that the parent states $\ket{X^{\nu=-1}_{0}}$, even though lead to gapped one-shot states with nonzero Chern numbers, do not have Chern numbers because they occupy gapless conduction bands plus a select number of local orbitals. 

We now apply the two rules given in \cref{sec:ground-state-rules} to discuss the one-shot ground states $\ket{X^{\nu=-1}_{1}}$ ($X=$VP, IVC, K-IVC, VP+K-IVC). 
VP, K-IVC and VP+K-IVC states satisfy the first rule, \ie two $f$-electrons tend to occupy a flavor flat-U(4) symmetry.
Here the two $f$-electrons are $f_{\RR1+\up}^\dagger f_{\RR2+\up}^\dagger$ or its flat-U(4) partners. 
We need to compare the quasi-particle energies of the three states to determine the ground states. 
The density matrices given by $\ket{\text{VP}^{\nu=-1}_0}$ are 
{\small
\begin{equation}\label{eq:density-nu=-1-VP}
O^f = \sigma_0 (\frac{\tau_z+\tau_0}2) (\frac{\spin_0 + \spin_z}2) 
+  (\frac{\sigma_0+\sigma_z}2) (\frac{\tau_0+\tau_z}2)(\frac{\spin_0 - \spin_z}2), \quad  
     O^c_{a\eta s,a'\eta's'}=\delta_{\eta\eta'} \delta_{ss'} O^c_{a\eta s,a'\eta s},\quad O^c_{a\eta s,a\eta s}=0, \quad 
O^{cf}=0\ . 
\end{equation}}
The density matrix $O^f$ is diagonal in spin.
The two spin blocks of $O^f$ are 
\begin{equation}
O^{f(s=\up)} = \sigma_0 (\frac{\tau_z+\tau_0}2),\qquad 
O^{f(s=\down)} = (\frac{\sigma_0+\sigma_z}2) (\frac{\tau_0+\tau_z}2)\ . 
\end{equation}
Substituting $O^{f(s)}$ into \cref{eq:MF-GM1GM2}, we obtain the two spin blocks of $H^{(\Gamma_1\oplus\Gamma_2)}(\kk=0)$ as 
\begin{equation}
  H^{(\Gamma_1\oplus\Gamma_2, s=\up)}_{\rm VP} =   -W_3 \sigma_0\tau_0 + M \sigma_x \tau_0
  - \frac{J}2 \tau_z \frac{\sigma_0\tau_z}2 \tau_z 
  - \frac{J}2 \sigma_z \frac{\sigma_0\tau_z}2 \sigma_z
 = -W_3 \sigma_0\tau_0 + M \sigma_x \tau_0  - \frac{J}2 \sigma_0 \tau_z\ ,
\end{equation}
and
\begin{equation}
  H^{(\Gamma_1\oplus\Gamma_2, s=\down)}_{\rm VP} =   -W_3 \sigma_0\tau_0 + M \sigma_x \tau_0
  - \frac{J}2 \tau_z X \tau_z 
  - \frac{J}2 \sigma_z X \sigma_z
 = -W_3 \sigma_0\tau_0 + M\sigma_x \tau_0 - \frac{J}2 \mrm{diag}([1,-1,-1,-1])\ ,
\end{equation}
respectively, where $X = (\frac{\sigma_0+\sigma_z}2) (\frac{\tau_0+\tau_z}2) - \frac12\sigma_0\tau_0 = \frac12 \mrm{diag}([1,-1,-1,-1])$. 
Here the subscript ``VP'' indicates that the corresponding density matrices are generated by the parent VP state. 
The lowest three levels at $\kk=0$ of the VP state are 
\begin{equation}
-W_3 - \sqrt{M^2+\frac{J^2}4},\qquad -W_3 -M - \frac{J}2,\qquad -W_3 + M - \frac{J}2 \ ,
\end{equation}
which come from the $s=\down,\eta=+$, $s=\up,\eta=+$, and $s=\up,\eta=+$ flavors, respectively. 
(We have made use of $J/2>M$.) 

Applying a flat-U(4) rotation $e^{-i\frac{\pi}2 \UF_{x0} }$ to the order parameter, we obtain the order parameters of the K-IVC state as 
\begin{equation}
    O^{f(s=\up)} = \frac12 \sigma_0\tau_0 - \frac12 \sigma_y\tau_y ,\qquad 
    O^{f(s=\down)} = \begin{pmatrix}
    0 & 0 & 0 & 1\\
    0 & -1& 0 & 0\\
    0 & 0 &-1 & 0\\
    1 & 0 & 0 & 0
\end{pmatrix} \ .
\end{equation}
Substituting $O^{f(s)}$ into \cref{eq:MF-GM1GM2}, we obtain the two spin blocks of $H^{(\Gamma_1\oplus\Gamma_2)}(\kk=0)$ as 
\begin{equation}
H^{(\Gamma_1\oplus\Gamma_2, s=\up)}_{\rm K-IVC}= -W_3 \sigma_0\tau_0 + M \sigma_x \tau_0  - \frac{J}2 \sigma_y \tau_y,\qquad
H^{(\Gamma_1\oplus\Gamma_2, s=\down)}_{\rm K-IVC} = -W_3 \sigma_0\tau_0 + M\sigma_x \tau_0 +  \frac{J}2 
\begin{pmatrix}
    0 & 0 & 0 & 1\\
    0 & 1& 0 & 0\\
    0 & 0 &1 & 0\\
    1 & 0 & 0 & 0
\end{pmatrix} \ ,
\end{equation}
where the subscript ``K-IVC'' indicates that the corresponding density matrices are generated by the parent K-IVC state. 
The lowest three levels at $\kk=0$ of the K-IVC state are all $-W_3 - \sqrt{M^2+\frac{J^2}4}$, where  two come from the $s=\up$ flavor and one comes from the $s=\down$ flavor. 
Clearly, the K-IVC state has lower quasi-particle energy than the VP state.
In the end we consider the one-shot HF mean field Hamiltonian spanned by the $\Gamma_1\oplus\Gamma_2$ for the VP+K-IVC state 
\begin{equation}
H^{(\Gamma_1\oplus\Gamma_2, s=\up)}_{\rm VP+K-IVC}= -W_3 + M  \sigma_x \tau_0 - \frac{J}2 \sigma_y \tau_y,\qquad
H^{(\Gamma_1\oplus\Gamma_2, s=\down)}_{\rm VP+K-IVC} = -W_3 + M\sigma_x\tau_0  
    - \frac{J}2 \mrm{diag}([1,-1,-1,-1]) \ ,
\end{equation}
where the subscript ``VP+K-IVC'' indicates that the corresponding density matrices are generated by the parent mixed (VP+K-IVC) state. 
The lowest three levels at $\kk=0$ of the VP+K-IVC state are also all $-W_3 - \sqrt{M^2+\frac{J^2}4}$, which are same as those of the K-IVC state. 
Thus the rules in \cref{sec:ground-state-rules} cannot tell us whether K-IVC or VP+K-IVC has a lower energy. 
In fact, the one-shot and self-consistent HF energies of the two states are indeed very close to each other (\cref{tab:HFEnergy}). 

Taking into account the bands at $\kk=0$ contributed by the $\Gamma_3$ basis still cannot distinguish the energies of the K-IVC and the VP+K-IVC states.
Following a similar calculation as in \cref{sec:GS-nu=0-analytical}, one will find that the energy levels from the $\Gamma_3$ basis are the same for both states. 
The energy difference must be contributed by the bands at finite $\kk$. 

As shown in \cref{fig:HFbands_nu1}, one-shot and self-consistent HF bands of all the different states at $\nu=-1$ have a feature in common: There is a set of flat bands above the zero energy. 
Such flat bands are also observed in one of our previous studies \cite{TBG5} but have not been understood. 
(See the particle excitation spectra in Fig. 11 of Ref.~\cite{TBG5}.)
We now can explain the origin of the flatness through our topological heavy fermion model.
We consider the k$\cdot$p expansion of the one-shot mean-field Hamiltonian. 
For simplicity, here we mainly focus on the VP state. 
Following the same procedure we have done to obtain the one-shot mean field Hamiltonian at $\nu=0$ (\cref{eq:HMF-nu=0-VP}), we obtain
{\small
\begin{equation}
H_{\rm VP}^{\rm(MF)} (\kk) \approx \begin{pmatrix}
- W_1\sigma_0\tau_0\spin_0   &  v_\star( k_x \sigma_0 \tau_z + ik_y\sigma_z\tau_0 )\spin_0  & \gamma \sigma_0 \tau_0 \spin_0 + v_\star'( k_x \sigma_x\tau_z + k_y\sigma_y \tau_0 ) \spin_0 \\
v_\star( k_x \sigma_0 \tau_z - ik_y\sigma_z\tau_0 )\spin_0 & - W_3\sigma_0\tau_0\spin_0 + M \sigma_x\tau_0\spin_0 - \frac{J}2 \sigma_0\tau_z\spin_0 & 0 \\
\gamma \sigma_0\tau_0 \spin_0 + v_\star'( k_x \sigma_x\tau_z + k_y  \sigma_y\tau_0)\spin_0 & 0 & - (U_1+6U_2)\sigma_0\tau_0\spin_0 - U_1( O^f - \frac12 \sigma_0\tau_0\spin_0)  
\end{pmatrix} .
\end{equation}}%
The density matrix $O^f$ is given in \cref{eq:density-nu=-1-VP}. 
Comparing it to the one-shot mean field Hamiltonian at $\nu=0$ (\cref{eq:HMF-nu=0-VP}), there are three additional terms, \ie $- W_1\sigma_0\tau_0\spin_0 $, $- W_3\sigma_0\tau_0\spin_0$, and $- (U_1+6U_2)\sigma_0\tau_0\spin_0$ in the three diagonal blocks. 
These three terms come from the Hartree channel terms $\nu_f W_1$, $\nu_f W_3$, $\nu_f (U_1+6U_2)$ of $\ovl{H}_W$ (\cref{eq:HW-MF}) and $\ovl{H}_U$ (\cref{eq:HU-MF}) and only shift the energies of the three blocks. 
Without the energy shift and the couplings ($\gamma,v_\star'$) between $c$- and $f$-electrobs, the $c$-bands (the first two blocks) would have a quadratic touching at zero energy and the $f$-levels (the third block) have energies $\pm U_1/2$, as illustrated by the red bands in Fig. 3(a) in the main text. 
Using the parameters obtained at $w_0/w_1=0.8$ (\cref{tab:HIparameters}), the average energy shift of the $c$-electron bands (the first two blocks) is $-(W_1+W_3)/2 \approx -47$meV, and the energy shift of the $f$-electron bands is $-(U_1+6U_2)\approx -72$meV. 
Thus, the relative energy shift of $f$-electrons with respect to $c$-electrons is given by $\delta E \approx (-72 + 47) \mrm{meV} \approx - 25$meV and is approximately $-U_1/2 \approx 29$meV. 
That means, if we turn off the hybridizations, the upper branch of the $f$-electron levels will be shifted to the quadratic touching point of the $c$-electrons, as shown by the red bands in Fig. 3(b) in the main text. 
Turning on the hybridizations will gap out the quadratic touching point, then the upper branch of $f$-electron bands form an isolated set of flat bands. 

\subsubsection{Filling \texorpdfstring{$\nu=-2$}{nu=-2}}
\begin{figure}[t]
\centering
\includegraphics[width=0.7\linewidth]{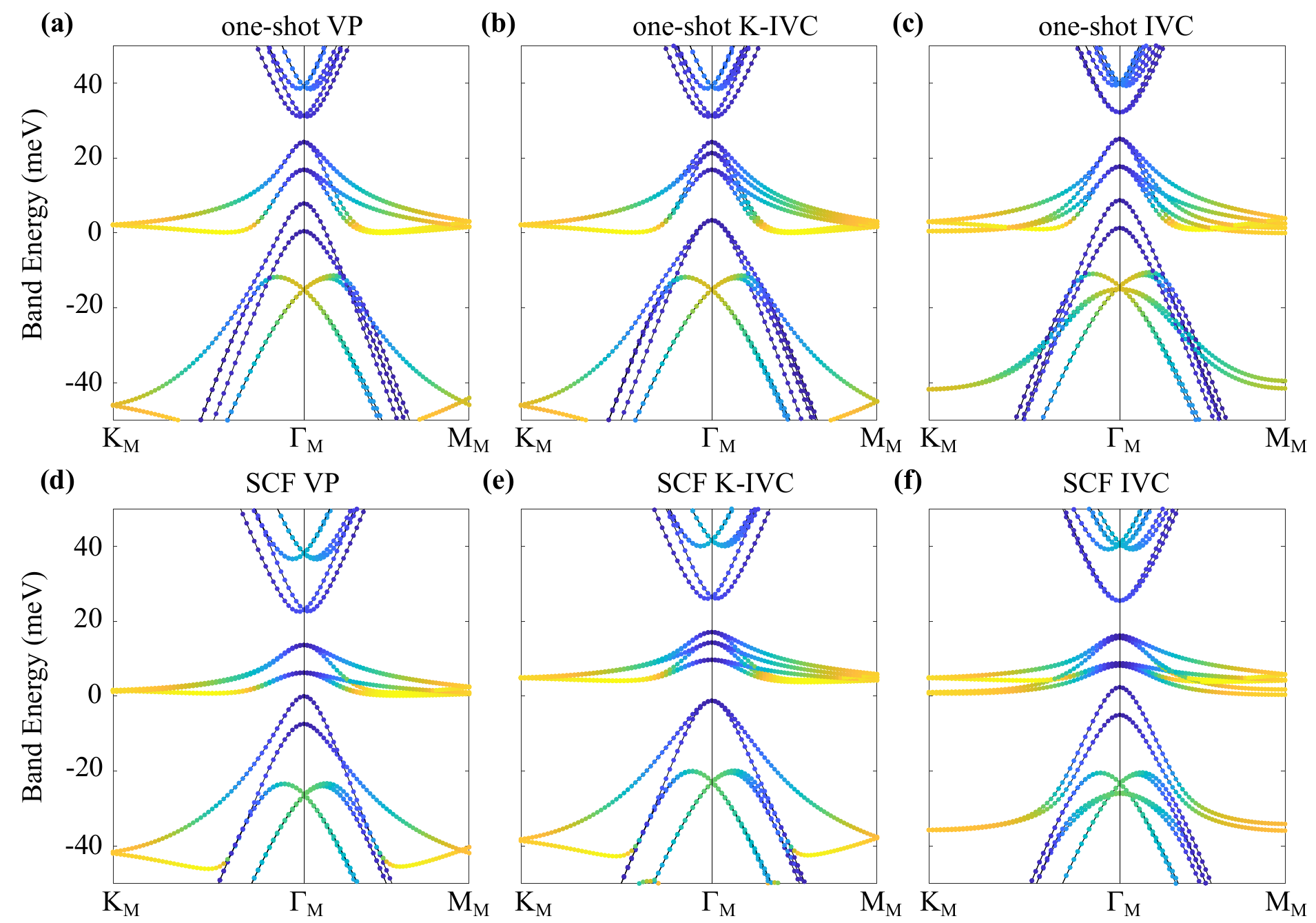}
\caption{HF band structures of correlated insulator phases at $\nu=-2$.
(a), (b), (c) are the one-shot HF band structures of the VP, K-IVC, and IVC phases, respectively.
(d), (e), (f) are the self-consistent HF band structures of the VP, K-IVC, and IVC phases, respectively.
The color represents the composition of the energy bands, where yellow corresponds to the local orbitals and blue corresponds to the conduction bands. 
We have chosen $w_0/w_1=0.8$ in the calculation. Other parameters of the single-particle and interaction hamiltonians are given in \cref{tab:H0parameters,tab:HIparameters}. 
\label{fig:HFbands_nu2} 
}
\end{figure}

We consider the parent wave function of valley polarized (VP) state at $\nu=-2$ as 
\begin{equation}
\ket{\mrm{VP}_0^{\nu=-2}} = \prod_{\RR} f_{\RR 1 + \up}^\dagger f_{\RR 2 + \up}^\dagger \ket{\mrm{FS}}\ , 
\end{equation}
where $\ket{\rm FS}$ is the Fermi sea state of the conduction bands at the charge neutrality point (\cref{eq:FS-def}).
The parent wave functions for the IVC and K-IVC states are rotated from $\ket{\mrm{VP}_0^{\nu=-1}}$ by the chiral-U(4) (\cref{eq:chiralU4}) and flat-U(4) (\cref{eq:flatU4}) symmetries as 
{\small
\begin{equation} \label{eq:IVC-parent-n2}
\ket{\mrm{IVC}_0^{\nu=-2}} = e^{-i\frac{\pi}2 \UC_{x0} } \ket{\mrm{VP}_0^{\nu=-1}}
= \prod_{\RR} \frac1{2} (f_{\RR 1 + \up}^\dagger -i f_{\RR 2 - \up}^\dagger)  ( - i f_{\RR 1 - \up}^\dagger + f_{\RR 2 + \up}^\dagger)  \ket{\rm FS}\ ,  
\end{equation}}
and 
{\small
\begin{equation} \label{eq:KIVC-parent-n2}
\ket{\text{K-IVC}_0^{\nu=-2}} = e^{-i\frac{\pi}2 \UF_{x0} } \ket{\mrm{VP}_0^{\nu=-1}}
= \prod_{\RR} \frac1{2} (f_{\RR 1 + \up}^\dagger + f_{\RR 2 - \up}^\dagger)  ( - f_{\RR 1 - \up}^\dagger + f_{\RR 2 + \up}^\dagger)  \ket{\rm FS}\ ,  
\end{equation}}
respectively.
Feeding the parent states into the HF loops, we obtain the HF band structures and energies as shown in \cref{fig:HFbands_nu2} and \cref{tab:HFEnergy}, respectively. 
We find that the K-IVC state has the lowest energy. 

We now apply the two rules given in \cref{sec:ground-state-rules} to discuss the one-shot states, \ie  $\ket{X^{\nu=-2}_{1}}$ ($X=$VP, IVC, K-IVC), which are Fock states occupying the one-shot HF bands (\cref{fig:HFbands_nu2}) to the filling $\nu=-2$. 
The VP and K-IVC states satisfy the first rule, \ie two $f$-electrons tend to occupy a flavor flat-U(4) symmetry.
Here the two $f$-electrons are $f_{\RR1+\up}^\dagger f_{\RR2+\up}^\dagger$ or its flat-U(4) partners. 
We need to compare the quasi-particle energies of the three states to determine the ground states. 
The density matrices given by $\ket{\text{VP}^{\nu=-2}_0}$ are 
\begin{equation}
O^f = \sigma_0 (\frac{\tau_z+\tau_0}2) (\frac{\spin_0 + \spin_z}2),\qquad 
     O^c_{a\eta s,a'\eta's'}=\delta_{\eta\eta'} \delta_{ss'} O^c_{a\eta s,a'\eta s},\qquad O^c_{a\eta s,a\eta s}=0,  \qquad 
     O^{cf}=0\ . 
\end{equation}
According to \cref{eq:MF-GM1GM2} and the density matrices, the one-shot HF mean field Hamiltonian spanned by the $\Gamma_1\oplus\Gamma_2$ states at $\kk=0$, \ie $H^{(\Gamma_1\oplus\Gamma_2)}$, is diagonal in the spin index. 
The spin-up and down blocks are given by
\begin{equation}
H^{(\Gamma_1\oplus\Gamma_2, s=\up)}_{\rm VP}= -2 W_3\sigma_0\tau_0 + M  \sigma_x \tau_0 - \frac{J}2 \sigma_0\tau_z,\qquad
H^{(\Gamma_1\oplus\Gamma_2, s=\down)}_{\rm VP} = -2 W_3 \sigma_0\tau_0 + M\sigma_x\tau_0 + \frac{J}2 \sigma_0\tau_0 \ ,
\end{equation}
where the subscript ``VP'' indicates that the corresponding density matrices are generated by parent VP state. 
The lowest two levels at $\kk=0$ of the VP state are 
\begin{equation}
-2 W_3 -M - \frac{J}2,\qquad  -2W_3+ M - \frac{J}2 \ ,
\end{equation}
both of which come from the $s=\up$ flavor. 
(We have making use of $J/2>M$.) 
Applying a flat-U(4) rotation $e^{-i\frac{\pi}2 \UF_{x0} }$ to the order parameter, we obtain the one-shot HF mean field Hamiltonian spanned by the $\Gamma_1\oplus\Gamma_2$ (\cref{eq:MF-GM1GM2}) for the K-IVC state 
\begin{equation}
H^{(\Gamma_1\oplus\Gamma_2, s=\up)}_{\rm K-IVC}= -2W_3 \sigma_0\tau_0 + M \sigma_x\tau_0   - \frac{J}2 \sigma_y\tau_y,\qquad
H^{(\Gamma_1\oplus\Gamma_2, s=\down)}_{\rm K-IVC} = -2W_3 \sigma_0\tau_0 + M \sigma_x\tau_0  + \frac{J}2 \sigma_0\tau_0 \ ,
\end{equation}
where the subscript ``K-IVC'' indicates that the corresponding density matrices are generated by parent VP state. 
The lowest two levels at $\kk=0$ of the K-IVC state are 
\begin{equation}
-2W_3 - \sqrt{M^2+\frac{J^2}4},\qquad 
-2W_3 - \sqrt{M^2+\frac{J^2}4}\ ,
\end{equation}
both of which come from the $s=\up$ flavor. 
Clearly, the K-IVC state, where the $M\sigma_x\tau_0$ and $\frac{J}2\sigma_y\tau_y$ terms in the spin-up sector anti-commute with each other,  has lower quasi-particle energy than the VP state, where the $M\sigma_x\tau_0$ and $\frac{J}2\sigma_0\tau_z$ terms in the spin-up sector commute with each other. 


\subsubsection{Filling \texorpdfstring{$\nu=-3$}{nu=-3}}

\begin{figure}[t]
\centering
\includegraphics[width=0.9\linewidth]{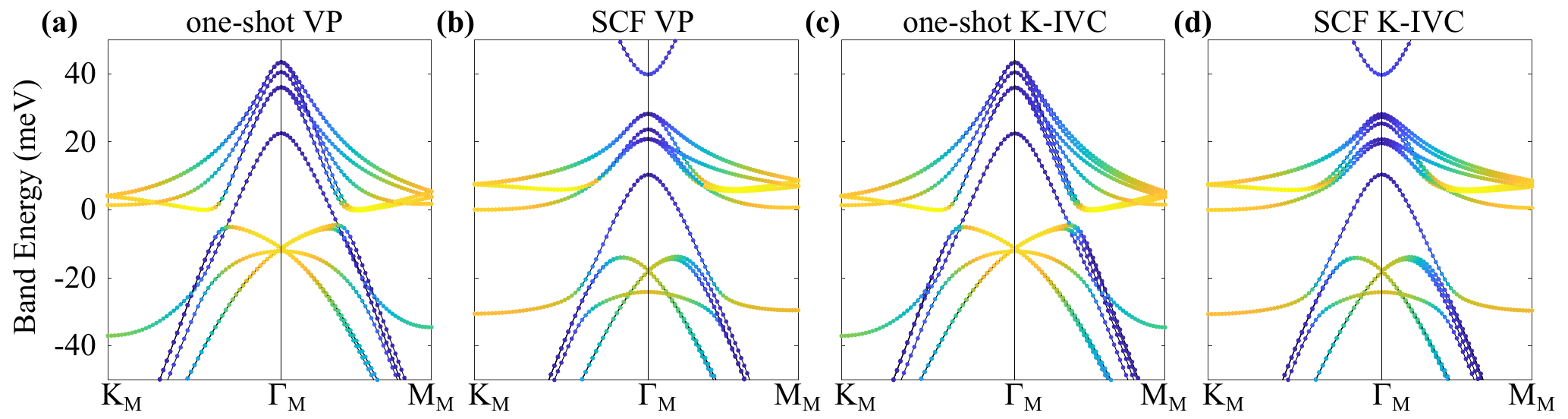}
\caption{HF band structures of correlated insulator phases at $\nu=-3$.
(a), (b) are the one-shot HF band structures of the VP and K-IVC phases, respectively.
(c), (d) are the self-consistent HF band structures of the VP and K-IVC phases, respectively.
The color represents the composition of the energy bands, where yellow corresponds to the local orbitals and blue corresponds to the conduction bands. 
We have chosen $w_0/w_1=0.8$ in the calculation. Other parameters of the single-particle and interaction hamiltonians are given in \cref{tab:H0parameters,tab:HIparameters}. 
\label{fig:HFbands_nu3} 
}
\end{figure}

We consider the parent wave function of valley polarized (VP) and spin polarized state at $\nu=-3$ as 
\begin{equation}
\ket{\mrm{VP}_0^{\nu=-3}} = \prod_{\RR} f_{\RR 1 + \up}^\dagger \ket{\mrm{FS}}\ , 
\end{equation}
where $\ket{\rm FS}$ is the Fermi sea state of the conduction bands at the charge neutrality point.
The parent wave functions for the IVC and K-IVC states are rotated from $\ket{\mrm{VP}_0^{\nu=-3}}$ by the chiral-U(4) (\cref{eq:chiralU4}) and flat-U(4) (\cref{eq:flatU4}) symmetries as 
{\small
\begin{equation}
\ket{\mrm{IVC}_0^{\nu=-3}} = e^{-i\frac{\pi}2 \UC_{x0} } \ket{\mrm{VP}_0^{\nu=-3}}
= \prod_{\RR} \frac1{\sqrt2} (f_{\RR 1 + \up}^\dagger -i f_{\RR 2 - \up}^\dagger)  \ket{\rm FS}\ ,  
\end{equation}}
and 
{\small
\begin{equation}
\ket{\text{K-IVC}_0^{\nu=-3}} = e^{-i\frac{\pi}2 \UF_{x0} } \ket{\mrm{VP}_0^{\nu=-3}}
= \prod_{\RR} \frac1{\sqrt2} (f_{\RR 1 + \up}^\dagger + f_{\RR 2 - \up}^\dagger)  \ket{\rm FS}\ ,  
\end{equation}}
respectively.
We can see that $\ket{\mrm{IVC}_0^{\nu=-3}}$ and $\ket{\text{K-IVC}_0^{\nu=-3}} $ are related by a valley-U(1) rotation, which adds the factor $i$ to the $f$-electron ($f_{\RR 2 - \up}^\dagger$) in the valley $\eta=-$.  
(One should notice that, at the filling $\nu=-2$, the K-IVC (\cref{eq:KIVC-parent-n2}) and IVC (\cref{eq:IVC-parent-n2}) parent states are not related by a valley-U(1) rotation.  If one applies the same rotation as above, one will obtain $\prod_{\RR} \frac12 (f_{\RR1+\up}^\dagger + f_{\RR2-\up}^\dagger ) (f_{\RR1-\up}^\dagger + f_{\RR2+\up}^\dagger) \ket{\text{FS}}$, which is still different from the  K-IVC  parent state (\cref{eq:KIVC-parent-n2}).)
Thus the two states must have the same band structure and the same energy. 
Hence in the following we will only discuss the VP and the K-IVC states. 
Feeding the parent states into the HF loops, we obtain the HF band structures and energies as shown in \cref{fig:HFbands_nu3} and \cref{tab:HFEnergy}, respectively. 
We find that the VP state has the lowest energy. 

All the one-shot states, \ie $\ket{X^{\nu=-3}_{1}}$ ($X=$VP, K-IVC), and self-consistent states, \ie $\ket{X^{\nu=-3}_{\infty}}$, are Fock states occupying the HF bands (\cref{fig:HFbands_nu1}) up to the filling $\nu=-3$.
Even though they do not have indirect gaps, they do have direct gaps at every momentum. 
Hence Chern numbers are still well defined for the direct gaps. 
As will be explained in \cref{sec:Chern}, all these states have the Chern number $C=1$. 
It is worth mentioning that the parent states $\ket{X^{\nu=-3}_{1}}$, even though lead to one-shot states with nonzero Chern numbers, do not have Chern numbers because they occupy gapless conduction bands plus one local orbital. 

We now apply the two rules given in \cref{sec:ground-state-rules} to discuss the ground states. 
We need to compare the quasi-particle energies of the two states to determine the ground states. 
The density matrices given by $\ket{\text{VP}^{\nu=-3}_0}$ are 
\begin{equation}
O^f = (\frac{\sigma_0+\sigma_z}2) (\frac{\tau_z+\tau_0}2) (\frac{\spin_0 + \spin_z}2),\qquad
O^c_{a\eta s,a'\eta's'}=\delta_{\eta\eta'} \delta_{ss'} O^c_{a\eta s,a'\eta s},\qquad O^c_{a\eta s,a\eta s}=0, \qquad 
O^{cf}=0\ . 
\end{equation}
According to \cref{eq:MF-GM1GM2} and the density matrices, the one-shot HF mean field Hamiltonian spanned by the $\Gamma_1\oplus\Gamma_2$ states at $\kk=0$, \ie $H^{(\Gamma_1\oplus\Gamma_2)}$, is diagonal in the spin index. 
The spin-up and down blocks of $\Gamma_1\oplus\Gamma_2$ states are given by 
\begin{equation}
H^{(\Gamma_1\oplus\Gamma_2, s=\up)}_{\rm VP}= -3 W_3 + M \tau_0 \sigma_x  - \frac{J}2 \mrm{diag}([1,-1,-1,-1]),\qquad
H^{(\Gamma_1\oplus\Gamma_2, s=\down)}_{\rm VP} = -3 W_3 + M\sigma_x + \frac{J}2\sigma_0\ ,
\end{equation}
where the subscript ``VP'' indicates that the input density matrices are generated by the parent VP state. 
The lowest level at $\kk=0$ of the VP state is 
\begin{equation}
-3 W_3 -\sqrt{M^2 + \frac{J^2}4} ,
\end{equation}
which comes from the $s=\up$ flavor. 
Applying a flat-U(4) rotation $e^{-i\frac{\pi}2 \UF_{x0} }$ to the order parameter, we obtain the one-shot HF mean field Hamiltonian spanned by the $\Gamma_1\oplus\Gamma_2$ (\cref{eq:MF-GM1GM2}) for the K-IVC state 
\begin{equation}
H^{(\Gamma_1\oplus\Gamma_2, s=\up)}_{\rm K-IVC}= -2W_3 + M \tau_0 \sigma_x  + \frac{J}2 
\begin{pmatrix}
    0 & 0 & 0 & 1\\
    0 & 1& 0 & 0\\
    0 & 0 &1 & 0\\
    1 & 0 & 0 & 0
\end{pmatrix} ,\qquad
H^{(\Gamma_1\oplus\Gamma_2, s=\down)}_{\rm K-IVC} = -2W_3 + M\sigma_x  + \frac{J}2\sigma_0 \ ,
\end{equation}
where the subscript ``K-IVC'' indicates that the input density matrices are generated by the parent K-IVC state. 
The lowest level at $\kk=0$ of the K-IVC state is still $-3 W_3 -\sqrt{M^2 + \frac{J^2}4}$. 
Thus the second rule given in \cref{sec:ground-state-rules} cannot tell us which has lower energy. 
In fact, the one-shot and self-consistent HF energies of the two states are indeed very close to each other (\cref{tab:HFEnergy}). 
As discussed in the end of \cref{sec:GS-nu=-1}, the energy difference between the two states must be contributed by bands at finite $\kk$. 

\subsection{Chern numbers of the ground states}\label{sec:Chern}

We first consider the Chern numbers of the insulator ground states at the filling $\nu=0$. 
Since the Chern numbers are stable against perturbations, we can simplify the problem by (i) looking at the one-shot mean-field Hamiltonian instead of the self-consistent mean-field Hamiltonian and (ii) deforming the one-shot mean-field Hamiltonian adiabatically. 
As shown in previous subsections, flat-U(4) is a good approximation of the Hamiltonian. 
Tuning $M$ to zero, which makes flat-U(4) an exact symmetry of the considered Hamiltonian, will not close the gap of the one-shot VP ($2|M-J/2|$), gap of one-shot K-IVC ($2\sqrt{M^2+J^2/4}$), and gaps of their U(4) partner states, provided that $M<J/2$. 
Therefore, the VP and K-IVC states must have the same Chern number since they are related by flat-U(4) rotation. 
We rewrite the kp expansion of the one-shot mean-field Hamiltonian for the VP state (\cref{eq:HMF-nu=0-VP}) as 
\begin{equation}
H^{({\rm MF})}_{\rm VP}(\kk) = 
\begin{pmatrix}
A(\kk)         & S(\kk) \\
S^\dagger(\kk) & B 
\end{pmatrix} \otimes \spin_0
\end{equation}
with 
\begin{equation} 
A(\kk) = \begin{pmatrix}
0  &  v_\star( k_x \tau_z \sigma_0 + ik_y\tau_0 \sigma_z)   \\
v_\star( k_x \tau_z\sigma_0 - ik_y\tau_0 \sigma_z) & M \tau_0\sigma_x - \frac{J}2 \tau_z\sigma_0 
\end{pmatrix},\qquad 
S(\kk) = \begin{pmatrix}
    \gamma \sigma_0 \tau_0 + v_\star'(k_x\sigma_x \tau_z + k_y\sigma_y \tau_0) \\
    0
\end{pmatrix}\ ,
\end{equation}
$B = - \frac{U_1}2 \sigma_0 \tau_z$. Since $U_1$ is large, we can integrate out the $f$-orbital levels to derive a four-by-four Hamiltonian to describe the low energy bands around $\kk=0$.
Applying a second order perturbation theory, or a Schrieffer-Wolf transformation that decouples $f$- and $c$-electrons, we obtain
\begin{equation} \label{eq:VP-nu=0-MF-cond}
H^{({\rm eff})}_{\rm VP}(\kk) \approx 
A(\kk) + S(\kk) B^{-1} S^\dagger(\kk) = 
\begin{pmatrix}
 \frac{2\gamma^2}{U_1} \tau_z\sigma_0 + \frac{4\gamma v_\star'}{U_1} (k_x \tau_0 \sigma_x + k_y \tau_z \sigma_y)  &  v_\star( k_x \tau_z \sigma_0 + ik_y\tau_0 \sigma_z)   \\
v_\star( k_x \tau_z\sigma_0 - ik_y\tau_0 \sigma_z) & M \tau_0\sigma_x - \frac{J}2 \tau_z\sigma_0 
\end{pmatrix} \otimes \spin_0 \ . 
\end{equation}
Since the ``integrated out'' $f$-orbitals are topologically trivial, there should be no worry about missing topological bands at higher energies. 
Without closing the gap, we can continuously change the parameters as $2\gamma^2/U \to m$, $v_\star'\to 0$, $M\to 0$, $J/2 \to -m$ with $m>0$ such that the Hamiltonian becomes
\begin{equation}
m \zeta_z \tau_z \sigma_0 + v_\star k_x \zeta_x \tau_z \sigma_0 - v_\star k_y \zeta_y \tau_0 \sigma_z\ .
\end{equation}
Here $\zeta_{x,y,z}$ are Pauli matrices introduced for the $\Gamma_3$ ($\zeta_z=1$) and the $\Gamma_1\oplus \Gamma_2$ ($\zeta_z=-1$) subspaces. 
We can view this Hamiltonian as four independent gapped Dirac points in the eigenspaces of $\tau_z$ and $\sigma_z$. 
To be concrete, we use $\eta=\pm1$ and $(-1)^{\alpha-1}$ ($\alpha=1,2$) to represent the eigenvalues of $\tau_z$ and $\sigma_z$, respectively. 
Then the Dirac Hamiltonian can be written as 
\begin{equation}
m \eta \zeta_z + v_\star k_x \eta \zeta_x  - v_\star k_y (-1)^{\alpha-1} \zeta_y .
\end{equation}
The Chern number contributed by each Dirac point is given by $- \frac12 \eta\cdot\eta\cdot (-1)^{\alpha-1} = -\frac12 (-1)^{\alpha-1}$.
The total Chern number is hence zero.

We then consider the Chern numbers of the insulator states at other fillings.
We can simplify the problem by deforming the one-shot mean-field Hamiltonian adiabatically in the sense that the {\it direct gap} at each $\kk$ does not close. 
We first let $M\to 0$ to recover the flat-U(4) symmetry. 
The states do not close their gaps ($2|J/2-M|$ or $\sqrt{M^2+J^2/4}$) in this process because $M$ is small compared to $J/2$. 
Thus all the phases related by flat-U(4) rotations must have the same topology.
For a set of flat-U(4) related states, we only need to calculate the Chern number of the simplest one of them, \ie the valley, spin, and orbital polarized state. 
We hence consider the parent wave function
\begin{equation}\label{eq:trial-chern}
    \ket{{\rm trial}} = \prod_{\RR} \prod_{\alpha\eta s} (f_{\RR \alpha\eta s}^\dagger)^{n_{\alpha\eta s}} \ket{\mrm{FS}}\ ,
\end{equation}
where $n_{\alpha\eta s}=1$ or 0 are the occupation numbers of the $\alpha$-th ($\alpha=1,2$) $f$-orbitals in the valley $\eta$ ($=\pm$) and spin $s$ ($=\up\down$) sector. 
The density matrix and filling of local orbitals are given by 
\begin{equation}
    O_{\alpha\eta s, \alpha'\eta' s'}^f = \delta_{\alpha,\alpha'} \delta_{\eta,\eta'} \delta_{s,s'} n_{\alpha\eta s} ,\qquad  
    \nu_f = -4 + \sum_{\alpha\eta s} n_{\alpha\eta s},\qquad 
    \nu_c = 0\ . 
\end{equation}
The k$\cdot$p expansion of the one-shot mean-field Hamiltonian (from \cref{eq:H0-def,eq:HU-MF,eq:HW-MF,eq:HJ-MF,eq:HV-MF}) for the parent state reads 
{\small
\begin{equation} 
H^{\rm(MF)} (\kk) \approx \begin{pmatrix}
\nu_f W_1  &  v_\star( k_x \sigma_0 \tau_z + ik_y\sigma_z\tau_0 )\spin_0  & \gamma \sigma_0 \tau_0 + v_\star'( k_x \sigma_x\tau_z + k_y\sigma_y \tau_0 )\spin_0 \\
v_\star( k_x \sigma_0 \tau_z - ik_y\sigma_z\tau_0 )\spin_0 & \nu_f W_3 + M \sigma_x\tau_0\spin_0 - J (O^f-\frac12 \sigma_0\tau_0\spin_0) & 0 \\
\gamma \sigma_0\tau_0\spin_0 + v_\star'( k_x \sigma_x\tau_z + k_y  \sigma_y\tau_0)\spin_0 & 0 & 
    \nu_f(U_1+6U_2) - U_1 (O^f - \frac12\sigma_0\tau_0\spin_0) 
\end{pmatrix} \ .
\end{equation}}
We further deform the Hamiltonian $H^{\rm(MF)} (\kk)$ adiabatically to simplify the problem. 
First, we can continuously turn off the $\nu_f W_1$, $\nu_f W_3$, $\nu_f W_1(U_1+6U_2)$ terms without changing the direct gap at $\kk=0$.
A simple argument to justify this process is the following.
As mentioned in \cref{sec:ground-state-rules}, for all the fillings the first block ($\Gamma_3$ representation from conduction bands) and the third block ($\Gamma_3$ representation from local orbitals) are at high energies and the levels closest to the gap are always from the second block ($
\Gamma_1\oplus\Gamma_2$ representation from the conduction bands), \ie $\nu_f W_3 \pm M \pm J/2$. 
In the process of turning off the three terms, the first block and the third block continue to be at high energies and do not enter the low energy physics around $\kk=0$.
This is confirmed by the calculated levels in in the following (\cref{eq:chern-level-1,eq:chern-level-0}). 
Second, we tune $v_\star'\to 0$, $M\to 0$, which is also adiabatic. 
The deformed Hamiltonian is diagonal in the sublattice ($\sigma_z$), valley ($\tau_z$), and spin ($\spin_z$) indices, \ie 
\begin{equation} 
H^{(\mrm{MF}, \alpha,\eta,s)} (\kk) \approx \begin{pmatrix}
0  &  v_\star( \eta k_x  + i (-1)^{\alpha-1} k_y)  & \gamma \\
v_\star( \eta k_x - i (-1)^{\alpha-1} k_y ) & - J (n_{\alpha\eta s}-\frac12) & 0 \\
\gamma  & 0 & - U_1 (n_{\alpha\eta s} - \frac12) 
\end{pmatrix} \ .
\end{equation}
For $n_{\alpha\eta s}=1$, given $|\frac{U_1}4 \pm \sqrt{\gamma^2+U_1^2/4}|>J/2$, the energy levels from low to high are 
\begin{equation}\label{eq:chern-level-1}
-\frac{U_1}4 - \sqrt{\gamma^2 + \frac{U_1^2}{16} },\qquad 
-\frac{J}2,\qquad
-\frac{U_1}4 + \sqrt{\gamma^2 + \frac{U_1^2}{16} }\ . 
\end{equation}
For $n_{\alpha\eta s}=0$, the energy levels from low to high are 
\begin{equation}\label{eq:chern-level-0}
\frac{U_1}4 - \sqrt{\gamma^2 + \frac{U_1^2}{16}  },\qquad 
\frac{J}2,\qquad
\frac{U_1}4 + \sqrt{\gamma^2 + \frac{U_1^2}{16}  }\ . 
\end{equation}
We can choose the chemical potential at zero: 
For $n_{\alpha\eta s}=1$, the $-\frac{U_1}4 - \sqrt{\gamma^2 + \frac{U_1^2}{16} }$ and $-J/2$ levels are occupied, while $-\frac{U_1}4 + \sqrt{\gamma^2 + \frac{U_1^2}{16} }$ is empty;
For $n_{\alpha\eta s}=0$, the $\frac{U_1}4 - \sqrt{\gamma^2 + \frac{U_1^2}{16} }$ level is occupied, while the $J/2$ and $\frac{U_1}4 + \sqrt{\gamma^2 + \frac{U_1^2}{16} }$ levels are empty. 
The total number of occupied levels (subtracting 12 contributed by $c$-electrons) equals to $\nu_f$. 
Now we integrate out the third block (local orbitals), as we did in \cref{eq:VP-nu=0-MF-cond}, to derive an effective Dirac Hamiltonian.
For $n_{\alpha\eta s}=1$, there is 
\begin{equation} 
H^{({\rm eff},\alpha,\eta, s)}(\kk) \approx 
\begin{pmatrix}
 \frac{2\gamma^2}{U_1}  &  v_\star( k_x \eta \sigma_0 + i(-1)^{\alpha-1} k_y\tau_0 )   \\
v_\star( k_x \eta \sigma_0 - i(-1)^{\alpha-1} k_y) & - \frac{J}2 
\end{pmatrix} \ .
\end{equation}
Tuning $\frac{2\gamma^2}{U_1} \to m$, $\frac{J}2 \to -m$ ($m>0$), the above Hamiltonian is adiabatically deformed to $m \xi_z + v_\star \eta k_x \xi_x - v_\star (-1)^{\alpha-1} \xi_y$.
Hence the contributed Chern number by this Hamiltonian is $- \frac12 \eta (-1)^{\alpha-1}$.
For $n_{\alpha\eta s}=0$, there is 
\begin{equation} 
H^{({\rm eff},\alpha,\eta, s)}(\kk) \approx 
\begin{pmatrix}
 - \frac{2\gamma^2}{U_1}  &  v_\star( k_x \eta \sigma_0 + i(-1)^{\alpha-1} k_y\tau_0 )  \\
v_\star( k_x \eta \sigma_0 - i(-1)^{\alpha-1} k_y) & + \frac{J}2 
\end{pmatrix} \ .
\end{equation}
Tuning $\frac{2\gamma^2}{U_1} \to m$, $\frac{J}2 \to -m$ ($m>0$), the above Hamiltonian is adiabatically deformed to $-m \xi_z + v_\star \eta k_x \xi_x - v_\star (-1)^{\alpha-1} \xi_y$.
Hence the contributed Chern number by this Hamiltonian is $\frac12 \eta (-1)^{\alpha-1}$.
Therefore, for the parent state of the form \cref{eq:trial-chern}, the total Chern number contributed by the $n_{\alpha\eta s}=1$ and $n_{\alpha\eta s}=0$ flavors is given by 
\begin{equation} \label{eq:chern-number}
C = \sum_{\alpha\eta s } \pare{ -\frac12\eta (-1)^{\alpha-1} n_{\alpha\eta s} 
    + \frac12\eta (-1)^{\alpha-1} (1-n_{\alpha\eta s}) } 
= - \sum_{\alpha \eta s} \eta (-1)^{\alpha-1} n_{\alpha\eta s}\ .
\end{equation}
According to the $C_{3z}$ representation matrix of the $f$-orbitals, $D^f(C_{3z}) = e^{\frac{2\pi}3 \tau_z \sigma_z }$ (\cref{eq:Df}), the factor $\eta (-1)^{\alpha-1}$ is nothing but the angular momentum of the corresponding $f$-orbital. 

Substituting \cref{eq:chern-number} into the VP states at the fillings $\nu=0,-1,-2,-3$, we find their Chern numbers as 
\begin{equation}
    C_{\nu=0}=0,\qquad C_{\nu=-1}=-1,\qquad C_{\nu=-2}=0,\qquad C_{\nu=-3}=-1\ .
\end{equation}
The flat-U(4) related K-IVC and VP+K-IVC states have the same Chern numbers. 
These results are same as Ref.~\cite{Biao-TBG4}.

\section{More discussions}


\paragraph{Insights for the heavy fermion physics from experiments} --- 
Our model fits precisely with the experimental facts of  Ref.~\cite{rozen2020entropic} which writes {\it ``The correlated state features an unusual combination of seemingly contradictory properties, some associated with itinerant electrons - such as the absence of a thermodynamic gap, metallicity and a Dirac-like compressibility - and others associated with localized moments, such as a large entropy and its disappearance under a magnetic field.''}

\paragraph{Charge-2 excitations} --- 
As an example of the improved understanding of the MATBG physics through our model, as shown in Ref.~\cite{TBG5}, there is a high-energy ``flat-band'' charge-2 collective mode upon the ground state at $\nu=0$ (see Fig. 4 of Ref.~\cite{TBG5}).
Since its energy is about 120meV $\sim 2U_1$, it is likely that the charge-2 collective mode is just an excitation of two $f$-electrons at the same site.

\paragraph{Position-dependence of STM spectrum and Landau level quantization} --- 
One key feature of our heavy fermion model is that the quasi-particle excitation has a minimal band gap at the $\Gamma_M$ point, and the corresponding wavefunctions are contributed by $c$-electrons. 
This feature should be reflected in position dependence of the spectrum: At an AA-stacking site, the local spectrum (density of states) is mainly contributed by $f$-electrons and hence should have strong weight at the large gap ($\sim U$) states but less weight at the minimal gap ($\sim J$) states. On the contrary, at an AB-stacking site, where c-electron dominates, the local spectrum should have more weight at the minimal gap ($\sim J$) states. In addition, since $c$-electrons are more delocalized, their response to magnetic field will be much more significant than that of $f$-electrons. As a result, the Landau level (LL) quantization in the presence of a magnetic field should be easier to be observed at AB-stacking sites. 

The position-dependences of local spectrum (density of states) and LLs have been seen in STM experiments, e.g., Refs.~\cite{choi2020tracing,wong_cascade_2020}. 
Comparing Fig. 1c and Extended Fig. 2a of Ref.~\cite{choi2020tracing} one can see that, in the same sample, when the large gap ($\sim U$) is well developed, there are additional peaks inside the large gap ($\sim U$) at AB-stacking sites but none at AA-stacking sites.
This phenomenon is consistent with the above analysis.
Ref.~\cite{choi2020tracing} also found that LLs at AA-stacking sites are harder to resolve than those at AB-stacking sites, which are also consistent the analysis above. 
Comparing Fig. 1c and Fig. 3a of Ref.~\cite{wong_cascade_2020}, one can see the same behavior: Spectrum at AB-stacking sites have additional peaks inside the large gap ($\sim U$). (But Fig. 1c and Fig. 3a of Ref.~\cite{wong_cascade_2020} are from two samples. We did not find spectra at both AA and AB sites of the same sample in Ref.~\cite{wong_cascade_2020}.)

The quasi-particle spectra shown in Fig. 2 of the main text also explain the Landau fans observed in transport experiments \cite{wu_chern_2020,saito2020,das2020symmetry,park2020flavour}. 
Around integer fillings $\nu\ge 0$ ($\nu\le 0$), the quasi-particles (quasi-holes) are contributed by c-electrons and will form Landau levels in the presence of magnetic field, while quasi-holes (quasi-particles) are contributed by f-electrons and will form Landau levels. Therefore, Landau fans are observed at $\nu=n+\delta$ ($\nu=- n-\delta$) for $n=0,1,2,3$ and $0<\delta\ll 1$.

\paragraph{Ground states at fillings $\nu=\pm 3$} --- 
Transport experiments \cite{yankowitz2019tuning,lu2019superconductors} observed resistivity peaks at the fillings $\nu=\pm3$, suggesting existence of strongly correlated phases. 
Then, STM \cite{nuckolls_chern_2020,choi2020tracing} and transport \cite{wu_chern_2020,saito2020,das2020symmetry,park2020flavour} experiments found that a finite magnetic field can stabilize a correlated Chern insulator phase with Chern numbers $\nu=\pm1$ at $\nu=\pm3$. These experiments suggest that the Chern insulator phase is at least a metastable state slightly above the actual ground state. 

A perturbation \cite{Biao-TBG4} and an ED calculation \cite{TBG6} by some of the authors found that the ground states at $\nu=\pm3$ would be correlated Chern insulator with Chern number $\pm1$ if the so-called flat-metric-condition is imposed to the model. These results are also consistent analysis in Ref.~\cite{bultinck_ground_2020}. 
However, with realistic parameters, there is no indirect gap between the quasi-hole and electron bands - even though there is a direct gap at each momentum \cite{TBG5}. 
Thus, with realistic parameters, the correlated Chern insulator phase becomes unstable \cite{TBG5}. 
The gapless spectrum obtained in the current work \cref{fig:HFbands_nu3} is fully consistent with the results in Refs.~\cite{TBG5,TBG6}. 

DMRG studies \cite{kang_nonabelian_2020,soejima2020efficient} found that, using realistic parameters, gapless $C_{2z}T$-symmetric nematic or gapped $C_{2z}T$-symmetric stripe states may have lower energies than the Chern insulator states at $\nu=\pm3$. 
They may provide an explaination to the absence of Chern insulator at zero magnetic field. 
We leave HF calculations of such states for future studies. 

\end{document}